\documentclass[twocolumn,aps,prd,amsmath,amssymb,longbibliography]{revtex4-1}
\makeatletter                           
\@ifclasswith{revtex4-1}{draft}{        
    \usepackage[notref,notcite]{showkeys}
    
}{}
\makeatother

\usepackage[utf8]{inputenc}
\usepackage{amsmath} 
\usepackage{amsfonts} 
\usepackage{amssymb}
\usepackage{epsfig}
\usepackage{graphics}
\usepackage{graphicx}\graphicspath{{figs/}}
\usepackage{titlesec}
\usepackage{mathtools}
\usepackage{environ}
\usepackage[english]{babel}
\usepackage[inline]{enumitem}
\usepackage{tikz}
\usetikzlibrary{positioning}
\usetikzlibrary{arrows}

\usepackage[colorlinks=true]{hyperref}
\hypersetup{urlcolor=black,linkcolor=black,citecolor=black}
\usepackage[toc,page]{appendix}

\makeatletter
\renewcommand*\env@matrix[1][c]{\hskip -\arraycolsep
  \let\@ifnextchar\new@ifnextchar
  \array{*\c@MaxMatrixCols #1}}
\makeatother

\titleformat{\section}[hang]{\Large\bfseries}{\thesection}{2ex}{}
\titleformat{\subsection}[hang]{\large\bfseries}{\thesubsection}{2ex}{}
\titlespacing{\subsection}{0pt}{20pt}{10pt}[0pt]
\titleformat{\subsubsection}[hang]{\large\bfseries}{\thesubsubsection}{1ex}{}
\titlespacing{\subsubsection}{0pt}{20pt}{10pt}[0pt]
\titlespacing{\paragraph}{0pt}{15pt}{5pt}[0pt]
\titleformat{\paragraph}[block]{\normalsize\bfseries}{\theparagraph}{1ex}{}

\renewcommand{\thesection}{\arabic{section}}
\renewcommand{\thesubsection}{\thesection.\arabic{subsection}}
\renewcommand{\thesubsubsection}{\thesubsection.\arabic{subsubsection}}
\renewcommand{\theparagraph}{\thesubsubsection.\arabic{paragraph}}

\makeatletter
\renewcommand{\p@subsection}{}
\renewcommand{\p@subsubsection}{}
\makeatother

\usepackage[mathscr]{euscript}
\usepackage{braket}
\usepackage{bm}

\usepackage{makecell,tabularx}
\setcellgapes{3pt}

\usepackage[disable]{todonotes} 

\newcommand{\tr}{\mathrm{tr}}

\newcommand{\dif}{\,\mathrm{d}}

\newcommand{\del}{\partial}
\newcommand{\up}{\ket{\uparrow}}
\newcommand{\down}{\ket{\downarrow}}
\newcommand{\trho}{\tilde{\rho}}

\newcommand{\Shat}{\hat{S}}
\renewcommand{\epsilon}{\varepsilon}
\newcommand{\eps}{\varepsilon}
\newcommand{\vp}{\varphi}

\DeclareMathOperator{\diag}{diag}

\let\Re\relax
\DeclareMathOperator{\Re}{Re}
\let\Im\relax
\DeclareMathOperator{\Im}{Im}
\DeclareMathOperator{\TO}{T}

\renewcommand{\d}{\mathop{}\!\mathrm{d}}

\newcommand{\nn}{\nonumber}
\newcommand{\subt}[1]{_{\text{#1}}}

\definecolor{refkey}{rgb}{0,0,1}
\definecolor{labelkey}{rgb}{0,1,0}

\numberwithin{equation}{subsection}

\usepackage{mathtools}

\begin{document}
\title[ ]{\Huge The probabilistic world II: \\
\Huge Quantum mechanics\\
from\\
classical statistics
\vspace{1cm}}

\author{C. Wetterich}
\affiliation{Institut  f\"ur Theoretische Physik\\
Universit\"at Heidelberg\\
Philosophenweg 16, D-69120 Heidelberg
\vspace{1cm}}

\begin{abstract}
This work discusses simple examples how quantum systems are obtained as
subsystems of classical statistical systems. For a single qubit with arbitrary
Hamiltonian and for the quantum particle in a harmonic potential we provide
explicitly all steps how these quantum systems follow from an overall
''classical" probability distribution for events at all times. This overall
probability distribution is the analogue of Feynman's functional integral for
quantum mechanics or for the functional integral defining a quantum field
theory. In our case the action and associated weight factor are real, however,
defining a classical probabilistic system. Nevertheless, a unitary
time-evolution of wave functions can be realized for suitable systems, in
particular probabilistic automata. Based on these insights we discuss novel
aspects for correlated computing not requiring the extreme isolation of quantum
computers. A simple neuromorphic computer based on neurons in an active or quiet
state within a probabilistic environment can learn the unitary transformations
of an entangled two-qubit system. Our explicit constructions constitute a proof
that no-go theorems for the embedding of quantum mechanics in classical
statistics are circumvented. We show in detail how subsystems of classical
statistical systems can explain various ``quantum mysteries". Conceptually our
approach is a straightforward derivation starting from an overall probability
distribution without invoking non-locality, acausality, contextuality, many
worlds or other additional concepts. All quantum laws follow directly from the
standard properties of classical probabilities.
\end{abstract}

\maketitle

\newpage\null\thispagestyle{empty}\newpage
\newpage\null\thispagestyle{empty}\newpage

\tableofcontents
\clearpage
\addtocontents{toc}{\protect\null\vspace*{2mm}}
\section[The classical and the quantum world]{The classical and the\\quantum
world}\label{sec:The_classical_and_the_quantum_world}

Why is the world described by quantum mechanics? The answer to this basic
question involves three central ingredients~\cite{CWPW}:
\begin{enumerate}[label={(\arabic*)}] \item The physical description of the
universe and its laws is based on probabilities. One may imagine one ``overall
probability distribution" for all possible events that could happen at different
times and locations in our universe. Given the state of the universe at a given
time, the probability for most particular events in our rich and complex
evolving universe is neither very close to one nor to zero. Then a physical
description is not predictive. It will not tell us if on planet earth a certain
bee will visit a certain flower around noon on a certain sunny summer day in a
garden in Heidelberg. Nevertheless, the classical statistical laws of
probabilities can develop strong predictive power for certain questions. This
concerns correlations of events, for example sequences of events. If a raindrop
has been found at different moments of a time sequence at positions which allow
to construct its trajectory and velocity up to a certain time $t$, the
conditional probability to find it at the next time step $t+\Delta t$ at a given
position may be very close to one or zero. In this case a physicist will predict
the position of the drop at $t+\Delta t$ and a physical law for the falling
raindrop may be established. The only assumption that we will use is the
description of the universe by an overall probability distribution and the
standard classical laws for probabilities~\cite{KOL}. Quantum mechanics and its
``axioms" follow once the concept of evolution is introduced and the evolution
is unitary. Probabilistic realism~\cite{CWPW} does not view probabilities as an
epistemic description for the lack of knowledge of an observer about some
ontological deterministic reality. The overall probability distribution is
rather the basic conceptual setting for the description of the universe, similar
to the functional integral for quantum field theory.

\item The second ingredient is a time structure of the probabilistic
system~\cite{CWPT}. We assume that a set of ``basis events" can be ordered in
some discrete or continuous variable that we call time. The overall probability
distribution assigns probabilities to these basis events. We further assume
"locality in time" in the sense that the probability distribution can be written
as a product of time-local factors which each involve basis events only at two
neighboring times. A simple example is the two-dimensional Ising
model~\cite{LENZ, ISI, KBI} with next-neighbor interactions. The basis events
are the configurations of Ising spins on the sites of a two-dimensional lattice.
Ising spins take values $\pm1$ and can be associated with yes/no-decisions or
bits in information theory~\cite{SHA}, or occupation numbers of
fermions~\cite{CWFIM}. We may define sequences of hypersurfaces which each
divide the two-dimensional lattice into the present (on the hypersurface), the
past and future. The choice of the time-hypersurfaces is not unique. We take one
such that the interactions between the Ising spins only involve spins on two
neighboring hypersurfaces. This implements the time-locality structure. The
time-locality structure is not a very particular choice but rather common to
many classical statistical systems. One can define a time-local probabilistic
information by a suitable sum over events in the past and future. Time-locality
permits the notion of evolution according to the simple question: Given the
time-local probabilistic information at a certain time $t$, what will be the
time-local probabilistic information at the next time step $t+\Delta t$? Most
parts of the quantum formalism, namely wave functions or the density matrix
encoding the time-local probabilistic information, operators for observables and
the evolution operator emerge from the answer to this question. The
non-commutative structures between operators characteristic for quantum
mechanics are well known for classical statistical systems once they are
investigated by the transfer matrix formalism~\cite{TM, MS, FU}.

\item Within the large family of classical statistical systems with time
locality the specific property which singles out quantum systems is the unitary
evolution. For many classical statistical systems much of the initial
time-local information is lost as time progresses. For the example of the Ising
model we may specify the initial time-local probabilistic information at a given
time $t_0$, typically on a boundary. As time increases (towards the bulk), the
time-local probabilistic information will approach an equilibrium distribution.
The rate how fast the more detailed initial information is lost is given by the
correlation length. In contrast, for classical statistical systems describing
quantum systems the initial information is never lost. Simple examples for
``classical" probabilistic systems of this type are probabilistic cellular
automata. For cellular automata~\cite{ULA, JVN, ZUS, HED, GAR, RICH, AMPA, HPP,
LIRO, TOOM, DKT, WOLF, VICH, PREDU, CREU, TOMA, HOOFT2, TH, GTH, ELZE, FLN} the
updating of the bit-configuration in a local cell is influenced only by a few
neighboring cells. Probabilistic cellular automata are defined by a probability
distribution over initial configurations. For probabilistic automata a
deterministic updating rule guarantees that the initial information is not lost.
The preservation of this time-local probabilistic information is the basis for
the unitary evolution in quantum mechanics. The deterministic updating maps the
probability for a given configuration in the next time step to the probability
for the configuration which obtains by the updating. For any probability
distribution over initial configurations this defines the overall probability
distribution for events at all times. All probabilistic automata are actually
quantum systems. Very often it is possible to introduce a complex structure,
such that quantum mechanics appears in the familiar form with a complex wave
function or density matrix.
\end{enumerate}

If physicists want to describe our highly complex universe, with a rich
dynamical evolution of structures seen everywhere, they better employ an overall
probabilistic distribution with a time-local structure and unitary
evolution~\cite{CWGENS}. Without a unitary evolution most initial probabilistic
information would be lost as time progresses, and physicists could only describe
some equilibrium state which has not much to do with our universe. The need for
a unitary evolution is the need for quantum mechanics, answering the question
why we describe the world by quantum mechanics. It is actually sufficient that a
large enough subsystem follows a unitary evolution. The remaining part of the
time-local probabilistic information can then be seen as an environment for the
subsystem which may approach some type of equilibrium. The complex evolution of
our world is then described by the subsystem.

In the language of modern quantum field theory we may give a short answer to the
question: what is quantum mechanics? The basic description of the world is based
on a euclidean functional integral which involves the overall probability
distribution for fields or configurations of infinitely many bits. Quantum
physics is the projection to the part of the local-time subsystem which follows
a unitary evolution. This projection is a quantum field theory in the operator
formalism. Quantum mechanics for a few particles or a few qubits follows for
appropriate subsystems of the local-time subsystem or quantum field theory.

From this general conceptual setting which is explained in detail in the first
part of this work~\cite{CWPW}, there remains still a long road to go before one
understands the properties of a quantum particle in a potential. The reason is
that a particle is not a simple object. The historic view has been that
particles are simple basic objects, while complexity can be understood, at least
in principle, from the interactions of particles. Modern quantum field theory
has inverted this view. The quantum field theory for fundamental particles
describes infinitely many interacting degrees of freedom. One first needs to
find a vacuum state which is a highly complex object -- a prime example being
the theory of quantum chromodynamics for the strong interactions~\cite{WIL,
GATLA}. Particles are seen as excitations of this vacuum. Even a single particle
involves infinitely many degrees of freedom. The particle properties are no
longer assumed as fundamental. They rather depend on properties of the vacuum. A
good example is the mass of the electron which is due to the vacuum expectation
value of the Higgs scalar. Actually, in very early cosmology (before the
electroweak phase transition) this expectation value vanishes, and so does the
electron mass. Quite generally, in cosmology the vacuum corresponds to the
dynamical cosmological background solution and therefore depends itself on time.

We should therefore not be surprised if no simple classical statistical system
is found for a quantum particle in a potential, or for a few interacting
qubits. The classical statistical systems describing fully these simple quantum
systems typically involve infinitely many degrees of freedom, similar to the
particle in quantum field theory. Nevertheless, reduced quantum features can
often be found for classical probability distributions involving only a few
degrees of freedom. An example are the realization of discrete subsets of
unitary transformations for a few qubits by generalized Ising models of a few
classical bits or Ising spins.

The first part of this work~\cite{CWPW} has developed the general probabilistic
view of the world and the concepts of evolution and time. This leads to the
quantum formalism for classical statistics~\cite{CWQF}. The local factors
describing
the overall probability distribution for systems with a time-local structure are
closely related to the step evolution operator, which is a normalized version of
the transfer matrix. The overall probability distribution can be seen as a
generalized local chain consisting of a product of local factors. The particular
case of unique jump chains realizes a unitary evolution. This case corresponds
to probabilistic automata.

In this first part we have developed probabilistic automata which are equivalent
to discretized quantum field theories for fermions in one time and one space
dimension. For the particular case of free massless fermions the continuum limit
of the discrete formulation can be taken and corresponds indeed to the standard
quantum field theory of massless free Dirac, Weyl or Majorana fermions in two
dimensions. We have established a time- and space-translation invariant vacuum
state which respects particle-antiparticle symmetry. All excitations of this
vacuum have positive energy, in close analogy to the half-filled Dirac sea in
particle physics.

Our discrete analysis of the probabilistic automaton for free massless fermions
reveals that already the vacuum state corresponds to a highly non-trivial
overall probability distribution. Single particle excitations can be constructed
by applying fermionic creation and annihilation operators on this vacuum state.
We emphasize that in our approach the basic simple object is directly the
quantum field theory, while one-particle quantum mechanics is realized for
particular subsystems in a complex setting. This subsystem involves infinitely
many degrees of freedom of the overall system. Even though the model is very
simple, it is striking how all the concepts of quantum field theory and quantum
mechanics emerge in a very natural way from the overall probability
distribution. All laws and axioms follow from the standard classical statistical
properties of probabilities, without any further assumptions.

In the first part of this work we also have constructed rather simple
probabilistic cellular automata which are equivalent to two-dimensional quantum
field theories for fermions with interactions. They correspond to generalized
Thirring or Gross-Neveu models~\cite{THI, KLA, GN, WWE, AAR, FAIV} in a
particular discretization. In principle, the path to one-particle quantum
mechanics is straightforward to follow. One has to establish the vacuum state
for these models and to investigate the properties of the one-particle
excitations. In practice, this task is rather complex, however. In the presence
of interactions the construction of a particle-antiparticle symmetric vacuum
state for which all excitations have positive energy is a rather complicated
issue. This influences the form of the possible continuum limit. One expects
that important renormalization effects from quantum fluctuations distinguish the
true continuum limit from a ``naive continuum limit".

In the present part of this work we approach the construction of one-particle
quantum mechanics from the opposite end, starting with only a few classical
bits. We keep in mind the basic observation that a quantum particle is a
subsystem of a much more complex system, i.e. the quantum field theory. The same
holds for the quantum mechanics of a certain number of qubits. Qubits can be
seen as a focus on restricted sets of states of a quantum particle. Qubits are
in turn subsystems of the quantum mechanics for particles. In principle, we
expect that even a single qubit involves infinitely many degrees of freedom of
the overall system. In this part of our work we approach the limit of infinitely
many degrees of freedom by establishing fundamental concepts leading to quantum
mechanics for only a small number of classical bits or Ising spins. We will see
how continuous quantum mechanics can emerge in the limit of infinitely many
classical bits.

In more detail, we will start from classical probability distributions for only
a few discrete degrees of freedom or Ising spins on a given time layer. They can
already describe the quantum mechanics of a single qubit for which only a
restricted set of unitary transformations is allowed. We will extend this to the
full quantum mechanics of a single qubit. This involves indeed infinitely many
``classical Ising spins". We discuss entanglement for classical statistical
systems describing two qubits and continue this approach in the direction of
several qubits.

This ``bottom-up" approach for a few qubits has two important advantages. First,
the classical statistical systems are very simple. This allows us to follow very
explicitly how the characteristic quantum properties as non-commuting operators
or entanglement arise for suitable subsystems of the classical statistical
generalized Ising models. The so-called ``quantum paradoxes" find explicit
solutions in our classical statistical setting, demonstrating how no go theorems
for the embedding of quantum mechanics in classical statistics are circumvented.

The second advantage is the direct contact with quantum computing or
neuromorphic computing. This highlights the crucial importance of correlations
for these types of computing, and points towards more general forms of
``correlated computing". We will show that for two qubits classical statistical
systems can realize arbitrary unitary transformations. We demonstrate this by a
``model neuromorphic computer" based on spiking neurons. It learns how to
perform the complete set of the well known basic quantum gates. Suitable
sequences of these gates result in arbitrary unitary transformations for the
two-qubit quantum subsystem. This demonstrates that the performance of certain
quantum tasks does not need the high degree of isolation of a quantum system
often assumed to be necessary. These tasks could be performed under ``human
conditions", for example by our brain. On the other hand, an extension to full
quantum operations for many qubits requires a highly complex control of
correlations. This underlines the great prospects of ``real" quantum computers
for which the nature of atoms as quantum objects guarantees the correlations
necessary for quantum computing. One may envisage the possibility that
intermediate forms of correlated computing, which do not perform arbitrary
unitary transformations for entangled qubits, could still be used by macroscopic
systems without extreme isolation, as the human brain.

In summary, quantum mechanics and ``classical'' probabilistic systems are in a
much closer relation than commonly realized. In short, quantum systems are
particular types of subsystems of general ``classical'' probabilistic systems.
The general properties of subsystems and their relation to quantum mechanics are
the central topic of this work. We will see how all the ``mysterious''
properties of quantum systems arise in a natural way from the generic properties
of subsystems. The correlations of subsystems with their environment play an
important role in this respect, leading to many features familiar from quantum
mechanics. These features are not realized for the often considered uncorrelated
subsystems. The presence of various quantum features in classical statistical
systems has been proposed in different settings in
refs.~\cite{LHAR,KIR,JBAR,FUSH,CSD1,BURO,CSD2,YY1,YY2,CCDA}. We advocate here the
viewpoint that \emph{all} quantum features actually arise from suitable
classical probabilistic systems~\cite{CWICS, CWIS, CWB}.

In sect.\,\ref{sec:quantum_subsystems} we discuss a first simple discrete
quantum system for a single qubit. It is based on a local chain for three Ising
spins at every time-layer $t$. Already this simple system shows many features of
quantum mechanics, as the whole formalism and particle-wave duality. We recall
in this section several key concepts of this work, as the classical wave
function and density matrix, the step evolution operator, or the status of
observables and associated operators. These concepts are described in detail in
ref.~\cite{CWPW}. We proceed in
sect.\,\ref{sec:entanglement_in_classical_and_quantum_statistics} to entangled
systems, both entangled quantum systems and entangled classical probabilistic
systems. Entanglement is not a property particular to quantum mechanics. We
construct explicitly entangled classical probabilistic systems which lead to
entangled two-qubit quantum subsystems.

In sect.\,\ref{sec:continuous_classical_variables} we take the limit of
continuous variables for the description of a classical probabilistic system. It
obtains for an infinite number of Ising spins or yes/no decisions. All
properties follow from the case of discrete variables by taking a suitable
limit. There is no practical difference between continuous variables and a very
large number of discrete variables. In this respect the continuum description is
rather a matter of convenience. Nevertheless, the continuum limit often shows
universal features which lead to important simplifications. The equivalence of
continuous variables with an infinite number of discrete variables is at the
basis of an important property of the one-qubit quantum system. The quantum
system has an infinity of observables with only two possible measurement values.
These are given by the quantum spin in arbitrary directions. The yes/no
decisions associated to continuous classical variables can be mapped to the
two-level observables in the quantum subsystem.

In sect.\,\ref{sec:quantum_mechanics} we address continuous quantum mechanics.
We first discuss the dynamics of a single qubit with an arbitrary time-dependent
Hamiltonian. It is based on a classical statistical system with a probability
distribution depending on continuous variables. A continuous set of yes/no
questions is mapped to a continuous set of quantum observables corresponding to
the quantum spin in different directions. The possible measurement values are
discrete, as given by the eigenvalues of the associated quantum operators. The
classical overall probability distribution realizes both sides of quantum
mechanics: the continuous wave function and the discrete observables. As it
should be, the quantum operators for spins in different directions do not
commute. Their expectation values obey the uncertainty relations of quantum
mechanics. This is related to the presence of ''quantum constraints" for the
subsystem which enforce correlations between the spin in different directions.
In this section we also construct a simple probabilistic automaton which
describes a quantum particle in a harmonic potential. It is based on the
classical statistical Liouville equation in phase space for a particle with two
colors. Suitable initial conditions lead to color oscillations with periods
predicted by the equidistant spectrum of the Hamiltonian of the quantum
subsystem. This underlines the usefulness of the quantum description for the
understanding of the dynamics of classical statistical systems.

In sect.\,\ref{sec:classical_and_quantum_computing} we turn to a possible use of
our setting for computing. Classical and quantum computing are treated in the
same general setting of probabilistic computing as different limiting cases.
Many intermediate cases between the two limits could lead to new powerful
computational structures. In particular, we address artificial neural networks
and neuromorphic computing within our general setting and ask if computers
constructed according to these principles, or even biological systems as the
human brain, could perform quantum operations. We provide examples where this is
the case for simple systems of spiking neurons. Since these systems are
``classical'', this demonstrates in a very direct way that there are no
conceptual boundaries between classical probabilistic systems and quantum
systems. 

In sect.\,\ref{sec:conditional_probabilities_4_7} we turn to the important topic
of conditional probabilities and their relations to sequences of measurements.
Most of the questions that humans ask about Nature invoke conditional
probabilities, of the type ``if an experimental setting is prepared, what will
be the probability for a certain outcome under this condition''. Conditional
probabilities are closely related to different types of measurements. In
particular, one has to think about the notion of ``ideal measurements'' for
subsystems. The ``reduction of the wave function'' turns out to be a convenient
mathematical tool for the description of conditional probabilities, rather than
a physical process. The concepts of conditional probabilities and ideal
measurements for subsystems play an important role for our discussion of the
``paradoxes'' of quantum mechanics in
sect.\,\ref{sec:the_paradoxes_of_quantum_mechanics}. There we address Bell's
inequalities, the Kochen-Specker no-go theorem and the Einstein-Podolski-Rosen
paradox. They all find a natural explanation in our ``classical'' probabilistic
setting.

We conclude in
sect.~\ref{sec:Embedding_quantum_mechanics_in_classical_statistics} by a short
overview of the embedding of quantum mechanics in classical statistical systems.
In particular, we address a list of often asked questions about the origin of
various aspects of quantum mechanics. We provide short answers how these quantum
features emerge from classical statistics if one focuses on appropriate
subsystems.

\section{Qubit automaton}
\label{sec:quantum_subsystems}

Quantum mechanics is realized for local subsystems with unitary evolution. 
For a given quantum state, as characterized by the quantum density matrix
\(\rho(t)\), or wave function \(\psi(t)\) for the special case of a pure quantum
state, only the local probabilistic information at $t$ is used. We will express
$\rho(t)$ in terms of the time-local probability distribution $\{p_\tau(t)\}$
for three classical Ising spins. The quantum subsystem typically does not use
all the local information contained
in $\{p_\tau(t)\}$. 
A few particular expectation values or classical correlations of the Ising spins
\(s_\gamma(t)\) specify the subsystem. The evolution law of the subsystem is
inherited from the evolution law of the underlying local chain. It describes a
linear unitary evolution of the density matrix, such that no information
contained in the quantum subsystem is lost. Our quantum subsystem admits a
complex structure. In the complex formulation the density matrix is Hermitian
and normalized,
\begin{align}\label{QM1}
\rho^\dagger (t) = \rho(t), && tr \rho(t) = 1.
\end{align}
An important property is the positivity of the density matrix, i.e. the property
that all its eigenvalues are positive or zero.

In the present section we concentrate on quantum mechanics for a single
qubit. A simple local chain with three classical Ising spins realizes already
many characteristic features of quantum mechanics, as non-commuting operators
for observables, the quantum rule for the computation of expectation values,
discrete measurement values corresponding to the spectrum of operators, the
uncertainty principle, unitary evolution and complex structure.

\subsection{Discrete qubit chain}
\label{sec:discrete_qubit_chain}

Let us consider a simple automaton for three classical bits or Ising spins
$s_k=\pm1$. With probabilistic initial conditions the overall probability
distribution is given by a local chain with three Ising spins \(s_k(t)\) or
\(s_k(m)\), \(k=1,2,3\), at every discrete position \(m\). 
The discrete qubit chain is a unique jump chain for which each orthogonal step
evolution operator maps \(s_k = s_k(m)\) to \(s_k'=s_k(m+1)\). 
The order of these operators in the chain is left arbitrary. We employ six basis
operators and products thereof,
\begin{align}\label{trafos}
\begin{split}
T_{12}:\ s_1' = s_2,\ \quad s_2' = -s_1,\\
T_{23}:\ s_2' = s_3,\ \quad s_3' = -s_2, \\
T_{31}:\ s_3' = s_1,\ \quad s_1' = -s_3, \\
T_{1}:\ s_2' = -s_2, \quad s_3' = -s_3, \\
T_{2}:\ s_1' = -s_1, \quad s_3' = -s_3, \\
T_{3}:\ s_1' = -s_1, \quad s_2' = -s_2. 
\end{split}
\end{align}
The Ising spins not listed explicitly remain invariant. The first three
transformations correspond to \(\pi/2\) rotations of the spin \(s_k\) in
different ``directions", where \(k=1,2,3\) may be associated to three
``coordinate directions", say \(x,y,z\). The three last transformations are
combined reflections of two spins. We also admit all products of the six
transformations (\ref{trafos}). The transformations form a discrete group. 

The unique jump operators \(\hat{S}(m)\) may differ for different $m$. The
different transformations do not commute, such that for \(\hat{S}(m)\)
depending on \(m\) the order of the matrices according to \(m\) matters for the
overall probability distribution and the expectation values of local
observables. A given sequence of \(\hat{S}(m)\) could correspond to a
deterministic classical computer with three bits \(s_k\). This is realized if
the initial state is a fixed spin configuration. In contrast, we will consider
here probabilistic initial conditions by specifying at some initial $m$, say
$m=0$, the probabilities $p_\tau$ for each configuration $\tau$. This defines a
probabilistic automaton. We will restrict the initial probability distribution
to obey a certain ``quantum constraint''.
The layers $m$ in the local chain may be a time sequence, but they could also
label any other sequence, for example an order in space, or layers in a neural
network. 
We will see that the discrete qubit chain can also be viewed as an embryonic
quantum computer.

For three bits there are eight classical states, \(\tau = 1,...,8\), that we may
label by eight different spin configurations, e.g. in the order $(- - -),$ $(- -
+),$ $(- + -),$ $(- + +),$ $(+ - -),$ $(+ - +),$ $(+ + -),$ $(+ + +)$ for
\(\tau\) from 1 to 8. The eight time-local probabilities \(p_\tau(m)\) are the
probabilities for these configurations on the layer $m$.
The expectation values of the three spins follow the basic probabilistic rule
\begin{equation}
\rho_k (m) = <s_k(m)> = \sum_\tau p_\tau(m) (S_k)_\tau,
\label{QM3}
\end{equation}
with \((S_k)_\tau\) the value of the spin observable in the state \(\tau\), e.g.
\begin{align}
\begin{split}
(S_1)_\tau &= (-1,-1,-1,-1,\ 1,\ 1,\ 1,\ 1),\\
(S_2)_\tau &= (-1,-1,\ 1,\ 1,-1,-1,\ 1,\ 1),\\
(S_3)_\tau &= (-1,\ 1,-1,\ 1,-1,\ 1,-1,\ 1).
\end{split}
\label{QM4}
\end{align}

\subsection{Classical wave function and step\\evolution operator}
\label{sec:classical_wave_function_and_step_evolution_operator}

A convenient formalism for probabilistic automata is based on the real classical
wave function. Its components $q_\tau(m)$ obey
\begin{equation}
\label{AA2}
p_\tau(m)=q_\tau^2(m)\ .
\end{equation}
For arbitrary $q_\tau$ the probabilities $p_\tau$ are positive, $p_\tau\geq0$.
The normalization of the probability distribution is guaranteed if $q_\tau$ are
the components of a unit vector. Any change of the probability distribution with
time results in a simple rotation of this unit vector. The step evolution
operator $\Shat(m)$ performs this rotation
\begin{equation}
\label{AB2}
q_\tau(m+1)=\Shat_{\tau\rho}(m)q_\rho(m)\ .
\end{equation}
It is therefore an orthogonal matrix -- in our case an $8\times8$ matrix.

A classical density matrix $\rho'(m)$ can be constructed as a bilinear of the
classical wave function, with elements
\begin{equation}
\label{AC2}
\rho'_{\tau\rho}(m)=q_\tau(m)q_\rho(m)\ .
\end{equation}
We can express the expectation values~\eqref{QM3} in terms of the classical
density matrix \(\rho'(m)\), which is a real \(8\times8\) matrix, as 
\begin{equation}\label{QM5}
\rho_k(m) = \langle s_k(m) \rangle =tr\left( \hat{S}_k \rho'(m)\right) ,
\end{equation}
with diagonal classical spin operators
\begin{equation}\label{QM6}
(\hat{S}_k)_{\tau \rho} = (S_k)_\tau \delta_{\tau \rho}.
\end{equation}
Only the diagonal elements \(p_\tau (m) = \rho'_{\tau\tau}(m)\) contribute in
this expression. 
The classical spin operators commute among themselves, but do not commute with
the step evolution operator, except for those spins that remain invariant under
a given transformation \(T_a\). A similar expression in terms of the classical
wave function is the analogue of the quantum law
\begin{equation}
\label{AD2}
\langle s_k(m)\rangle=q_\tau(m)\big(\Shat_k\big)_{\tau\rho}q_\rho(m)\ .
\end{equation}

The overall probability distribution for probabilistic automata is rather easily
visualized. For each initial configuration $\tau_0$ we can construct the
trajectory by applying the updating rule of the automaton. This trajectory is
the sequence of configurations reached by the updating. Each point on the
trajectory has the same probability, given by the initial probability
$p_{\tau_0}(0)$. Every configuration of spins $\{s_k(m)\}$ belongs to a unique
trajectory. In this way one assigns overall probabilities to all configurations
$\{s_k(m)\}$ which can be constructed from the spins at all $m$. The step
evolution operators~\eqref{AB2} are constructed in order to realize this overall
probability distribution.

For all probabilistic automata the step evolution operators are unique jump
matrices. They have in each row and column precisely one element equal to one
or minus one, and all other elements zero. For each one of the
transformations~\eqref{trafos} we can construct an associated step evolution
operator. This is done by following how each one of the eight configurations
$\tau$ is mapped by the operation of $T$ to a new configuration. We do not need
the explicit form of the step evolution operators $\Shat(m)$ for the present
purpose and refer to ref.~\cite{CWPW} for their explicit construction for cellular
automata.

The unique jump step operators transform the
local probabilities among themselves as a limiting case of a Markov chain
without loss of information. This transformation reproduces for the expectation
values \(\rho_k = \langle s_k \rangle\) the same transformation as for the spins
\(s_k\), e.g. for \( \hat{S}(m)\) corresponding to \(T_{12}\) one has
\(\rho_1(m+1)=\rho_2(m)\), \(\rho_2(m+1) = -\rho_1(m)\),
\(\rho_3(m+1)=\rho_3(m)\). (There should be no confusion between expectation
values \(\rho_k(m)\) and elements of the classical density matrix \(\rho_{\tau
\rho}' (m)\).) We also do not need here the explicit form of the overall
probability distribution. It is sufficient to realize that it exists in order to
see that we deal with a classical statistical system. For a detailed discussion
of the realization of the overall probability distribution as a constrained
generalized Ising model we refer to ref.~\cite{CWPW}.

\subsection{Quantum subsystem}

It is a key property of many quantum systems that they are subsystems of more
extended classical probabilistic systems. The resulting incomplete statistics is
the origin of the uncertainty relation and the non-commuting operator structure.
We discuss the quantum subsystem for the discrete qubit chain here. We could
already interpret the discrete qubit chain as a type of discrete quantum
mechanics with real wave functions. This quantum system is somewhat boring since
all operators commute and everything looks as a trivial reformulation of simple
classical properties. We will show that the restriction to a subsystem can
change these properties profoundly, leading to complex discrete quantum
mechanics with non-commuting operators for the spin observables. On the one
hand, the map to the subsystem discards part of the information contained in the
time-local probability distribution $\{p_\tau(m)\}$ for the discrete qubit
chain. The information retained for the subsystem exceeds, however, the one in
a classical probability distribution for two or less classical bits. We can
still compute all three expectation values $\rho_k(m)$ from the information
contained in the subsystem.

\paragraph*{Time-local subsystem}

The quantum subsystem is based on the three expectation values \(\rho_k(m)\).
For every given $m$ it is a time-local subsystem. It is also a simple form of a
subsystem based on correlations. (See ref~\cite{CWPW} for a general discussion of
subsystems based on correlations.) The three values \(\rho_k(m)\) are the only
information used by and available to the subsystem. The evolution of the
discrete qubit chain transforms the expectation values among themselves and thus
the subsystem is closed under the evolution. The subsystem uses only part of
the local probabilistic information in the form of three particular combinations
of local probabilities,
\begin{align}\label{QM7}
\begin{split}
\rho_1 = -p_1-p_2-p_3-p_4+p_5+p_6+p_7+p_8\ ,\\
\rho_2 = -p_1-p_2+p_3+p_4-p_5-p_6+p_7+p_8\ ,\\
\rho_3 = -p_1+p_2-p_3+p_4-p_5+p_6-p_7+p_8\ .
\end{split}
\end{align}

\paragraph*{Quantum density matrix and quantum operators}

We collect the probabilistic information for the quantum subsystem in the form
of a Hermitian \(2\times 2\) matrix~\cite{CWB}
\begin{align}\label{QM8}
\rho = \frac{1}{2}(1+\rho_k \tau_k), && \rho^\dagger=\rho , && tr \ \rho = 1.
\end{align} 
Hermiticity follows for real \(\rho_k\) from the hermiticity of the Pauli
matrices \(\tau_k\), and \(tr \ \rho = 1\) follows from \(tr \ \tau_k = 0\). 
This matrix is the quantum density matrix describing the subsystem, provided
that it is a positive matrix, see below in sect.\,\ref{sec:quantum_condition}.

We introduce three Hermitian quantum operators \(S_k\) for the three ``Cartesian
directions'' of the qubit, given by the Pauli matrices,
\begin{align}\label{QM9}
S_k = \tau_k, && S_k^\dagger = S_k.
\end{align}
In terms of these quantum operators we can compute for every $m$ the expectation
values of the classical spins from the density matrix,
\begin{equation}
\langle s_k \rangle = \rho_k = tr \ (\rho S_k).
\end{equation}
This follows from \(tr(\tau_k \tau_l)=2 \delta_{kl}\), \(tr \ \tau_k = 0\),
\begin{equation}
\rho_k = \frac{1}{2} tr\left\{ (1+\rho_l \tau_l) \tau_k \right\} = \frac{1}{2}
\rho_l \delta_{lk}\ tr \ 1 .
\end{equation}
We identify the three components of the quantum spin or qubit with the three
classical Ising spins \(s_k\), 
\begin{equation}\label{QM12}
\langle S_k \rangle_q = \langle s_k \rangle_{cl}.
\end{equation}
Here\(\langle s_k \rangle_{cl}\) is computed according to the classical rule,
while \(\langle S_k \rangle_q\) is computed according to the quantum rule which
associates to every observable \(A\) an Hermitian operator and computes the
expectation value from the density matrix
\begin{equation}\label{QM13}
\langle A \rangle = tr(\rho A).
\end{equation} 
We emphasize that with the identification (\ref{QM12}) the quantum rule
(\ref{QM13}) is no independent new rule or axiom. 
It follows directly from the classical probabilistic definition of expectation
values.

The classical spin operators \(\hat{S}_k\) in eq. (\ref{QM6}) and the quantum
spin operators \(S_k\) in eq. (\ref{QM9}) are different objects. 
The classical spin operators \(S_k\) are real diagonal \(8\times 8\) matrices
and commute. 
The quantum spin operators $\hat{S}_k$ are Hermitian \(2\times 2\) matrices that
do not commute,
\begin{equation}
\label{QM13A}
\left[S_k,S_l\right]=2i \varepsilon_{klm}S_m .
\end{equation}
For distinction, we use a hat for classical operators and no hat for quantum
operators. The map to the subsystem maps commuting ''classical" operators to
non-commuting quantum operators~\cite{CWNC}.

\paragraph*{Particle-wave duality}
Already in this very simple form we see the particle-wave duality of quantum
mechanics. 
The possible measurement values of the quantum spin components are \(\pm 1\), as
given by the possible measurement values of the three classical Ising spins
\(s_k\). 
The possible measurement values of the quantum spin are the eigenvalues of the
spin operators \(S_k\). 
The quantum rule states that the possible measurement values of an observable
are given by the spectrum of the associated operator. This is not a new rule or
axiom, but follows from the association with the classical Ising spins. 
The discreteness of the possible measurement values is the ``particle side" of
particle-wave duality.

The ``wave-side" is the continuous character of the time-local probabilistic
information. 
The probabilities \(p_\tau\), and therefore the expectation values \(\rho_k\) in
eq.(\ref{QM7}), are continuous. 
The density matrix is continuous as well. The density matrix \(\rho\) is a
``pure state density matrix" if it obeys the condition
\begin{equation}\label{QM15}
\rho^2 = \rho.
\end{equation}
In this case \(\rho\) can be composed as a product of the pure state wave
function \(\psi\) and its complex conjugate \(\psi^*\) according to
\begin{equation}\label{QM16}
\rho_{\alpha \beta} = \psi_\alpha \psi_\beta^*.
\end{equation}
The wave function is a normalized two component vector, \(\psi^\dagger \psi =
1\), which is an element of Hilbert space. 
The overall phase of \(\psi\) plays no role since it does not appear in the
density matrix (\ref{QM16}). All the wave-aspects of quantum mechanics are
associated to the continuous character of the local probabilistic information.

For the particular case of a pure quantum state the rule (\ref{QM13}) for the
expectation value of an observable takes the form familiar from quantum
mechanics
\begin{equation}
\langle A \rangle = tr(\rho A) = \rho_{\beta\alpha} A_{\alpha\beta} =
\psi_\alpha^* A_{\alpha\beta} \psi_\beta.
\end{equation}
It may be written in the conventional bra-ket notation as
\begin{equation}
\langle A\rangle = \psi^\dagger A \psi = \bra{\psi} A \ket{\psi}.
\end{equation}
We see that already for the simple one qubit subsystem the rules of quantum
mechanics emerge in a natural way.

\subsection{Incomplete statistics}

The operators for the quantum spins do not commute. This is no accident or
result of some particular choice. It is a direct consequence of the quantum
subsystem being characterized by incomplete statistics~\cite{CWICS, CWIS}.
Incomplete statistics is defined here in the sense that the statistical
information in the subsystem is not sufficient to compute classical correlation
functions for all observables.

\paragraph*{Quantum subsystem and environment}

\begin{figure}[t!]
	\includegraphics[scale=0.35]{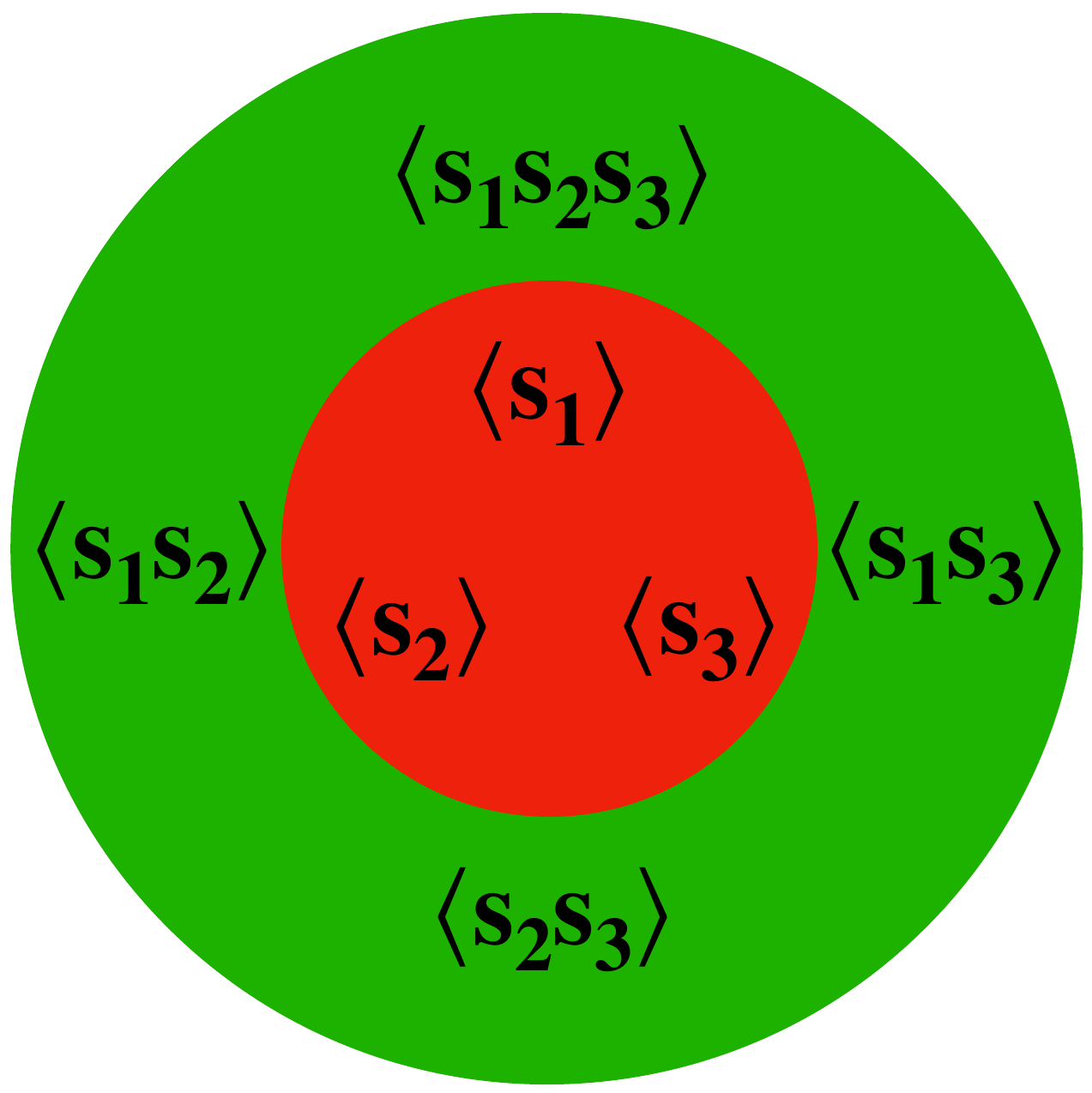}
	\caption{Schematic embedding of the quantum subsystem within the classical
statistical system in the space of correlation functions. The inner region (red)
comprises the quantum subsystem, and the outer region (green) constitutes the
environment. In contrast, a ``classical subsystem'' would eliminate $s_{1}$,
consisting of the correlations $s_{2}$, $s_{3}$ and $s_{2}s_{3}$. The quantum
subsystem is clearly not of this type.}\label{fig:3} 
\end{figure}

The quantum subsystem is characterized by the three expectation values \(\langle
s_k\rangle\). All other classical correlation functions of the Ising spins, as
\(\langle s_k s_l\rangle\) or \(\langle s_1 s_2 s_3\rangle\), belong to the
``environment". This is depicted in Fig.\,\ref{fig:3}. The quantum subsystem can
be seen as a submanifold in the manifold of all classical correlation functions
of Ising spins at a given $m$. In terms of the local probabilities \(p_\tau\)
the quantum subsystem is a three-dimensional submanifold of the
seven-dimensional manifold of the independent local probabilities, specified by
the relations (\ref{QM7}). The other four independent combinations of \(p_\tau\)
specify the environment, but are not relevant for the quantum subsystem. For a
given \( (\rho_1,\rho_2,\rho_3)\) all local probability distributions leading to
the same \(\rho\) describe the same quantum subsystem. The map from the local
probability distribution to the quantum subsystem is not invertible. It
``forgets" the probabilistic information pertaining to the environment.

\paragraph*{Classical correlation functions}

The classical two-point and three-point correlation functions belong to the
environment, and not to the quantum subsystem. They cannot be computed from the
probabilistic information of the quantum subsystem. For example, one has
\begin{equation}\label{QM19}
\langle s_1 s_2\rangle = p_1 + p_2 - p_3 - p_4 -p_5 - p_6 + p_7 + p_8.
\end{equation}
This linear combination cannot be expressed in terms of \(\rho_k\).
Classical correlations are ``inaccessible" for the quantum subsystem, since
their computation needs information about the environment beyond the subsystem.
This is ``incomplete statistics". For incomplete statistics the probabilistic
information is sufficient for the computation of expectation values of a certain
number of observables, but insufficient for the computation of all classical
correlation functions for these observables. For more general systems of
incomplete statistics some of the correlation functions may belong to the
subsystem, but not all of them. We will encounter this case for two qubits in
sect.\,\ref{sec:two-qubit_quantum_systems}.

\paragraph*{Incomplete statistics and non commuting operators}

For an expression of expectation values by eq.\,(\ref{QM13}) not all operators
for observables of the incomplete statistical system can commute. This is the
basic origin for the non-commutativity of the quantum spin operators for the
discrete qubit chain.

If two quantum operators \(A\) and \(B\) commute, \([A,B]=0\), also the product
\(C = A B = B A\) is a valid quantum operator that commutes with \(A\) and
\(B\). The expectation values of $A$, $B$
and $C$ are independent real numbers that have to be part of the probabilistic
information in the quantum subsystem. More precisely, $\langle C\rangle$ is
restricted by the values of $\langle A\rangle$ and $\langle B\rangle$, but not
computable in terms of $\langle A\rangle$ and $\langle B\rangle$ except for the
particular limiting cases $\langle A\rangle = \pm 1$, $\langle B\rangle = \pm
1$. If we associate $A$ with $s_1$ and $B$ with $s_2$ we have one more quantum
observable $D$ associated with $s_3$. Since $\langle D\rangle$ cannot be
expressed in terms of $\langle A\rangle$,$\langle B\rangle$ and $\langle
C\rangle$, such a system would need at least the probabilistic information given
by four real numbers. This is more than available by a $2\times 2$ Hermitian
normalized density matrix. The assumption $[A,B]=0$ leads to a contradiction.
One concludes that the operators representing the three classical spins $s_k$ in
the subsystem cannot commute. This holds for every pair of quantum operators
$S_k$.

It is interesting to consider for an extended setting the particular case where
the quantum correlation $\langle AB\rangle_q$ of two commuting quantum
observables equals the classical correlation $\langle AB\rangle_{cl}$ for two
classical observables $A$ and $B$ whose expectation values are used for the
definition of the quantum subsystems, i.e. $\langle A\rangle_q = \langle
A\rangle_{cl}$, $\langle B\rangle_q = \langle B\rangle_{cl}.$ While this is not
the general case, we will discuss in sect.\,\ref{sec:correlation_map} an
interesting ``correlation map" where this is realized. In this case one has
\begin{equation}\label{QM20}
\langle AB\rangle_q = tr(\rho A B) = \langle AB\rangle_{cl} = \sum_\tau p_\tau
A_\tau B_\tau.
\end{equation}
This identity can hold only for commuting quantum operators. Indeed, for any two
commuting operators there exists a basis where both are diagonal,
\begin{align}\label{QM21}
A_{\alpha\beta} = A_\alpha \delta_{\alpha\beta}, && B_{\alpha\beta} =
B_{\alpha}\delta_{\alpha\beta}, 
\end{align}
with \(A_\alpha\) and \(B_\alpha\) given by possible measurement values of the
observables. In this basis one has
\begin{equation}\label{QM22}
\langle AB\rangle_q =\sum_\alpha \rho_{\alpha\alpha} A_\alpha B_\alpha,
\end{equation}
which corresponds precisely to the classical expectation value \(AB_{cl}\),
provided that the diagonal elements \(\rho_{\alpha\alpha}\) can be associated
with probabilities of a subsystem of the classical system.

More precisely, for two-level observables $A$ and $B$ with possible measurement
values $\pm1$ the ``simultaneous probability" $p_{++}$ for finding $A=+1$ and
$B=+1$ is computable as an appropriate combination of diagonal elements
$\rho_{\alpha\alpha}$. This also holds for the other simultaneous probabilities
$p_{+-}$, $p_{-+}$ and $p_{--}$. The same simultaneous probabilities are
computable from the classical probabilities $p_\tau$. The relation \eqref{QM20}
requires that all simultaneous probabilities are the same in the quantum
subsystem and the classical statistical system. On the other hand, simultaneous
probabilities are not available for the quantum system if two associated
operators do not commute. (An exception may be states for which $\langle
[A,B]\rangle$ vanishes.) The two operators $A$ and $B$ cannot be diagonalized
simultaneously. In a basis where $A$ is diagonal, linear combinations of the
positive semidefinite diagonal elements $\rho_{\alpha\alpha}$ can be employed to
define the probabilities to find $A=1$, or $A=-1$. Similar probabilities can be
computed for $B$ in a basis where $B$ is diagonal. There is no way, however, to
extract simultaneous probabilities.

We conclude the following properties:
If the classical correlation function $\langle AB\rangle_{cl}$ is part of the
probabilistic information of the quantum subsystem, the associated quantum
operators $A$ and $B$ have to commute. Inversely, if $A$ and $B$ do not commute,
the classical correlation function is not available for the quantum subsystem
and therefore belongs to the environment. If $A$ and $B$ commute, the classical
correlation function can belong to the quantum subsystem but does not need to.
It may also be part of the environment. This issue depends on the precise
implementation of the quantum subsystem. 

\subsection{Quantum condition}
\label{sec:quantum_condition}

In order to realize a quantum subsystem the three expectation values $\rho_k =
\langle s_k\rangle_{cl}$ have to obey an inequality
\begin{equation}\label{QC1}
\sum_k \rho_k^2 \leq 1.
\end{equation}
This ``quantum constraint" or ``quantum condition" arises from the requirement
that the quantum density matrix $\rho$ is a positive matrix. Pure quantum states
require the ``pure state condition"
\begin{equation}\label{QC2}
\rho_k\rho_k = 1,
\end{equation}
while mixed states obey
\begin{equation}\label{QC2A}
\rho_k\rho_k < 1.
\end{equation}
The quantum subsystem can therefore not be realized for arbitrary time-local
probabilities $\{p_\tau(m)\}$, but only for a submanifold defined by
eq.~\eqref{QC1}. We will see that the quantum constraint is preserved by the
evolution. It has important consequences for the expectation values in the
quantum subsystem.

The quantum constraint arises here as a condition for the realization of a
subsystem with closed time evolution. There are some analogues with restricted
classical probability distributions which induce certain quantum
features~\cite{CF, HA, FU2, SPE, HSP, BRS}. Our general idea is that suitable
subsystems are selected by the evolution dynamics of the overall probability
distribution in a context of infinitely many degrees of freedom, similar to
isolated atoms in a quantum field theory. This dynamical selection can impose
the quantum constraint. For the purpose of our example we simply postulate the
quantum constraint.

\paragraph*{Pure state condition}

Consider first the pure state condition \eqref{QC2}. For a pure quantum state
one needs the condition \eqref{QM15}. We write the definition \eqref{QM8} of the
quantum subsystem as
\begin{equation}\label{QC3}
\rho = \frac{1}{2} \rho_\mu \tau_\mu,
\end{equation}
where we employ
\begin{align}\label{QC4}
\tau_0 = 1, && \rho_0 = 1,
\end{align}
and that the sum over $\mu$ extends form zero to three.
The condition $\rho^2 = \rho$ amounts to
\begin{equation}\label{QC6}
\frac{1}{4}(\rho_\mu \tau_\mu)(\rho_\nu \tau_\nu) = \frac{1}{8}\rho_\mu \rho_\nu
\{\tau_\mu \tau_\nu\} = \frac{1}{2} \rho_\mu \tau_\mu.
\end{equation}
With $\{\tau_k,\tau_l\} = 2\delta_{kl}$, $\{\tau_k,\tau_0\} = 2\tau_k$,
$\{\tau_0,\tau_0\} = 2$ the condition \eqref{QC6} becomes
\begin{equation}
\frac{1}{4}(1+\rho_k \rho_k) + \frac{1}{2} \rho_k \tau_k = \frac{1}{2} +
\frac{1}{2} \rho_k \tau_k,
\end{equation}
which indeed requires the condition \eqref{QC2}. Inversely, eq.~\eqref{QC2}
implies a pure state density matrix $\rho^2 = \rho$.

\paragraph*{Positive eigenvalues of density matrix}

For a pure quantum state the two eigenvalues of $\rho$ are $\lambda_1 = 1$,
$\lambda_2 = 0$. In general, the positivity of $\rho$ requires $\lambda_1 \geq
0$, $\lambda_2 \geq 0$. From
\begin{align}
\tr(\rho) = \lambda_1 + \lambda_2 = 1, && \det(\rho) = \lambda_1 \lambda_2,
\end{align}
we conclude that $\rho$ is a positive matrix if $\det(\rho) \geq 0$. Computing
from eq.\,\eqref{QC3}
\begin{equation}
\det(\rho) = \frac{1}{4}(1-\rho_k\rho_k),
\end{equation}
the condition $\det(\rho)\geq 0$ indeed coincides with the quantum constraint
\eqref{QC1}. The boundary value $\det(\rho) = 0$ is realized for the pure state
condition \eqref{QC2}, as appropriate since one eigenvalue of $\rho$ vanishes.
We conclude that mixed quantum states with positive $\rho$ not obeying
$\rho^2=\rho$ require the inequality \eqref{QC2A}.

\paragraph*{Bloch sphere}

\begin{figure}[t!]
	\includegraphics[scale=0.25]{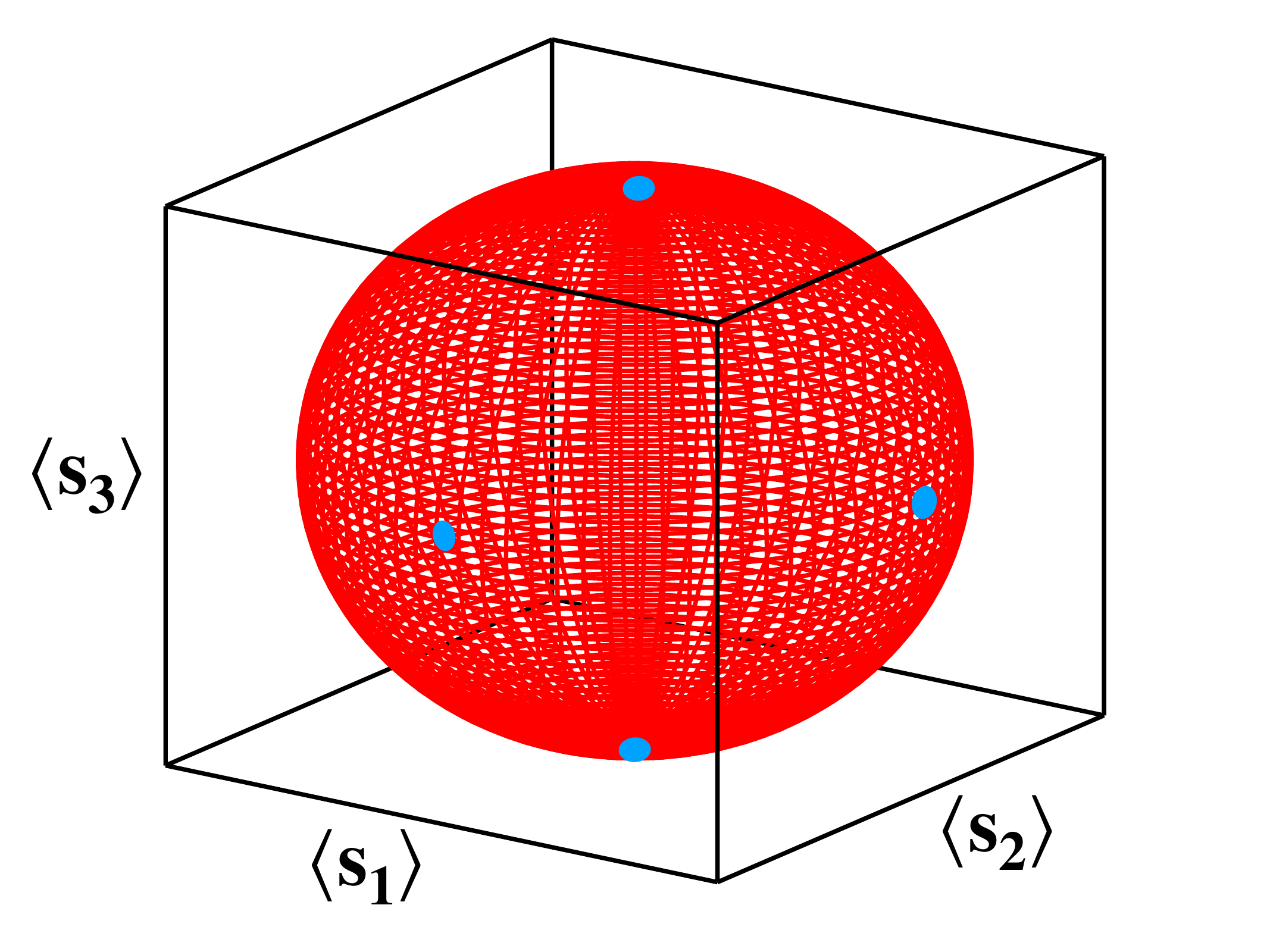}
	\caption{Quantum condition. For a quantum subsystem the expectation values
$s_{z}$ must be inside or on the Bloch sphere. Points on the Bloch sphere are
pure quantum states. Points outside the Bloch sphere correspond to classical
statistical probability distributions that do not realize a quantum subsystem.
Corners of the cube have $|s_{k}|=1$ for all $k$ and are not compatible with the
quantum subsystem. The Bloch sphere touches the cube at the points indicated at
the center of its surfaces.}\label{fig:4} 
\end{figure}

The quantum condition is visualized in Fig.\,\ref{fig:4}. Pure quantum states
are points on the Bloch sphere with
$\braket{s_1}^2+\braket{s_2}^2+\braket{s_3}^2 = 1$. The mixed quantum states
correspond to points inside the Bloch sphere.

\paragraph*{Uncertainty relation}

The most general classical probability distributions for three Ising spins can
realize arbitrary values $\braket{s_k}$ in the interval $-1 \leq \braket{s_k}
\leq 1$. These correspond to all points inside the cube in Fig.~\ref{fig:4}.
Points inside the cube but outside the Bloch sphere are valid classical
probability distributions, but the associated probability distributions do not
admit a quantum subsystem. Quantum subsystems can therefore be only realized by
a subfamily of classical probability distributions. The non-invertible map from
the classical probability distribution $\{p_\tau\}$ to the matrix $\rho$ can be
defined by eq.~\eqref{QC3} for arbitrary $\{p_\tau\}$. Only for a submanifold of
$\{p_\tau\}$ the matrix $\rho$ describes a valid positive quantum density
matrix, however.

As an example we consider the limiting classical distribution for which $p_\tau$
differs from zero only for the particular state with $s_1 =s_2=1$, $s_3 = -1$.
This translates to $\braket{s_1} = \braket{s_2} = 1$, $\braket{s_3}=-1$ and
corresponds to one of the corners of the cube in Fig. (\ref{fig:4}). With
$\rho_k \rho_k = 3$ this classical probability distribution violates the quantum
constraint \eqref{QC1}. Indeed, no valid quantum state can realize
simultaneously fixed values for the quantum spin in all directions.

More generally, the uncertainty relation of quantum mechanics follows directly
from the quantum constraint. Indeed, for a positive Hermitian normalized density
matrix $\rho$ the formulation of quantum mechanics can be applied and induces
the uncertainty relation. We can see directly from the quantum condition
\eqref{QC1} that a sharp value $\braket{s_1}=\pm 1$ requires a vanishing
expectation value for the spins in the two other Cartesian directions,
$\braket{s_2} = \braket{s_3} = 0$. Two spins cannot have simultaneously sharp
values, as well known in quantum mechanics from the commutation
relation~\eqref{QM13A} for the associated operators.

\subsection{Unitary evolution}
\label{sec:Unitary_evolution}

So far we have discussed how to extract a local quantum density matrix $\rho(m)$
from a classical probability distribution $\{p_\tau (m)\}$ at a given position
$m$ in the local chain. We can identify $m$ with time, $t=m\eps$, with $\eps$
the time interval for a discrete formulation of the quantum evolution. For the
discrete qubit chain \eqref{trafos} the probability distribution at $t$ is
mapped to the probability distribution at $t+\epsilon$. Indeed, the unique jump
operation corresponding to one of the discrete transformations \eqref{trafos}
maps every spin configuration $\sigma$ at $t$ to precisely one configuration
$\tau(\sigma)$ at $t+\epsilon$. The probabilities $p_\tau(t+\epsilon)$ obtain from
$p_\tau(t)$ by the simple relation
\begin{equation}
\label{QE0}
p_\tau(t+\eps)=p_{\sigma(\tau)}(t)\ ,
\end{equation}
with $\sigma(\tau)$ the inverse of the map $\tau(\sigma)$. In other words, the
probability of a configuration $\tau$ at $t+\eps$ equals the probability for the
configuration $\sigma(\tau)$ at $t$ from which it originates by the updating
rule of the automaton.
 From $\{p_\tau(t+\epsilon)\}$ we can
compute $\rho(t+\epsilon)$ by eqs.\,\eqref{QM8}, \eqref{QM5}.

\paragraph*{Discrete quantum evolution operator}

The question arises if $\rho(t+\epsilon)$ is again a positive quantum density
matrix if $\rho(t)$ obeys the quantum constraint, and if the change from
$\rho(t)$ to $\rho(t+\epsilon)$ follows the unitary evolution law of quantum
mechanics.
We will see that both properties hold. The quantum evolution of a density matrix
is given by the unitary quantum evolution operator $U(t+\epsilon,t)$
\begin{equation}\label{QE1}
\rho(t+\epsilon) = U(t+\epsilon,t) \rho(t) U^\dagger(t+\epsilon,t).
\end{equation}
For pure states, this is equivalent to the unitary evolution of the wave
function, 
\begin{equation}\label{QE2}
\psi(t+\epsilon) = U(t+\epsilon,t)\psi(t).
\end{equation}
Any mixed state quantum density matrix can be represented as a linear
combination of pure state density matrices $\rho^{(a)}$
\begin{equation}\label{QE3}
\rho = \sum_a w_a \rho^{(a)},
\end{equation}
with $(\rho^{(a)})^2 = \rho^{(a)}$. The pure state density matrices $\rho^{(a)}$
can be written in terms of wave functions $\psi^{\alpha}$,
\begin{equation}\label{QE4}
\rho_{\alpha \beta}^{(a)} = \psi_\alpha^{(a)} \psi_\beta^{(a)*},
\end{equation}
for which the evolution is given by eq.~\eqref{QE1}. Since the evolution
equation is linear in $\rho^{(a)}$ it also holds for linear combinations of
$\rho^{(a)}$.
The positive coefficients $w_a \geq 0$ can be interpreted as probabilities to
find a given pure state $a$. With eqs.~\eqref{QE3}, \eqref{QE4} eq.~\eqref{QE1}
follows from eq.~\eqref{QE2}.

We are interested here in discrete time steps from $t$ to $t+\epsilon$, where
the distance $\epsilon$ between two neighboring time points is always the same.
We therefore use the abbreviated notation
\begin{align}\label{QE5}
U(t)=U(t+\epsilon,t), && U^\dagger(t) U(t) = 1.
\end{align}
The unitary $2\times 2$ matrices $U(t)$ are the discrete evolution operators.

\paragraph*{Unitary evolution for discrete qubit chain}

Consider as a particular transformation $T_{31}$, that acts on the expectation
values $\rho_k = \braket{s_k}$ as 
\begin{align}\label{QE6}
\rho_3(t+\epsilon) = \rho_1(t), && \rho_1(t+\epsilon) = -\rho_3(t), &&
\rho_2(t+\epsilon) = \rho_2(t).
\end{align}
This corresponds to a unitary transformation in the quantum subsystem, given by
the unitary matrix
\begin{equation}\label{QE7}
U_{31} - \frac{1}{\sqrt{2}} 
\begin{pmatrix}
1 & 1 \\
-1 & 1
\end{pmatrix}.
\end{equation}\label{QE8}
Indeed, one verifies
\begin{align}\label{QE9}
&\ \frac{1}{2} U_{31}(t)
\begin{pmatrix}[c]
1+\rho_3(t)\phantom{00} & \rho_1(t) - i\rho_2(t) \\[2mm]
\rho_1(t) + i \rho_2(t)\phantom{00}& 1-\rho_3(t)
\end{pmatrix}
U_{31}^{\dagger}(t)
\nonumber \\
&= \frac{1}{2}
\begin{pmatrix}[c]
1+\rho_1(t)\phantom{00} & -\rho_3(t) - i\rho_2(t) \\[2mm]
-\rho_3(t) + i \rho_2(t)\phantom{00} & 1-\rho_1(t)
\end{pmatrix} \\
&= \frac{1}{2}
\begin{pmatrix}[c]
1+\rho_3(t+\epsilon)\phantom{00} & \rho_1(t+\epsilon) - i\rho_2(t+\epsilon)
\\[2mm]
\rho_1(t+\epsilon) + i \rho_2(t+\epsilon)\phantom{00} & 1-\rho_3(t+\epsilon)
\end{pmatrix}\nonumber\ , 
\end{align}
in accordance with eq.~\eqref{QE6}. The unique jump operation $T_{31}$ acting on
the probability distribution for the classical bits is reflected as a unitary
transformation for the qubit.

The other unique jump operators in eq.~\eqref{trafos} also act as unitary
transformations on the quantum density matrix, with discrete evolution operators
given by 
\begin{align*}
U_{12} = \frac{1}{\sqrt{2}}
\begin{pmatrix}
1+i & 0\\
0 & 1-i
\end{pmatrix},
&& U_{23} = \frac{1}{\sqrt{2}}
\begin{pmatrix}
1 & i \\
i & 1
\end{pmatrix},
\end{align*}
\begin{align}\label{QE10}
U_1 = 
\begin{pmatrix}
0 & i \\
i & 0
\end{pmatrix},
&& U_2 = 
\begin{pmatrix}
0 & 1 \\
-1 & 0
\end{pmatrix},
&& U_3 = 
\begin{pmatrix}
i & 0 \\
0 & -i
\end{pmatrix}.
\end{align}
The overall phase of $U$ is arbitrary since it drops out in the transformation
\eqref{QE1}. We observe that $U_{23} = \exp(i\pi \tau_1/4)$, $U_{31} =
\exp(i\pi\tau_2 /4)$ and $U_{12} = \exp(i\pi\tau_3 /4)$ induce rotations by
$\pi/2$ for the vector $(\rho_1, \rho_2, \rho_3)$ in the planes indicated by the
indices. The operators $U_k = i \tau_k$ induce rotations by $\pi$ around the
$k$-axis for the Cartesian spin directions, equivalent to simultaneous
reflections of two spin directions.

Not all unique jump operators on the classical probability distributions lead to
unitary transformations for the quantum subsystem. As an example, consider the
unique jump operation $p_1 \leftrightarrow p_3$, $p_2 \leftrightarrow p_4$. It
corresponds to a conditional change of $s_2$. If $s_1 = -1$ the sign of $s_2$
flips, $s_2' = - s_2$, while for $s_1 = 1$ the spin $s_2$ remains unchanged.
This transformation leaves $\rho_1$ and $\rho_3$ invariant, while $\rho_2$
changes to $\rho_2' = p_1 + p_2 - p_3 - p_4 - p_5 - p_6 + p_7 + p_8$. The
combination $\rho_2'$ cannot be expressed in terms of $\rho_1$, $\rho_2$,
$\rho_3$. For realizing a unitary transformation on the quantum subsystem it is
necessary that the density matrix at $t + \epsilon$ can be expressed in terms of
the density matrix at $t$. This is not the case for the above conditional spin
flip. Another type of unique jump operation that does not correspond to a
unitary quantum evolution is the reflection of an odd number of classical Ising
spins. For example, $s_2 \to - s_2$ results in complex conjugation of the
quantum density matrix, $\rho \to \rho^*$, rather than a unitary transformation
of $\rho$.

\paragraph*{Sequences of unitary evolution steps}

For the discrete qubit chain one may choose arbitrary sequences of unique jump
operations \eqref{trafos}. On the level of the quantum subsystem this is reflected
by a sequence of unitary operations, e.g.
\begin{align}\label{QE11}
&\rho(t + 3\epsilon) =\nonumber\\
&U_a(t+2\epsilon) U_b(t+\epsilon) U_c (t) \rho (t)
U_c^\dagger (t) U_b^\dagger (t+\epsilon) U_a^\dagger(t+2\epsilon)\ .
\end{align}
Such transformations are elements of a discrete group that is generated by two
basis transformations, say $T_{31}$ and $T_{12}$. On the level of unitary
transformations of the quantum subsystem this is the group generated by $U_{31}$
and $U_{12}$, with matrices differing only by an overall phase.

We note the identities
\begin{align}
U_{31}^2 = U_2 && U_{12}^2 = U_3, && U_{23}^2 = U_1,
\end{align}
which correspond to a sequence of two identical $\pi/2$-rotations producing a
$\pi$-rotation around the same axis. The inverse $\pi/2$-rotations obey
\begin{align}
U_{13} &= - U_{31} U_2 =- U_2 U_{31} = U_{31}^\dagger, \nonumber \\
U_{21} &= - U_{12} U_3 =- U_3 U_{12} = U_{12}^\dagger, \nonumber \\
U_{32} &= - U_{23} U_1 =- U_1 U_{23} = U_{23}^\dagger.
\end{align}
We finally observe
\begin{equation}
U_{23} = U_{12} U_{31} U_{21},
\end{equation}
such that two basic transformations induced by $U_{31}$ and $U_{12}$ generate
the complete discrete group. The discrete qubit chains can realize arbitrary
sequences of unitary transformations belonging to this discrete group.

\paragraph*{Quantum computing}

What is quantum computing? Quantum computing is based on a stepwise evolution of
a quantum system. For simplicity we consider equidistant time steps $t$,
$t+\epsilon$, $t + 2\epsilon$ and so on. A given computational step maps the
probabilistic information of a quantum system at time $t$ to the one at time $t
+\epsilon$. The discrete unitary transformations of the density matrix
\eqref{QE1} are called gates. In case of pure states the gates $U(t)$ act on the
wave function \eqref{QE2}. We concentrate on the formulation in terms of the
density matrix
\begin{equation}
\rho(t +\epsilon) = U(t)\rho(t)U^\dagger (t),
\end{equation}
from which eq.\,\eqref{QE2} can be derived as a special case. A quantum
computation consists of a sequence of quantum gates, corresponding to matrix
multiplication of unitary matrices according to eq.~\eqref{QE11}. In this way
the input in form of $\rho(t_{in})$ is transformed to the output in form of
$\rho(t_f)$, where it can be read out by measurements.

The discrete bit chain \eqref{trafos} can be viewed as a quantum computer. It is
a very simple one since it can only perform a rather limited set of gates,
corresponding to the discrete group discussed above. Nevertheless, it can
perform a set of quantum operations by simple deterministic manipulation of
classical bits. The discrete subgroup generated by the $\pi/2$-rotations of the
vector of classical Ising spins $(s_1,s_2,s_3)$ are only a small subgroup of the
general deterministic operations for three classical spins. The latter
correspond to the group of permutations for eight elements, corresponding to the
eight states $\tau$.

One may ask what is particular about the quantum operations realized by three
classical spins. The particularity arises from the quantum constraint
\eqref{QC1}. The classical Ising spins or bits do not all have well determined
values $s_k = \pm 1$, as for classical computing. Only the three independent
probabilities to find the values one or zero are available for a given bit. The
probabilities for the possible states of three bits, corresponding to $p_\tau$,
are not needed. Many probability distributions for the states of three bits lead
to the same expectation values $\braket{s_k} = \rho_k$. On the other hand,
knowledge of the probability distribution for one spin, say $\braket{s_1}$,
entails information on the two other spins. For example, if $\braket{s_1} = \pm
1$, one knows $\braket{s_2} = \braket{s_3} = 0$.

\paragraph*{Complete unitary transformations}

In quantum computing it is well known that if a system can perform a suitable
set of basis gates it can perform the complete set of all unitary
transformations by a suitable sequence of the basis gates. For the two basis
gates for a one-qubit system one usually takes the Hadamard gate $U_H$ and the
rotation gate $U_T$, 
\begin{align}
U_H = \frac{1}{\sqrt{2}}
\begin{pmatrix}
1&1\\
1&-1
\end{pmatrix}
&& U_T =
\begin{pmatrix}
1 & 0\\
0&e^{i\pi/4}
\end{pmatrix}.
\label{eq:4.2.46}
\end{align}
For the Hadamard gate one has
\begin{align}
U_H:\ \rho_1(t+\epsilon) &=\rho_3(t),\quad \rho_3(t+\epsilon) = \rho_1(t),
\nonumber\\
\rho_2(t+\epsilon) &= -\rho_2(t),
\end{align}
while the $T$-gate amounts to
\begin{align}
U_T:\ \rho_1(t+\epsilon) &= \frac{1}{\sqrt{2}} (\rho_1(t)
-\rho_2(t)),\nonumber\\
\rho_2(t+\epsilon) &= \frac{1}{\sqrt{2}} (\rho_1(t) +\rho_2(t)),\nonumber\\
\rho_3(t+\epsilon) &= \rho_3(t).
\end{align}
An arbitrary unitary matrix $U$ can be approximated with any wanted precision by
a sequence of factors $U_H$ and $U_T$.

The Hadamard gate can be realized by a deterministic operation on classical
bits, $s_1 \leftrightarrow s_3$, $s_2 \to -s_2$. The matrix $U_H$ is a product
of the rotation matrices discussed above,
\begin{equation}
U_H = -i U_{31} U_1.
\end{equation}
It can be realized by the corresponding combination of updatings~\eqref{trafos}
of the probabilistic automaton. The rotation gate cannot be obtained by unique
jump operations. If we could represent it as a product of the unitary matrices
of the discrete group generated by $\pi/2$-rotations, these transformations
would generate arbitrary unitary transformations by suitable products. This is
obviously not possible for the finite discrete group.

\paragraph*{General unitary transformations}

The rotation gate requires a change of the classical probability
distribution $\{p_\tau\}$ that does not correspond to a unique jump operation.
Since every quantum density matrix $\rho(t)$ can be realized by some probability
distribution $\{p_\tau(t)\}$ according to eqs.\,\eqref{QM7}, \eqref{QM8},
suitable changes of probability distributions that realize the rotation gate do
exist. This extends to arbitrary unitary transformations of the one-qubit
density matrices. Any arbitrary unitary quantum evolution can be realized by
suitable evolutions of time-local probability distributions. The issue is not a
question of principle, but rather if possible concrete realizations of the
required changes of probability distributions are available.

While it is not possible to realize the $T$-gate by a simple automaton acting on
three classical bits, it is possible to realize it in more extended classical
statistical systems. The probability distributions for three classical Ising
spins could perhaps be realized by three suitably correlated probabilistic bits
($p$-bits)~\cite{CSD1}. This would permit to perform transformations of these
probability distributions, possibly conserving automatically the quantum
constraint. We will discuss in sect.~\ref{sec:classical_and_quantum_computing}
artificial neural networks or neuromorphic computers that can learn to perform
changes of the classical probability distribution which realize the $T$-gate.

\paragraph*{Unitary evolution and quantum condition}

A unitary quantum evolution and the quantum condition \eqref{QC1} are in close
correspondence. Unitary transformations act as rotations on the three component
vector $(\rho_1,\rho_2,\rho_3)$. They therefore preserve the ``purity"
\begin{equation}
P = \rho_k \rho_k.
\end{equation}
In particular, a pure quantum state with $P=1$ remains a pure quantum state
after the transformation. More generally, if $\rho(t)$ obeys the quantum
constraint $P\leq 1$, this is also the case for $\rho(t+\epsilon)$.

On the other hand, the possibility to perform arbitrary unitary evolution steps
requires the quantum condition \eqref{QC1}. For points outside the Bloch sphere
in Fig.~\ref{fig:4}, for which $P>1$, arbitrary rotated points do not lie within
the cube. In other words, a general rotation of the vector of expectation values
$(\braket{s_1},\braket{s_2},\braket{s_3})$ is no longer a set of allowed
expectation values. Some of the $|\braket{s_k}|$ would have to be larger than
one, which is not possible. If the dynamics is such that arbitrary unitary
transformations are possible for a simple qubit quantum subsystem, the
probability distributions have to obey the quantum condition.

\paragraph*{Correlated computing}

Even though the discrete qubit chain cannot perform arbitrary unitary
transformations of a single qubit, it realizes already a key property of quantum
computing, namely correlated computing. In a quantum computer the different
Cartesian spin directions are not independent but obey strong correlations. If
one changes one spin direction $\langle s_k\rangle$, one necessarily influences
simultaneously the other two. This extends to several qubits in an entangled
state. Manipulating one qubit immediately affects the other qubits. This use of
correlations is a key feature of quantum computing which enhances its power as
compared to a classical computer. For a classical computer changing one bit
$s_k$ does not necessarily affect other bits.

The reason for this ``global effect" of a change in a single quantity (e.g.
single qubit) resides in strong correlations. Our embryonic quantum computer is
a simple model for the understanding of this ``correlated computing". Indeed, the
probability distributions $\{p_\tau(t)\}$ for the classical spins which are
compatible with the quantum constraint~\eqref{QC1} all describe states with
strong correlations between the different spins. Probability distributions for
which two (or three) of the classical spins are uncorrelated can be written in a
suitable product form. This product form is not compatible with the quantum
constraint. We conclude that the quantum constraint enforces correlations. Once
these correlations are realized for some initial state they will be preserved by
the unitary evolution.

\subsection{Probabilistic observables}
\label{sec:probabilistic_observables}

The time-local probabilistic information of the quantum subsystem is given by
expectation values $\rho_k(m)$. The vector $\rho_k$ or the associated density
matrix specify the state of the system at a given time $t$. The transition to
the subsystem entails important conceptual changes for the status of
observables.

Observables have no longer fixed values for every state of the subsystem. They
become ``probabilistic observables" for which only probabilities to find a given
possible measurement value are given for any state of the
subsystem~\cite{BINE,VONE1,MIS,ALP,HOL,SISU,BEBU,BEBU2,BUG,STUBU}. This
change of character of the observables is not a fundamental change -- the
possible measurement values are not changed by the transition to the subsystem.
We only deal with restricted information available for the observables in the
subsystem. Since realistic quantum systems are typically subsystems of systems
with infinitely many degrees of freedom it is essential to understand the
concept of probabilistic observables for subsystems. Many quantum features
emerge from the map to a subsystem. We have discussed the concept of
probabilistic observables in detail in the first part of this work~\cite{CWPW}.

Consider the three Ising spins $s_k$. Within the subsystem they are time-local
system observables whose expectation values $\braket{s_k}$ can be 
computed from the probabilistic information of the subsystem. The latter is
given by the three system variables $\rho_k$ that define the 
density matrix. These system observables have associated local-observable
operators $\hat{S}_k = \tau_k$. The possible measurement values
$\lambda_{\pm}^{(k)} = \pm 1$ correspond to the eigenvalues of the operators
$\hat{S}_k$. Together with the probabilities $w_\pm^{(k)}$ to find for $s_k$
the value $\lambda_{\pm}^{(k)}$ they specify probabilistic observables. These
probabilities are given by
\begin{equation}
\label{PQ01}
w_\pm^{(k)} = \frac{1}{2} \left( 1 \pm \rho_k \right)\,.
\end{equation}
They are computable from the system variables $\rho_k$. Due to the quantum
constraint $\rho_k \rho_k \leq 1$ at most
one of the spins can have a sharp value, however. This requires the state of the
subsystem to be a particular pure quantum state, namely an
eigenstate to the corresponding operator $\hat{S}_k$. Thus one has genuinely
probabilistic observables which cannot all take simultaneously sharp values. 
The quantum subsystem admits no microstates for which all system observables
have sharp values. 

One may question about other possible system observables. The spin operators in
arbitrary directions,
\begin{equation}
\label{PQ02}
\hat{S}(e) = e_k \tau_k\,, \quad e_k e_k = 1\ ,
\end{equation}
obey the criteria for local-observable operators~\cite{CWPW}. The question is if
there are measurement procedures that identify 
probabilistic observables $s(e_k)$ for which the possible outcomes are the
values $\pm 1$, and for which the probabilities $w_\pm(e_k)$
are given by
\begin{equation}
\label{PQ03}
w_\pm(e_k) = \frac{1}{2} (1 + e_k \rho_k)\,.
\end{equation}
If yes, these are system observables. We will discuss in
sect.~\ref{sec:classical_ising_spins_and_quantum_spin} a setting for which
the observable $s(e_k)$ are associated to yes/no decisions in a classical
statistical setting. In this case we are guaranteed that they
are system observables of the quantum subsystem.

\subsection{Bit-quantum map}\label{sec:bit_quantum_map}

A bit-quantum map is a map from the local probabilistic information for
classical Ising spins or bits to the density matrix for qubits. It maps a
``classical'' probabilistic system to a quantum subsystem. This map is
compatible with the local structure associated to time and evolution. It maps a
time-local subsystem to a quantum subsystem at the same time $t$. In general, a
bit-quantum map is a map from the classical density matrix $\rho'(t)$ to a
quantum density matrix $\rho(t)$. In our case it is a map from the time-local
probability distribution $\{p_\tau(t)\}$ to the density matrix of the subsystem.
The bit-quantum map can be generalized from a finite set of classical Ising
spins to continuous variables.

For the present one-qubit quantum system realized by the discrete qubit chain
the bit-quantum map is given by eq.\,\eqref{QM8}, with coefficients $\rho_k(t)$
expressed in terms of the probabilities $p_\tau(t)$ by eq.\,\eqref{QM7}. This
map is ``complete'' in the sense that for every quantum density matrix $\rho(t)$
one can find a local probability distribution $\{p_\tau(t)\}$ such that the
bit-quantum map realizes this density matrix. The bit-quantum map is not an
isomorphism. Many different probability distributions $\{p_\tau(t)\}$ realize
the same quantum density matrix $\rho(t)$. For the particular case of a density
matrix for a pure quantum state the bit-quantum map is a map between the eight
classical probabilities and the complex normalized two-component quantum wave
function. A map of this type is also considered in ref.~\cite{YY1}.

\begin{figure}
\includegraphics{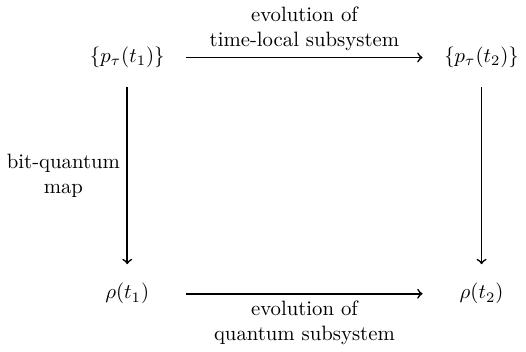}
\caption{Evolution of quantum subsystem induced by evolution of ``classical''
time-local subsystem.}
\label{fig:BQ}
\end{figure}

The bit-quantum map transports the time evolution of the time-local subsystem to
the time evolution of the quantum subsystem. In our case, a time evolution of
the probabilities $p_\tau(t)$ results in a time evolution of $\rho(t)$, as shown
in Fig.\,\ref{fig:BQ}.
This requires the time evolution of the time-local subsystem to be compatible
with the bit-quantum map. If $\{p_\tau(t_1)\}$ obeys the quantum constraints,
this has to hold for $\{p_\tau(t_2)\}$ as well. For the discrete qubit chain as
a probabilistic automaton with updatings~\eqref{trafos} the evolution of the
quantum subsystem is indeed unitary, with discrete evolution
operators~\eqref{QE7},~\eqref{QE10}.

More generally, a unitary quantum evolution requires particular properties for
the evolution of the probability distribution $\{p_\tau(t)\}$. Consider two
different Hermitian, normalized, and positive matrices $\rho_1$ and $\rho_2$
that are related by a unitary transformation $U$,
\begin{equation}
\rho_2 = U \rho_1 U^\dagger.
\label{eq:BQ1}
\end{equation}
If $\{p_\tau(t_1)\}$ is mapped to $\rho_1(t_1)$, and $\{p_\tau(t_2)\}$ to
$\rho_2(t_2)$, the unitary quantum evolution operator $U(t_2,t_1)$ is given by
eq.\,\eqref{eq:BQ1}. An arbitrary evolution of $\{p_\tau(t)\}$ defines the
evolution of a hermitean normalized matrix $\rho(t)$ according to
eq.\,\eqref{QM8}. In the general case, however, $\rho(t_2)$ needs not to be
related to $\rho(t_1)$ by a unitary transformation \eqref{eq:BQ1}. 

A necessary condition for a unitary transformation is that $\{p_\tau(t)\}$ obeys
the quantum constraint for all $t$. The quantum constraint ensures positivity of
the associated density matrix. Since a unitary evolution preserves the
eigenvalues of $\rho(t)$, a violation of the quantum constraint cannot be
compatible with a unitary evolution for which $\rho(t)$ remains positive for all
$t$. As a sufficient condition for a unitary evolution of $\rho(t)$ we may state
that the evolution of $\{p_\tau(t)\}$ must be such that all eigenvalues of
$\rho(t)$ are invariant. Two hermitean matrices with the same eigenvalues can
indeed be related by a unitary transformation \eqref{eq:BQ1}. In particular, if
the evolution of the ``classical'' time-local subsystem is such that every
$\{p_\tau(t_1)\}$ representing a pure state density matrix $\rho(t_1)$ evolves
at $t_2$ to a distribution representing another (unique) pure state density
matrix $\rho(t_2)$, the quantum evolution has to be unitary.

It should be clear by this short discussion that the time evolution of classical
probabilistic systems can generate by the quantum-bit map an evolution law for
the quantum subsystem that is not unitary. In particular, it can describe
phenomena as decoherence or syncoherence for which pure quantum states evolve to
mixed quantum states and vice versa. We will discuss in
sec.\,\ref{sec:dynamic_selection_of_quantum_subsystems} the general reason why
Nature selects the unitary quantum evolution among the many other possible
evolution laws.

For a complete bit-quantum map an arbitrary unitary evolution of the quantum
subsystem can be realized by a suitable evolution of the time-local
``classical'' subsystem. For every $U$ in eq.\,\eqref{eq:BQ1} one obtains from
$\rho_1$ a given $\rho_2$, for which a probability distribution
$\{p_\tau(t_2)\}$ exists by virtue of completeness. Since $\{p_\tau(t_2)\}$
realizing $\rho_2$ is not unique, the ``classical evolution law'' for
$\{p_\tau(t)\}$ realizing a given unitary quantum evolution is not unique.

\subsection{Simple quantum system from classical statistics}

In summary of this chapter we have constructed a simple discrete quantum system
from a classical statistical setting. It is described by a complex Hermitian and
positive density matrix, or by a complex wave function in case of a pure state.
Its time evolution performs a restricted set of unitary transformations. This
quantum system can be regarded as a restricted one-qubit quantum computer. On
the formal side the three Cartesian spin observables are represented by
non-commuting quantum operators. This non-commutativity reflects the incomplete
statistics of the subsystem. The eigenvalues of the quantum operators coincide
with the possible measurement values. The spin observables are probabilistic
observables which cannot have simultaneous sharp values. The quantum mechanical
uncertainty is realized. The classical correlation function for the spin
observables is not accessible for the quantum subsystem. Our classical
probabilistic model constitutes a simple example for the emergence of quantum
mechanics from classical statistics~\cite{CWB}.

In our example all these quantum properties arise from the focus to a subsystem.
We should emphasize, however, that the map to a subsystem is not the only way
how incomplete statistics and non-commutative operators are realized for
classical statistical systems. Another origin can be ``statistical observables"
which characterize properties of the probabilistic information and have no
definite value for a given spin configuration. The momentum observable for
probabilistic cellular automata discussed in ref.~\cite{CWPW} is a good
example.

\section{Entanglement in classical and quantum statistics}
\label{sec:entanglement_in_classical_and_quantum_statistics}

Entanglement describes situations where two parts of a system are connected and
cannot be separated. The properties in one part depend on the properties of the
other part. The quantitative description of such situations is given by
correlation functions.
There is no conceptual difference between entanglement in classical statistics
and in quantum mechanics\,\cite{CWA}. In this chapter we will construct
explicitly probabilistic automata that realize the maximally entangled state of
a two-qubit quantum system.

\subsection{Entanglement in classical statistics and quantum mechanics}

A simple example of entanglement in a classical probabilistic system is a system
of two Ising spins $s_1$ and $s_2$ for which the probabilities for equal signs
of both spins vanish. The two spins are maximally anticorrelated. We denote by
$p_{++}$ the probability for $s_1 = s_2 = 1$, and by $p_{--}$ the one for the
state $s_1 = s_2=-1$. Similarly, we label the probabilities $p_{+-}$ for
$s_1=1$, $s_2 = -1$ and $p_{-+}$ for $s_1 = -1$, $s_2 = 1$. For a probability
distribution 
\begin{align}
p_{+-} = p_{-+} = \frac{1}{2}, && p_{++} = p_{--} = 0
\end{align}
one finds the correlation function
\begin{equation}
\braket{s_1 s_2} = -1,
\end{equation}
while the expectation values for both spins vanish 
\begin{align}
\braket{s_1} = 0, && \braket{s_2} = 0.
\end{align}

The interpretation is simple: the two spins necessarily have opposite signs.
Assume that a measurement of $s_1$ yields $s_1 = 1$, and the measurement is
ideal in the sense that it eliminates the possibilities to find $s_1 = -1$
without affecting the relative probabilities to find $s_2$. The
\textit{conditional probability} $p_{+-}^{(c)}$ to find $s_2 = -1$ after a
measurement $s_1 =1$ equals one in this case, while the conditional probability
$p_{++}^{(c)}$ to find $s_2 = 1$ after a measurement $s_1 = 1$ vanishes. One is
certain to find $s_2 = -1$ in a second measurement of $s_2$. We observe,
however, that this statement involves the notion of conditional probabilities
and ideal measurements which may not always be as simple as for the assumed
situation. We will discuss this issue in
sect.\,\ref{sec:conditional_probabilities_4_7}.

There is no need that the measurement of $s_1$ sends any ``signal" to $s_2$. For
example, the two spins may be separated by large distances, such that no light
signal can connect $s_1$ and $s_2$ for the time span relevant for the two
measurements. An example is the cosmic microwave background where $s_1$ and
$s_2$ may correspond to temperatures above or below the mean in two regions of
the sky at largely different angles. The two temperature differences or Ising
spins are correlated, even though no maximal anticorrelation will be found in
this case. No signal can connect the two regions at the time of the CMB-emission
or during the time span of the two measurements at different angles. At the time
of the CMB-emission the correlations on large relative angles are non-local.
They can be prepared by some causal physics in the past, however. We will
discuss the issue of causality much later in this work. What is already clear at
this simple level is the central statement:
In the presence of correlations a system cannot be divided into separate
independent parts. The whole is more than the sum of its parts. 

In the concept of probabilistic realism there exists one real world and the laws
are probabilistic. The reality is given by the probability distribution without
particular restrictions on its form. One may nevertheless introduce a restricted
concept of reality by calling real only those properties that occur with
probability one or extremely close to one. This is the approach used by
Einstein, Podolski and Rosen in ref.~\cite{EPR}. If we apply this restricted
concept of reality to the entangled situation above, it is the anticorrelation
between the two spins that is real. In contrast, the individual spin values are
not real in this restricted sense, since they have the value $+1$ or $-1$ with
probability one half. If one tries to divide the system artificially into
separated parts, and assigns ``restricted reality" to the spin values in each
part, one should not be surprised to encounter paradoxes. We will discuss the
issue in more detail in sect.\,\ref{sec:the_paradoxes_of_quantum_mechanics}.

In quantum mechanics the precise quantitative definition of the notion of
entanglement is under debate. An entangled state is typically a state that is
not a direct product state of two single spin states. The main notion is a
strong correlation between two individual spins. Consider a two qubit system in
a basis of eigenstates to the spins in the 3-direction $S_3^{(1)}$ and
$S_3^{(2)}$. A ``maximally entangled state" is given by
\begin{equation}\label{E1}
\psi_{\text{en}} = \frac{1}{\sqrt{2}}(\ket{\uparrow} \ket{\downarrow} -
\ket{\downarrow} \ket{\uparrow}),
\end{equation}
where $\ket{\uparrow}$ and $\ket{\downarrow}$ in the first position denote the
spin $S_3^{(1)} = 1$ or $-1$ of the first qubit, while the second position
indicates $S_3^{(2)} = +1,\ -1$. In the state $\psi_{\text{en}}$ the spins of
the two qubits are maximally anticorrelated in all directions
\begin{equation}\label{E2}
\braket{S_1^{(1)} S_1^{(2)}}= \braket{S_2^{(1)} S_2^{(2)}} = \braket{S_3^{(1)}
S_3^{(2)}} = -1,
\end{equation}
while all expectation values of spins vanish.
\begin{equation}\label{E3}
\braket{S_k^{(i)}} = 0.
\end{equation}
Furthermore, for $k\neq l$ one has
\begin{equation}\label{E3A}
\braket{S_k^{(1)} S_l^{(2)}} = 0.
\end{equation}
The problems with the precise definition of entanglement are connected to the
possibility of different choices of basis. Here we employ a fixed basis,
associated to the two individual quantum spins. 

It is often believed that entanglement is a characteristic feature of quantum
systems, not present in classical probabilistic systems. If quantum systems are
subsystems of classical statistical systems, however, all quantum features,
including the notion of entanglement, should be present for the classical
probabilistic systems. We will see that this is indeed the case.

\subsection{Two-qubit quantum systems}
\label{sec:two-qubit_quantum_systems}

We briefly recall here basic notions of two-qubit quantum systems. This fixes
the notation and specifies the relations that we want to implement by a
classical statistical system.

\paragraph*{Direct product basis}

A system of two quantum spins or qubits is a four-state system. Its density
matrix $\rho$ is a positive Hermitian $4 \times 4 $ matrix, normalized by $\tr
\rho = 1$.
Correspondingly, for a pure quantum state the wave function is a complex
four-component vector. We will use a basis of direct product states of wave
functions for single qubits, 
\begin{equation}
\psi = \begin{pmatrix}
\psi_1\\\psi_2\\\psi_3\\\psi_4
\end{pmatrix}
= \psi_1 \ket{\uparrow}\ket{\uparrow} + \psi_2 \ket{\uparrow}\ket{\downarrow} +
\psi_3 \ket{\downarrow}\ket{\uparrow} + \psi_4 \down\down,
\end{equation}
with $\psi_\alpha^* \psi_\alpha = 1$, $\alpha = 1...4$. A general direct product
state is given by
\begin{equation}
\psi_{\text{dp}} = (b_1 \up + b_2 \down)(c_1 \up + c_2 \down)\ ,
\end{equation}
or
\begin{align}\label{E6}
\psi_1 = b_1 c_1\ ,\quad \psi_2 = b_1 c_2\ ,\quad \psi_3 = b_2 c_1\ ,\quad
\psi_4 = b_2 c_2\ .
\end{align}
Pure states that do not obey the relations \eqref{E6} are called entangled. An
example is the maximally entangled state \eqref{E1} with
\begin{align}
\psi_2 = -\psi_3 = \frac{1}{\sqrt{2}}, && \psi_1 = \psi_4 = 0.
\end{align}

\paragraph*{Unitary transformations and the CNOT-gate}

Unitary transformations can transform direct product states into entangled
states and vice versa. A prominent example is the CNOT-gate
\begin{equation}\label{E8}
U_C = 
\begin{pmatrix}
\boldsymbol{1}_2 & 0 \\
0 & \tau_1
\end{pmatrix}
= U_C^\dagger.
\end{equation}
Starting with a direct product state
\begin{align}
\label{E8A}
\psi_{\text{dp}} &= \frac{1}{\sqrt{2}}(\up - \down)\down =
\frac{1}{\sqrt{2}}(\up\down-\down\down) \nonumber\\
&= \frac{1}{\sqrt{2}} (0,1,0,-1),
\end{align}
one obtains the maximally entangled state by multiplication with $U_C$
\begin{equation}
\label{E8B}
\psi_{\text{en}} = U_C \psi_{\text{dp}} = \frac{1}{\sqrt{2}} (0,1,-1,0).
\end{equation}
In quantum mechanics unitary transformations can be employed for a change of
basis. This demonstrates that the concept of entanglement needs some type of
selection of a basis that accounts for the notion of direct product states for
individual quantum spins.

Together with the Hadamard gate $U_H$ and rotation gate $U_T$ for single qubits,
the CNOT-gate $U_C$ forms a set of three basis matrices from which all unitary
matrices can be approximated arbitrarily closely by approximate sequences of
products of basis matrices. If we include CNOT-gates for arbitrary pairs of
qubits this statement generalizes to arbitrary unitary matrices for an arbitrary
number of qubits.

\paragraph*{Density matrix for two qubits}

The most general Hermitian $4\times 4$ matrix can be written in terms of sixteen
Hermitian matrices $L_{\mu\nu}$,
\begin{align}\label{E11}
\rho= \frac{1}{4} \rho_{\mu\nu} L_{\mu\nu}, && L_{\mu\nu} = \tau_\mu \otimes
\tau_\nu,
\end{align}
with
\begin{align}
\tau_\mu = (1,\tau_k), && \mu,\nu = 0...3, && k=1...3.
\end{align}
The normalization $\tr \rho = 1$ requires
\begin{equation}
\rho_{00} = 1.
\end{equation}
The matrix $L_{00}$ is the unit matrix, and the other $L_{\mu\nu}$ are the
fifteen generators $L_z$ of $SU(4)$.
The relation
\begin{equation}\label{E14}
\tr(L_{\mu\nu} L_{\sigma\lambda}) = 4 \delta_{\mu\rho} \delta_{\sigma\lambda}
\end{equation}
implies 
\begin{equation}
\tr(L_{\mu\nu} \rho) = \rho_{\mu\nu}.
\end{equation}
We observe
\begin{equation}
L_{\mu\nu}^2 = 1,
\end{equation}
and the eigenvalues of $L_z$ are $+1$ and $-1$.

We further need the quantum constraint which requires that all eigenvalues
$\lambda_i$ of $\rho$ obey $\lambda_i \geq 0$. We first discuss the condition
for pure quantum states, $\rho^2 = \rho$,
\begin{equation}
\frac{1}{16}(\rho_{\mu\nu} L_{\mu\nu})^2 = \frac{1}{32} \rho_{\mu\nu}
\rho_{\sigma\lambda} \{L_{\mu\nu}, L_{\sigma\lambda}\} = \frac{1}{4}
\rho_{\alpha\beta} L_{\alpha\beta}.
\end{equation}
With
\begin{equation}\label{E17}
\{L_{\mu\nu}, L_{\sigma\lambda}\} = 2d_{\mu\nu,\sigma\lambda,\alpha\beta}
L_{\alpha\beta},
\end{equation}
the constraint for pure states reads
\begin{equation}
\frac{1}{4} d_{\mu\nu,\sigma\lambda,\alpha\beta} \rho_{\mu\nu}
\rho_{\sigma\lambda} = \rho_{\alpha\beta}.
\end{equation}
This relation constrains the allowed values of $\rho_{\mu\nu}$ for which $\rho$
describes a pure quantum state. In particular, the relation
\begin{equation}
\tr \rho^2 = \tr \rho = 1
\end{equation}
implies with eq.~\eqref{E14} the condition
\begin{equation}
\rho_{\mu\nu} \rho_{\mu\nu} = 4.
\end{equation}

With the 15 generators of SO(4) denoted by $L_z$
\begin{align}
L_{00} = 1, && L_{\mu\nu} = L_z \text{ for } (\mu\nu)\neq(00)
\end{align} 
we can write the density matrix in a way analogous to the single qubit case
\begin{equation}\label{E22}
\rho = \frac{1}{4}(1 + \rho_z L_z).
\end{equation}
The pure state condition then requires
\begin{equation}\label{E23}
\rho_z \rho_z =3,
\end{equation}
in distinction to the single qubit case where $\rho_z\rho_z = 1$.
In this language eq.~\eqref{E17} reads
\begin{equation}
\{L_z,L_y\} = 2 \delta_{zy} + 2 d_{zyw} L_{w}
\end{equation}
and the pure state condition requires
\begin{equation}
d_{zyw} \rho_z \rho_y = 2 \rho_w,
\end{equation}
in addition to the constraint \eqref{E23}. From 
\begin{align}
\tr L_z = 0, && L_z^2=1,
\end{align}
we conclude that the spectrum of each $L_z$ has two eigenvalues $+1$ and two
eigenvalues $-1$.

The operators for the spin of the first and second qubit are given by 
\begin{align}
S_k^{(1)} = L_{k0} = \tau_k \otimes 1, && S_k^{(2)} = L_{0k} = 1 \otimes \tau_k.
\end{align}
The generators with two indices $k,l$ are products of single spin operators
\begin{equation}
L_{kl} = L_{k0} L_{0l}.
\end{equation}
This implies simple relations as (no sums over repeated indices here)
\begin{align}
L_{kl} L_{0l} = L_{k0}, && L_{kl} L_{k0} = L_{0l}.
\end{align}
The operators $L_{k0}$ and $L_{0l}$ commute
\begin{equation}
[L_{k0},L_{0l}] = 0.
\end{equation}
For given pairs $(k,l)$ all three generators $L_{k0}$, $L_{0l}$ and $L_{kl}$
commute.

\subsection{Classical probabilistic systems for two qubits}

This section presents explicit time-local classical probability distributions
which are mapped to a two-qubit quantum subsystem by a bit-quantum map. The
implementation of a quantum subsystem for two qubits by a classical probability
distribution for Ising spins is not unique. Different implementations correspond
to different bit-quantum maps.

\paragraph*{Average spin map}

A simple bit-quantum map is based on fifteen Ising spins $s_z$, one
corresponding to each generator $L_z$. With eigenvalues of $L_z$ being $\pm 1$
the possible measurement values of the quantum observables associated to $L_z$
coincide with the ones for the Ising spins $s_z$. Identifying $\rho_z$ with the
classical expectation values of $s_z$,
\begin{equation}\label{E31}
\rho_z = \braket{s_z},
\end{equation}
defines the bit-quantum map by eq.~\eqref{E22}. Only the average spins
$\braket{s_z}$ and no correlations are employed for this definition of the
quantum subsystem.

The ``average spin map" \eqref{E31} is a complete bit quantum map, since every
possible ensemble of eigenvalues $\braket{s_z}$ can be realized by suitable
classical probability distributions. As a direct consequence, arbitrary unitary
SU(4)-transformations of the density matrix can be realized by suitable changes
of classical probability distributions.

For the average spin map the CNOT-gate can be realized by a deterministic unique
jump operation. On the level of the coefficients $\rho_z = \rho_{\mu\nu}$ of the
density matrix \eqref{E11} the CNOT gate \eqref{E8} corresponds to the
transformation
\begin{align}
&
\begin{tabular}{c c c}
$\rho_{10} \leftrightarrow \rho_{11},$ & $\rho_{20} \leftrightarrow \rho_{21},$
& $\rho_{13} \leftrightarrow -\rho_{22},$ \\
$\rho_{02} \leftrightarrow \rho_{32},$ &$\rho_{03} \leftrightarrow \rho_{33},$ &
$\rho_{23} \leftrightarrow \rho_{12},$
\end{tabular}\nonumber \\
&\rho_{30},\ \rho_{01},\ \rho_{31} \text{ invariant.}
\label{eq:4.3.37}
\end{align}
It can be realized directly for a probabilistic automaton by the analogous
transformations between the Ising spins $s_z = s_{\mu\nu}$.

We can start with a classical probability distribution for the $15$ Ising spins
which realizes the direct product state~\eqref{E8A} and let the automaton
perform the updating equivalent to the CNOT-gate. This produces a probability
distribution corresponding to the maximally entangled state~\eqref{E8B}. Of
course, we could also directly construct a probability distribution which
realizes the entangled state~\eqref{E8B} for the quantum subsystem. This
setting constitutes a direct explicit example for a classical statistical system
which realizes a maximally entangled quantum state for a suitable subsystem. Not
only general entanglement, but also specific quantum entanglement can be found
in classical statistical systems.

\paragraph*{General bit-quantum maps}

For a general class of bit-quantum maps we consider Ising spins
$\sigma_{\mu\nu}$ that are not necessarily independent, and denote their
expectation values by
\begin{equation}
\chi_{\mu\nu} = \braket{\sigma_{\mu\nu}}.
\label{eq:GBQ1}
\end{equation}
We define the bit-quantum map by associating the quantum density matrix to these
expectation values
\begin{equation}
\rho = \frac{1}{4} \chi_{\mu\nu} L_{\mu\nu},
\label{eq:GBQ2}
\end{equation}
where
\begin{equation}
\sigma_{00} = 1,\quad \chi_{00} = 1.
\label{eq:GBQ3}
\end{equation}
In this case the parameters $\rho_{\mu\nu}$ characterizing the subsystem are
given by these expectation values
\begin{equation}
\rho_{\mu\nu} = \chi_{\mu\nu} = \braket{\sigma_{\mu\nu}}.
\label{eq:GBQ4}
\end{equation}
(The parameters $\rho_z = \rho_{\mu\nu}$ characterizing the subsystem should not
be confounded with the elements $\rho_{\alpha\beta}$ of the density matrix. In
most cases of interest the map from $\rho_z$ to $\rho_{\alpha\beta}$ is
invertible, such that both sets of parameters contain equivalently the
probabilistic information for the subsystem. This is the reason why we employ
the same symbol $\rho$.)

For the average spin map the Ising spins $\sigma_{\mu\nu}$ are independent
spins, $\sigma_{\mu\nu}= s_{\mu\nu}$. Since products of Ising spins are again
Ising spins, we can construct different bit-quantum maps by associating some of
the $\sigma_{\mu\nu}$ to products of two or more ``fundamental'' Ising spins. A
particularly important bit-quantum map of this type is the correlation map which
employs correlation functions of ``fundamental'' Ising spins.

\subsection{Correlation map}
\label{sec:correlation_map}

The correlation map~\cite{CWQCCB} is a bit-quantum map that maps probability
distributions for six classical Ising spins to a two-qubit quantum subsystem. It
is more economical than the average spin map in the sense that only six Ising
spins are used instead of fifteen. On the other hand, the probabilistic
information of the subsystem does not only involve the expectation values of
classical spins, but also some of the correlation functions. The correlation map
employs two sets of Cartesian Ising spins $s_k^{(1)}$ and $s_k^{(2)}$,
$k=1...3.$ They will be associated to the Cartesian directions of the two
quantum spins. It defines the quantum density matrix \eqref{E11} by
\begin{align}
\rho_{k0} = \braket{s_k^{(1)}}, && \rho_{0k} = \braket{s_k^{(2)}}, &&
\rho_{kl} = \braket{s_k^{(1)} s_l^{(2)}}.
\label{eq:GBQ5B}
\end{align}
Besides the six expectation values $\braket{s_k^{(i)}}$ it also employs nine
classical correlation functions $\braket{s_k^{(1)} s_l^{(2)}}$.

The product $s_k^{(1)} s_l^{(2)}$ can only take the values $\pm 1$ and is
therefore again an Ising spin.
We may consider it as a composite Ising spin
\begin{equation}
\sigma_{kl} = s_k^{(1)} s_l^{(2)}.
\label{eq:GBQ5}
\end{equation}
Using a four-component notation for the independent Ising spins with $s_0^{(i)}
= 1$, $s_\mu^{(i)} = (1,s_k^{(i)})$, we can write
\begin{equation}
\sigma_{\mu\nu} = s_\mu^{(1)} s_\nu^{(2)},\quad \chi_{\mu\nu} =
\braket{\sigma_{\mu\nu}} = \braket{s_\mu^{(1)} s_\nu^{(2)}},
\label{eq:GBQ6}
\end{equation}
with density matrix given by eq.\,\eqref{eq:GBQ2}.

In contrast to the average spin map, $\sigma_{kl} = s_k^{(1)} s_l^{(2)}$ is,
however, not an independent spin. Its expectation value is given by the
probability distribution for the six Ising spins $s_k^{(1)},\ s_k^{(2)}$. The
expectation values of $\sigma_{kl}$ and $s_k, s_l$ are therefore related. They
have to obey the restrictions for classical correlations, as the inequality for
all pairs $(k,l)$
\begin{equation}
-1 + |\braket{s_k^{(1)}} + \braket{s_l^{(2)}} | \leq \braket{s_k^{(1)}
s_l^{(2)}} \leq 1- |\braket{s_k^{(1)}} - \braket{s_l^{(2)}}|.
\label{eq:GBQ7}
\end{equation}
It is therefore not guaranteed a priori that the correlation map is a complete
bit-quantum map for which every positive density matrix can be realized. We
discuss the completeness of the correlation map for two qubits in
sect.~\ref{subsec:completeness_of_correlation}.

For the quantum system the expectation value for the operator $L_{kl}$ is given
by the quantum correlation function of the spin operators $S_k^{(1)}$ and
$S_l^{(2)}$,
\begin{align}
\braket{L_{kl}}_q &= \tr(\rho L_{kl}) = \tr (\rho S_k^{(1)} S_l^{(2)}) \nonumber
\\
&= \braket{S_k^{(1)} S_l^{(2)}}_q = \chi_{kl}.
\end{align}
For this particular set of correlation functions the quantum correlation and the
classical correlation coincide 
\begin{equation}
\braket{S_k^{(1)} S_l^{(2)}}_q = \braket{s_k^{(1)} s_l^{(2)}}_\mathrm{cl}.
\end{equation}
We observe that the correlation functions $\braket{S_k^{(1)} S_l^{(2)}}_q$ only
involve two commuting operators. The correlation functions for non-commuting
operators as $\braket{S_k^{(1)} S_l^{(1)}}_q$ are not expressed in terms of
classical correlation functions. Also the classical correlation functions
$\braket{s_k^{(1)} s_l^{(1)}}_\mathrm{cl}$ are not part of the probabilistic
information of the quantum subsystem. They belong to the environment, similar to
Fig.~\ref{fig:3}. Also the three-point and higher classical correlation
functions belong to the environment. 

The subsystem is still characterized by incomplete statistics, since only a
small part of the classical correlation functions is accessible for the
subsystem. The probabilistic information in the subsystem is sufficient for the
computation of the simultaneous or joint probabilities to find for $s_k^{(1)}$
and $s_l^{(2)}$ given pairs of values as $(1,-1)$ etc. It is insufficient for
the computation of joint probabilities for Ising spins corresponding to
different Cartesian directions of a single given quantum spin, as $s_k^{(1)}$
and $s_l^{(1)}$. We recall that the association between quantum correlations and
classical correlations is not a general property, but rather depends on the
particular bit-quantum map. No identification of classical and quantum
correlations is present for the average spin map. 

For the correlation map the deterministic operations on the classical Ising
spins are restricted to permutations among the 64 classical states $\tau$.
They can be performed by operations on the bits of a probabilistic automaton.
The CNOT-gate cannot be realized by these unique jump operations \cite{CWQCCB}.
The unique jump operations can still realize the unitary transformations
\eqref{QE7},\eqref{QE10} for each individual quantum spin. They are given by
$(U^{(1)} \otimes 1)$ and $(1 \otimes U^{(2)})$ respectively. 
Here the matrices $U^{(1)}$ and $U^{(2)}$ can be multiplied by arbitrary phases.
Another deterministic operation is the exchange between the two quantum spins,
as given by the ``swap operation" 
\begin{equation}
U_S =
\begin{pmatrix}
1 & 0 & 0 & 0 \\
0 & 0 & 1 & 0 \\
0 & 1 & 0 & 0 \\
0 & 0 & 0 & 1
\end{pmatrix}.
\end{equation} 
It is realized by a simultaneous exchange of the classical Ising spins
$s_k^{(1)} \leftrightarrow s_k^{(2)}$.

On the level of classical Ising spins an exchange of expectation values and
correlations
\begin{equation}
s_k^{(1)} \leftrightarrow s_k^{(1)} s_l^{(2)}\quad\mathrm{or}\quad 
s_k^{(2)} \leftrightarrow s_k^{(2)} s_l^{(1)}
\label{eq:BBA}
\end{equation}
can be achieved by a conditional jump: If $s_l^{(2)} = -1$, switch the sign of
$s_k^{(1)}$, or similar for the second switch in eq.\,\eqref{eq:BBA}. It is
difficult, however, to construct unitary quantum transformations with a switch
$\rho_{k0} \to \rho_{kl}$. The reason is that other classical correlations, as
$s_k^{(1)} s_{l'}^{(2)}$ for $l'\neq l$, transform into a three-point function,
$s_k^{(1)} s_l^{(2)} \to s_k^{(1)} s_{l'}^{(2)} s_l^{(2)}$, which is not part of
the probabilistic information for the quantum subsystem.

\subsection{Classical entanglement}

It is not difficult to simultaneously realize the maximal anticorrelation
\eqref{E2}, the vanishing expectation values \eqref{E3} and the vanishing
correlations \eqref{E3A} for $k\neq l$ with a suitable classical probability
distribution.
For six Ising spins $s_k^{(i)}$, $k=1...3$, $i = 1,2$, we have $2^6 = 64$ states
$\tau$, labeled by the configurations for six Ising spins. If $p_\tau$ vanishes
for all states for which any pair $k$ of spins $(s_k^{(1)},s_k^{(2)})$ has the
same signs, the system is maximally anticorrelated according to eq.~\eqref{E2}.
These vanishing probabilities concern 56 out of the 64 configurations. For the
remaining eight configurations the spins $s_k^{(1)}$ and $s_k^{(2)}$ have
opposite signs for each $k$. If the probabilities for these eight states are all
equal, one infers, in addition, the relations \eqref{E3} and \eqref{E3A}. 

\paragraph*{Classical probability distributions for maximally anticorrelated
states}

For the six Ising spins $s_k^{(i)}$ we can label the classical states by $\tau =
(\tau_1,\tau_2)$, where $\tau_1 = 1,...,8$ labels the eight configurations for
the triplet of spins $s_k^{(1)}$, and $\tau_2 = 1,...,8$ the ones of
$s_k^{(2)}$. Instead of $\tau_2$ we may equivalently use $\tilde{\tau}_2$ for
which the signs of all spins are switched as compared to $\tau_2$. For example,
$\tau_2 = (1,1,-1)$ corresponds to $\tilde{\tau_2} = (-1,-1,1)$. The
non-vanishing probabilities for a maximally anticorrelated state are given by
\begin{equation}
p(\tau_1,\tilde{\tau}_2) = p(\tau_1,\tilde{\tau}_2 = \tau_1) = \bar{p}(\tau_1).
\end{equation}
In other words, $p(\tau_1,\tau_2)$ differs from zero only if for each $k$ the
value of $s_k^{(2)}$ is opposite to $s_k^{(1)}$. The non-vanishing probabilities
$\bar{p}(\tau_1)$ are therefore labeled by the eight configurations of the first
triplet of Ising spins $s_k^{(1)}$. The expectation values of $s_k^{(1)}$ only
depend on $\tau_1$,
\begin{equation}
\braket{s_k^{(1)}} = \sum_{\tau_1,\tau_2} p(\tau_1,\tau_2) s_k^{(1)}(\tau_1) =
\sum_{\tau_1} \bar{p}(\tau_1) s_k^{(1)}(\tau_1).
\end{equation}
with $s_k^{(1)}(\tau_1)$ the value of the Ising spin $s_k^{(1)}$ in the state
$\tau_1$. For every $\tau_1$ the second triplet of spins $s_k^{(2)}$ has
opposite signs to $s_k^{(1)}$. We conclude for the maximally anticorrelated
systems that the expectation values of $s_k^{(1)}$ and $s_k^{(2)}$ are opposite
\begin{equation}
\braket{s_k^{(2)}} = - \braket{s_k^{(1)}},
\end{equation}
while the classical correlation functions are the same,
\begin{equation}
\braket{s_k^{(2)} s_l^{(2)}} = \braket{s_k^{(1)} s_l^{(1)}}.
\end{equation}
For arbitrary $\bar{p}(\tau_1)$ the maximal anticorrelation \eqref{E2} is
realized by the classical correlations between pairs of different spin triplets
in arbitrary Cartesian directions 
\begin{equation}
\braket{s_k^{(1)} s_k^{(2)}} = -1.
\label{new1}
\end{equation}

Probability distributions that additionally realize vanishing expectation values
\begin{equation}
\braket{s_k^{(i)}} = 0
\label{new2}
\end{equation} 
require three conditions on $\bar{p}(\tau_1)$, namely for each $k$
\begin{equation}
\sum_{\tau_1} \bar{p}(\tau_1)s_k^{(1)}(\tau_1) = 0.
\end{equation}
Together with the normalization one has four constraints on eight real positive
numbers. As an example for two different classical probability distributions
that realize eqs.~\eqref{E2},\eqref{E3} we
first take an equipartition for which all $\bar{p}(\tau_1)$ are equal and
$\braket{s_k^{(1)} s_l^{(1)}} = 0$ for $k \neq l$,
and second $\bar{p}(1,-1,-1) = \bar{p}(1,-1,1) = \bar{p}(-1,1,1) = 1/4$, for
which $\braket{s_1^{(1)} s_2^{(1)}} = -1$. 
 

If we want to realize, in addition, the vanishing correlations \eqref{E3A} for
$k\neq l$
\begin{equation}
\label{E42A}
\braket{s_k^{(1)} s_l^{(2)}} = 0 \quad \text{for } k \neq l\,,
\end{equation}
we need to impose three additional constraints. One has 
\begin{equation}\label{E42B}
\begin{split}
\braket{s_1^{(1)} s_2^{(2)}} &= \braket{s_2^{(1)} s_1^{(2)}} \\
& = \bar{p}_{+-+} + \bar{p}_{+--} + \bar{p}_{-++} + \bar{p}_{-+-} \\
& \quad -\bar{p}_{+++} - \bar{p}_{++-} - \bar{p}_{--+} - \bar{p}_{---}\,,
\end{split}
\end{equation} 
\begin{equation}\label{E42C}
\begin{split}
\braket{s_1^{(1)} s_3^{(2)}} &= \braket{s_3^{(1)} s_1^{(2)}} \\
& = \bar{p}_{++-} + \bar{p}_{+--} + \bar{p}_{-++} + \bar{p}_{--+} \\
& \quad -\bar{p}_{+++} - \bar{p}_{+-+} - \bar{p}_{-+-} - \bar{p}_{---}\,,
\end{split}
\end{equation}
and
\begin{equation}\label{E42D}
\begin{split}
\braket{s_2^{(1)} s_3^{(2)}} &= \braket{s_3^{(1)} s_2^{(2)}} \\
& = \bar{p}_{++-} + \bar{p}_{-+-} + \bar{p}_{+-+} + \bar{p}_{--+} \\
& \quad -\bar{p}_{+++} - \bar{p}_{-++} - \bar{p}_{+--} - \bar{p}_{---}\,,
\end{split}
\end{equation} 
where $\bar{p}_{+-+}$ is a shorthand for $\bar{p}(1,-1,1)$ etc. The general
family of classical probability
distributions that obeys simultaneously the relations \eqref{new1}, \eqref{new2}
and \eqref{E42A} is given by
\begin{equation}\label{E42E}
\begin{split}
\bar{p}_{+++} &= \bar{p}_{+--} = \bar{p}_{-+-} = \bar{p}_{--+} = \frac{1}{8} +
\Delta\,, \\
\bar{p}_{---} &= \bar{p}_{-++} = \bar{p}_{+-+} = \bar{p}_{++-} = \frac{1}{8} -
\Delta\,.
\end{split}
\end{equation}
Thus the classical probability distributions corresponding to the ``maximally
entangled classical state"
\eqref{new1}, \eqref{new2}, \eqref{E42A} is not unique. It is given by a one
parameter family, with 
$| \Delta | \leq 1/8$. All classical correlation functions depend on a single
parameter $\Delta$.

In analogy to the two quantum spins we may divide the system of six classical
spins into two parts. The
first part is composed of the triplet $s_k^{(1)}$ and the second part involves
the three Ising spins
$s_k^{(2)}$. ``Direct product states" are those for which the probability
distribution factorizes, 
\begin{equation}
\label{E43}
p(\tau_1,\tau_2) = p_1(\tau_1) p_2(\tau_2)\,.
\end{equation}
Probability distributions for which eq.\,\eqref{E43} is violated, as the
maximally anticorrelated states,
may be called entangled. The notion of ``entangled states" refers to the
probabilistic information encoded
in $\{ p_\tau \}$, not to properties of the spin configurations $\tau$. The
double use of the wording 
``state" is similar to quantum mechanics, where an ``entangled state" refers to
the probabilistic information,
while a ``two state system" counts the dimension of the wave function or the
number of independent basis states.

Similar to quantum mechanics, the notion of entanglement in classical
probabilistic systems needs the selection
of a basis. More generally, entanglement is a statement about relations or
correlations between two (or several)
parts of a system. It needs the specification of what the parts are. We
demonstrate this next by instructive
examples. 

\paragraph*{Fundamental and composite degrees of freedom}

In particle physics or condensed matter physics there is no sharp distinction
between fundamental and 
composite particles or between fundamental and composite degrees of freedom. For
the theory of strong 
interactions, the microscopic particles are quarks and gluons, while the
observed propagating particles are
mesons and baryons. The field for the mesons can be represented as a correlation
function for quarks and 
antiquarks. Fields for baryons are associated to three point correlations for
three quarks. Baryons are as
``real" as quarks, demonstrating in a simple striking way that sometimes
``restricted reality" concerns the
correlations, rather than the expectation values of ``fundamental observables". 

The partition function in condensed matter physics can often be expressed in
terms of different degrees of 
freedom. A variable transform can switch degrees of freedom, without affecting
the functional integral. The
notion of what is ``composite" or ``fundamental", what is a correlation or an
expectation value, depends on the
choice of the variables which are associated to ``fundamental degrees of
freedom".

\paragraph*{Different divisions into parts}

As we have seen before the notion of entanglement depends on the specification
of parts of the system. These
parts are often associated to different particles. For our example of quantum
entanglement the system consists
of two particles to which the two qubits are associated. For the classical
statistical counterpart the two triplets of 
Ising spins $s_k^{(1)}$ and $s_k^{(2)}$ have been associated to two different
particles. We call this division 
the ``two-particle basis". Entanglement concerns then the correlations between
the different particles 
$i = 1$ or $2$. 

For our classical statistical example with six Ising spins $s_k^{(i)}$ we can
order the degrees of freedom in 
a different way. The Ising spins may be associated to three different particles,
labeled by $k$. For each 
particle $k$ the ``internal degrees of freedom" are now labeled by $i$. We call
this assignment the 
``three-particle basis". As compared to the previous discussion the new
assignment exchanges the role of $k$
and $i$. In the two-particle basis $i$ labels the two particles, and $k$ the
internal degrees of freedom. 
In the three-particle basis the direct product states correspond to probability
distributions with three 
factors 
\begin{equation}
\label{E44}
p_\tau = p_{1,\sigma_1} p_{2,\sigma_2} p_{3,\sigma_3}\,,
\end{equation}
where $\sigma_k = 1\ldots 4$ denotes for each $k$ the four states $s_k^{(1)} =
s_k^{(2)} = 1$, 
$s_k^{(1)} = 1$, $s_k^{(2)} = -1$, $s_k^{(1)} = -1$, $s_k^{(2)} = 1$,
$s_k^{(1)} = s_k^{(2)} = -1$.
The maximally anticorrelated states \eqref{new1} in the two-state basis can be
direct product states
in the three-state basis. Indeed, if for each $k$ one has $p_{k++} = p_{k--} =
0$,
one finds maximal anticorrelation $\braket{s_k^{(1)} s_k^{(2)}} = -1$.
For these states one remains with three independent probabilities $p_{k+-}$,
with $p_{k-+} = 1 - p_{k+-}$.
They fix the expectation values
\begin{equation}
\label{E45}
\braket{s_k^{(1)}} = p_{k+-} - p_{k-+} = 2 p_{k+-} - 1 = - \braket{s_k^{(2)}}\,.
\end{equation}
Vanishing expectation values \eqref{E42A} obtain for $p_{k+-} = 1/2$. The direct
product form \eqref{E44} 
implies vanishing connected correlation functions for each pair of different
``particles", e.g. for $k \neq l$
one has 
\begin{equation}\label{E46}
\braket{s_k^{(i)} s_l^{(j)}}_c = \braket{s_k^{(i)} s_l^{(j)}} -
\braket{s_k^{(i)}} \braket{s_l^{(j)}} = 0\,.
\end{equation}
For this family of classical probability distributions the relations
\eqref{E42A} follow from eq.\,\eqref{new2}.
We conclude that out of the one-parameter family of probability distributions
\eqref{E42E} for maximally
entangled classical states only the one with $\Delta =0$ can be realized as a
direct product state \eqref{E44}.

On the other hand, direct product states in the two-particle basis can appear as
entangled states in the 
three-particle basis. For a direct product state in the two-particle basis one
has
\begin{equation}\label{E47}
\braket{s_k^{(1)}s_l^{(2)}} = \braket{s_k^{(1)}} \braket{s_l^{(2)}}\,,
\end{equation}
whereas a direct product state in the three-particle basis obeys
\begin{equation}
\label{E48}
\braket{s_k^{(i)}s_l^{(j)}} = \braket{s_k^{(i)}} \braket{s_l^{(j)}} \quad
\text{for } k\neq l\,.
\end{equation}
Consider a direct product state \eqref{E43} in the two-particle basis, with
$p_1(\tau_1)$ chosen such that
\begin{equation}
\label{E49}
\braket{s_1^{(1)} s_2^{(2)}} = -1, \quad \braket{s_1^{(1)}} = \braket{s_2^{(2)}}
= 0\,,
\end{equation}
and similarly for $p_2(\tau_2)$. The relation \eqref{E49} contradicts
eq.~\eqref{E48}, such that this state
can only be realized as an entangled state in the three-particle basis. The
notion of classical entanglement
depends on the division into parts or the basis for direct product states. There
is no difference in this
respect from quantum mechanics.

\paragraph*{Classical probabilities for quantum dices}


The maximally entangled quantum state for two qubits is sometimes associated
with a pair of two dice with 
mysterious properties. Whenever the first dice shows a number $\tau$, the second
dice shows a complementary
number $\bar{\tau}$. For example, we may take pairs of complementary numbers
$(\tau, \bar{\tau}) = (1,6), (2,5)$
and $(3,4)$. Otherwise the dice have unbiased probabilities, e.g. the
probability to find a number $\tau_1$ for dice one equals
$1/6$, and the probability for finding another number $\tau_2$ for dice two is
also given by $1/6$. No number
is preferred for one of the individual dice. There is widespread prejudice that
this mysterious behavior of the pair
of ``quantum dice" is not compatible with classical probabilistic systems. 

This prejudice is inappropriate. The only thing that cannot work is a direct
product state for the probability distribution of the two
dice. The classical states of dice one can be labeled with $\tau_1$, $\tau_1 =
1\ldots 6$, and similarly with $\tau_2$ for dice
two. The two numbers $(\tau_1,\tau_2)$ occur with probabilities
$p(\tau_1,\tau_2)$. For a direct product state,
\begin{equation}
\label{E50}
p(\tau_1,\tau_2) = p(\tau_1) p(\tau_2)\,,
\end{equation}
unbiased dice correspond to $p(\tau_1) = 1/6$ independent of $\tau_1$, and
similarly $p(\tau_2) = 1/6$. The probability for
any given pair $(\tau_1,\tau_2)$ equals $1/36$, in contrast to the behaviour of
the quantum dice. We conclude that the classical
probability distribution for the pair of quantum dice has to be entangled,
showing strong correlation between the two dice.

Indeed, we can realize the strong correlation by classical probabilities that
vanish whenever $\tau_2 \neq \bar{\tau}_1$, e.g.
\begin{equation}
\label{E51}
p(\tau_1,\tau_2 \neq \bar{\tau}_1) = 0\,.
\end{equation}
Nonzero probabilities occur only if $\tau_1 + \tau_2 = 7$. The six non-vanishing
probabilities may be assigned by
\begin{equation}
\label{E52}
\bar{p}(\tau_1) = p(\tau_1,\tau_2 = \bar{\tau}_1)\,.
\end{equation}
For $\bar{p}(\tau_1) = 1/6$ the two dice are unbiased, showing every number with
probability $1/6$.

In everyday life unbiased dice in a game will not show the correlation $p(\tau_1
+ \tau_2 \neq 7)=0$. Even if the correlation
would be prepared by the hands of a gifted player, the stochastic evolution of
the dice once they have left the hands of the 
player would destroy the correlation. This is somewhat analogous to decoherence
in quantum mechanics. One may imagine a different
evolution of the pair of correlated dice. For example, the could perform
rotations in vacuum such that $\tau_1+\tau_2=7$ is 
conserved. 

While the realization of such a system for dice may be very difficult, many
analogous systems can be found in nature.
For example, there may be conserved total angular momentum of two bodies. Assume
that a system of two bodies has initially zero 
total angular momentum
\begin{equation}
\label{E53}
L_k^{(1)} + L_l^{(2)} = 0\,, \quad k = 1\ldots3\,,
\end{equation}
and that the subsequent evolution preserves total angular momentum, such that
eq.~\eqref{E53} holds for all later times $t$. This
implies for the correlation functions for every $k$
\begin{equation}
\label{E54}
\braket{L_k^{(1)} L_k^{(2)}} = - \braket{ (L_k^{(1)})^2} = - \braket{
(L_k^{(2)})^2}\,.
\end{equation}
No particular direction may be preferred by the system, such that
\begin{equation}
\label{E55}
\braket{L_k^{(1)}} = \braket{L_k^{(2)}} = 0\,,
\end{equation}
as well as
\begin{equation}
\label{E56}
\braket{L_k^{(1)} L_l^{(2)}} = 0 \quad \text{for } k \neq l \,.
\end{equation}

If we assume further probability distributions with
\begin{equation}
\label{E57}
\braket{ (L_k^{(1)})^2} = \braket{ (L_k^{(2)})^2} = c_k^2\,, \quad c_k > 0\,,
\end{equation}
we can define 
\begin{equation}
\label{E58}
s_k^{(i)} = L_k^{(i)}/c_k\,.
\end{equation}
The relations \eqref{E54}--\eqref{E56} coincide with the relations \eqref{new1},
\eqref{new2}, \eqref{E42A} in this case.
It does not matter for these properties of correlation functions if we deal with
macroscopic bodies or the microscopic decay 
of a spinless particle into a pair of particles with spin. We also note that we
do not require that the angular momentum
of individual bodies or particles is conserved during the evolution. The
conservation of zero total angular momentum 
during the evolution is sufficient to guarantee eq.\,\eqref{E54} for arbitrary
$t$, including possible large distances between 
the bodies such that the correlation becomes non-local.


\paragraph*{Correlation map for maximally entangled quantum state}

Let us define a two-qubit quantum subsystem in terms of the probability
distribution for six classical Ising spins $s_k^{(1)}, s_k^{(2)}$ 
by the correlation map \eqref{eq:GBQ5B}. In this case the quantum correlations
\eqref{E2}, \eqref{E3}, \eqref{E3A} are directly given by
the classical correlations \eqref{new1}, \eqref{new2}, \eqref{E42A}. The family
of classical probability distributions \eqref{E42E} realises the maximally
entangled pure state \eqref{E1} for the quantum subsystem. This demonstrates by
direct construction
that entanglement is not an obstruction for obtaining quantum systems as
subsystems of classical probabilistic systems. 

Inversely, the probabilistic information contained in the quantum subsystem for
the maximally entangled state is sufficient to compute
the classical correlation functions \eqref{new1}, \eqref{new2}, \eqref{E42A}. It
also contains many relations among other classical 
functions since all can be computed in terms of a simple parameter $\Delta$ in
eq.\,\eqref{E42E}. One may wonder if the maximally entangled
quantum state contains information beyond the correlation functions
\eqref{new1}, \eqref{new2}, \eqref{E42A}. This is not the case. 
The maximally entangled correlation functions \eqref{new1}, \eqref{new2},
\eqref{E42A} impose restrictions on the possible classical
probability distribution that can realize them. These restrictions lead
precisely to eq.\,\eqref{E42E} and the corresponding relations
between classical correlations functions.


\subsection{Classical wave function and\\entanglement}
\label{sec:normalized_classical_wave_function}

The classical wave function\,\cite{CWQP} is a powerful tool for the discussion
of entanglement in classical probabilistic
systems. It provides for classical statistics a formulation in close analogy to
quantum mechanics. This makes the similarity between
quantum entanglement and classical entanglement particularly apparent. 

\paragraph*{Classical wave function and probabilities}

We define the classical wave function $q$ as a root of the
probability distribution 
\begin{equation}
\label{eq:E59}
p_\tau = q_\tau^2\,.
\end{equation}
This determines $q_\tau$ up to a sign $\sigma_\tau$,
\begin{equation}
\label{eq:E60}
q_\tau = \sigma_\tau \sqrt{p_\tau}\,, \quad \sigma_\tau = \pm 1\,.
\end{equation}
The normalization of the probabilities $\sum_\tau p_\tau = 1$ implies that $q$
is a unit vector,
\begin{equation}
\label{eq:E61}
q_\tau q_\tau = 1\,.
\end{equation}
Transformations of the probability distribution that preserve the normalization
are simply rotations of the normalized wave function.
This simplicity is an important advantage for many purposes. For an orthogonal
step evolution operator, as realized for probabilistic cellular automata, the
evolution directly performs such rotations. For general classical statistics a
linear evolution law involves a pair of independent wave functions. The wave
function~\eqref{eq:E59} can be constructed from this pair as the "normalized
classical wave functions"~\cite{CWIT, CWQF}.

Using the diagonal classical operators,
\begin{equation}
\label{eq:E66}
\hat{A}_{\tau \rho} = A_\tau \delta_{\tau \rho}\ ,
\end{equation}
one finds for the expectation value a relation similar to quantum mechanics
\begin{equation}
\label{eq:E65}
\braket{A} = q_\tau \hat{A}_{\tau \rho} q_\rho = \braket{q | \hat{A} | q}\,.
\end{equation}
The signs $\sigma_\tau$ drop out for diagonal classical operators.
Eq.~\eqref{eq:E65} reproduces directly the fundamental definition 
of expectation values (2.1.2) in classical statistics
\begin{equation}
\label{eq:E67}
\sum_{\tau, \rho} q_\tau \hat{A} q_\rho = \sum_\tau A_\tau q_\tau^2 = \sum_\tau
A_\tau p_\tau\,.
\end{equation}

\paragraph*{Classical entanglement}

In the formalism for classical wave functions we can directly implement concepts
familiar from quantum mechanics as direct product wave functions and entangled
wave functions. As in quantum mechanics, the notions of direct product and
entanglement depend on the definition of parts of the system and the adapted
choice of basis functions.

In the two-particle basis the six classical spin operators corresponding to the
Ising spins $s_k^{(1)}, s_k^{(2)}$ are represented as
\begin{equation}
\label{eq:E68}
\hat{S}_k^{(1)} = \hat{S}_k \otimes 1\,, \quad \hat{S}_k^{(2)} = 1 \otimes
\hat{S}_k\,,
\end{equation}
with diagonal $8 \times 8$ matrices $\hat{S}_k$ given by eqs.\,\eqref{QM6},
\eqref{QM4}. A direct product wave function takes the form 
\begin{equation}
\label{eq:E69}
q_\tau = q_{\tau_1 \tau_2} = q_{\tau_1}^{(1)} q_{\tau_2}^{(2)}\,,
\end{equation}
with 8-component unit vectors $q^{(1)}$ and $q^{(2)}$. For direct product wave
functions one has
\begin{equation}
\label{eq:E70}
p_\tau = p_{\tau_1}^{(1)} p_{\tau_2}^{(2)}, \quad p^{(1)}_{\tau_1} = (
q_{\tau_1}^{(1)} )^2,\quad p^{(2)}_{\tau_2} = ( q_{\tau_2}^{(2)} )^2,
\end{equation}
and with eq.\,\eqref{eq:E65}
\begin{equation}
\label{eq:E71}
\begin{split}
\braket{s_k^{(1)}} &= q_{\tau_1}^{(1)} \left(\hat{S}_k \right)_{\tau_1 \rho_1}
q_{\rho_1}^{(1)}, \\
\braket{s_k^{(2)}} &= q_{\tau_2}^{(2)} \left(\hat{S}_k \right)_{\tau_2 \rho_2}
q_{\rho_2}^{(2)}, \\
\braket{s_k^{(1)} s_l^{(2)}} &= \braket{s_k^{(1)}} \braket{s_l^{(2)}}.
\end{split}
\end{equation}
The probability distribution \eqref{E42E} for the classically entangled state
cannot be obtained from a direct product normalized
classical wave function.

A general entangled classical wave function can be represented as a linear
combination of direct product wave functions
\begin{equation}
\label{eq:E72}
q_\tau = \sum_a c_a q_{\tau_1}^{(a, 1)} q_{\tau_2}^{(a, 2)}\,.
\end{equation}
If we chose the direct product wave function orthogonal 
\begin{equation}
\label{eq:E73}
q_{\tau_1}^{(a, 1)} q_{\tau_1}^{(b, 1)} q_{\tau_2}^{(a, 2)} q_{\tau_2}^{(b, 2)}
= \delta_{a b}\,,
\end{equation}
the normalization reads
\begin{equation}
\label{eq:E74}
\sum_a c_a^2 = 1\,.
\end{equation}
Every probability distribution can be represented in this way as $p_\tau =
q_\tau^2$, including the one for the classically entangled
state \eqref{E42E}. We observe complete analogy with entanglement in quantum
mechanics.

\paragraph*{Three particle basis}

Following ref.\,\cite{CWQCCB} we can represent the classical entangled state for
$\Delta = 0$ by a direct product classical wave function. We
will generalize the setting and construct a classical probability distribution
for which the correlation map to the two-qubit
quantum subsystem yields a pure entangled state of the form
\begin{equation}
\label{eq:E75}
\psi = \begin{pmatrix}
0 \\ \cos(\vartheta) \\ \sin(\vartheta) \\ 0
\end{pmatrix}\,.
\end{equation}
The maximally entangled quantum state \eqref{E1} arises for $\vartheta =
-\pi/4$. The non-vanishing quantum expectation values
or equivalent classical expectation values and correlations are given by
\begin{equation}
\label{eq:E76}
\begin{split}
\rho_{30} &= -\rho_{03} = \cos^2(\vartheta)-\sin^2(\vartheta)\,,\quad
\rho_{33}=-1\,,\\
\rho_{11} &= \rho_{22} = 2\cos(\vartheta) \sin(\vartheta)\,.
\end{split}
\end{equation}

The construction of a probability distribution realizing the
properties~\eqref{eq:E76} is most easily done with the classical wave function
in the three particle basis. In the three-particle basis the classical spin
operators are represented as
\begin{equation}
\label{eq:E77}
\begin{split}
\hat{S}_1^{(i)} &= t^{(i)} \otimes 1 \otimes 1\,, \quad \hat{S}_2^{(i)} = 1
\otimes t^{(i)} \otimes 1 \,, \\
\hat{S}_3^{(i)} &= 1 \otimes 1 \otimes t^{(i)} \,,
\end{split}
\end{equation}
with diagonal $4 \times 4$ matrices,
\begin{equation}
\label{eq:E78}
t^{(1)} = \diag(1,1,-1,-1)\,, \quad t^{(2)} = \diag(1,-1,1,-1)\,.
\end{equation}
A direct product classical wave function takes the form
\begin{equation}
\label{eq:E79}
q_\tau = q_\alpha^{(1)} q_\beta^{(2)} q_\gamma^{(3)}\,,
\end{equation}
with normalized four-component vectors $q_\alpha^{(k)} q_\alpha^{(k)} = 1$. One
infers the expectation values
\begin{equation}
\label{eq:E80}
\rho_{k0} = \sum_\alpha t_\alpha^{(1)} \left( q_\alpha^{(k)} \right)^2\,, \quad 
\rho_{0k} = \sum_\alpha t_\alpha^{(2)} \left( q_\alpha^{(k)} \right)^2,
\end{equation}
with $t_\alpha^{(i)}$ the appropriate eigenvalues of $t^{(i)}$. For the
correlations one has
\begin{equation}
\label{eq:E81}
\begin{split}
\rho_{k k} &= \sum_\alpha t_\alpha^{(1)} t_\alpha^{(2)} \left( q_\alpha^{(k)}
\right)^2\,, \\
\rho_{k l} &= \rho_{k0} \rho_{0l} \quad \text{for } k\neq l\,. 
\end{split}
\end{equation}
Taking 
\begin{equation}
q^{(1)} = q^{(2)} = \begin{pmatrix}
a \\ b \\ b \\ a 
\end{pmatrix}\,, \quad q^{(3)} = \begin{pmatrix}
0 \\ \cos(\vartheta) \\ \sin(\vartheta) \\ 0
\end{pmatrix}\,,
\end{equation}
with
\begin{equation}
\label{eq:E83}
a = \frac{1}{2}\left( \cos(\vartheta)+\sin(\vartheta)\right)\,, \quad 
b = \frac{1}{2}\left( \cos(\vartheta)-\sin(\vartheta)\right)\,,
\end{equation}
one realizes the entangled state according to eq.~\eqref{eq:E76}. These examples
demonstrate in a straightforward way the construction of classical entangled
states which are mapped by the bit-quantum map to entangled quantum states.

\subsection{Bell's inequalities}
\label{sec:bells_inequalities}

Bell's inequalities \cite{BELL}, or the more general form of the CHSH
inequalities \cite{CHSH}, are identities for 
correlation functions in classical probabilistic systems. They become relevant
for quantum subsystems if parts of the
probabilistic information contained in the quantum subsystem is given by
classical correlation functions. This is the case
for the correlation map. In contrast, the average spin map employs no classical
correlation functions. In this case the 
generalized Bell's inequalities only concern the environment. They are
irrelevant for the quantum subsystem.

\paragraph*{Generalized Bell's inequalities and bit-quantum maps}

For the correlation map there is a set of quantum correlations, namely
$\rho_{kl}$, that is given by classical correlation functions.
As for any classical correlation function they have to obey the CHSH inequality.
Otherwise the correlation map cannot be a complete
bit-quantum map. If there would exist positive density matrices for which the
quantum correlations $\rho_{kl}$ violate the
CHSH-inequality, this set of density matrices cannot be obtained from classical
probability distributions. The only assumption for
the CHSH-inequality is the existence of some complete probability distribution
for which simultaneous probabilities for the 
two factors in the correlation are available, and that the classical correlation
is computed in the usual way using these simultaneous 
probabilities. We will show that the particular quantum correlations $\rho_{kl}$
obey the CHSH-inequality. No obstruction to the completeness of the 
correlation map arises from this side. It is important that the particular set
of quantum correlations $\rho_{kl}$ concerns correlations
for commuting quantum operators. There exist other quantum correlations which
violate the CHSH-inequality. They are not related to
classical correlation functions, such that no contradiction arises.

\paragraph*{CHSH-inequality}

For the relevant CHSH-inequality we employ two sets of classical Ising spins,
namely $A, A'$ from the triplet of spins
$s_k^{(1)}$, and $B, B'$ from $s_k^{(2)}$,
\begin{equation}
\label{eq:E84}
\begin{split}
A &= \pm s_k^{(1)}\,, \quad A' = \pm s_l^{(1)}\,,\\
B &= \pm s_m^{(2)}\,, \quad B' = \pm s_n^{(2)}\,.
\end{split}
\end{equation}
We define the combination 
\begin{equation}
\label{eq:E85}
\begin{split}
C &= AB + AB' + A'B - A'B' \\
&= A \left(B+B'\right) + A' \left(B-B' \right)\,.
\end{split}
\end{equation}
Since $B$ and $B'$ are Ising spins with possible values $\pm 1$, one has either
$B = B'$ or $B = -B'$. For $B' = B$ one has 
$C = 2AB$, such that $C$ can take the values $\pm 2$. For $B' = -B$ one finds $C
= 2A'B$. Again $C$ can only take the values
$\pm 2$. For any probability distribution the inequality $-2 \leq \braket{C}
\leq 2$, $\lvert\braket{C}\rvert \leq 2$, holds.
For a complete classical probability distribution the classical correlations
$\braket{AB}$ etc. can be computed from the same
probability distribution as used for $\braket{C}$. One concludes the
CHSH-inequality 
\begin{equation}
\label{eq:E86}
\lvert\braket{C}\rvert = \lvert \braket{AB} + \braket{AB'} + \braket{A'B}-
\braket{A'B'}\rvert \leq 2\,.
\end{equation}
Bell's inequalities are special cases of the more general CHSH inequality. We
observe that the completeness of the probabilistic 
information plays a central role. For the incomplete statistics of quantum
subsystems this completeness is not given, in general.
For this reason, quantum correlations need not to obey the CHSH-inequality.

\paragraph*{CHSH inequality for the correlation map}

For two qubits the maximally entangled state is often believed to lead to a
maximal violation of the CHSH inequality. One can verify
by explicit computation\,\cite{CWQCCB} that the quantum correlations \eqref{E2},
\eqref{E3A} obey the CHSH inequality. We can anticipate this finding
since we have already constructed an explicit classical probability distribution
\eqref{E42E} from which these correlations can be 
computed as classical correlations. They therefore have to obey the CHSH
inequality. This extends to the family of entangled state
\eqref{eq:E75}. A general proof that the correlation map is compatible with the
CHSH inequality has to establish the inequality
\begin{equation}
\label{eq:E87}
\lvert \rho_{km} + \rho_{kn} + \rho_{lm} - \rho_{ln} \lvert \leq 2\,,
\end{equation}
for all possible density matrices and arbitrary $k, l, m, n = 1\ldots 3$.

\subsection{Completeness of correlation map}
\label{subsec:completeness_of_correlation}

The correlation map is a complete bit-quantum map if for every positive density
matrix one can find at least one classical probability distribution for the two
triplets of Ising spins $s_k^{(1)}$ and $s_k^{(2)}$ such that the coefficients
$\rho_{\mu \nu}$ can be expressed in terms of classical expectation values and
correlations $\chi_{\mu\nu}$ in eq.\,\eqref{eq:GBQ6}. This requires the
inequality \eqref{eq:E87}, which involves four correlation functions. Further
inequalities that have to be obeyed for all positive density matrices arise from
the restriction 
\eqref{eq:GBQ7} for classical correlation functions
\begin{equation}
\label{eq:E88}
-1 + \lvert \chi_{k0} + \chi_{0l} \lvert \leq \chi_{kl} \leq 1 - \vert \chi_{k0}
- \chi_{0l} \vert\,.
\end{equation}
We will demonstrate that eq.\,\eqref{eq:E88} indeed holds for arbitrary pairs
$(k,l)$.

For a given pair $(k,l)$ the quantum operators $S_k^{(1)}$ and $S_l^{(2)}$
commute and can be diagonalized simultaneously. In the basis where both are
diagonal the positive diagonal elements of the density matrix can be associated
with probabilities: $p_{++}$ for the element corresponding to the eigenvalues
$+1$ of $S_k^{(1)}$ and $+1$ for $S_l^{(2)}$, $p_{+-}$ for the pair of
eigenvalues $(+1,-1)$ and so on. The four probabilities $(p_{++}, p_{+-},
p_{-+}, p_{--})$ form a normalized probability distribution, from which
$\braket{S_k^{(1)}}, \braket{S_l^{(2)}}$ and $\braket{S_k^{(1)} S_l^{(2)}}$ can
be computed according to the classical rule. As for any classical correlation
function the inequality \eqref{eq:GBQ7} holds, which coincides in this case with
eq.\,\eqref{eq:E88}. More in detail, one has
\begin{equation}
\label{eq:E89}
\begin{split}
\chi_{k0} &= p_{++} + p_{+-} - p_{-+} - p_{--}\,,\\
\chi_{0l} &= p_{++} - p_{+-} + p_{-+} - p_{--}\,,\\
\chi_{kl} &= p_{++} - p_{+-} - p_{-+} + p_{--}\,,
\end{split}
\end{equation}
from which eq.\,\eqref{eq:E88} follows directly. We conclude that 
the positivity of the quantum density matrix ensures that
the inequality \eqref{eq:E88} is indeed obeyed for arbitrary pairs
$(k,l)$. No obstruction to the completeness of the correlation map arises from
this type of inequalities. The positivity 
of the density matrix is crucial for this property. For two different pairs
$(k,l)$ the pairs
of operators are diagonal in two different bases. The probability distributions
$(p_{++}, p_{+-}, p_{-+}, p_{--})$ are different.
The positivity of the density matrix guarantees that the diagonal elements are
all positive semidefinite in an arbitrary basis, such
that they constitute indeed normalized probability distributions. The
normalization follows from $\tr \rho = 1$, 
which is independent of the choice of basis.

So far we have seen that no obstruction to the completeness of the correlation
map arises from the CHSH-inequality or from the
inequalities \eqref{eq:E88}. We also have found explicit classical probability
distributions for a family of entangled quantum 
states, including the maximally entangled state. These findings suggest that the
correlation map is complete.
They are not a proof, however, since obstructions on a higher level involving
six or more correlation functions could, in principle,
exist.

An analytic proof of completeness of the correlation map is not a simple task.
The classification of all possible inequalities for classical correlation
functions is cumbersome. We have not yet succeeded to find an analytic
expression for finding a probability distribution for an arbitrary density
matrix. The issue has been settled numerically in ref.\,\cite{PW}. For a very
large set of randomly chosen density matrices it has always been possible to
find an associated classical wave function \eqref{eq:E59}, and therefore a
probability distribution, for the classical time-local subsystem of six Ising
spins. We therefore consider the correlation map for two qubits as complete.
Arbitrary density matrices for two qubits can be obtained by the correlation map
from a probability distribution for six Ising spins. As a direct consequence, an
arbitrary unitary quantum evolution can be described by a suitable evolution of
probabilities in the time-local system. Only if we restrict the evolution to the
deterministic updatings of a probabilistic automaton, for which we are
guaranteed that an overall probability distribution exists, the possible unitary
transformations will be a restricted discrete subset.

\subsection{Many qubits}\label{sec:many_qubits}

The generalization to an arbitrary number $Q$ of qubits is rather
straightforward. The generators of SU($Q$) can be written as a direct product of
$Q$ factors
\begin{equation}
L_{\mu_1 \mu_2 ... \mu_Q} = \tau_{\mu_1} \otimes \tau_{\mu_2} \otimes
\tau_{\mu_3} \otimes ... \otimes \tau_{\mu_Q},
\label{eq:MQ1}
\end{equation}
and a general Hermitian normalized density matrix takes the form
\begin{equation}
\rho = 2^{-Q} \rho_{\mu_1 \mu_2 ... \mu_Q} L_{\mu_1 \mu_2 ... \mu_Q},
\label{eq:MQ2}
\end{equation}
with $L_{00...0} = 1$, $\rho_{00...0} = 1$. The $2^{2Q}$ real numbers
$\rho_{\mu_1 ... \mu_Q}$ correspond to the $2^Q \times 2^Q$ elements of the
matrix $\rho$. (Since $\rho^\dagger = \rho$, there are $2^{2Q}$ real independent
elements, where one element is fixed by $\tr \rho = 1$, corresponding to
$\rho_{00...0} = 1$.)

A general class of bit-quantum maps expresses the probabilistic information of
the quantum system, as encoded in $\rho_{\mu_1 ... \mu_Q}$, by expectation
values of $2^{2Q}-1$ Ising spins $\sigma_{\mu_1 ... \mu_Q}$
\begin{equation}
\rho_{\mu_1...\mu_Q} = \chi_{\mu_1...\mu_Q} = \braket{\sigma_{\mu_1...\mu_Q}},
\label{eq:MQ3}
\end{equation}
where $\sigma_{00...0} =1$. For the average spin map all
$\sigma_{\mu_1...\mu_Q}$ are independent Ising spins. Already for a rather
modest number of qubits, say $Q=20$, this requires a very high number of
$\approx2^{40}$ Ising spins.

The minimal correlation map for $Q$ qubits is much more economical, involving
only $3Q$ independent Ising spins $s_k^{(i)}$, $k=1...3$, $i=1...Q$. Composite
spins are formed as products
\begin{equation}
\sigma_{\mu_1 \mu_2 ... \mu_Q} = s_{\mu_1}^{(1)} s_{\mu_2}^{(2)} ...
s_{\mu_Q}^{(Q)}.
\label{eq:MQ4}
\end{equation}
With $s_0^{(i)}=1$ the composite spins with only one index $\mu_a$ different
from zero correspond to the ``fundamental'' Ising spins
\begin{equation}
\sigma_{00...k...0} = s_k^{(a)},
\label{eq:MQ5}
\end{equation}
where the index $k$ on the l.\,h.\,s.\ is at the position $a$. Similarly, if
only two indices $\mu_a$ and $\mu_b$ differ from zero, the expectation value
$\chi_{0...k_a...k_b...0}$ corresponds to a two-point correlation function
\begin{equation}
\chi_{0...k_a 0...k_b 0...0} = \braket{s_{k_a}^{(a)} s_{k_b}^{(b)}}.
\label{eq:MQ6}
\end{equation}
For $Q$-qubits the density matrix involves $n$-point functions of the Ising
spins with $n$ up to $Q$. The price to pay for the use of only a small number
$3Q$ of Ising spins is the need for rather high correlation functions for the
complete characterization of the quantum density matrix. For a pure state
density matrix the minimal correlation map expresses the $2^Q$-component wave
function for $Q$ qubits in terms of $2^{3Q}$ probabilities. A map of this type
is also discussed in ref.~\cite{YY2}.

The Cartesian directions of the $Q$ quantum spins $S_k^{(i)}$ can be associated
directly to the classical Ising spins, with
\begin{equation}
\braket{S_k^{(i)}}_\mathrm{q} = \braket{s_k^{(i)}}_\mathrm{cl}.
\label{eq:MQ7}
\end{equation}
This extends to all correlation functions which involve only different quantum
spins
\begin{equation}
\braket{S_{k_1}^{(i_1)} S_{k_2}^{(i_2)} ...\, S_{k_n}^{(i_n)}}_\mathrm{q} 
= \braket{s_{k_1}^{(i_1)} s_{k_2}^{(i_2)} ...\, s_{k_n}^{(i_n)}}_\mathrm{cl},
\label{eq:MQ8}
\end{equation}
where $i_1 \neq i_2 \neq ... i_n$. This can be seen easily from the form of the
operator associated so $S_k^{(i)}$,
\begin{equation}
\hat{S}_k^{(i)} = 1 \otimes 1 ...\otimes \tau_k \otimes 1 ...\otimes 1, 
\label{eq:MQ9}
\end{equation}
with $\tau_k$ at the position $i$. The quantum operators with different $i_1$
and $i_2$ all commute,
\begin{equation}
\left[ \hat{S}_{k_1}^{(i_1)}, \hat{S}_{k_2}^{(i_2)} \right] = 0 \quad
\textnormal{for } i_1 \neq i_2.
\label{eq:MQ10}
\end{equation}

Concerning completeness of the minimal correlation map for three qubits
ref.\,\cite{PW} has found a small subset of highly entangled density matrices,
in particular the GHZ-state~\cite{GHZ, MER2}, for which no classical probability
distributions exist which are mapped to these states. The minimal correlation
map is therefore not complete. For an extended correlation map explicit
classical probability distributions have been constructed which are mapped to
the GHZ-state. It is not known how many classical bits are needed for a complete
bit-quantum map for arbitrary $Q$. One may suspect that this number increases
faster than linear in $Q$. Nevertheless, special purpose ``Ising
machines''~\cite{YYHA, MMHA, IHI, WARO, CBGH, BBFU, DUKG, ARV} can generate and
manipulate probability distributions for a large number of classical Ising
spins. A possible alternative could be suitably correlated $p$-bits~\cite{CSD1}.
It is an interesting question for how many qubits general unitary
transformations can be realized by these types of probabilistic computing.

For the minimal correlation map a large number of observables has the same
expectation value in the quantum system and in the ``classical'' time-local
system -- namely all the correlations \eqref{eq:MQ8}. The quantum operators for
these observables do, in general, not commute. More precisely, two products of
spins for which at least one factor for a given spin has a different Cartesian
direction, are represented by non-commuting quantum operators. For those pairs
of observables the simultaneous probabilities to find a given combination of
their values $(++)$, $(+-)$, $(-+)$ and $(--)$ are not available in the quantum
subsystem. The quantum subsystem is again characterized by incomplete
statistics. In particular, it contains no information on $n$-point functions
with $n>Q$. The probabilistic information of the quantum subsystem is given by
$2^{2Q}-1$ real numbers. In contrast, the probability distribution
$\{p_\tau(t)\}$ for the time-local system of $3Q$ Ising spins has $2^{3Q}-1$
independent probabilities. Obviously, only a small part of this information is
available for the quantum subsystem. The number of classical Ising spins
increases for extended correlation maps~\cite{PW}.

For large $Q$ the complete information about the density matrix involves a large
number of real parameters, namely $2^{2Q}-1$. This is the reason why rather high
correlations are needed for its full characterization. There is no difference
between the quantum system and the classical system in this respect. Also for
the quantum system $2^{2Q}-1$ expectation values of observables are needed for a
full characterization of the density matrix. As an example one may take the
products of quantum spins in different Cartesian directions as on the
l.\,h.\,s.\ of eq.\,\eqref{eq:MQ8}. In the quantum case, the number of
independent real numbers gets reduced to $2^Q-2$ for pure states. Still, it
increases very rapidly with $Q$.

In practical applications for many qubits the complete information about the
density matrix or the wave function is neither available nor needed. The
question arises which part of the information actually matters for a given
problem. For example, for certain cases a Gaussian approximation for the
probability distribution may be sufficient
\begin{equation}
p[s] = \exp \left\{ -\frac{1}{2} A_{kl}^{ij} (s_k^{(i)} - \tilde{\chi}_k^{(i)})
(s_l^{(j)} - \tilde{\chi}_l^{(j)}) \right\}.
\label{eq:MQ11}
\end{equation}
For the minimal correlation map it involves $3Q$ numbers $\tilde{\chi}_k^{(i)}$
and $(3Q)^2$ coefficients $A_{kl}^{ij}$. This is much less than the $2^{2Q}-1$
independent elements of the density matrix. These elements can be computed for
given $\tilde{\chi}_k^{(i)}$ and $A_{kl}^{ij}$. In the next approximation one
may add in the exponent terms involving three or four Ising spins.

In summary, the extension of our simple bit-quantum maps for one and two qubits
to an arbitrary number $Q$ poses problems in two respects. The first concerns
the issue of completeness of the map. The second is the need to reproduce the
high number of independent elements of the quantum density matrix, either by a
high number of classical bits or high correlations of classical bits. On the
conceptual level it is important to note that there is no obstruction to the
construction of a complete bit-quantum map for an arbitrary number of $Q$-bits.
An example is the average spin map which associates to each
$\sigma_{\mu_1\mu_2\dots\mu_Q}$ an independent classical bit. For an arbitrary
density matrix for $Q$ qubits one can find time-local classical probability
distributions which are mapped to this density matrix.

On the practical side it seems doubtful if an all purpose quantum computer for
many entangled $Q$-bits can be based on probabilistic automata or similar
concepts. The use of real quantum particles or atoms as provided by nature as
subsystems of a suitable quantum field theory seems to offer much better
chances. An interesting issue remains to be explored: Are there useful forms of
correlated computing that are not full quantum computations but nevertheless
constrain correlations similar to the quantum constraint? Does nature use this
possibility for life? We briefly turn back to this question in
sect.~\ref{sec:classical_and_quantum_computing}.

\section{Continuous classical
variables}\label{sec:continuous_classical_variables}

Most classical probabilistic systems are formulated in terms of continuous
variables. The probability distribution $p(\varphi) \geq 0$ then depends on
points $\varphi$ of some continuous manifold. It is normalized by 
\begin{equation}\label{CV1}
\int_\varphi p(\varphi) = 1,
\end{equation}
where $\int_\varphi = \int \dif \varphi$ denotes the integration over the
manifold, which may be multi-dimensional. As compared to the previous discussion
with Ising spins, the discrete classical states or spin configurations $\tau$
are replaced by the points $\varphi$. Every point $\varphi$ denotes a classical
state. Since a continuous variable can be associated to an infinite set of
discrete variables, the classical statistical systems for continuous variables
can be viewed as a limiting case for discrete variables. Observables are real
functions of $\varphi$. The expectation value of an
observable $A(\varphi)$ is given by 
\begin{equation}\label{CV2}
\braket{A} = \int_\varphi p(\varphi)A(\varphi).
\end{equation}

The use of continuous classical variables brings us closer to the quantum
particle for which we have argued that it involves an infinite number of degrees
of freedom. If we allow for an arbitrary orthogonal evolution of the classical
wave function we can find a bit-quantum map which maps this to the evolution of
a quantum particle in a potential according to the usual Schrödinger equation.
This orthogonal evolution is not a unique jump evolution and therefore cannot be
realized directly by a probabilistic automaton. The existence of an overall
probability distribution for events at all times is then not guaranteed. This
general orthogonal evolution would have to obtain by mapping a quantum field
theory to a suitable one-particle subsystem. For the particular case of a
harmonic potential a probabilistic automaton which is mapped to the quantum
particle will be presented in
sect.~\ref{subsec:quantum_particle_in_harmonic_potential}.

\subsection{Continuous variables and Ising spins}
\label{sec:continuous_variables_and_ising_spins}

We begin by a discussion of the relation between continuous variables and Ising
spins. This may seem somewhat trivial and pedantic. In the context of quantum
mechanics it encodes, however, a crucial aspect. The Ising spins provide for
discrete observables associated to yes/no decisions of the type: is a particle
present in a certain region of space or not. The discreteness of these
observables is associated to the ``particle side" of particle-wave duality. In
the context of quantum spins in arbitrary directions it explains why the
possible measurement values of the spin in an arbitrary direction are discrete.

The association of classical Ising spins and continuous classical variables
usually proceeds by some type of ``binning". For example, $\varphi$ may denote
the position of a single particle. A most efficient binning divides the space
into a finite number of bins that do not overlap and cover the whole space. The
yes/no question associated to an Ising spin asks if the particle is in a given
bin or not. Some of the Ising spins may be composite, i.e. products of other
Ising spins. We take $s_j = 1$ if the particle is in the bin $j$, and $s_j = -1$
if it is not.

\paragraph*{Ising spins and most efficient binning of a circle}
As an example, we take $\varphi$ to be a point on a circle or an angle, $-\pi
\leq \varphi \leq \pi$, with endpoints of the interval identified. A first Ising
spin is associated to the question if a particle is in the right half of the
circle, $\cos \varphi \geq 0$, or in the left half of the circle, $\cos \varphi
\leq 0$. The corresponding Ising spin observable is
\begin{equation}\label{CV3}
s_1(\varphi) = \Theta(cos\varphi)-\Theta(-\cos \varphi).
\end{equation}
(We may define spin variables such that $s_1(\pi/2) = 1$ and $s_1(-\pi/2) = -1$.
The precise definition does not matter for expectation values since the points
$\varphi = \pm \pi/2$ are of measure zero in the corresponding integrals.) A
second Ising spin may distinguish between the upper and lower halves of the
circle
\begin{equation}\label{CV4}
s_2(\varphi) = \Theta(\sin \varphi) - \Theta(\sin \varphi).
\end{equation}
We can employ the two spins to define four bins
\begin{alignat}{4}\label{CV5}
&I: & s_1 &= 1, & s_2 &= 1 : & 0&<\varphi<\pi/2 \nonumber\\
&II: & s_1 &= 1, & s_2 &= -1 : & \ -\pi/2&<\varphi<0 \nonumber\\
&III:\ & s_1 &= -1, & s_2 &= 1 : & \pi/2&<\varphi<\pi \nonumber\\
&IV: & s_1 &= -1, & \ s_2 &= -1 : & -\pi&<\varphi<-\pi/2.
\end{alignat} 
We could further subdivide the bins by additional yes/no decisions or Ising
spins, making the bins narrower and narrower. In the limit of infinitely many
Ising spins the size of the bins shrinks to zero and a given point $\varphi$ can
be resolved arbitrarily accurately. This procedure corresponds to the
representation of real numbers in terms of bits on a computer. It is a type of
``most efficient binning" since $M$ spins are sufficient for $2^M$ bins. We see
the direct association of the bins with the classical states $\tau$ discussed
previously.
 
\paragraph*{Overlapping Ising spins on a circle}

Another family of Ising spins associates to each angle $\psi$ a half-circle and
asks if $\varphi$ is within this half circle or not. This association is
employed for the description of a quantum spin in an arbitrary direction. The
corresponding expression for this family of Ising spin observables $s(\psi)$ is
given by 
\begin{equation}\label{CV6}
s(\psi;\varphi) = \Theta(\cos(\varphi - \psi)) - \Theta(-\cos(\varphi-\psi)).
\end{equation}
Here the range of $\psi$ is restricted to the half-circle
\begin{equation}\label{CV7}
-\pi/2 <\psi<\pi/2,
\end{equation}
since the other half-circle is already covered by the opposite value of the spin
observable, $s(\psi-\pi;\varphi) = -s(\psi;\varphi)$. Instead of the angle
$\psi$ we may use a two-component unit vector $e = (e_1,e_2)$, $e_1^2 +e_2^2 =
1$, and similarly employ a unit vector $f$ for $\varphi$ 

\begin{equation}\label{CV8}
\begin{tabular}{l l}
$e_1 = \cos \psi$, & $e_2 = \sin \psi$, \\
$f_1 = \cos \varphi$, & $f_2 = \sin \varphi$.
\end{tabular}
\end{equation}
With these definitions the Ising spin observables involve the scalar product $e
f = e_k f_k$, $k=1,2$,
\begin{equation}\label{CV9}
s(e;f) = \Theta(ef) - \Theta(-ef).
\end{equation}
The two spins $s_1$ and $s_2$ in eqs.~ \eqref{CV3},\eqref{CV4} belong to this
family for unit vectors $e=(1,0)$ and $e=(0,1)$,
\begin{align}\label{CV10}
s_1(f) = s((1,0);f), && s_2(f) = s_2((0,1);f).
\end{align}
Finer binning by using more spins $s(e)$ is less efficient than the previous
case. For example, we may add $s_+ = s(e = (1/\sqrt{2},1/\sqrt{2}))$ for half-
spheres in the direction of a diagonal. Using different values of $s_+$ we can
subdivide the bins II and III in eq.~\eqref{CV5}, but not the intervals I and
IV. This occurs since the interval $0<\varphi<\pi/2$ with $s_1 = s_2 = 1$
automatically has $s_+ = 1$. For positive $f_1(s_1 = 1)$ and positive
$f_2(s_2=1)$ one has $(f_1+f_2)/\sqrt{2}>0$ and therefore $s_+=1$. For
subdividing the Intervals I and IV we need an additional Ising spin associated
to $e = (1/\sqrt{2})(1,-1))$. For dividing the circle into eight equal bins we
therefore need four Ising spins instead of three for the most efficient binning.
Nevertheless, in the limit of infinitely many spins every point $\varphi$ can be
resolved arbitrarily accurately. We will see that the family of Ising spins
\eqref{CV9} is characteristic for quantum systems.

In contrast to the most effective binning the bins defined by eq.~\eqref{CV6}
overlap. A given point $\varphi$ can belong to a large number of bins. A
particle at a given $\varphi$ can be ``seen" by many detectors based on the
yes/no decision \eqref{CV6}. For a precise location of a particle at a given
point $\varphi$ one has to specify a large number of values of Ising spins,
going to infinity if the precision is to be sharply determined. This contrasts
to the most efficient binning for which a single detector $s_j$ can decide if
the particle is at the precise position associated to it.

\paragraph*{Ising spins on spheres and $\mathbb{R}^d$}

The family of Ising spin observables $s(e)$ in eq.~\eqref{CV9} is easily
extended to unit spheres. In this case $e$ and $f$ become $(d+1)$-component unit
vectors. We can also define these Ising spins for $\varphi \in \mathbb{R}^d$. In
this case we replace the unit vector $f$ by $\varphi$, e.g.
\begin{equation}\label{CV11}
s(e;\varphi) = \Theta(e\varphi)-\Theta(-e\varphi).
\end{equation}
We may equivalently use eq.~\eqref{CV9}, with 
\begin{align}\label{CV12}
\varphi_k = r f_k, && r^2 = \varphi_k\varphi_k.
\end{align}
We observe that this binning only concerns the angular direction. Each bin still
contains points with an arbitrary value of $r$. For resolving points on
$\mathbb{R}^d$ one would need an additional binning of the radial coordinate
$r$.

For the description of a single qubit we are no longer restricted to the three
Cartesian spins $s_k$. One may use an infinite number of classical Ising spins
$s(e_k)$. For a fixed classical state denoted by $\vp$ or $f_k$ as a point on
the sphere these Ising spins have values $\pm1$ according to eq.~\eqref{CV11}.
One can construct~\cite{CWPO} a bit-quantum map from a classical probability
distribution $p[\vp]$ to the quantum system for a single qubit such that the
classical spin $s(e_k)$ is mapped to the quantum spin in the direction $e_k$.
This entails that for an arbitrary direction $e_k$ the quantum spin can only
have the possible measurement values $\pm1$. One can construct the associated
quantum spin operators $\hat s(e_k)$ which have eigenvalues $\pm1$. We will
report this construction in chapter~\ref{sec:quantum_mechanics}.

\subsection{Quantum clock system}\label{sec:quantum_clock_system}

The quantum clock system is a simple example of a probabilistic automaton or a
unique jump local chain for a single periodic continuous variable $\varphi$,
with $-\pi<\varphi<\pi$ similar to
sect.~\ref{sec:continuous_variables_and_ising_spins}. The step evolution
operator is a unique jump operator
\begin{equation}\label{CV25}
\hat{S}_{\varphi' \varphi}(t) = \delta_{\varphi' , \varphi+\Delta \alpha}.
\end{equation}
A state $\varphi$ at $t$ necessarily changes to $\varphi +\Delta \alpha$ at
$t+\epsilon$. Correspondingly, one finds for the local probability distributions
\begin{equation}\label{CV26}
p(t+\epsilon,\varphi) = p(t,\varphi - \Delta\alpha).
\end{equation}
In the continuous limit $\epsilon \to 0$ this yields the evolution
equation
\begin{align}\label{CV27}
\del_t p(t,\varphi) = - \omega \del_\varphi p(t,\varphi), && \omega =
\frac{\Delta \alpha}{\epsilon}\ .
\end{align}
This can be seen as the continuum limit of a discrete automaton with discrete
points $\vp$ on a circle and discrete time steps.

\paragraph*{Quantum clocks}

For the quantum system of a single qubit the expectation value of the spin in an
arbitrary direction is computable in terms of the expectation values of the spin
in the three Cartesian directions. The latter fix the density matrix, and the
density matrix determines the expectation values of all quantum observables. If
one wants to construct a classical system that is mapped by the bit-quantum map
to a one qubit quantum system, it has to share this property for the relation of
expectation values of spins in arbitrary directions. We will construct a
bit-quantum map to the one-qubit quantum subsystem in
chapter~\ref{sec:quantum_mechanics}. Here we develop a somewhat simpler system
of a quantum clock for which expectation values of spins in arbitrary angles can
be computed from the expectation values in two Cartesian directions.

The probability distribution $p(\vp)$ permits the computation of the expectation
value of an arbitrary spin $s(\psi)$ with an angle $\psi$. Here we define
$s(\psi)$ by the relation~\eqref{CV6}, resulting in
\begin{equation}
\label{QQ1}
\langle s(\psi)\rangle=\int\text{d}\vp\,
p(\vp)\big[\theta\big(\cos(\vp-\psi)\big) -
\theta\big(-\cos(\vp-\psi)\big)\big]\ .
\end{equation}
We will realize an initial value $\langle s(\psi)\rangle=\cos\psi$ by a suitable
initial probability distribution $p(\vp)$. For later $t$ this can be followed by
the evolution~\eqref{CV26},~\eqref{CV27}. Indeed if we choose at some initial
time $t=0$ the particular probability distribution
\begin{equation}\label{CV28}
p(\varphi) = \frac{1}{2} \cos \varphi \ \Theta(\cos \varphi)\ ,
\end{equation}
the expectation value of the Ising spin in the $\psi$-direction \eqref{CV6} is
given by 
\begin{equation}\label{CV29}
\braket{s(\psi)} = \cos \psi.
\end{equation} 
This follows from the simple angular integration
\begin{align}\label{CV30}
&\braket{s(\psi)} = \frac{1}{2} \int_\varphi \cos(\varphi)
\Theta(\cos(\varphi))\nonumber\\ 
& \quad \quad \quad \quad \quad \times
[\Theta(\cos(\varphi-\psi))-\Theta(-\cos(\varphi-\psi))] \nonumber\\
&= \frac{1}{2} \int_{- \frac{\pi}{2}}^{ \frac{\pi}{2}} \dif \varphi
\cos(\varphi)[\Theta(\cos(\varphi- \psi))-\Theta(-\cos(\varphi-\psi))] \nonumber
\\
&= \frac{1}{2} \left[ \int_{\psi-\pi/2}^{\pi/2} \dif \varphi \cos \varphi -
\int_{-\pi/2}^{\psi-\pi/2} \dif\varphi \cos \varphi \right] \nonumber \\
&= \cos \psi.
\end{align}
The probability distribution describes an eigenstate of the spin in the
direction $\psi=\pi/2$, $\langle s(\pi/2)\rangle=1$. Eq.~\eqref{CV30} describes
the expectation value of a quantum spin in a direction that has an angle
$\psi-\pi/2$ with respect to the direction of the spin for which the system is
in an eigenstate.

A shift of the probability distribution \eqref{CV28} by a constant angle
$\beta$, 
\begin{equation}\label{CV31}
p_\beta (\varphi) = \frac{1}{2} \cos(\varphi-\beta) \Theta(\cos(\varphi-\beta)),
\end{equation}
results by a shift in the angle $\psi$
\begin{equation}\label{CV32}
\braket{s(\psi)}_\beta = \cos(\psi-\beta).
\end{equation}
For the evolution equation \eqref{CV27} one concludes that $p(t,\varphi)$
depends only on the combination $\varphi-\omega t$. For the initial distribution
\eqref{CV31} one infers the time-local probability distribution
\begin{equation}\label{CV33}
p(t,\varphi) = \frac{1}{2} \cos(\varphi - \omega t - \beta) \Theta (\cos(\varphi
- \omega t - \beta)).
\end{equation}
The expectation value of $s(\psi)$ rotates correspondingly
\begin{equation}\label{CV34}
\braket{s(t;\psi)}_\beta = \cos(\psi - \omega t - \beta).
\end{equation}
We conclude that the quantum clock is a probabilistic clock for which the
maximum of the expectation values of the spins $s(\psi)$ can be used as a
pointer. The expectation values of spins in arbitrary directions are fixed by
their angle to the pointer direction. They rotate together with the pointer.

Instead of the angles $\varphi$, $\psi$ and $\beta$ we may also use two
component unit vectors $f=(f_1,f_3)$, $e = (e_1,e_3)$, $\rho = (\rho_1,\rho_3)$,
\begin{equation}\label{CV35}
\begin{tabular}{c c}
$e_1 = \cos \psi,$ & $e_3=\sin \psi,$ \\
$f_1 = \cos \varphi,$ & $f_3 = \sin \varphi,$ \\
$\rho_1 = \cos \beta,$ & $\rho_3 = \sin \beta.$
\end{tabular}
\end{equation}
The initial probability distribution \eqref{CV31} for $t=0$ reads in this
representation
\begin{align}\label{CV36}
p(\rho;f) = \frac{1}{2} (\rho f)\Theta (\rho f), && \rho_k\rho_k = 1
\end{align}
the spins are given by
\begin{equation}\label{CV37}
s(e;f) = \Theta(ef)-\Theta (-(ef)),
\end{equation}
and the expectation values obey
\begin{equation}\label{CV38}
\braket{s(e)}_\rho = (\rho e).
\end{equation}
The appearance of the scalar products makes the invariance under simultaneous
rotations of $\rho$, $f$ and $e$ apparent. For the time evolution one has
\begin{align}\label{CV39}
\rho_1 (t) = \cos(\beta + \omega t), && \rho_3(t) = \sin(\beta + \omega t).
\end{align}

\paragraph*{Quantum subsystem}

The expectation values of the Cartesian spins $s_1 = s(e=(1,0))$ and $s_3 =
s(e=(0,1))$ are given by
\begin{align}\label{CV40}
\braket{s_1} = \rho_1, && \braket{s_3} = \rho_3.
\end{align}
We can define a quantum subsystem based on these two expectation values, with
density matrix
\begin{equation}\label{CV41}
\rho = \frac{1}{2} (1 + \rho_1 \tau_1 + \rho_3 \tau_3).
\end{equation}
This is the density matrix for a two-component quantum spin. The third spin
direction $s_2$ is absent. The density matrix \eqref{CV41} is real and
symmetric. It is a pure state density matrix, since $\rho_1^2 + \rho_3^2 =1$.
The quantum operator for the spin in the direction $e = (e_1,e_3)$ is given by
\begin{equation}\label{CV42}
S(e) = e_1 \tau_1 + e_3 \tau_3.
\end{equation}
The quantum rule,
\begin{equation}\label{CV43}
\braket{S(e)} = \tr \{\rho S(e)\} = \rho_1 e_1 +\rho_3 e_3,
\end{equation}
yields the same result as eq.~\eqref{CV38}. We therefore can identify the
quantum spin in an arbitrary direction $e$ with the classical spin $s(e)$ in the
same direction. The eigenvalues of the operators $S(e)$ are $\pm 1$,
corresponding to the possible measurement values of the classical Ising spins.
The expectation values can be evaluated equivalently with the classical rule or
the quantum rule \eqref{CV43}.

In contrast to the quantum subsystem discussed in
sect.~\ref{sec:quantum_subsystems} the identification of quantum spin directions
with classical Ising spins holds for arbitrary spin directions, not only for the
Cartesian spins. This involves the infinitely many classical Ising spins
associated to the continuous variable $\varphi$ by eq.~\eqref{CV37}. 

\paragraph*{Unitary evolution}

The deterministic unique jump operations \eqref{CV25} can realize arbitrary
rotations in the (1-3)-plane as unitary transformations. On the classical level
a rotation on the circle,
\begin{equation}\label{CV43A}
\varphi' = \varphi - \gamma,
\end{equation}
corresponds to
\begin{align}
f_1' &= \cos \gamma \ f_1 + \sin \gamma \ f_3 \nonumber \\
f_3' &= \cos \gamma \ f_3 - \sin \gamma \ f_1.
\end{align}
A unique jump operation transforms a probability distribution $p(\rho;f)$ at $t$
to $p(\rho;f')$ at $t+\epsilon$. Using the same variables at $t+\epsilon$ and
$t$ the transformation amounts to $p(\rho';f)$ at $t+\epsilon$ with
\begin{align}\label{CV43C}
\rho_1' &= \cos \gamma \ \rho_1 - \sin \gamma \ \rho_3 \nonumber\\
\rho_3' &= \cos \gamma \ \rho_3 + \sin \gamma \ \rho_1.
\end{align}
The expectation values of the Cartesian spins and therefore the entries of the
quantum density matrix are given by eq.~\eqref{CV43C} as well.

On the level of the quantum density matrix the unitary transformation
\begin{equation}\label{CV44}
\rho' = exp\left( \frac{i\gamma \tau_2}{2}\right) \rho \ exp \left(-
\frac{i\gamma \tau_2}{2}\right)
\end{equation}
rotates by an angle $\gamma$ in the 1-3 plane and realizes eq.~\eqref{CV43C}
\begin{align}\label{CV45}
\rho_1' &= \cos \gamma \rho_1 - \sin \gamma \rho_3 = \cos(\beta+\gamma),
\nonumber \\
\rho_3' &= \cos \gamma \rho_3 + \sin \gamma \rho_1 = \sin(\beta+\gamma).
\end{align}
The evolution \eqref{CV39},
\begin{align}\label{CV46}
\rho(t) &= \frac{1}{2}(1+\rho_1(t)\tau_1 + \rho_3(t)\tau_3) \nonumber \\
&= U(t)\rho(0)U^\dagger(t) \nonumber \\
&= U(t)( \frac{1}{2}(1+\cos \beta \tau_1 + \sin \beta \tau_3))U^\dagger(t,)
\end{align}
is realized by
\begin{equation}\label{CV47}
U(t) = \exp \left( \frac{i\omega t}{2} \tau_2\right).
\end{equation}

The quantum subsystem obeys a unitary evolution law. In particular, we can
consider infinitesimal time steps $\epsilon \to 0$. In this case one finds the
von Neumann equation
\begin{equation}\label{CV48}
\del_t \rho = \del_t U U^\dagger \rho + \rho U \del_t U^\dagger = -i[H,\rho],
\end{equation} 
with hermitean Hamiltonian
\begin{equation}\label{CV49}
H = - \frac{1}{2} \omega \tau_2.
\end{equation}
This remains a real evolution equation since $-iH$ is a real antisymmetric
matrix, and $U$ therefore an orthogonal matrix. With
\begin{equation}
\rho_1(t) = \cos \beta(t),\quad \rho_3(t) = \sin \beta(t),
\label{eq:CW1}
\end{equation}
the solution of the von-Neumann equation \eqref{CV48} reads indeed
\begin{equation}
\beta(t) = \beta_0 + \omega t.
\label{eq:CW2}
\end{equation}

In summary, we have mapped the classical statistical quantum clock system for a
continuous classical variable $\vp$ and associated classical Ising spins
$s(\psi)$ to a quantum subsystem which corresponds to a type of two-dimensional
qubit. The unitary transformation of the quantum subsystem form the abelian
group $\text{U}(1)$ or $\text{SO}(2)$. We encounter a particular case of ``real
quantum mechanics" with a real symmetric density matrix. A complex formulation
could be realized by doubling the degrees of freedom. We will generalize in
sect.~\ref{sec:quantum_mechanics} this system to a full qubit with unitary
transformations forming the group $\text{SU}(2)$. This will automatically induce
the usual complex formulation of quantum mechanics. We emphasize that the
quantum clock system does not only realize the unitary evolution of the density
matrix which is rather easy to achieve. It also relates the discrete possible
measurement values of the quantum spin in an arbitrary direction to a yes/no
question of the classical statistical system. The quantum rule for possible
measurement values finds a direct root in the properties of classical
observables. This aspect of particle-wave duality is directly realized.

\paragraph*{General probability distributions for quantum clocks}

The realization of the quantum clock system by the unique jump operation
\eqref{CV25} with initial classical probability distribution \eqref{CV31}
belongs to a wide class of possible classical probabilistic systems. The
probability distributions may depend on additional variables, $p(t;\varphi;y)$.
It is sufficient that for every $t$ these distributions obey
\begin{equation}\label{CV50}
\int_y p(t;\varphi;y) = p_{\beta(t)}(\varphi),
\end{equation}
with $p_{\beta(t)}$ given by eq.~\eqref{CV31} for suitable $\beta(t)$.
Since the classical Ising spins $s(e)$ depend on $\varphi$ and are independent
of $y$, the relation \eqref{CV32} holds, with $\beta(t)$ defining $\rho(t)$ in
eqs.~\eqref{CV36},~\eqref{CV38}. If the relation \eqref{CV50} holds for $t=0$,
many different unique jump operations can ensure this relation for arbitrary
$t$. As a particular example, the unique jump operation may be given by
eq.~\eqref{CV25} with $y$ left invariant.
General $\beta (t)$ correspond to time dependent $\omega(t) = \del_t \beta(t)$.
As a particular case we may consider continuous variables $\varphi \in
\mathbb{R}^2$, with $\varphi_k = r f_k$. The Ising spins are independent of $r$,
which can be associated with the additional variable $y$.

\subsection{Deterministic evolution with\\continuous
variables}\label{sec:deterministic_evolution_for_continuous_variables}

The concept of probabilistic automata can be taken over directly to continuous
classical variables. This can be seen as the limit of infinitely many classical
bits. For a given discrete time step $\eps$ the updating describes an invertible
map among the continuous variables $\vp$. One often can take the continuous limit
$\eps\to0$. For the time-local probability distribution this results in a first
order non-linear differential evolution equation. It involves the derivatives of
the probability distribution with respect to $\vp$ in linear order. The
Liouville equation for the probability distribution in phase space for a
classical particle in a potential can be described in this way. We discuss a
generalization that can describe a quantum particle in a potential as a suitable
subsystem. For a harmonic potential this can be realized by a probabilistic
automaton.

\paragraph*{Unique jump operations for continuous variables}

Unique jump operations map every variable $\varphi$ at $t$ to a unique variable
$\varphi' = f(\varphi;t)$ at $t+\epsilon$. This is a deterministic evolution in
the space of variables, that we may denote as
\begin{equation}\label{CV51}
\varphi(t+\epsilon) = f(\varphi;t).
\end{equation}
It translates directly to the $t$-dependence of the time-local probability
distributions,
\begin{equation}\label{CV52}
p(t+\epsilon; f(\varphi;t)) = p(t;\varphi)\ .
\end{equation}
We will consider invertible transformations $f(\varphi)$ here, such that
\begin{equation}\label{CV54}
p(t+\epsilon;\varphi) = p(t;f^{-1}(\varphi;t))\ .
\end{equation}
The corresponding step evolution operator reads
\begin{equation}\label{CV55}
\hat{S}(t;\varphi',\varphi) = \delta(\varphi', f(\varphi;t)).
\end{equation}

\paragraph*{Differential evolution equations with classical \\ variables}

The deterministic evolution equation \eqref{CV51} admits a continuum limit if
the transformation $f(\varphi)$ is sufficiently smooth. In this case it turns to
a differential equation
\begin{align}\label{CV56}
\del_t \varphi(t) &= \frac{1}{2\epsilon} (\varphi(t+\epsilon)-
\varphi(t-\epsilon)) \nonumber \\
&= \frac{1}{2\epsilon} [ f(\varphi(t);t) - f^{-1}(\varphi(t);t-\epsilon)]
\nonumber \\
&= \mathcal{D}(\varphi(t);t),
\end{align}
with $\mathcal{D}(\varphi(t);t)$ a suitable operator defined by the second line.
For $\epsilon \to 0$ and smooth $f$ one finds that $\mathcal{D}$ is simply a
function of $\vp$. In particular, for
\begin{align}\label{CV57}
f(\varphi(t);t) &= \varphi(t)+ \epsilon g(\varphi(t);t), \nonumber \\
f^{-1}(\varphi(t);t) &= \varphi(t) - \epsilon g(\varphi(t);t),
\end{align}
one has
\begin{equation}\label{CV58}
\mathcal{D}(\varphi(t),t) = \frac{1}{2} [g(\varphi(t);t) +
g(\varphi(t);t-\epsilon)].
\end{equation}
If the $t$-dependence of $g$ is smooth, one can identify
$\mathcal{D}(\varphi(t);t) = g(\varphi(t);t)$.

The generalization to a multi-component classical continuous $\vp_k$ is
straightforward. The evolution equation
\begin{equation}\label{CV59}
\del_t \varphi_k = \mathcal{D}_k(\varphi;t)
\end{equation}
is a first order differential equation that is, in general, not linear. It is
local in time since only $\varphi(t)$ appears on the r.h.s, such that for given
$\mathcal{D}$ one can compute $\varphi(t+\epsilon)$ from $\varphi(t)$ without
any additional information. This is the situation encountered in many classical
deterministic systems with continuous variables.

The resulting time evolution of the time-local probability distribution follows from
eq.~\eqref{CV54},
\begin{equation}\label{CV60}
\del_t p(t;\varphi) = -\mathcal{D}_k(\varphi) \frac{\del}{\del \varphi_k}
p(t;\varphi).
\end{equation}
Here we employ for the definition of $\mathcal{D}_k$ a notation where $\vp$ can
be a multi-component vector
\begin{align}\label{CV61}
\del_t p(t;\varphi) &= \frac{1}{2\epsilon} [p(t+\epsilon;\varphi) -
p(t-\epsilon;\varphi)] \nonumber \\
&= \frac{1}{2\epsilon} [p(t;f^{-1}(\varphi;t)) - p(t; f(\varphi;t-\epsilon))]
\nonumber \\
&= \frac{1}{2\epsilon} [p(t;\varphi-\epsilon g(\varphi,t)) - p(t;\varphi +
\epsilon g(\varphi;t-\epsilon))] \nonumber \\
&= - \frac{1}{2}(g(\varphi;t) + g(\varphi;t-\epsilon)) \del_\varphi
p(t;\varphi).
\end{align}
For sufficiently smooth $g$ we take the continuum limit~\eqref{CV58} for which
$\mathcal{D}_k(\vp)$ becomes a function of $\vp$.
The evolution equation \eqref{CV60} holds for all differentiable local
probability distributions. It constitutes a non-linear partial differential
equation for the time evolution of the time-local probability distribution.

We recall that all classical statistical systems with continuous variables of
this type can be considered as probabilistic cellular automata. The overall
probability distribution as a fundamental quantity for the description of a
probabilistic world for all times is therefore well defined.

\paragraph*{Liouville equation}

As an example, we may consider a simple classical particle in a potential. The
variables are points in phase space, $\varphi = (x_k,p_k)$, $k=1...3$. The
deterministic equations of motion are Newton's equation, such that
eq.~\eqref{CV59} reads
\begin{align}
\label{eq:NE}
\del_t x_k = \frac{p_k}{m}, && \del_t p_k = - \frac{\del V}{\del x_k},
\end{align}
where $V(x)$ is the potential and $m$ is the particle mass. The resulting
evolution equation for the time-local probability distribution
$w(\vec{x},\vec{p})$,
\begin{equation}
\label{eq:LLB2}
\del_t w = - \frac{p_k}{m} \frac{\del w}{\del x_k} + \frac{\del V}{\del x_k} \frac{\del
w}{\del p_k},
\end{equation}
is the Liouville equation
for free particles in a potential. For a $\delta$-distribution of
$w(\vec{x},\vec{p})$ one recovers Newton's equations \eqref{eq:NE}. For more
general $w(\vec{x},\vec{p})$ one observes a broadening of wave packets similar
to quantum mechanics\,\cite{CWQP,VOL}.

\subsection{Classical wave function and quantum particles}
\label{sec:classical_wave_function_and_quantum_particles}

One may introduce a classical wave function
$\phi_\mathrm{c}(\vec{x},\vec{p})$
\begin{equation}
w(\vec{x},\vec{p}) = \phi_\mathrm{c}^2(\vec{x},\vec{p})\ .
\label{eq:LLA}
\end{equation}
Due to the particular structure of the Liouville operator it obeys the same
differential equation as the probability
distribution\,\cite{CWQP,CWQPCG,CWQPPS},
\begin{equation}
\partial_t \phi_\mathrm{c}(x,p) = -\hat{L} \phi_\mathrm{c}(x,p),\quad \hat{L} =
\frac{p}{m} \partial_x - \frac{\partial V}{\partial x} \partial_p\ .
\label{eq:LLB}
\end{equation}
The description in terms of a classical wave function shares important features
with the Hilbert space formulation of classical mechanics by Koopman\,\cite{KOP}
and von Neumann\,\cite{VNE}. This probabilistic view on classical mechanics has
triggered many interesting formal developments\,\cite{MAU,GORE,MAMA,GOMAS,NKO},
with connection to the work of Wigner\,\cite{WIG} and Moyal\,\cite{MOJ}.

There is no need that the evolution equation for the classical wave function in
phase space and associated probability distribution is given precisely by
eq.\,\eqref{eq:LLB}, \eqref{eq:LLB2}. We have discussed this in the
introduction or ref.~\cite{CWPW} for the example of rain drops. Interesting
experiments show quantum features in the statistical motion of classical
droplets\,\cite{COFO,EFMC}. For a suitable modification of the r.\,h.\,s.\ of
eq.\,\eqref{eq:LLB} one obtains the precise probabilistic motion of quantum
particles in a potential, including phenomena as
tunneling\,\cite{CWQP,CWQPCG,CWQPPS}. One can also obtain zwitters\,\cite{CWZWI}
-- particles between classical particles and quantum particles.

For the evolution of a classical wave function $\phi_c$ and associated
probability distribution in phase space $w$ which can be mapped to the evolution
of a quantum particle according to the Schrödinger equation, the operator
$\hat{L}$ in eq.\,\eqref{eq:LLB} is replaced\,\cite{CWQP,CWQPPS} by
$\hat{L}_\mathrm{W}$
\begin{equation}
\hat{L}_\mathrm{W} = \frac{p}{m} \partial_x + iV\left( x + \frac{i}{2}
\partial_p \right) - iV \left( x-\frac{i}{2} \partial_p \right)\ .
\label{eq:LLC}
\end{equation}
We note that $\hat{L}_\mathrm{W}$ is a real operator despite the complex
formulation. An appropriate coarse graining by taking a subtrace of the
classical density matrix constructed from $\phi_c$ yields a subsystem for which the
complex wave function obeys the Schrödinger equation\,\cite{CWQPCG} for the
potential $V$. This coarse graining defines the bit-quantum map to a subsystem for a
quantum particle. It associates to a classical probability distribution
$w(\vec{x},\vec{p})$ a complex density matrix $\rho(x,x')$ or a complex wave
function $\psi(x)$.

The generator $\hat L_W$ induces a rotation of the classical wave function, such
that no information is lost by the evolution
\begin{equation}
\label{4.75A}
\partial_t\phi_c=-\hat L_W\phi_c\ .
\end{equation}
We have seen in eq.~\eqref{CV60}, however, that the deterministic updating of a
probabilistic automaton results in a generator that is linear in the derivatives
$\partial/\partial x_k$ and $\partial/\partial p_k$. For a potential $V(x)$
which involves more than two powers of $x$ this is not the case for $\hat L_W$
in eq.~\eqref{eq:LLC}. We conclude that eq.~\eqref{4.75A} describes a rotation
of $\phi_c$ which induces a corresponding change in $w(\vec x,\vec p)$. This
rotation is not realized, however, by the simple updating rule~\eqref{CV51}. It
is not described by a probabilistic automaton and therefore the existence of an
overall probability distribution is not guaranteed. Nevertheless, the
evolution~\eqref{4.75A} describes a perfectly valid evolution of the time-local
probability distribution $w(t;\vec x,\vec p)$. If this evolution can be obtained
for a suitable subsystem from some overall probability distribution, this
overall probability distribution can describe a quantum particle in an arbitrary
potential $V(\vec x)$. If one has the freedom to use arbitrary orthogonal step
evolution operators for the evolution of the time-local probabilistic
information rather complex quantum field theories can be described in this
way~\cite{CWCPMW, CWQFTCS}.

For the special case of a harmonic potential,
\begin{equation}
\label{4.75B}
V(\vec x)=\frac12\beta_{kl}x_k x_l\ ,
\end{equation}
one has
\begin{equation}
\label{4.75C}
iV\left(\vec x+\frac{i}{2}\vec{\frac{\partial}{\partial p}}\right)-iV\left(\vec
x-\frac{i}{2}\vec{\frac{\partial}{\partial p}}\right) =
\beta_{kl}x_l\frac{\partial}{\partial p_k}\ .
\end{equation}
This coincides with the Liouville equation~\eqref{eq:LLB}. We conclude that a
quantum particle in a harmonic potential can be described as an appropriate
subsystem of a probabilistic automaton. The particular quantum features as the
discrete energy spectrum arise from the quantum constraint which requires that
the density matrix of the quantum subsystem has to be a positive matrix. We will
describe the quantum particle in a harmonic potential in
sect.~\ref{subsec:quantum_particle_in_harmonic_potential}.

\section{Quantum mechanics}
\label{sec:quantum_mechanics}

In this section we discuss probabilistic automata that realize all features of
quantum mechanics. We start with quantum mechanics for a two-state system or a
single qubit. The quantum spin in an arbitrary direction is associated to a
corresponding classical Ising spin. The deterministic evolution for the
automaton results for the quantum subsystem in the unitary evolution according
to the von-Neumann equation for the quantum density matrix. Suitable updating
rules can realize any arbitrary Hamiltonian for a single qubit. Quantum
mechanics for a single qubit is the extension of the quantum clock system to
rotations in three-dimensional space or on the two-dimensional sphere.

As a second example of a probabilistic automaton which describes a known quantum
system as a subsystem we discuss the quantum particle in a harmonic potential.
Both for the single qubit system and the quantum particle in a harmonic
potential all rules and properties of quantum mechanics follow directly from the
classical probability laws. The key ingredients are the identification of a
suitable subsystem, and constraints on the probability distribution for the
automaton which ensure a positive density matrix for the subsystem.

\subsection{Classical Ising spins and quantum
spin}\label{sec:classical_ising_spins_and_quantum_spin}

We first consider a given site on the local chain $m$ or time $t$. The
classical variables $\varphi$ are points in $\mathbb{R}^3$, $\varphi =
(\varphi_1, \varphi_2 , \varphi_3)$. We could restrict $\vp$ to be a unit
vector. We keep here an arbitrary length in order to demonstrate that large
classes of different classical probability distributions can lead to the same
quantum subsystem. We define Ising spin observables in an arbitrary direction $e
= (e_1,e_2,e_3)$, $e_k e_k =1$, similar to
eq.~\eqref{CV11}
\begin{equation}\label{Q1}
s(e) = \Theta(\varphi e) - \Theta(-\varphi e).
\end{equation}
They take the value $+1$ if the scalar product $\varphi e = \varphi_k e_k$ is
positive, and the value $-1$ otherwise. We also generalize the family of
probability distributions \eqref{CV36},
\begin{align}\label{Q2}
p(\rho) = \bar{p}(r)(\varphi\rho) \Theta(\varphi \rho), && r^2 = \varphi_k
\varphi_k,
\end{align}
with $\bar{p}(r) \geq 0$ arbitrary as long as it obeys the normalization
condition
\begin{equation}\label{Q3}
\int \dif^3 \varphi p(\rho) = 1.
\end{equation}
The different members of this family are labeled by a unit vector $\rho$,
\begin{equation}\label{Q4}
\rho_k \rho_k =1.
\end{equation}
While eq.~\eqref{Q2} comprises a large class of different probability
distributions, it remains nevertheless a small subset of the most general
time-local probability distributions. The particular form~\eqref{Q2} realizes
the quantum constraint.

\paragraph*{Expectation values of classical Ising spins}

For the probability distributions \eqref{Q2} the expectation values of the Ising
spins obey
\begin{equation}\label{Q5}
\braket{s(e)}_\rho = e \rho.
\end{equation}
In order to show this important relation we need to establish the integral
\begin{equation}\label{Q6}
\braket{s(e)}_\rho = \int \dif^3 \varphi \bar{p}(r)(\rho \varphi)\Theta(\rho
\varphi) [\Theta(\varphi e) - \Theta(-\varphi e)] = \rho e.
\end{equation}
We observe that the integral \eqref{Q6} is invariant under simultaneous
rotations of $\varphi,\rho$ and $e$, since only invariant scalar products are
involved. Without loss of generality we can choose 
\begin{align}
\rho = (1,0,0), && e = (e_1,0,e_3),
\end{align}
and proof the relation
\begin{align}\label{Q8}
& \braket{s(e)}_\rho = \nonumber \\ 
& \int \dif^3 \varphi \bar{p}(r) \varphi_1 \Theta (\varphi_1) [\Theta(\varphi_1
e_1 + \varphi_3 e_3)-\Theta(-\varphi_1 e_1 - \varphi_3 e_3)] \nonumber \\
& = e_1.
\end{align}

We can perform the $\varphi_2$-integration
\begin{align}\label{Q9}
\int \dif \varphi_2 \bar{p}(r) = H(R), && R^2 = \varphi_1^2 +\varphi_3^2.
\end{align}
The normalization condition \eqref{Q3} implies
\begin{equation}
\int \dif \varphi_1 \dif \varphi_3 H(R) \varphi_1 \Theta(\varphi_1) = 1.
\end{equation}
With
\begin{equation}
\varphi_1 = R \cos \alpha, \quad \varphi_3 = R \sin \alpha,
\end{equation}
this yields
\begin{align}\label{Q12}
&\int \dif R R^2 H(R) \int \dif \alpha \cos \alpha \ \Theta(\cos \alpha)\nn\\
&=2\int \dif R R^2 H(R) = 1,
\end{align}
in accordance with the normalization \eqref{Q3}.

Using furthermore
\begin{align}
e_1 = \cos \psi, \quad e_3 = \sin \psi,
\end{align}
the insertion of eqs.~\eqref{Q9}, \eqref{Q12} into eq.~\eqref{Q8} yields with
eq.~\eqref{CV30}
\begin{align}
& \braket{s(e)}_\rho =\nonumber \\ 
& \frac{1}{2} \int \dif\alpha \cos \alpha \Theta(\cos \alpha)
[\Theta(\cos(\alpha-\psi))- \Theta(-(\cos(\alpha-\psi))] \nonumber \\
& = \cos \psi,
\end{align}
confirming eq.~\eqref{Q8} and therefore establishing eq.~\eqref{Q5}.
We observe that the number of components of $\varphi_k$ plays no role in this
proof since for $k > 3$ the l.h.s. of eq.~\eqref{Q9} is replaced by an
integration over all components except $\varphi_1$ and $\varphi_3$.

\paragraph*{Quantum subsystem}

We next define the quantum subsystem as a bit-quantum map from the family of
probability distributions~\eqref{Q2} to the density matrix for a single
qubit~\cite{CWQCCB}. For this purpose we evaluate the relation \eqref{Q5} for
three ``Cartesian spins"
\begin{alignat}{2}
s_1 &= s(e=(1,0,0)), &\quad s_2 &= s(e=(0,1,0)), \nonumber \\
s_3 &= s(e=(0,0,1)), & &
\end{alignat}
with expectation values
\begin{equation}
\braket{s_k} = \rho_k.
\end{equation}
We employ these expectation values for the definition of the density matrix
$\rho$ of the quantum subsystem
\begin{equation}
\label{5.17}
\rho = \frac{1}{2} (1+\rho_k \tau_k)\ .
\end{equation}
With eq.~\eqref{Q4} the quantum constraint is obeyed and $\rho$ is a positive
matrix. Since the three components $\rho_k$ form a unit vector we actually deal
with a pure quantum state for a single qubit or two-state quantum system. As
usual, one can construct from $\rho$ the complex two-component pure state wave
function $\psi$.

The quantum spin operators in the direction $e$ are given by
\begin{equation}
\label{5.18}
S(e) = e_k \tau_k.
\end{equation}
According to the quantum rule their expectation values read
\begin{equation}\label{Q19}
\braket{S(e)} = \tr \{\rho S(e)\} = e_k \rho_k .
\end{equation}
This coincides with the expectation values of the classical Ising spins $s(e)$
according to eq.~\eqref{Q5}. We can identify the classical Ising spins $s(e)$
with the quantum spins $S(e)$ in the same direction. For both the possible
measurement values are $\pm 1$, according to the eigenvalues of the quantum
operators $S(e)$. The expectation value can equivalently be evaluated with the
classical rule \eqref{Q6} or the quantum rule \eqref{Q19}. 

The information in the quantum subsystem is sufficient for the computation of
all the infinitely many spin observables in the different directions. These
correspond to the infinitely many classical Ising spins that are defined for a
classical probabilistic system with continuous variables. None of the classical
correlation functions for the Ising spins is computable from the information of
the quantum subsystem. These classical correlation functions depend on the
specific choice of $\bar{p}(r)$. They cannot be expressed in terms of the three
numbers $\rho_k$ that characterize the quantum subsystem. 

\addtocontents{toc}{\protect\newpage}
\addtocontents{toc}{\vspace*{4.5em}}

\subsection{Unitary evolution for one-qubit\\quantum system}
\label{sec:unitary_evolution_4_5_2}

We next define updatings of the probabilistic automaton which realize unitary
transformations for the quantum subsystem. We consider finite time steps $\eps$,
for which a deterministic updating of the continuous classical variables is
given by
\begin{equation}
\label{5.19A}
\vp_k\to\vp_k+\eps g_k\ .
\end{equation}
We want to find out which $g_k$ lead to a unitary transformation of the quantum
subsystem. For this purpose we establish that an $\text{SO}(3)$-rotation in the
space of the continuous variables $\varphi$, namely
\begin{equation}
\label{5.19B}
\vp_k\to R_{kl}\vp_l\ ,\quad R^TR=1\ ,\quad \text{det}R=1\ ,
\end{equation}
induces a unitary evolution of the quantum subsystem. We may start with a
rotation in the (1-3) plane between the variables $\varphi_1$ and $\varphi_3$,
keeping $\varphi_2$ fixed,
\begin{align}
\label{5.20A}
\varphi_1' &= \cos \gamma \varphi_1 + \sin \gamma \varphi_3 \nonumber \\
\varphi_3' &= \cos \gamma \varphi_3 - \sin \gamma \varphi_1.
\end{align} 
The probability distribution~\eqref{Q2} is transformed to a new member of this
family, according to
\begin{equation}
\label{5.20A}
p(t+\eps,\vp';\rho)=p(t,\vp;\rho)=p(t+\eps,\vp,\rho')
\end{equation}
with $\rho'$ defined by
\begin{equation}
\label{5.20B}
\rho\vp'=\rho'\vp=\rho R\vp\ .
\end{equation}
For the transformation~\eqref{5.20A} this results for $\rho$ in the
transformation~\eqref{CV43C}. On the level of the quantum density matrix for the
subsystem we recover the unitary transformation~\eqref{CV44}. Indeed, rotations
around a given axis form a quantum clock system. The component $\rho_2$ remains
invariant under rotations around the $2$-axis.

Rotations around one of the other Cartesian axes replaces $\tau_2$ in
eq.~\eqref{CV43A} by $\tau_1$ or $\tau_3$. More generally, a rotation by
$\gamma$ around an arbitrary axis with direction given by a unit vector $b$
results for the quantum subsystem in the unitary transformation
\begin{equation}
\label{5.21}
U(b) = \exp \left\{ \frac{i\gamma}{2}(\tau_k b_k)\right\}.
\end{equation}
We conclude that the updatings~\eqref{5.19A} which are compatible with the
evolution of the subsystem are the rotations~\eqref{5.19B} around an arbitrary
axis. The probabilistic automaton can be defined by a sequence of rotations
around different axes. This results for the quantum subsystem in the
corresponding sequence of unitary transformation. As compared to the discrete
qubit chain in sect.~\ref{sec:quantum_subsystems}, the transition to continuous
variables allows us to realize arbitrary sequences of unitary transformations.

A given discrete rotation and associated unitary transformation can be described
for the quantum subsystem by a Hamiltonian that is constant between $t$ and
$t+\eps$,
\begin{equation}
\label{5.21A}
U(b)=\exp\big(-i\eps H(b)\big)\ ,
\end{equation}
with
\begin{equation}
\label{5.22}
H(b)=-\frac{\omega}{2}\tau_kb_k\ ,\quad \omega=\frac{\gamma}{\eps}\ .
\end{equation}

If $b$ and $\gamma$ change only smoothly with time we can take the continuum
limit $\eps\to0$. The discrete evolution equation turns then to the von Neumann
equation for the density matrix
\begin{equation}
\label{5.22A}
i\partial_t\rho=\big[H,\rho\big]\ .
\end{equation}
For a pure state this yields for the associated complex wave function $\psi$ the
Schrödinger equation
\begin{equation}
\label{5.22B}
i\partial_t\psi=H\psi\ .
\end{equation}
Here both $\omega$ and $b_k$ can depend on $t$, such that the most general time
evolution of single qubit quantum mechanics can be implemented by a suitable
updating prescription for a probabilistic automaton.

Taking things together, we have found a classical probabilistic system for which
a suitable subsystem realizes all features of quantum mechanics for a two-state
system. This is an example for the embedding of quantum mechanics in a classical
probabilistic setting. More in detail, the overall probability distribution is
the one for a probabilistic automaton. The map from the overall probability
distribution to the quantum subsystem proceeds in two steps. One first defines
the time-local subsystem for the automaton, as characterized by the classical
wave function or the associated time-local probability distribution. From there
the bit-quantum map maps a suitable family of time-local probability
distributions~\eqref{Q2} to the quantum subsystem. Our construction can be
verified numerically by initializing a probabilistic automaton with a
probability distribution over the states $\vp$ obeying the quantum constraint.
The automaton performs the rotations~\eqref{5.19B}. One can determine the
expectation values $\langle s(e)\rangle$ of the classical Ising spins~\eqref{Q1}
at any time $t$ from the time-local probability distribution for the automaton.
One the quantum side one constructs the initial density matrix~\eqref{5.17} from
the initial probability distribution of the automaton. One performs the unitary
transformation~\eqref{5.21} corresponding to the rotations~\eqref{5.19B} of the
automaton. The expectation values of the quantum spin with operator $S(e)$ given
by eq.~\eqref{5.18} can be evaluated for any $t$ by the usual quantum
rule~\eqref{Q19}. They will coincide with the expectation values $\langle
s(e)\rangle$ of the corresponding classical Ising spins.

\subsection{Time reversal and complex conjugation}

If we revert the time direction the l.h.s. of the von-Neumann equation,
\begin{equation}\label{Q23}
\del_t \rho = -i [H,\rho],
\end{equation}
changes sign. In the time reverted system the positive direction points from the
site $m$ to the site $m-1$ on the local chain.
In other words, the time reversal transforms the von-Neumann equation and the
Hamiltonian according to
\begin{align}\label{Q24}
T: && \del_t\rho = i[H,\rho], && H \to - H.
\end{align}
The von-Neumann equation is a complex equation and we can take its complex
conjugate:
\begin{align}\label{Q25}
C: && \del_t \rho^* = i[H^*,\rho^*], && H\to - H^*.
\end{align}
For the second expression in eqs.~\eqref{Q24}, \eqref{Q25} we perform the
transformation by a transformation of the Hamiltonian, keeping the structure
\eqref{Q23} of the von-Neumann equation fixed. We observe that for $H \to - H_*$
the term $\sim b_2$ is invariant. This reflects that the evolution for rotations
in the (1-3)-plane involves only real quantities.

Finally, a reflection at the (1-3) plane changes the sign of $\rho_2$, such that
$\rho$ is replaced by $\tilde{\rho}$ with opposite sign of $\rho_2$. Keeping the
form of the von Neumann equation fixed this changes $H \to \tilde{H}$, where
$\tilde{H}$ obtains from $H$ by changing the sign of $b_2$.
\begin{align}
P_2: && \del_t \trho = -i[H, \trho], && H \to \tilde{H}.
\end{align} 
From eq.~\eqref{5.22} we conclude
\begin{equation}
\tilde{H} = H^*.
\end{equation}
As a result, one finds for the combination $CP_2$
\begin{equation}
CP_2: H \to -H.
\end{equation}
This is the same as for time reversal. The von-Neumann equation is invariant
under the combined transformation $CP_2 T$.

The transformation $C$ is the analogue of charge conjugation in particle
physics which involves a complex conjugation. Similarly, the transformation
$P_2$ is a particular version of the parity transformation. We may define parity
as the reflection $P=P_1 P_2 P_3$, with $P_1$ a reflection at the (2-3) plane,
and $P_3$ a reflection at the (1-2) plane. Acting on $\rho_k$ the three
reflections commute. The combination $P_1 P_3$, i.e. $\rho_1 \to -\rho_1$,
$\rho_3 \to -\rho_3$ is a rotation in the (1-3) plane. Thus $P_2$ is equivalent
to $P$ up to a rotation. We conclude that the quantum subsystem is invariant
under a type of CPT-transformation, similar to the situation in particle
physics. Complex conjugation is directly linked to the discrete transformation
$P_2 T$. This reveals a relation between the complex structure in quantum
mechanics and discrete reflections in time.

\subsection{Quantum mechanics in continuous time}
\label{sec:quantum_mechanics_in_continuous_time}

So far we have mainly described discrete quantum mechanics for which the
evolution is described in discrete time steps. A unitary step evolution operator
maps wave function and density matrix at time $t$ to a subsequent time
$t+\varepsilon$. In general, we use quantum mechanics in a continuous version,
with dynamics described by the Schrödinger- or von-Neumann-equation. This
corresponds to the continuum limit $\varepsilon\to 0$ at fixed time intervals
$\Delta t$. We have taken this continuum limit for the examples of the preceding
sections and generalize it here.

\paragraph*{Hamilton operator}

Discrete quantum mechanics is formulated with discrete evolution steps between
$t$ and $t+\epsilon$, given by a unitary matrix $U(t)$. This can be translated
to a Hamiltonian formulation which can be employed for a possible continuum
limit for $\eps\to0$. The evolution is then described by the continuous
Schrödinger equation
\begin{equation}
i \partial_t \psi(t) = H(t) \psi(t).
\label{eq:QC6}
\end{equation}
For the definition of the Hamiltonian one uses
\begin{equation}
U(t) = e^{-i\epsilon H(t)}\ ,\quad H(t) = \frac{i}{\epsilon} \ln
U(t),\quad H^\dagger = H\ .
\label{eq:QC7}
\end{equation}
For the definition \eqref{eq:QC7} even a real (e.g. orthogonal) evolution
operator can yield a complex matrix $-iH$, such that a complex wave
function is needed \cite{GTH}. The solution of eq.\,\eqref{eq:QC7} may not be
unique and require sometimes some effort to be found
\cite{GTH1,GIK,TH,EL1,ELZE,EL2,EL3,GTH2,EL4}.

Since the direct construction~\eqref{eq:QC7} of the Hamiltonian $H$ can
sometimes be cumbersome in practice, one may define an approximation $\bar{H}$
which coincides with $H$ in the continuum limit. For this purpose one defines
for the discrete evolution the operator $G(t)$,
\begin{equation}
G(t) = \frac{i}{2\epsilon} [ U(t) - U^\dagger(t-\epsilon) ]\ .
\label{eq:QC1}
\end{equation}
It results in a discrete Schrödinger equation for the quantum wave function
\begin{equation}
\frac{i}{2\epsilon} [ \psi(t+\epsilon) - \psi(t-\epsilon) ] = G(t) \psi(t).
\label{eq:QC2}
\end{equation}
Splitting $G$ into an Hermitian and anti-Hermitian part
($\bar{H}^\dagger=\bar{H}$, $J^\dagger=J$),
\begin{equation}
G(t) = \bar{H}(t) + iJ(t),
\label{eq:QC3}
\end{equation}
yields
\begin{equation}
\bar{H}(t) = \frac{i}{4\epsilon} [ U(t) + U(t-\epsilon) - U^\dagger(t) -
U^\dagger(t-\epsilon) ],
\label{eq:QC4}
\end{equation}
and
\begin{equation}
J(t) = \frac{1}{4\epsilon} [ U(t) - U(t-\epsilon) + U^\dagger(t) -
U^\dagger(t-\epsilon) ].
\label{eq:QC5}
\end{equation}
A consistent continuum limit for $\epsilon\to 0$ requires that $J(t)$ vanishes
in this limit.

In the continuum limit $\bar{H}(t)$ and $H(t)$ coincide, as can be seen by
expanding
\begin{equation}
U(t) = 1 - i\epsilon H(t) - \frac{\epsilon^2}{2} H^2(t) + ...,
\label{eq:QC8}
\end{equation}
for which one finds
\begin{align}
\begin{split}
H(t) &= \frac{1}{2} [ H(t) + H(t-\epsilon) ] +
\mathcal{O}(\epsilon^2), \\
J(t) &= -\frac{\epsilon}{4} [ H^2(t) - H^2(t-\epsilon) ] +
\mathcal{O}(\epsilon^3).
\end{split}
\label{eq:QC9}
\end{align}
In the continuum limit we can write
\begin{equation}
H(t) = i\partial_t U(t) U^\dagger(t).
\label{eq:QC10}
\end{equation}

\paragraph*{Quantum systems from motion in internal space}

In the continuum limit both the quantum clock system in
sect.\,\ref{sec:quantum_clock_system} and the one qubit quantum mechanics in
sect.\,\ref{sec:classical_ising_spins_and_quantum_spin},
\ref{sec:unitary_evolution_4_5_2} can be associated to the motion in an
appropriate geometry. We may associate this geometry to some type of "internal
space". A quantum particle with spin would then be described by its motion in
ordinary (external) space plus a motion in internal space. The focus on the spin
leading to qubits forgets about the motion in external space and only retains
the motion in internal space. This motion uses a probabilistic description by a
Liouville-type equation that can be associated to a very simple deterministic
Newton-type equation. For particular choices of the probability distributions
solving the Liouville-type equation the probabilistic information has the
properties which allow a reduction to a simple quantum subsystem.

The quantum clock system can be associated to the motion on a circle with
constant velocity. For a particle at a sharp position on the circle or at a
sharp angle $\beta$ the deterministic motion is simply
\begin{equation}
\beta(t) = \beta_0 + \omega t.
\label{eq:QC12}
\end{equation}
The probability distribution associated to a particle centered around $\beta$ is
given by $p_\beta(\varphi)$. It could be realized by a probabilistic
distribution of initial conditions for sharp particles, but there is actually no
need for this in our genuinely probabilistic description of the world. The
evolution of the probability distribution obeys the Liouville-type equation
\eqref{CV27}
\begin{equation}
\partial_t p_\beta(\varphi) = -\omega \partial_\varphi p_\beta(\varphi).
\label{eq:QC13}
\end{equation}
For this particularly simple motion no phase-space description with momentum is
needed as for the usual Liouville equation -- the evolution of
$p_\beta(\varphi)$ is closed.

The general solution of eq.\,\eqref{eq:QC13} reads
\begin{equation}
p_\beta(\varphi;t) = p_\beta(\varphi-\omega t).
\label{eq:QC14}
\end{equation}
For an initial condition for which $p_\beta(\varphi;t_0)$ depends only on
$\varphi-\beta_0$ this yields
\begin{equation}
p_\beta(\varphi;t) = p_\beta(\varphi-\beta(t))
\label{eq:QC15}
\end{equation}
with $\beta(t)$ given by eq.\,\eqref{eq:QC12}. If the initial condition is given
more specifically by the particular form \eqref{CV31},
\begin{equation}
p_\beta(\varphi;t) = \frac{1}{2} \cos(\varphi-\beta_0)
\Theta(\cos(\varphi-\beta_0)),
\label{eq:QC16}
\end{equation}
the probabilistic information allows for a map to a quantum subsystem from which
all expectation values of the Ising spins \eqref{CV6} in different directions
can be computed. The quantum subsystem provides for a conceptually much simpler
description of the time evolution of the expectation values of Ising spins.
Instead of following the evolution of a whole function $p(\varphi;t)$, it is now
sufficient to investigate the evolution of a real two component wave function
$\psi(t)$ or real $2\times 2$ density matrix \eqref{CV48} with the Hamiltonian
\eqref{CV49}. We emphasize that the map to the simple quantum subsystem is only
possible for the particular initial condition \eqref{eq:QC16}.

The one-qubit quantum system corresponds to the time evolution of the
probability distribution for a particle moving on a sphere. For a given constant
unit vector $b$ the probability distribution differs from zero for a rotating
half-sphere whose direction is perpendicular to $b$. This half-sphere rotates
with angular frequency $\omega$ around the axis $b$. We can imagine particles on
this half-sphere with trajectories rotating around the axis $b$. The
corresponding Liouville-type equation describes the associated rotation of the
probability distribution. Again, a particular initial condition for this
probability distribution is needed in order to allow for the construction of a
quantum subsystem. If $\gamma$ and $\omega$ depend on $t$, the direction and
frequency of the rotations of all particles change with $t$. As compared with
the classical Liouville-type evolution of probability distributions for
particles on a sphere, the quantum evolution with a arbitrary Hamiltonian $H(t)$
is an important simplification. Still the Liouville-type probabilistic
description of ``classical particles'' is useful for an understanding of the
origin of the quantum rules from the basic rules of ``classical statistics''.

The ``internal space" and the particular form of the probability distribution
which obeys the quantum constraint may not appear very natural. We do not
consider these constructions as fundamental. On the fundamental side a particle
is an excitation of the vacuum of a quantum field theory, and a qubit is a
subsystem of a particle. Our constructions should rather serve as concrete
examples that it is possible to obtain quantum subsystems from a classical
probabilistic setting. Any direct explicit construction demonstrates that "no go
theorems" for the embedding of quantum mechanics in classical statistics are
circumvented. The main reason is the incomplete statistics of the quantum
subsystem for which classical correlation functions are not accessible. In
addition, our construction highlights an important point, namely that even the
quantum mechanics of a single qubit needs infinitely many classical bits or
classical continuous variables. The central reason is that a quantum spin in an
arbitrary direction involves infinitely many observables with a discrete
spectrum, corresponding to the infinity of possible directions.

\paragraph*{Single fermion}

The one-qubit quantum system can also describe a single fermionic excitation. A
single fermion is a two-level quantum system, the two states corresponding to
the occupation number one or zero. It can therefore be described by a qubit.
For the wave function of a pure state we can employ a basis of an empty and an
occupied state,
\begin{equation}
\ket{1} = \begin{pmatrix}
1 \\ 0
\end{pmatrix},\quad 
\ket{0} = \begin{pmatrix}
0 \\ 1
\end{pmatrix}\ .
\label{eq:QC19}
\end{equation}
For $\gamma=(0,0,-1)$ the Hamiltonian,
\begin{equation}
H = \frac{\omega}{2} \begin{pmatrix}
1 & 0 \\ 0 & -1
\end{pmatrix}
= \omega \left(n-\frac{1}{2}\right)\ ,
\label{eq:QC17}
\end{equation}
can be expressed in terms of the occupation number operator $n$,
\begin{equation}
n = \begin{pmatrix}
1 & 0 \\ 0 & 0
\end{pmatrix}.
\label{eq:QC18}
\end{equation}

We may introduce fermionic annihilation and creation operators
\begin{equation}
a = \begin{pmatrix}
0 & 0 \\ 1 & 0
\end{pmatrix},\quad
a^\dagger = \begin{pmatrix}
0 & 1 \\ 0 & 0
\end{pmatrix},
\label{eq:QC20}
\end{equation}
with
\begin{equation}
a\ket{0} = 0,\quad a\ket{1} = \ket{0},\quad a^\dagger \ket{0} = \ket{1},\quad
a^\dagger \ket{1} = 0,
\label{eq:QC21}
\end{equation}
and
\begin{equation}
n = a^\dagger a,\quad \{ a^\dagger, a \} = 1.
\label{eq:QC22}
\end{equation}
The Hermitian linear combinations of $a$ and $a^\dagger$ are expressed by the
quantum spin operators
\begin{equation}
a + a^\dagger = S_1 = \tau_1,\quad i(a-a^\dagger) = S_2 = \tau_2.
\label{eq:QC23}
\end{equation}
For a given initial state the evolution of their expectation values can be
computed from the Schrödinger or von-Neumann equation.

\subsection{Quantum particle in harmonic\\potential}
\label{subsec:quantum_particle_in_harmonic_potential}

In this section we construct a probabilistic automaton which realizes a quantum
subsystem for a quantum particle in a harmonic potential. Our starting point is
the Liouville equation for a classical particle in a harmonic potential. The
evolution is given by a unique jump step evolution operator. The overall
probability distribution exists as for all probabilistic automata. The subsystem
is described by a complex wave function which obeys the Schrödinger equation for
a quantum particle in the same potential as the one for the classical particle.
We express the quantum observables which are represented by the familiar
non-commuting quantum operators for position, momentum and energy as observables
for the classical probabilistic automaton. For energy and momentum these are
statistical observables.

The harmonic potential is special since particles with arbitrary initial
conditions oscillate with the same frequency $\omega = \sqrt{c/m}$. We therefore
expect periodicity in the evolution of the probability distribution with
frequency $\omega$. The map to the quantum subsystem will reveal that suitable
initial probability distributions lead to a periodic evolution with frequencies
$n\omega$ for every integer $n$. This corresponds to the equidistant energy
spectrum of the quantum Hamiltonian. We will also discuss briefly the Liouville
equation for an anharmonic potential. The quantum formulation again predicts the
existence of periodic evolution for suitable initial probability distributions.

\paragraph*{Liouville equation for the classical wave function}

Let us consider a probabilistic classical particle in a harmonic potential,
\begin{equation}
\label{PP1}
V=\frac{c}{2}z_kz_k\ .
\end{equation}
We can describe the time evolution of the probability in phase space $w(z,p)$ in
terms of the Liouville equation for the classical wave function $\phi_c(z,p)$
\begin{align}
\label{PP2}
\partial_t\phi_c=&\,-\hat L\phi_c\ ,\quad w(z,p)=\phi_c^2(z,p)\ ,\nn\\
\hat L=&\,\frac{p_k}{m}\frac{\partial}{\partial
z_k}-cz_k\frac{\partial}{\partial p_k}\ .
\end{align}
(As compared to eq.~\eqref{CV37} we have replaced $\vec{x}\to\vec{z}$ and use
the specific potential~\eqref{PP1}.) The formulation in terms of the classical
wave function allows us to perform a Fourier transform to a "double position
basis", with $z=(x+y)/2$,
\begin{equation}
\label{PP3}
\tilde\psi_c(x,y)=\int_pe^{ip(x-y)}\phi_c\left(\frac{x+y}{2},p\right) =
\tilde\psi^*_c(y,x)\ .
\end{equation}
In this basis the time evolution equation~\eqref{PP2} takes the form
\begin{align}
\label{PP4}
\partial_t\tilde\psi_c=&\,-i(H_Q-\tilde H_Q)\tilde\psi_c\ ,\nn\\
H_Q=&\,-\frac{1}{2m}\frac{\partial}{\partial x_k}\frac{\partial}{\partial x_k} +
\frac{c}{2}x_kx_k\ ,\nn\\
\tilde H_Q=&\,-\frac{1}{2m}\frac{\partial}{\partial y_k}
\frac{\partial}{\partial y_k} + \frac{c}{2}y_ky_k\ .
\end{align}
This is already suggestive for the quantum particle in a harmonic potential.

\paragraph*{Quantum subsystem}

We next restrict the classical wave function $\tilde\psi_c$ to a direct product
form
\begin{equation}
\label{PP5}
\tilde\psi_c(x,y)=\psi_Q(x)\psi_Q^*(y)\ .
\end{equation}
This restriction encodes the quantum constraint which will allow the map to a
quantum subsystem. The direct product form is consistent with the evolution if
$\psi_Q(t,x)$ obeys the Schrödinger equation
\begin{equation}
\label{PP6}
i\partial_t\psi_Q(t,x)=H_Q\psi_Q(t,x)\ .
\end{equation}
The normalization of $\tilde\psi_c(x,y)$ is guaranteed by
\begin{equation}
\label{PP7}
\int_x\psi_Q^*(x)\psi_Q(x)=1\ .
\end{equation}
For $\tilde\psi_c$ obeying the quantum constraint~\eqref{PP5} we can perform the
map to a quantum subsystem by ``integrating" over the $y$-position. For this
purpose one defines the classical density matrix for the pure state classical
wave function~\eqref{PP3}
\begin{equation}
\label{PP8}
\rho_c(x,y;x',y')=\tilde\psi_c(x,y)\tilde\psi_c^*(x',y')\ .
\end{equation}
The coarse graining
\begin{equation}
\label{PP9}
\rho_Q(x,x')=\int_y\rho_c(x,y;x',y)\ ,
\end{equation}
leads for the quantum constraint~\eqref{PP5} to the pure state quantum density
matrix
\begin{equation}
\label{PP10}
\rho_Q(x,x')=\psi_Q(x)\psi_Q^*(x')\ .
\end{equation}
We recognize $\psi_Q(t,x)$ as the complex wave function of the quantum
subsystem. With eqs.~\eqref{PP6},~\eqref{PP4} its evolution equation is the
Schrödinger equation for a quantum particle in the harmonic
potential~\eqref{PP1}.

To summarize, the classical probabilistic Liouville equation for a classical
particle in a harmonic potential is mapped to a quantum subsystem for a quantum
particle in the same harmonic potential. This map requires as a condition or
quantum constraint the factorized form~\eqref{PP5} of the classical wave
function. We may translate the quantum condition~\eqref{PP5} to the equivalent
condition for the classical wave function in phase space ($r=x-y$)
\begin{equation}
\label{PP11}
\phi_c(z,p)=\int_re^{-ipr}\psi_Q\left(z+\frac{r}{2}\right)
\psi_Q^*\left(z-\frac{r}{2}\right)\ .
\end{equation}
The associated classical probability distribution $w(z,p)$ follows by
$w(z,p)=\phi_c^2(z,p)$. This family of probability distributions covers only a
part of the most general $w(z,p)$. It is parameterized by the quantum wave
function $\psi_Q(x)$.

One can now take over all results for a quantum particle in a harmonic
potential, as the discrete equidistant spectrum of a quantum energy observable
associated to the operator $H_Q$, the conserved quantum angular momentum and so
on. Every result for the quantum wave function $\psi_Q(x)$ can be taken over to
the corresponding probability distribution $w(z,p)$ by use of eq.~\eqref{PP11}.
The equivalence can be verified directly by following the evolution of $w(z,p)$
according to the Liouville equation.

In summary, we have here a simple example how a rather standard evolution of a
classical probability distribution can describe the unitary quantum evolution of
an appropriate subsystem. This phenomenon has been observed first in the
investigation of the evolution of correlation functions in simple classical
field theories~\cite{CWQMTE}.

\paragraph*{Quantum operators as statistical observables in classical
statistical systems}

We may also employ the quantum position and momentum operators
\begin{equation}
\label{PP12}
\hat X_{Q,k}=x_k\delta(x-x')\ ,\quad \hat P_{Q,k}=-i\frac{\partial}{\partial
x_k}\delta(x-x')\ .
\end{equation}
They obey the usual commutation relation,
\begin{equation}
\label{PP13}
\big[\hat X_{Q,k},\hat P_{Q,l}\big]=i\delta_{kl}\ ,
\end{equation}
and
\begin{equation}
\label{PP14}
H_Q=\frac{1}{2m}\hat P_{Q,k}\hat P_{Q,k} + \frac{c}{2}\hat X_{Q,k}
\hat X_{Q,k}\ .
\end{equation}
The action on the classical wave function $\phi_c(z,p)$ is given by
\begin{align}
\label{PP15}
\hat X_{Q,k}\phi_c(z,p)=&\,\left( z_k+\frac{i}{2}\frac{\partial}{\partial p_k}
\right)\phi_c(z,p)\ ,\nn\\
\hat P_{Q,k}\phi_c(z,p)=&\,\left( p_k-\frac{i}{2}\frac{\partial}{\partial z_k}
\right)\phi_c(z,p)\ .
\end{align}
For a real classical wave function they are not well defined.

In order to have observables associated to $\hat X_Q$ and $\hat P_Q$ we can
extend our model by adding a discrete property to the particle. For example, the
particle may be red or green. This remains a perfectly valid setting for a
probabilistic classical particle. The probability distribution in phase space
has now an additional index, $w_i(z,p)$, with $i=1$ for the green particle and
$i=2$ for the red one. The same holds for the classical wave function
$\phi_{c,i}(z,p)$. The Liouville equation does not affect the internal index
$i$. This setting allows for the introduction of a simple complex structure
defined by the complex wave function
\begin{equation}
\label{PP16}
\phi_c(z,p)=\phi_{c,1}(z,p)+i\phi_{c,i}(z,p)\ .
\end{equation}
The steps to the quantum subsystem can be done in a similar way as before. In
eq.~\eqref{PP3} $\tilde\psi_c(x,y)$ is no longer automatically identified with
$\tilde\psi_c^*(y,x)$. The quantum constraint can now be extended to a larger
class
\begin{equation}
\label{PP17}
\tilde\psi_c(x,y)=\psi_Q(x)\tilde\psi_Q^*(y)\ ,
\end{equation}
where $\tilde\psi_Q(y)$ is any normalized solution of the Schrödinger equation,
\begin{equation}
\label{PP18}
i\partial_t\tilde\psi_Q(t,y)=\tilde H_Q\tilde\psi_Q(t,y)\ ,
\end{equation}
and no longer related to $\psi_Q(y)$. The complex wave function $\phi_c(z,p)$
which obeys the quantum constraint replaces in eq.~\eqref{PP11} the last factor
$\psi_Q^*\big(z-\frac{r}{2}\big)$ by $\tilde\psi_Q^*\big(z-\frac{r}{2}\big)$.

For a complex classical wave function the operator expressions~\eqref{PP15} have
a well defined meaning. We can evaluate mean values as
\begin{equation}
\label{PP19}
\langle X_{Q,k}\rangle=\int_{z,p}\phi_c^*(z,p)\left(
z_k+\frac{i}{2}\frac{\partial}{\partial p_k} \right)\phi_c(z,p)\ .
\end{equation}
The expectation value of $X_Q$ evaluated from the classical wave function in
phase space by eq.~\eqref{PP19} agrees with the one evaluated from the quantum
subsystem, provided that $\phi_c(z,p)$ obeys the quantum
constraint~\eqref{PP17}. We observe that $X_Q$ differs from the classical
position observable for which the piece $(i/2)(\partial/\partial p_k)$ is
absent. Due to the momentum derivative it has no fixed value for the points
$(z,p)$ in phase space. It is rather a statistical observable which reflects in
parts properties of the probabilistic information.

The expectation values of arbitrary functions of $X_Q$ and $P_Q$, defined by
Hermitian functions of the associated operators, $\hat F(\hat X_Q,\hat P_Q)=\hat
F^\dagger(\hat X_Q,\hat P_Q)$, can be evaluated from the classical wave function
\begin{equation}
\label{PP20}
\langle\hat F(\hat X_Q,\hat P_Q)\rangle = \int_{z,p}\phi_c^*(z,p) \hat F(\hat
X_Q,\hat P_Q)\phi_c(z,p)\ .
\end{equation}
In particular, this concerns the quantum energy operator $H_Q$,
\begin{align}
\label{PP21}
H_Q=&\,\frac{1}{2m}\left( p_kp_k-\frac14\frac{\partial}{\partial
z_k}\frac{\partial}{\partial z_k} - ip_k\frac{\partial}{\partial z_k}
\right)\nn\\
&+\frac{c}{2}\left( z_kz_k-\frac14\frac{\partial}{\partial
p_k}\frac{\partial}{\partial p_k} + iz_k\frac{\partial}{\partial p_k} \right)\ .
\end{align}
This operator commutes with the Liouville operator~\eqref{PP2}
\begin{equation}
\label{PP22}
\big[\hat L, H_Q\big]=0\ .
\end{equation}
This concludes the construction of an embedding of the quantum subsystem for a
quantum particle in a harmonic potential into a classical statistical model.

\paragraph*{Quantum results for classical statistical system}

One may ask if in turn the existence of the quantum subsystem can be useful for
the understanding of the probabilistic evolution of a classical particle with
two colors. The answer is positive. The quantum subsystem allows us to identify
particular initial probability distributions which follow periodic oscillations
between the red and green particle. This periodicity is a typical quantum
feature not present for general classical statistical systems. For the harmonic
potential the new features concern the oscillations with frequencies $n\omega$,
$n>1$.

Another important point is the discovery of new conserved observables as the
quantum energy $H_Q$. The observable~\eqref{PP21} is defined for arbitrary
complex classical wave functions $\phi_c(z,p)$. The commutation
relation~\eqref{PP22} does not involve the quantum constraint. Thus $H_Q$ is a
conserved quantity for arbitrary complex $\phi_c$, and similar for $\tilde H_Q$.
For classical wave functions obeying the quantum constraint the possible
measurement values of $H_Q$ exhibit the familiar discrete equidistant spectrum.
Similar remarks apply to the quantum angular momentum. One easily constructs
complex classical wave functions $\phi_c$ which are eigenfunctions for these
observables. They follow from the eigenfunctions in the quantum system by
eq.~\eqref{PP11}. The corresponding periodic evolution in time for the classical
probabilities $w_i(t,z,p)$ may not easily be guessed without the insight of the
quantum subsystem.

Let us focus on complex classical wave functions which obey the quantum
constraint, with $\psi_Q$ an eigenstate of $H_Q$ with energy $E_n$, and
$\tilde\psi_Q$ and eigenstate of $\tilde H_Q$ with energy $\tilde E_{n'}$,
\begin{align}
\label{PP23}
\psi_Q=&\,\psi_n(x)\exp\big(-iE_nt\big)\ ,\quad
\tilde\psi_Q=\psi_{n'}(y)\exp\big(-i\tilde E_{n'}t\big)\ ,\nn\\
E_n=&\,\left(n+\frac12\right)\omega\ ,\quad \tilde
E_{n'}=\left(n'+\frac12\right)\omega\
,\quad \omega^2=\frac{c}{m}\ .
\end{align}
These classical wave functions show a periodic evolution
\begin{align}
\label{PP24}
&\phi_c(z,p)=\nn\\
&\int_re^{-ipr}\psi_n\left(z+\frac{r}{2}\right)
\psi_{n'}^*\left(z-\frac{r}{2}\right) \exp\left\{-i\omega(n-n')t\right\}\ .
\end{align}
This periodicity is taken over to the classical probability distribution in
phase space $w_i(z,p)$. One can explicitly start with a probability distribution
in phase space at $t=0$ according to eq.~\eqref{PP24}, with
$w_1(z,p)=\big(\text{Re}\phi_c(z,p)\big)^2$,
$w_2(z,p)=\big(\text{Im}\phi_c(z,p)\big)^2$ and follow the evolution according
to the Liouville equation. This will reveal periodic oscillations between the
two colors, reflecting the properties of the quantum subsystem.

One may ask more generally if the local-time probability distribution for the
two-color classical particle in a harmonic potential admits solutions with a
periodic time evolution. For this purpose we write the Liouville equation for
the complex wave function $\phi_c(z,p)$ in a Hamiltonian form
\begin{align}
\label{PP25}
i\partial_t\phi_c(t;z,p)=&\,H_L\phi_c(t;z,p)\ ,\nn\\
H_L=&\,-i\hat L=-i\frac{p_k}{m}\frac{\partial}{\partial z_k}+icz_k
\frac{\partial}{\partial p_k} = H_L^\dagger\ .
\end{align}
A periodic evolution is realized for eigenstates of $H_L$. With eq.~\eqref{PP4}
\begin{equation}
\label{PP26}
H_L=H_Q-\tilde H_Q\ ,
\end{equation}
the spectrum of eigenvalues $E_L$ of $H_L$ follows from the discrete spectrum of
the quantum Hamiltonian,
\begin{equation}
\label{PP27}
E_L(n,n')=\omega(n-n')\ .
\end{equation}
Thus the periods of the possible oscillations are fixed by the eigenvalues of
the quantum energy operator. The corresponding oscillating probability
distributions are determined by the quantum
eigenstates~\eqref{PP23},~\eqref{PP24}.
The quantum subsystem is central for identifying and understanding the
periodic probability distributions!

\paragraph*{Quantum system for Liouville equation with\\anharmonic potential}

The Liouville equation for arbitrary potentials constitutes a probabilistic
automaton and describes therefore a quantum system. In general, this quantum
system may differ from the one of a quantum particle in a potential. Still,
there exists a Hamiltonian with a spectrum of eigenvalues. The corresponding
eigenfunctions will evolve periodically.

We demonstrate this for simplicity for one space dimension, with Liouville
equation
\begin{equation}
\label{eq:ANN1}
\partial_t\phi_c = -\hat L\phi_c = \left( -\frac{p}{m}\partial_z + (cz+bz^3)
\partial_p \right)\phi_c\,.
\end{equation}
Performing the Fourier transformation~\eqref{PP3} one arrives at
\begin{align}
\label{eq:ANN2}
i\partial_t\tilde\psi_c =&\, H_L\tilde\psi_c\,,\nonumber\\
H_L =&\, -\frac{1}{2m}(\partial_x^2-\partial_y^2) + \frac{c}{2}(x^2-y^2)
\nonumber\\
&+ \frac{b}{8}(x^2-y^2)(x+y)^2\,.
\end{align}
The Hamiltonian $H_L$ is unbounded, similar to the harmonic potential. It can,
however, no longer be decomposed into two pieces involving only the coordinates
$x$ or $y$, respectively. The anharmonic piece $~\sim b$ couples the motion in
the $x$- and $y$-directions. Determining the spectrum of $H_L$ and its
eigenfunctions will reveal periodic probability distributions.

\subsection{Dynamical selection of quantum\\subsystems}
\label{sec:dynamic_selection_of_quantum_subsystems}

Quantum systems are ubiquitous in Nature. All what we observe is governed by
quantum mechanics. In our concept of a probabilistic world quantum systems are
particular subsystems of more general ``classical'' probabilistic systems. This
raises the question: ``what singles out quantum systems?''. Is the formulation
of quantum systems just a particular choice of structures between observables
that we use for the description of the world, and the associated choice of an
overall probability distribution? Or are quantum systems singled out dynamically
by the time evolution over many time steps, even if some initial time-local
probability distribution does not describe a quantum system? In this section we
argue that quantum systems are indeed selected by the dynamical evolution in the
large time limit. If initial conditions are set in the infinite past, the
distance to the present involves infinite time. Only quantum systems ``survive''
in this limit.

Our setting of a probabilistic world not only contains the possibility of
quantum systems. It could give a fundamental explanation why our world is
described by quantum physics. ``Classical'' probabilistic systems describe the
overall probabilistic system of the whole Universe. The time-local subsystem at
``finite time'', separated from the initial time in the infinite past by an
infinite time interval, contains a quantum subsystem for which the probabilistic
information is preserved. All relevant dynamics is related to the probabilistic
information of this quantum subsystem. A possible environment of the quantum
subsystem plays no longer a role. In turn, time-local quantum systems can have
subsystems for which the probabilistic information in the subsystem is not
conserved. Such subsystems are not quantum systems, but more general
probabilistic systems. The notions of quantum systems and ``classical''
probabilistic systems are intrinsically related. Which aspect matters depends on
the subsystem under consideration.

\paragraph*{Conservation of information}

General quantum systems have the property that they are time-local subsystems
for which the initial probabilistic information is preserved. This corresponds
to an orthogonal step evolution operator. In the presence of an appropriate
complex structure the evolution is unitary. This property of an unitary or
orthogonal evolution does not have to hold for the complete time-local
subsystem. It is sufficient that it holds for an appropriate closed subsystem.

Consider the evolution of time-local subsystems with a step evolution operator
that is not orthogonal. The step evolution operator may have a set of maximal
eigenvalues $|\lambda_i|=1$, and another set of eigenvalues $\lambda_j$ with
$|\lambda_j|<1$. Expanding the classical wave function in eigenfunctions of the
step evolution operator, all eigenfunctions to eigenvalues $|\lambda_j|<1$ will
approach zero as time progresses. Only the eigenfunctions to the maximal
eigenvalues survive for infinite time. This reduces the time-local system to a
subsystem for which the step evolution operator becomes orthogonal. The dynamics
therefore selects systems for which the information in the classical wave
functions is preserved~\cite{CWPW}.

This dynamical selection leads to a subsystem for which all eigenvalues of the
step evolution operator obey $|\lambda_i|=1$. There are two possible outcomes.
Either one has $\lambda_i =1$ for all eigenvalues. In this case the time-local
subsystem approaches some type of equilibrium state which is static in the sense
that the classical wave functions and density matrix become independent of $t$.
The evolution stops sufficiently far away from the boundaries. For boundaries in
the infinite past and future this leads to a world without evolution. For the
second alternative some of the eigenvalues differ from one, $\lambda\neq 1$,
while $|\lambda_i|=1$. The eigenvalues are characterized by non-trivial phases,
$\lambda_i = e^{i\alpha_i}$. In this case one observes a non-trivial evolution
even arbitrarily far away from the boundaries. At this point we may formulate a
simple postulate: The presence of our world is characterized by evolution. This
is meant in the sense of a non-trivial evolution, with some phases $\alpha_i\neq
0$. Strictly speaking, this is not a postulate about the structure of a
probabilistic description of the world. Since we know that structures among
observables and associated overall probability distributions with a non-trivial
evolution of the time-local subsystem exist, the postulate is rather a decision
for the choice of these structures for an efficient description of the world.

\paragraph*{Quantum systems and general information\\preserving systems}

Our postulate selects for the present world time-local subsystems for which the
local probabilistic information in the classical wave functions and density
matrix is preserved. These subsystems follow an orthogonal evolution. There are
many such systems that we may not immediately associate with quantum
systems. All unique jump step evolution operators have this property. This
includes all discrete cellular automata and all systems characterized by
deterministic evolution equations for a classical particle in the phase space of
position and momentum. In fact, all those systems can be viewed as discrete
quantum subsystems in a real formulation. If, in addition, a complex structure
exists which is compatible with the evolution, the usual complex formulation of
quantum mechanics can be implemented. The unitary transformation guarantees the
existence of a Hermitian Hamiltonian, even though it may sometimes be difficult
to find its explicit form. Furthermore, some of these discrete quantum systems
may not admit a smooth continuum limit.

Nothing prevents us from choosing a description of the world with step evolution
operators that are unique jump operators. For such a description the evolution
is deterministic. The probabilistic aspects enter only through the probabilistic
boundary condition. All eigenvalues of the step evolution operator obey
$\lambda_i = e^{i\alpha_i}$, and we choose systems with some $\alpha_i\neq 0$.
For our description of a continuous clock system or a classical probabilistic
system for the one-qubit quantum systems arbitrary time-local probability
distributions $p(\varphi;t)$ obey our postulate and follow an orthogonal
evolution. The question is then raised if there exists some dynamical selection
process that leads for a subsystem to the particular shape of
$p_\beta(\varphi;t)$ or $p(\varphi,\rho;t)$ given by eq.\,\eqref{CV31} or
eq.\,\eqref{Q2}, that allows for the formulation of simple quantum subsystems.

As a first important observation we notice that every deterministic or unitary
evolution formally preserves the initial information completely, while in
practice part of the information is lost. An example is the approach to a
thermal equilibrium state for a system of a great number of interacting
classical particles. The preserved information is shuffled to $n$-point
functions with very high $n$, while the $n$-point functions with low $n$ all
reach their thermal equilibrium values. We may sharpen our postulate in the
sense that we focus on overall probability distributions for which the present
shows a non-trivial evolution of expectation values, propagators, or $n$-point
correlation functions with low $n$. This restriction favors a dynamical
selection of quantum subsystems
in the common sense for which periodic behavior becomes, in principle,
observable.

\paragraph*{Dynamical selection of atoms}

For the bottom-up approach followed in this part of our investigation the simple
question ``why do we observe identical atoms following a quantum evolution''
remains an open issue. A possible answer by dynamical selection would have to be
on the level of subsystems for individual atoms. We believe that the answer to
this question is of a more global nature by the dynamical selection of a quantum
field theory. The fact that the parameters for all atoms, as the fine structure
constant or the ratio of electron to proton mass, are precisely the same for all
atoms, and all atoms in a given quantum state are identical, points to the
global answer in terms of a quantum field theory. If a quantum field theory and
a corresponding vacuum are selected by the evolution from the infinite past to
the present, all excitations as elementary particles or atoms are indeed
identical.

At the end, our proposal for an explanation of the ubiquitous quantum systems in
our world states that quantum field theories are well suited for the
organization of the probabilistic information in our world. They are robust due
to universal properties of their long-distance behavior\,\cite{CWGEO}. Quantum
field theories contain as subsystems identical single atoms, or the single
quantum spins.

\subsection{Particle-wave duality}
\label{sec:particle_wave_duality}

In the beginning of quantum mechanics particle-wave duality was considered as a
great mystery. Light from a very distant star passes through the lenses of a
telescope according to the laws of wave propagation. If the intensity is very
low, single photons can be counted as hits of particles in light-detectors. How
can an object be simultaneously a discrete particle and a continuous wave? In
our probabilistic description of the world the answer is very simple. Many
observables correspond to discrete yes/no-decisions. Does a particle detector
fire or not? Such two-level observables or Ising spins have discrete possible
measurement values: yes or no, $+1$ or $-1$. This is the particle side of
events.

On the other hand, dynamics and evolution are described by the propagation of
probabilistic information. This allows one to compute at every time the
probabilities to find $+1$ or $-1$ for a two-level observable. The probabilistic
information is encoded in the form of classical or quantum wave functions, the
density matrix or the probability distribution. All these objects are
continuous, given by real or complex numbers that depend on $t$. Furthermore,
the wave function and the density matrix obey a linear evolution law. This holds
both for classical and for quantum wave functions, and the corresponding
classical or quantum density matrices. For a linear evolution law the
superposition principle for possible solutions holds, as typical for the
propagation of waves. Particle-wave duality deals with discrete possible
outcomes of observations whose probabilities can be predicted by a linear
evolution law for continuous probability waves. The probability waves are
probability amplitudes, with probabilities given by a quadratic expression of
the amplitudes.

We may recall at this occasion the one-qubit quantum system of
sect.\,\ref{sec:classical_ising_spins_and_quantum_spin},
\ref{sec:unitary_evolution_4_5_2}. The quantum spins in different directions
correspond to discrete yes/no decisions if an event belongs to the associated
hemisphere or not. The evolution of the probabilistic information is given by the
Schrödinger equation for a continuous wave function.

Our probabilistic setting addresses also another apparent ``mystery''. Particles may
be located in small space regions. A very high resolution photon detector either
detects a particle or not. On the other hand, waves are typically much more
extended objects. Already the wave propagation inside the telescope involves
characteristic length scales of the size of the telescope, for example for
interference. In our picture there is no contradiction between very localized
observables (particles) and a much more extended character of the probabilistic
information and its evolution (waves).

\section{Correlated computing}
\label{sec:classical_and_quantum_computing}

Computing consists of a sequence of computational steps. Discrete
time steps transform the state of the system at $t$ to the state of the system
at $t+\varepsilon$. The formalism described in the present work is suitable for
a general description of computing. ``Time'' orders here the sequence of
computation steps and needs not to be identified with physical time. Each step
performs a particular operation on the state of the system, which consists of a
particular configuration for a sequence of bits or qubits. Since we describe
here qubits in terms of classical bits we can develop a unified approach to
classical and quantum computing.

As we have emphasized already, the crucial feature of quantum computing is the
large amount of correlations between the associated classical bits. These
correlations result from the quantum constraint of positivity of the density
matrix for the quantum subsystem. There may exist intermediate forms of
correlated computing which impose constraints on the probabilistic information
of the time-local subsystem leading to correlations among the classical bits.
These constraints may be weaker than the full quantum constraint. As a result,
such an intermediate system will not be able to perform the most general quantum
operations on many qubits. We explore here systems that only can perform parts
of the quantum operations or only quantum operations for a few qubits. On the
other hand, our systems do not involve a very large number of classical bits.

In particular, we are interested in the question if classical probabilistic
systems which are not under the extreme conditions of isolation of a quantum
computer, as for example artificial neural networks, neuromorphic computers or
the brain, can learn the changes of classical probability distributions needed
for the performance of certain quantum tasks. In this case we no longer deal
with simple probabilistic automata. For probabilistic automata the deterministic
updating of the probabilistic information severely restricts their capabilities.
More general rotations of the classical wave function beyond unique jump
operations offer a large spectrum of new possibilities. Our results establish
that such a learning is indeed possible.

In order to collect a few first examples for correlated computing with a not too
large number of classical bits we focus here on systems that perform simple
unitary quantum operations. The fact that quantum operations are performed
guarantees that the classical bits are indeed highly correlated. The field of
correlated computing for which the realization of certain constraints enforces
correlations for classical bits is much larger than a restriction to quantum
operations. This is a wide area that needs to be explored!

\subsection{Deterministic and probabilistic\\computing}
\label{sec:Deterministic_and_probabilistic_computing}

Standard or ``classical'' computing is deterministic. The state $\tau$ at a
given time $t$ corresponds to one specific configuration of bits or Ising spins
$\rho_0$. Here bits can be identified with fermionic occupation numbers $n$
taking values one or zero, and therefore with Ising spins, by $n=(s+1)/2$. At
any given $t$ the normalized classical wave function for deterministic
computation is a $\delta$-function, $q_\rho(t) = \delta_{\rho,\rho_0}$. A
deterministic operation changes the bit configuration $\rho_0$ to a new bit
configuration $\tau_0 = \bar{\tau}(\rho_0)$. A specific computational operation
corresponds to a specific function $\bar{\tau}(\rho)$. Correspondingly, the
normalized wave function after this computational step becomes
$q_\tau(t+\varepsilon) = \delta_{\tau,\tau_0} =
\delta_{\tau,\bar{\tau}(\rho_0)}$. The corresponding step evolution operator
$\hat{S}(t)$ is a unique jump operator. This process can be repeated for the
next computational step from $t+\varepsilon$ to $t+2\varepsilon$. Classical
computing corresponds to a deterministic automaton. The formulation with a
normalized wave function and step evolution operator describes the result of a
sequence of operations on arbitrary input states $\rho_0$.

\paragraph*{Probabilistic computing}

Probabilistic computing arises on two levels. First, the input state may be
given by a probability distribution over initial configurations. In this case
the step evolution operators $\hat{S}(t)$ remain unique jump operators and the
sequence of operations remains the same as for deterministic computing. Only the
initial wave function $q(t_\mathrm{in})$ is no longer a
$\delta$-function, but rather some general unit vector. This is the setting that
we call a probabilistic automaton.

Second, the computational operations may become probabilistic themselves. In
this case the step evolution operators $\hat{S}(t)$ are no longer unique jump
operators. We may distinguish two cases. For the first, the step evolution
operator is an orthogonal matrix. In this case, no information is lost during
the evolution. The corresponding maps in the space of probability distributions
are not easy to realize in practice. In our example, they will be learned by
artificial neural networks or neuromorphic computers. For the second case where
$\hat S$ is not orthogonal a general formalism employs the evolution of the
classical density matrix $\rho'(t)$, from which the probabilities for bit
configurations at every step $t$ can be extracted~\cite{CWIT}. In this case the
information is at least partly lost during the evolution. In the long time limit
the system is expected to equilibrate at least partly.

Even for non-orthogonal step evolution operators in many cases the probabilities
$p_\tau(t)$ at $t$ are sufficient for a determination of the probabilities
$p_\tau(t+\varepsilon)$ at the next computation step. In this case the
normalized classical wave function $q(t)$, with $q_\tau^2=p_\tau$, is a useful
concept for describing the probabilistic state at every stage of the
computation. It offers the important advantage that a rotation of $q(t)$ keeps
easily control of the normalization of the probability distribution due to
$p_\tau(t)=q_\tau^2(t)$. The computational operation from $t$ to $t+\varepsilon$
is then specified by an effective orthogonal step evolution operator. Every step
of the calculation rotates the normalized classical wave function $q(t)$. At
first sight, this looks rather similar to an orthogonal step evolution operator.
The important difference is, however, that the effective step evolution operator
can now depend on the wave function. The linearity of the evolution equation is
lost.

Formulated in terms of the normalized wave function $q(t)$ the general form of
probabilistic computing shares already many aspects of quantum computing. For
general probabilistic computing the evolution law is not always linear, however.
The effective step evolution operator $\tilde{S}(t)$, which transforms
$q(t+\varepsilon)$ to $q(t)$,
\begin{equation}
q(t+\varepsilon) = \tilde{S}(t) q(t),\quad \tilde{S}^\mathrm{T}(t) \tilde{S}(t)
= 1,
\label{eq:C01}
\end{equation}
is orthogonal, but it may depend on $q(t)$. This is an important difference to
quantum computing. We will in the following discuss several interesting cases
where quantum computing is realized as a special case of more general
probabilistic computing.

\paragraph*{Error propagation}

A direct field of application for probabilistic computing is a systematic
description of error propagation in classical computing. Due to errors, the
effective step evolution operator $\hat{S}(t)$ is not precisely a unique jump
operator. For a good computer it will produce ``wrong'' configurations at
$t+\varepsilon$ only with small probabilities. This changes the zero elements in
the unique jump step evolution operator to small non-zero entries. Error
propagation investigates how such small entries can produce a substantial
cumulative effect by products of many effective step evolution operators,
corresponding to many computational steps. Furthermore, the input configuration
may contain errors. This corresponds to a deviation of the input wave function
$q(t_\mathrm{in})$ from a $\delta$-function.

\paragraph*{Quantum computing}

Quantum computing\,\cite{BEN,MAN,FEY,DEU} is a particular form of probabilistic
computing. In this case the density matrix is a positive Hermitian matrix, and
the step evolution operator $\hat{S}(t)$ is replaced by the unitary evolution
operator $U(t+\varepsilon,t)$, that we denote here by $U(t)$. These are the only
particular features.

We will not discuss in this work all the fascinating developments of performing
quantum computing with atoms, photons or qubits in solids. 
(For some developments close to our topic see
refs.\,\cite{AOR,SFN,AAHE,SLL,MNI}.)
Since we have understood how quantum systems can arise as subsystems of general
probabilistic systems, we explore here to what extent the operations of quantum
computing can be performed by the evolution of ``classical'' statistical
systems. The Ising spins whose expectation values define the quantum subsystem
can now be macroscopic two-level observables, as neurons in an active or quiet
state. There is no need for small isolated subsystems or low temperatures. On
the conceptual side, the realization of quantum operations by classical
statistical systems will shed additional light on the embedding of quantum
systems within general probabilistic systems.

\subsection{Quantum computing by probabilistic automata}
\label{sec:Quantum_computing_by_probabilistic_cellular_automata}

We have seen in sect.\,\ref{sec:quantum_subsystems} that certain unitary quantum
operations can be realized as deterministic operations on classical spin
configurations. This typically concerns a discrete subgroup of the general
unitary transformations. For the example in sect.\,\ref{sec:quantum_subsystems}
the discrete qubit chain employs an automaton consisting of three Ising spins.
It realizes a discrete subgroup of the SU(2)-transformations for one-qubit
quantum system. As we have discussed in sect.\,\ref{sec:Unitary_evolution} this
subgroup can be associated to $\pi/2$-rotations around the Cartesian axes.

\paragraph*{Correlations between Ising spins}

The quantum aspects of this simple ``quantum computer'' are due to the quantum
constraint $\sum_k \rho_k^2 \leq 1$, $\rho_k = \braket{s_k}$. This forbids a
deterministic initial state. For any specific spin configuration the expectation
values for a sharp state coincide with the values of the spins in this
configuration, and therefore $\sum_k \rho_k^2 = 3$. This contradicts the quantum
constraint. States respecting the quantum constraint are necessarily
probabilistic. We therefore deal with probabilistic automata. At first sight the
probabilistic input state may only look as a loss of precision of the
computation. What is new, however, are the correlations between the three
classical Ising spins. Given expectation values of two of the spins constrain
the possible expectation value of the third spin. 

Consider an initial state which is a pure quantum state, $\sum_k \rho_k^2 = 1$.
This will remain a pure state for all steps of the computation. If some
algorithm leads to $\rho_1(t) = \rho_2(t) = 0$ at some step in the evolution,
one automatically knows $\rho_3(t) = \pm 1$. This type of correlation enables
one to influence the state of all spins by acting only on a subset of spins.
Such a behavior is a characteristic of quantum computations.

\paragraph*{Icosahedron}

One may ask which other non-abelian subgroups of the unitary
SU(2)-transformations for a single qubit can be realized by a probabilistic
automaton. The maximal discrete subgroup of SU(2) is the symmetry group of the
icosahedron. It can be realized by six classical bits labeled here by
$s_k^\pm$. Their expectation values generate the quantum density matrix by
\begin{align}
\begin{split}
\braket{s_{1\pm}} &= a\rho_1 \pm b\rho_3, \\
\braket{s_{2\pm}} &= a\rho_2 \pm b\rho_1, \\
\braket{s_{3\pm}} &= a\rho_3 \pm b\rho_2,
\end{split}
\label{eq:C02}
\end{align}
where
\begin{equation}
a = \left( \frac{1 + \sqrt{5}}{2\sqrt{5}} \right)^\frac{1}{2},\quad b = \left(
\frac{2}{5 + \sqrt{5}} \right)^\frac{1}{2}
\label{eq:C03}
\end{equation}
with
\begin{equation}
a^2 + b^2 = 1,\quad b = \frac{2a}{1+\sqrt{5}}.
\label{eq:C04}
\end{equation}

The expectation values of the six classical Ising spins $s_{k\pm}$ coincide with
the expectation values of quantum spins $S_{k\pm}$ in particular directions,
namely
\begin{align}
\begin{split}
S_{1\pm} = (a,0,\pm b), \\
S_{2\pm} = (\pm b,a,0), \\
S_{3\pm} = (0,\pm b,a).
\end{split}
\label{eq:C05}
\end{align}
The associated operators are
\begin{equation}
\hat{S}_{k\pm} = a\tau_k \pm b\tilde{\tau}_k,
\label{eq:C06}
\end{equation}
where $\tilde{\tau}_3 = \tau_2$, $\tilde{\tau}_2 = \tau_1$, $\tilde{\tau}_1 =
\tau_3$. Six quantum spins \eqref{eq:C05} correspond to six corners of the
icosahedron on the Bloch sphere, the other six corners being given by the
opposite values of these spins. The twelve corners of the icosahedron give
already a reasonable approximation of the sphere.

The particular feature of quantum computing consists again in the correlations
between the spins. Besides the constraint $\sum_k \rho_k^2 = 1$, there are
additional quantum constraints since six expectation values are given by three
numbers $\rho_k$. For example, the relation
\begin{equation}
\rho_1 = \frac{1}{2a} \left( \braket{s_{1+}} + \braket{s_{1-}} \right)
	= \frac{1}{2b} \left( \braket{s_{2+}} - \braket{s_{2-}} \right)
\label{eq:C07}
\end{equation}
implies the constraint
\begin{equation}
\braket{s_{2+}} - \braket{s_{2-}} = \frac{2}{1+\sqrt{5}} \left( \braket{s_{1+}}
+ \braket{s_{1-}} \right).
\label{eq:C08}
\end{equation}
With two similar constraints for the differences $\braket{s_{1+}} -
\braket{s_{1-}}$ and $\braket{s_{3+}} - \braket{s_{3-}}$, any change of the
expectation value of one of the classical Ising spins is necessarily accompanied
by changes for other spins. 

The operations of the probabilistic automaton realizing the icosahedron subgroup
of the unitary quantum transformations of the density matrix $\rho$ are
permutations of the classical bits or Ising spins. Only those are permitted that
respect the quantum constraint. These are precisely the $2\pi/5$ rotations
around appropriate axes which leave the icosahedron invariant, and compositions
thereof. The corresponding unitary transformations of the quantum density matrix
can be performed by simple bit permutations of a classical computer. If an
algorithm aims at exploiting the correlations due to the quantum constraints,
the initial state has to be prepared in order to obey these constraints. For the
following computational steps the constraints will be preserved automatically.

Already for a single qubit we observe the general tendency: A more dense
subgroup of the unitary transformation can be realized by a larger number of
classical Ising spins. In turn, the system has to be initialized with a larger
number of quantum constraints.

\paragraph*{Two qubits}

For a quantum system with two qubits the relevant group of transformations is
SU(4). Similar to the case of a single qubit, one may investigate which
permutations of classical bit configurations can realize an appropriate
non-abelian discrete subgroup of SU(4). It is not known to us which subgroups
realize the CNOT-transformation. This is not crucial, however, since other
discrete transformations can transform direct product states into entangled
states. For the case of two qubits the construction of discrete subgroups of
$\text{SU}(4)$ which can be realized by a finite number of classical spins is
already rather complex.

The six-dimensional manifold spanned by the wave functions for pure two-qubit
quantum states corresponds to SU(4)/SU(3)$\times$U(1)\,\cite{BH}. (The four
complex components of the two-qubit wave function correspond to eight real
numbers. The normalization imposes a first constraint, and the overall phase is
irrelevant, leaving six independent real numbers.) For an arbitrary pure state a
particular triplet $(\sigma_1,\sigma_2,\sigma_3)$ of commuting observables with
eigenvalues $\pm 1$ has sharp values, with $\sigma_3 = \sigma_1 \sigma_2$. For
example, in the state $q_0 = (1,0,0,0)$ one has $\sigma_1^{(0)} = S_3^{(1)}$,
$\sigma_2^{(0)} = S_3^{(2)}$, $\sigma_3^{(0)} = S_3^{(1)} S_3^{(2)}$, with
$\braket{\sigma_k^{(0)}} = 1$. After an SU(4) transformation, $q = Uq_0$, the
two-level observables with sharp values $\sigma_k =1$ in the state $q$ are
$\sigma_k = U \sigma_k^{(0)} U^\dagger$. 

The manifold of all two-level quantum observables is the eight dimensional
homogeneous space SU(4)/SU(2)$\times$SU(2)$\times$U(1). It corresponds to the
unitary transformations of a particular spin operator, say $\hat{S}_3^{(1)}$.
Out of the spin-operators associated to these two-level observables a given pure
state selects two commuting ones corresponding to $\sigma_1$ and $\sigma_2$,
with $\sigma_3$ the product of the two. These three have a sharp value $+1$. The
six-dimensional manifold of pure states corresponds therefore to the possible
embeddings of the three sharp observables $\sigma_1$, $\sigma_2$, $\sigma_3$
into the eight-dimensional manifold of two-level observables. The
transformations of the discrete subgroup of SU(4) act both in the
eight-dimensional space of possible two-level observables and in the
six-dimensional space of possible embeddings of $(\sigma_1, \sigma_2,
\sigma_3)$. In other words, the action of SU(4) in the eight-dimensional space
of observables is such that each triplet of commuting sharp observables is
mapped to a new triplet of commuting sharp observables.

\paragraph*{Probabilistic automata for two qubits}

We want to know which discrete subgroups of $\text{SU}(4)$ can be realized by
unique jump operations for classical spins. This will depend on the number of
independent classical spins used, or on the bit-quantum map employed. We have
already seen that the CNOT-gate~\eqref{eq:4.3.37} can be realized by the average
spin map, but not by the correlation map.

A possible strategy for realizing discrete quantum rotations by permutations of
classical Ising spins selects first a discrete subgroup of SU(4). The action of
its elements on the quantum spin operator $\hat{S}_3^{(1)}$ produces a discrete
set of two-level quantum observables. One associates to each point of this set a
classical Ising spin. Here, a change of sign is not counted as a new variable,
but rather as a change of the value of the two-level observable. The action of
the unitary quantum transformation of the discrete subgroup of $\text{SU}(4)$
can then be realized by the corresponding permutations of classical bits. If all
the classical bits are independent this corresponds to the average spin map. If
some of the spins can be represented as products of other spins, as for the
correlation map, this induces additional relations. These relations may or may
not be compatible with the chosen subgroup of $\text{SU}(4)$.

The association between classical Ising spins and quantum spins is possible
provided that the expectation values of the classical Ising spins coincide with
the expectation values of the associated quantum spins. This constitutes the
quantum constraint. In a pure state all expectation values of the discrete set
of quantum spins generated by the discrete subgroup of SU(4) are fixed in terms
of the six parameters characterizing the pure state wave function. The pure
state quantum constraint requires for the classical probability distribution
that all expectation values of the associated Ising spins take the same value.
It is sufficient to realize this quantum constraint for the probability
distribution of the initial state. It is then preserved by the Ising spin
permutations that correspond to the discrete unitary transformation. Similar to
the case of the icosahedron for a single qubit, the quantum constraint induces
many correlations between the classical Ising spins.

A unitary quantum operation transforms the expectation values of two-level
quantum observables with associated quantum operators
\begin{align}
\begin{split}
\braket{A'} &= \tr \left\{ \hat{A}\rho(t+\varepsilon) \right\} \\
&= \tr \left\{ \hat{A} U \rho(t) U^{-1} \right\} = \tr \left\{
\hat{A}_\mathrm{H}(t) \rho(t) \right\},
\end{split}
\label{eq:LL1}
\end{align}
with unitary evolution operator $U = U(t+\varepsilon,t)$, and Heisenberg
operator
\begin{equation}
\hat{A}_\mathrm{H}(t) = U^{-1} \hat{A} U.
\label{eq:LL2}
\end{equation}
If all classical expectation values $\rho_{\mu\nu} = \chi_{\mu\nu}$ of Ising
spins, that are used for the definition of the quantum density matrix
\eqref{eq:GBQ2}, are transformed in the same way as the transformation from
$\braket{A} = \braket{A(t)}$ to $\braket{A'} = \braket{A(t+\varepsilon)}$ in
eq.\,\eqref{eq:LL1}, the corresponding $U$ can be realized by a change of the
time-local probability distribution. For a deterministic change it has to be
realized by a map between bit configurations $\tau \to \tau'$.

As an example, let us consider the unitary transformation 
\begin{equation}
U_\mathrm{D3} = \begin{pmatrix}
1 & 0 & 0 & 0 \\ 
0 & 1 & 0 & 0 \\ 
0 & 0 & -1 & 0 \\ 
0 & 0 & 0 & 1
\end{pmatrix},\quad
U_\mathrm{D3}^2 = 1.
\label{eq:LL3}
\end{equation}
It leaves the quantum spin operators $S_3^{(1)}$ and $S_3^{(2)}$ invariant. Its
action on $S_{1,2}^{(1),(2)}$ produces products of spin operators
\begin{align}
\begin{split}
U_\mathrm{D3}^\dagger S_1^{(1)} U_\mathrm{D3} &= -S_1^{(1)} S_3^{(2)}, \quad
U_\mathrm{D3}^\dagger S_2^{(1)} U_\mathrm{D3} = -S_2^{(1)} S_3^{(2)}, \\
U_\mathrm{D3}^\dagger S_1^{(2)} U_\mathrm{D3} &= S_1^{(2)} S_3^{(1)}, \quad
U_\mathrm{D3}^\dagger S_2^{(2)} U_\mathrm{D3} = S_2^{(2)} S_3^{(1)}.
\end{split}
\label{eq:LL4}
\end{align}
The corresponding changes of classical spin expectation values are
\begin{align}
\begin{split}
\rho_{10} \leftrightarrow -\rho_{13}&, \quad \rho_{20} \leftrightarrow
-\rho_{23}, \\
\rho_{01} \leftrightarrow \rho_{31}&,\quad \rho_{02} \leftrightarrow \rho_{32}.
\end{split}
\label{eq:LL5}
\end{align}
The remaining four quantities defining the density matrix, namely $\rho_{11}$,
$\rho_{12}$, $\rho_{21}$ and $\rho_{22}$, correspond to the quantum expectation
values of the product of commuting quantum spin operators $\braket{S_1^{(1)}
S_1^{(2)}}$, $\braket{S_1^{(1)} S_2^{(2)}}$, $\braket{S_2^{(1)} S_1^{(2)}}$ and
$\braket{S_2^{(1)} S_2^{(2)}}$, respectively. They transform under the
D3-transformation as
\begin{align}
\begin{split}
U_\mathrm{D3}^\dagger S_1^{(1)} S_1^{(2)} U_\mathrm{D3} &= -S_2^{(1)} S_2^{(2)},
\\
U_\mathrm{D3}^\dagger S_1^{(1)} S_2^{(2)} U_\mathrm{D3} &= S_2^{(1)} S_1^{(2)},
\end{split}
\label{eq:LL6}
\end{align}
corresponding to the map
\begin{equation}
\rho_{11} \leftrightarrow -\rho_{22},\quad \rho_{12} \leftrightarrow \rho_{21}.
\label{eq:LL7}
\end{equation}

\paragraph*{Average spin map}

The map \eqref{eq:LL5}, \eqref{eq:LL7}, with invariant $\rho_{30}$, $\rho_{03}$,
$\rho_{33}$, can be performed by spin exchanges and changes of sign for the
fifteen spins of the average spin map \eqref{E31}. In contrast, it cannot be
performed by the correlation map. For the correlation map with six classical
spins $s_k^{(1)}$, $s_k^{(2)}$, and correlations $\rho_{kl} = \braket{s_k^{(1)}
s_l^{(2)}}$, the transformation \eqref{eq:LL5} can be achieved by conditional
jumps. Also the transformation \eqref{eq:LL6} is achieved by a simple spin
exchange $s_1^{(1)} \to -s_2^{(1)}$, $s_2^{(1)} \to s_1^{(1)}$, $s_1^{(2)} \to
s_2^{(2)}$, $s_2^{(2)} \to -s_1^{(2)}$. This would, however, further change the
quantities appearing in eq.\,\eqref{eq:LL5}. There seems to be no transformation
of Ising spin configurations which realizes both eq.\,\eqref{eq:LL5} and
\eqref{eq:LL7} simultaneously, such that the unitary transformation
\eqref{eq:LL3} cannot be performed by deterministic operations for the
correlation map. The situation is similar to the CNOT gate \eqref{eq:4.3.37}.

We can employ the $\pi$-rotation around the 3-axis of spin one for realizing
\begin{equation}
U = \diag (1,1,-1,-1) = \tau_3 \otimes 1 = -i \left( U_3^{(1)} \otimes 1
\right).
\label{eq:LL7b}
\end{equation}
The overall phase of a transformation acting on a single spin does not matter,
such that single spin operations are also represented by $e^{i\varphi} U^{(1)}
\otimes 1$ and $e^{i\varphi} 1 \otimes U^{(2)}$, with arbitrary phases. The
$\pi$-rotation around the three-axis of spin two realizes 
\begin{equation}
U = \diag (1,-1,1,-1) = 1 \otimes \tau_3 = -i\left( 1\otimes U_3^{(2)} \right),
\label{eq:LL8}
\end{equation}
and the combination of $\pi$-rotations around the three-axis for both spins
gives
\begin{equation}
U = \diag (1,-1,-1,1) = \tau_3 \otimes \tau_3.
\label{eq:LL9}
\end{equation}
For the average spin map such transformations can be combined with the
transformation D3 in eq.~\eqref{eq:LL3}. Together with a free overall phase of
the unitary matrices, arbitrary diagonal $U$ with elements $\pm 1$ can be
realized by deterministic maps of spins.

Already at this state it becomes clear that the average spin map can realize a
rather dense set of unitary transformations of the two-qubit quantum system by
the simple deterministic updating of a probabilistic automaton.

\paragraph*{Phases in quantum computing}

Phases in quantum wave functions or in unitary operations play an important role
in quantum computing. If one wants to implement operations of quantum computing
by changes of probability distributions for classical bits one has to account
for these phases.
It is instructive to see how different phases in unitary quantum operators
correspond to different unique jump classical operators for the average spin
map. Let us consider the matrix
\begin{equation}
U^\dagger = \begin{pmatrix}
0 & a & 0 & 0 \\ 
0 & 0 & 0 & b \\ 
c & 0 & 0 & 0 \\ 
0 & 0 & d & 0
\end{pmatrix},\quad
U = \begin{pmatrix}
0 & 0 & c^* & 0 \\ 
a^* & 0 & 0 & 0 \\ 
0 & 0 & 0 & d^* \\ 
0 & b^* & 0 & 0
\end{pmatrix}.
\label{eq:M1}
\end{equation}
For $|a|=|b|=|c|=|d|=1$ this is a unitary matrix, $U^\dagger U = 1$. One has
\begin{equation}
\left( U^\dagger \right)^2 = \begin{pmatrix}
0 & 0 & 0 & ab \\ 
0 & 0 & bd & 0 \\ 
0 & ac & 0 & 0 \\ 
cd & 0 & 0 & 0
\end{pmatrix},\quad
\left( U^\dagger \right)^4 = abcd\, 1_4,
\label{eq:M2}
\end{equation}
such that $U^4$ is unity up to an irrelevant overall phase. We have chosen this
matrix such that it rotates for arbitrary phases $a$, $b$, $c$, $d$ the
three-components of the two quantum spins 
\begin{equation}
\left( \hat{S}_3^{(1)} \right)' = U^\dagger \hat{S}_3^{(1)} U =
\hat{S}_3^{(2)},\quad 
\left( \hat{S}_3^{(2)} \right)' = U^\dagger \hat{S}_3^{(2)} U =
-\hat{S}_3^{(1)}.
\label{eq:M3}
\end{equation}
For the other spins one finds
\begin{align}
\label{eq:M4}
\left( \hat{S}_1^{(1)} \right)' = U^\dagger \hat{S}_1^{(1)} U &= 
	\begin{pmatrix}
	 0 & ab^* & 0 & 0 \\ 
	 a^*b & 0 & 0 & 0 \\ 
	 0 & 0 & 0 & cd^* \\ 
	 0 & 0 & c^*d & 0
	 \end{pmatrix} ,\nonumber \\ 
\left( \hat{S}_2^{(1)} \right)' = U^\dagger \hat{S}_2^{(1)} U &= 
	\begin{pmatrix}
	 0 & -iab^* & 0 & 0 \\ 
	 ia^*b & 0 & 0 & 0 \\ 
	 0 & 0 & 0 & -icd^* \\ 
	 0 & 0 & ic^*d & 0
	 \end{pmatrix} ,\nonumber \\
\left( \hat{S}_1^{(2)} \right)' = U^\dagger \hat{S}_1^{(2)} U &= 
	\begin{pmatrix}
	0 & 0 & ac^* & 0 \\ 
	0 & 0 & 0 & bd^* \\ 
	a^*c & 0 & 0 & 0 \\ 
	0 & b^*d & 0 & 0
	 \end{pmatrix} , \\
\left( \hat{S}_2^{(2)} \right)' = U^\dagger \hat{S}_2^{(2)} U &= 
	\begin{pmatrix}
	0 & 0 & iac^* & 0 \\ 
	0 & 0 & 0 & ibd^* \\ 
	-ia^*c & 0 & 0 & 0 \\ 
	0 & -ib^*d & 0 & 0
	 \end{pmatrix} .\nonumber
\end{align}

By the choice of different phases $a$, $b$, $c$, $d$ we can realize different
transformations. For $a=b=c=d=1$ one has
\begin{align}
\begin{split}
\left( \hat{S}_1^{(1)} \right)' &= \hat{S}_1^{(2)},\quad 
\left( \hat{S}_2^{(1)} \right)' = \hat{S}_2^{(2)}, \\
\left( \hat{S}_1^{(2)} \right)' &= \hat{S}_1^{(1)},\quad
\left( \hat{S}_2^{(2)} \right)' = -\hat{S}_2^{(1)}.
\end{split}
\label{eq:M5}
\end{align}
On the other hand, for $a=1$, $b=i$, $c=1$, $d=i$ one finds
\begin{align}
\begin{split}
\left( \hat{S}_1^{(1)} \right)' &= \hat{S}_2^{(2)},\quad 
\left( \hat{S}_2^{(1)} \right)' = -\hat{S}_1^{(2)}, \\
\left( \hat{S}_1^{(2)} \right)' &= \hat{S}_1^{(1)},\quad
\left( \hat{S}_2^{(2)} \right)' = -\hat{S}_2^{(1)}.
\end{split}
\label{eq:M6}
\end{align}

A different type of transformation is realized for $a=b=c=1$, $d=-1$. Quantum
spins transform now into correlation functions,
\begin{align}
\begin{split}
\left( \hat{S}_1^{(1)} \right)' &= \hat{S}_1^{(2)} \hat{S}_3^{(1)},\quad 
\left( \hat{S}_2^{(1)} \right)' = \hat{S}_2^{(2)} \hat{S}_3^{(1)}, \\
\left( \hat{S}_1^{(2)} \right)' &= \hat{S}_1^{(1)} \hat{S}_3^{(2)},\quad
\left( \hat{S}_2^{(2)} \right)' = -\hat{S}_2^{(1)} \hat{S}_3^{(2)}.
\end{split}
\label{eq:M7}
\end{align}
Correspondingly, these correlations transform as
\begin{align}
\begin{split}
\left( \hat{S}_1^{(2)} \hat{S}_3^{(1)} \right)' &= \hat{S}_1^{(1)},\quad 
\left( \hat{S}_2^{(2)} \hat{S}_3^{(1)} \right)' = -\hat{S}_2^{(1)}, \\
\left( \hat{S}_1^{(1)} \hat{S}_3^{(2)} \right)' &= -\hat{S}_1^{(2)},\quad 
\left( \hat{S}_2^{(1)} \hat{S}_3^{(2)} \right)' = -\hat{S}_2^{(2)}.
\end{split}
\label{eq:M8}
\end{align}
The other five correlation functions employed in the correlation map transform
as
\begin{equation}
\left( \hat{S}_3^{(1)} \hat{S}_3^{(2)} \right)' = -\hat{S}_3^{(1)}
\hat{S}_3^{(2)},
\label{eq:M9}
\end{equation}
and
\begin{align}
\begin{split}
\left( \hat{S}_1^{(1)} \hat{S}_1^{(2)} \right)' &= \hat{S}_2^{(1)}
\hat{S}_2^{(2)},\quad 
\left( \hat{S}_2^{(1)} \hat{S}_2^{(2)} \right)' = -\hat{S}_1^{(1)}
\hat{S}_1^{(2)}, \\
\left( \hat{S}_1^{(1)} \hat{S}_2^{(2)} \right)' &= \hat{S}_1^{(1)}
\hat{S}_2^{(2)},\quad 
\left( \hat{S}_2^{(1)} \hat{S}_1^{(2)} \right)' = -\hat{S}_2^{(1)}
\hat{S}_1^{(2)}. 
\end{split}
\label{eq:M10}
\end{align}

All these transformations are realized for the average spin map by simple
exchanges and sign changes of spins. Different phases of the quantum operator
\eqref{eq:M1} clearly correspond to different deterministic classical
operations. 
This is a simple demonstration that there is no contradiction between the
importance of phases in quantum computing and implementations of quantum
computing by manipulations of classical bits or the associated probability
distributions.

\paragraph*{Density of unitary transformations}

For the average spin map with fifteen classical bits the covering of
SU(4)/SU(2)$\times$SU(2)$\times$U(1) by fifteen discrete points remains rather
sparse. Nevertheless, quite a substantial number of discrete
SU(4)-transformations can be performed by products of the unitary matrices
discussed so far, including the CNOT gate \eqref{eq:4.3.37}. These
transformations can transform direct product states to entangled states and vice
versa. The correlation map uses less spins, but also permits for far less
deterministic operations realizing unitary quantum changes. For all other
unitary transformations one has to employ truly probabilistic changes of the
probability distribution for the classical Ising spins.

If one aims for a dense set of discrete unitary quantum transformations
performed by deterministic bit operations one is interested in the other
direction, by using more classical spins, similar to the icosahedron for the
single qubit case. The issue of the most dense subset of unitary transformations
is related to the topic of maximal non-abelian discrete subgroups of
SU(4)\,\cite{HH}. 

\paragraph*{Many qubits}

For a larger number $Q$ of qubits rather large non-abelian subgroups of
SU($2^Q$) exist and can be realized by the deterministic operations of
probabilistic cellular automata. The prize to pay is a rapidly increasing number
of classical bits. For the average spin map one employs $2^{2Q}-1$ Ising spins,
one for each independent $\rho_{\mu_1 ... \mu_Q}$ characterizing the density
matrix. Even denser discrete subgroups of SU($2^Q$) employ even more Ising
spins. These numbers increase very rapidly with $Q$. For practical computations
one will have to make a compromise between the density of deterministic
operations realizing quantum operations, and the number of necessary classical
bits. Nevertheless, an investigation of the non-abelian discrete subgroups for
large $Q$ would be interesting from the conceptual side.

\paragraph*{Restricted unitary computing}

The possibility to realize a discrete subgroup of unitary transformations for a
certain number of qubits by the deterministic updating of classical bits may be
called ``restricted unitary computing". In contrast to classical computing it is
a form of probabilistic computing. One has to prepare an initial probability
distribution for the configurations of classical bits. This initial probability
distribution has to realize the ``quantum constraints" which ensure the necessary
correlations between the classical bits. We therefore deal with a form of
correlated computing. The terrain of possible algorithms which could solve
computational tasks by use of restricted unitary transformations, in particular
the use of entanglement, is essentially unexplored. (See refs.~\cite{LTP, CWEL,
NP, ALL, ICJ, KWC, LBL, MMW, BRH} for related ideas.)

The prize to pay for the use of correlations for computational tasks is the
preparation of an initial probability distribution. A straightforward way could
run repeatedly or in parallel over many initial bit-configurations, and evaluate
expectation values for observables at the end by sampling with weights given by
the initial probability distribution $\{p_\tau(t_0)\}$. Acceptable
$\{p_\tau(t_0)\}$ should all obey the quantum constraints. Still there will be
many different $\{p_\tau(t_0)\}$ compatible with these constraints.

A typical algorithm could construct a two-level observable $A(t_f)$ with
possible values $A(t_f)=\pm1$. At the final step of the computation its
expectation value will be either positive or negative, depending on the initial
$\{p_\tau(t_0)\}$. The determination of the expectation value $\langle
A(t_f)\rangle$ needs the sampling over initial bit-configurations. The outcome
$\langle A(t_f)\rangle$ can decide between two classes of initial probability
distributions. Generalizing to several $A(t_f)$ can be employed for
classification problems.

\subsection{Artificial neural networks}\label{sec:neural_networks}

Unitary quantum operations can be performed with a much smaller number of
classical bits if one employs genuinely probabilistic updating instead of
probabilistic automata. We have argued that for two qubits the correlation
map is complete. Suitable changes of the probability distribution $p(t)$ to
$p(t+\epsilon)$ can therefore induce any arbitrary unitary transformation $U(t)$
of the density matrix for the two-qubit the quantum subsystem. The question arises
how to find the required changes of the probability distribution in practice.
We pursue here the idea that a system can learn the required change of the
probability distribution, and explore the possibility of quantum computation by
artificial neural networks or neuromorphic computing. Several ideas for
realizing aspects of quantum computations by artificial neural networks (ANN)
can be found in
refs.\,\cite{LPLS,SNSSW,CARLTRO,CATRO,GADUA,KIQB,JBC,CNI,JOBC,CTMA}. Our focus
here is a complete quantum computation by ANN. We want to know if a classical
ANN (not a quantum neural network) can learn quantum operations. Our positive
answer raises the question if the deep neural networks used for artificial
intelligence~\cite{HOT,HTF,LCBH,GBC} could possibly make use of this capacity.

We investigate the ``learning of quantum operations" in two steps. The first
step concerns the learning of the change of expectation values required for a
unitary quantum transformation. At this stage there is no difference between the
average spin map or the correlation map, since we do not specify if the
expectation values concern basic Ising spins or include correlations as
expectation values of composite Ising spins. For two qubits we treat the fifteen
quantities $\rho_{\mu \nu}$ simply as real numbers whose change has to be
learned. For the second step we focus on the correlation map. In this step the
$\rho_{\mu \nu}$ are realized as expectation values and correlations of Ising
spins in some stochastic process. This second step is discussed in
sect.\,\ref{sec:neuromorphic_computing}. In the present section we concentrate
on the first, following ref.\,\cite{PEME}.

It is our aim to construct an ANN which learns the three basis gates for two
qubits, namely the Hadamard gate, the rotation gate and the CNOT-gate. After the
training the connections of the network are optimized such that it can perform
these three tasks for an arbitrary initial density matrix. Subsequently, we can
employ these learned transformations in order to realize arbitrary unitary
transformations by sequences of the basis gates. No new learning is necessary
for performing arbitrary sequences. In this sense the ANN realizes after the
learning part important aspects of a two-qubit quantum computer. It can
transform density matrices, but it does not yet realize the connection between
elements of the density matrix and expectation values of classical spins. From
the view of particle-wave duality it realizes the continuous wave aspects of the
probabilistic information, but not the particle aspects. A full two-qubit
quantum computer which also realizes the particle aspects of quantum mechanics
will be constructed in the next section.

\paragraph*{Quantumness gate}

As a first requirement, the neural network has to learn that it deals with a
quantum system expressed by a density matrix $\rho(t)$. In particular, it has
to learn the quantum constraints that guarantee the positivity of $\rho(t)$. We
may call the process that establishes a density matrix $\rho(t_0)$ as an initial
state for a quantum operation a ``quantumness gate".

There are different ways to realize quantumness gates. We describe here the
setting in ref.\,\cite{PEME} for two qubits. It works with an artificial neural
network (ANN), for which 64 real artificial neurons in a layer can be ordered
such that they represent a real $8 \times 8$-matrix. The input $8 \times 8$
matrix $A(t\subt{in})$ is arbitrary, i.e. the input is $64$ arbitrary real
numbers. The task of the quantumness gate consists of transforming $A(t)$ to a
representation of a positive Hermitian $4 \times 4$ density matrix $\rho(t)$ for
two qubits. In this way the quantumness gate prepares the quantum constraints.
 
Every complex $4 \times 4$ matrix $C = C_R + i C_I$, with real $4 \times 4$
matrices $C_R$ and $C_I$, has a real representation given by the real $8\times8$
matrix $\bar C$,
\begin{equation}
\label{NW1}
\bar{C} = \begin{pmatrix}
C_R & -C_I \\
C_I & C_R
\end{pmatrix} = 1_2 \otimes C_R + I_2 \otimes C_I\,,
\end{equation}
with $I_2 = -i \tau_2$. The real matrix product $\bar{C}_1 \bar{C}_2$ is
isomorphic to the complex matrix product $C_1 C_2$. The first step associates to
a given input matrix $A$ a representation $\bar{C}$ of a complex matrix by
\begin{equation}
\label{NW2}
\tilde{A} = -I A I\,, \quad I = \begin{pmatrix}
0 & -1_4 \\
1_4 & 0
\end{pmatrix},
\end{equation}
and 
\begin{equation}
\label{NW3}
\bar{C} = \frac{1}{2}\left( A + \tilde{A} \right)\,.
\end{equation}
The matrix $\bar{C}$ has the structure \eqref{NW1}, and we can associate to it a
complex matrix $C$.

The second step constructs from the complex $4 \times 4$ matrix $C$ a positive
Hermitian normalized density matrix
\begin{equation}
\label{NW4}
\rho = \frac{C C^\dagger}{\tr \left\{ C C^\dagger \right\}} = \rho_R + i
\rho_I\,.
\end{equation}
The associated real representation is the $8 \times 8$ matrix 
\begin{equation}
\label{NW5}
\bar{\rho} = 1_2 \otimes \rho_R + I_2 \otimes \rho_I\,.
\end{equation}
It can again be represented by particular values of the 64 real neurons. Taking
things together, the quantumness gate learns how to map every input matrix
$A(t\subt{in})$ to an input density matrix $\bar{\rho}(t\subt{in})$. This can be
used as an initial state for a sequence of quantum operations. A quantumness
gate is required if the quantum computation consists in processing the initial
information stored in an initial density matrix. Different ``preparations" of
initial density matrices are conceivable. In our construction the 64 real
numbers specifying $A(t_\text{in})$ are mapped to 15 real numbers specifying
$\rho(t_\text{in})$. There could also be quantum algorithms that start by a
fixed density matrix, and provide the information to be processed by a number of
``initial gates" acting on this fixed density matrix. In this case no
quantumness gate is necessary.

Constructing an ANN that learns the quantumness gate for the initialization of
the computation is not a very hard task. We may shortcut this by performing
analytically the map from $A(t\subt{in})$ to $\bar\rho(t\subt{in})$. Our main
emphasis is the processing of the information in $\bar\rho(t\subt{in})$.

\paragraph*{Learning unitary transformations}

\begin{figure}[t!]
\includegraphics{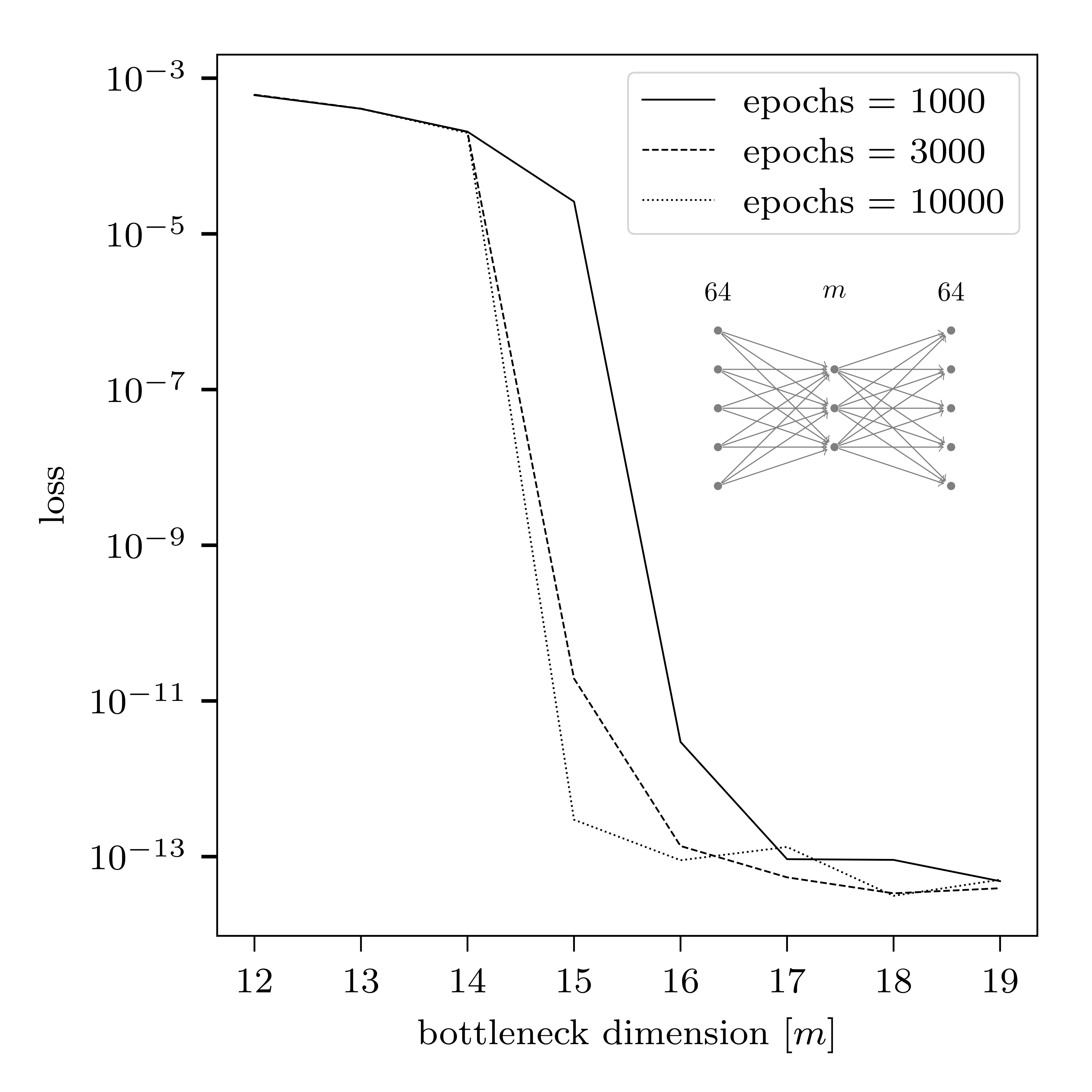}
\caption{Learning the CNOT-gate for two qubits. Loss function $C_\mathrm{l}$
after 1000, 3000 and 10\,000 epochs of training. The plot as a function of
bottleneck dimension $m$ shows that the ANN can learn the unitary transformation
of the CNOT-gate only for $m\geq 15$. The number $m=15$ corresponds to the
number of independent elements of the density matrix for two qubits. The figure
is taken from ref.\,\cite{PEME}.}
\label{fig:loss_function}
\end{figure}

Arbitrary unitary transformations for two qubits can be composed of three basis
gates: the Hadamard gate $U_H$ and the 
rotation gate $U_T$ in eq.\,\eqref{eq:4.2.46} acting on a simple qubit, and the
CNOT-gate $U_C$ in eq.\,\eqref{E8} connecting the
two qubits. If the ANN can learn to perform these three basis gates, it can
perform arbitrary unitary transformations by suitable 
sequences of these gates. A given gate transforms
\begin{equation}
\label{NW6}
\rho(t+\epsilon) = U(t)\rho(t)U^\dagger(t)\,, \quad \bar{\rho}(t+\epsilon) =
\bar{U}(t)\bar{\rho}(t)\bar{U}^{-1}(t)\,,
\end{equation}
with $\bar{\rho}(t)$ and $\bar{\rho}(t+\epsilon)$ the real representations of
$\rho(t)$ and $\rho(t+\epsilon)$. The task for the
ANN is therefore to learn how to transform $\bar{\rho}(t)$ by the unitary
transformations $U_H$, $U_T$ and $U_C$.

Ref.\,\cite{PEME} uses a small ANN consisting of three layers. The first layer
contains 64 real neurons and represents the input matrix 
$\bar{\rho}(t)$. The intermediate layer with $m$ real neurons, typically $m$
much smaller than 64, constitutes a ``bottleneck". 
The third layer has again 64 real neurons that parametrize the output matrix
$B(t+\epsilon)$. Without learning, the output matrix
$B(t+\epsilon)$ is an arbitrary real $8\times 8$ matrix. The learning consists
in adapting the connections between the neurons in 
the different layers such that the output matrix $B(t+\epsilon)$ equals
$\bar{\rho}(t+\epsilon)$. The loss function to be
minimized employs the Frobenius norm $|| B(t+\epsilon) - \bar{\rho}(t+\epsilon)
||$.

The ANN is trained by a sample of $N$ arbitrarily chosen input matrices
$\bar{\rho}_i(t)$, for which $B_i(t+\epsilon)$ results as a map
involving parameters specifying the connections between neurons. For each
$\bar{\rho}_i(t)$ the matrix $\bar{\rho}_i(t+\epsilon)$
is computed analytically for the particular unitary transformation to be
learned. The loss function is defined as
\begin{equation}
\label{NW7}
C_\mathrm{l} = \frac{1}{N} \sum_{i=1}^N || B_i(t+\epsilon)
-\bar{\rho}_i(t+\epsilon) ||^2\,.
\end{equation}
It depends on the parameters specifying the connections between neurons for a
given step in the training. At the next training step
the procedure is repeated with different parameters specifying the connections.
By comparison of the resulting loss, the connection
parameters are adapted in order to minimize the loss function. (For details see
ref.\,\cite{PEME}.)

Fig.\,\ref{fig:loss_function} shows the loss function for different numbers of
training steps (epochs)\,\cite{PEME}. The result is plotted as 
a function of the bottleneck dimension $m$. One observes successful training for
$m \geq 15$. We could use this result in order to
establish the minimal number of real quantities needed to store the necessary
information. The number fifteen coincides with the
number of real parameters specifying the density matrix for two qubits. The ANN
can learn how to combine 64 real numbers into 15 numbers
containing the relevant information for the given task.

\paragraph*{Sequence of unitary transformations}

The training is stopped after a certain number of epochs. The parameters of the
connections between neurons, that specify the map
$\bar{\rho}(t) \to B(t+\epsilon)$, are now kept at the fixed values that have
been learned in order to bring $B(t+\epsilon)$
close to $\bar{\rho}(t+\epsilon)$. For a given unitary gate the resulting
approximation to the map 
$\bar{\rho}(t) \to \bar{U}(t) \bar{\rho}(t) \bar{U}^{-1}(t)$ can now be applied
to arbitrary input density matrices $\bar{\rho}(t)$.
After having learned the three parameter sets for $U_H$, $U_T$ and $U_C$, the
trained system should be able to perform sequences of 
unitary transformations. For this purpose, the output matrix $B(t+\epsilon)$ is
used as the input density matrix $\bar{\rho}'(t+\epsilon)$ for the next
computational step from $t+\epsilon$ to $t+2\epsilon$, with $B(t+2\epsilon) =
\bar{U}(t+\epsilon) \bar{\rho}(t+\epsilon) \bar{U}^{-1}(t+\epsilon)$. Since
$B(t+\epsilon)$ is not exactly equal to $\bar{\rho}(t+\epsilon)$ there will be a
small
error in the matrix product $U(t+\epsilon)U(t)$. A typical computation uses
different gates $U(t+\epsilon)$ and $U(t)$. Different
basis gates do not commute. The process can be repeated for an arbitrary
sequence of unitary transformations. 

As a quantitative test for the quality of the learned unitary transformations
one can perform some given sequence of $n$ unitary
transformations. On the one hand one computes $\bar{\rho}(t+n\epsilon)$
analytically for this sequence. On the other hand the ANN
for a sequence of learned unitary transformations produces $B(t+n\epsilon)$.
Comparison of $B(t+n\epsilon)$ and $\bar{\rho}(t+n\epsilon)$
allows one to quantify the error of $n$ computational steps.

\begin{figure}[t!]
\includegraphics{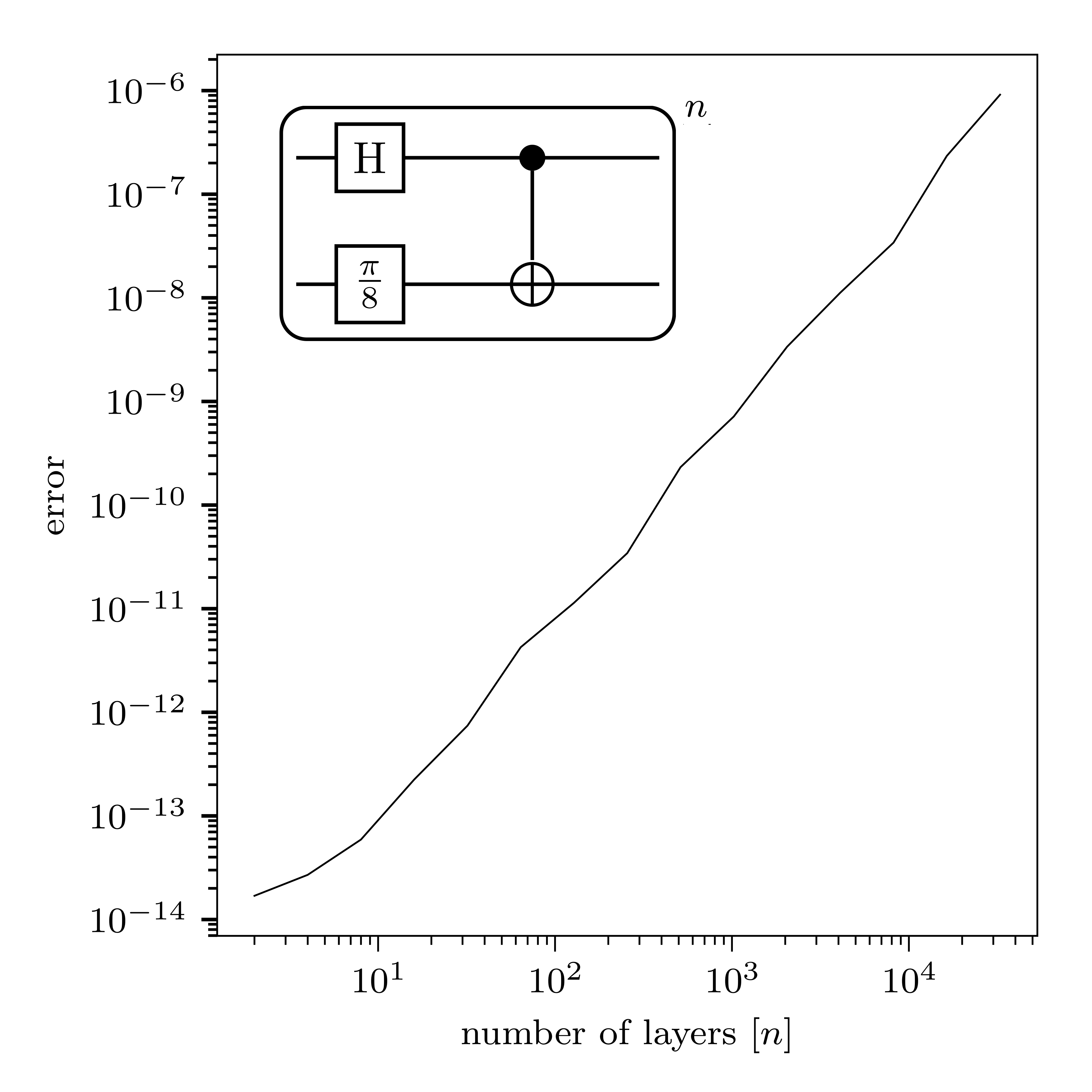}
\caption{Iteration of unitary gates. An alternating sequence of CNOT-gates and a
combination of Hadamard and rotation gates is applied to an initial density
matrix after the ANN has learned these operations. We plot the mean square error
between the final numerically computed and the analytic density matrix after $n$
steps of iteration. The error remains modest even after $10^4$ iterations. For
this high number of iterations the unitary SU(4)-transformations are already
covered very densely. The figure is taken from ref.\,\cite{PEME}.}
\label{fig:mean_square_error}
\end{figure}

Fig.\,\ref{fig:mean_square_error} shows the mean square error after $n$
computational steps or layers\,\cite{PEME}. Even after more
than $10^4$ layers the error remains small. The specific sequence used
alternates the CNOT-gate with a combination of Hadamard and
rotation gates
\begin{equation}
\label{NW8}
U = U_C U_{HR}\,, \quad U_{HR} = U_{H1} U_{R2}\,,
\end{equation}
where
\begin{equation}
\label{NW9}
U_{H1} = U_H \otimes 1\,, \quad U_{R2} = 1 \otimes U_T\,.
\end{equation}
Repeating $U$ many times explores the SU(4)-transformations very densely. One of
the products $U^n$ for $n$ between 1 and $2^{15}$ comes
very close to any arbitrary SU(4)-matrix. This demonstrates that after learning
the three basis gates the ANN can perform arbitrary
unitary transformations for two qubits. 

\subsection{Neuromorphic computing}
\label{sec:neuromorphic_computing}

In the preceding sect.\,\ref{sec:neural_networks} we have demonstrated how
arbitrary unitary quantum transformations could be performed by suitable changes
of expectation values. In the present section we realize these expectation
values in suitable classical probabilistic systems. The classical bits or Ising
spins are given by ``neurons" in an active or quiet state of a small neuromorphic
computer, which mimics in a rough way the dynamics of real neurons in a brain.
The probability distribution for the configurations of these neurons is
implemented by some stochastic time evolution. This permits us to compute
expectation values for the Ising spins by taking time averages over this
stochastic time evolution. We discuss here the implementation of ref.~\cite{PW} --
for related approaches see refs.~\cite{CZPA, MECAR, CZBAU}.

A subset of six expectation values of Ising spins defines the density matrix for
the quantum subsystem by the correlation map. Learning consists in adapting the
connections between the neurons. We train the system to learn initially the
quantum correlations by a quantumness gate. Subsequently, it learns how to
perform unitary transformations for the two-qubit quantum system. This classical
probabilistic system can therefore be regarded as a small quantum computer for
two qubits. It realizes both aspects of particle-wave duality of quantum
mechanics. The particle side of discrete observables is realized by the neurons
which can be observed to be in an active state or not. The wave side corresponds
to the continuous probability distribution for the configurations of neurons.

This system learns how to perform the changes of these probability distributions
which preserve the information without being unique jump operations. The step
evolution operator is orthogonal without being a unique jump matrix. More
precisely, the system has a subsystem for which the step evolution is
orthogonal, and this subsystem is mapped to the two-qubit quantum system. Albeit
rather simple, this two-qubit quantum computer demonstrates that quantum
operations can be performed by classical probabilistic systems. No extreme
isolation preventing the qubits from decoherence is needed. Our brain could
learn the simple two-qubit quantum operations if they would be of use for
performing certain tasks. Our system has not been implemented by hardware and
remains so far a theoretical neuromorphic quantum computer. There seems to be no
major obstacle for a hardware implementation.

In ref.~\cite{PW} the correlation map for two qubits has been implemented in
neuromorphic computing\,\cite{BBNM,PBBSM,PJTM,FSGH,ASM,THA,JPBSM,DBKB,RJP}. The
six classical Ising spins $s_k^{(1)}$, $s_k^{(2)}$ correspond to active ($s=1$)
or silent ($s=-1$) stages of six particular, but randomly chosen neurons. These
neurons are embedded in an environment of many other neurons that provide for
stochastic dynamics in the time evolution of the six selected neurons. The Ising
spins $s_k^{(i)}$ are ``macroscopic two-level observables" that ``measure" at
any time $\tau$ if a given neuron is active or silent.

The detailed stochastic dynamics used for the results below can be found in
ref.\,\cite{PW}. What is important for the present summary is only that the
neuron $j = (k,i)$ takes the value $s_j(\tau) = 1$ for some part of the time
$\tau_j^+$ during a ``measurement" period $T$. For the rest of the time,
$\tau_j^- = T - \tau_j^+$, it assumes the value $s_j(\tau) = -1$. Expectation
values can be determined by time averages
\begin{align}
\label{NC1}
\braket{s_j} &= \frac{1}{T} \int_0^T \dif \tau\, s_j(\tau) = \frac{\tau_j^+ -
\tau_j^-}{T} \nonumber \\
&= \frac{2\tau_j^+}{T} - 1\,.
\end{align}
With $\tau_{jl}^{++}$ the time interval when $s_j(\tau) = s_l(\tau) = 1$,
$\tau_{jl}^{+-}$ the interval with $s_j(\tau)=1$, $s_l(\tau)=-1$, and similarly
for $\tau_{jl}^{-+}$ with opposite sign, and $\tau_{jl}^{--}$ for both signs
negative, the correlations are given by
\begin{align}
\label{NC2}
\braket{s_j s_l} &= \frac{1}{T} \int_0^T \dif \tau\, s_j(\tau) s_l(\tau)
\nonumber \\
&= \frac{2 \left( \tau_{jl}^{++} + \tau_{jl}^{--} \right)}{T} -1\,.
\end{align}
Thus the expectation values and correlations needed for the construction of the
density matrix by the correlation map can be measured directly.

Denoting the relevant expectation values by $\sigma_{\mu \nu}$,
\begin{equation}
\label{NC3}
\sigma_{k0} = \braket{s_k^{(1)}}, \quad \sigma_{0k} = \braket{s_k^{(2)}}, \quad
\sigma_{kl} = \braket{s_k^{(1)}s_l^{(2)}},
\end{equation}
and identifying for the density matrix by $\rho_{\mu \nu} = \sigma_{\mu \nu}$,
\begin{equation}
\label{NC4}
\rho = \frac{1}{4} \sigma_{\mu \nu} L_{\mu \nu}\,,
\end{equation}
where $\rho_{00} = \sigma_{00} = 1$, the stochastic evolution during the
measurement time $T$ defines a quantum density matrix
if the expectation values $\sigma_{\mu \nu}$ obey the quantum constraints. This
construction applies for many stochastic systems.
The neurons may be the ones in a neuromorphic computer or in a biological system
as our brain. The neurons can also be used as abstract
quantities for suitable two-level observables in many other stochastic systems. 

A given measurement period corresponds to a given step $t$ in the computation.
This defines the expectation values $\sigma_{\mu \nu}(t)$ 
and the density matrix $\rho(t)$. For the next step at $t+\epsilon$ one may
change the parameters determining the stochastic evolution
and do again measurements for a time period $T$. This defines
$\sigma_{\mu\nu}(t+\epsilon)$ and $\rho(t+\epsilon)$. The change of 
the parameters of the stochastic evolution can again be done by learning. Thus
the parameters defining the map form $\rho(t)$ to
$B(t+\epsilon)$ in sect.~\ref{sec:neural_networks} are replaced here by the
parameters of the stochastic evolution. One may construct
the same neural network as in sect.~\ref{sec:neural_networks} and use the same
training for the learning of the basis gates for unitary
transformations of the quantum density matrix. 

We present here only a simple task for the learning process, namely how the
stochastic dynamics can learn the density matrix for a given
quantum state. The learning consists in adapting the parameters of the
stochastic evolution such that the expectation values \eqref{NC3}
yield by eq.\,\eqref{NC4} the quantum density matrix which is the goal for the
learning. This step can be viewed as a quantumness gate 
preparing the initial density matrix for following computational steps.

\begin{figure}[t!]
\includegraphics[width=\linewidth]{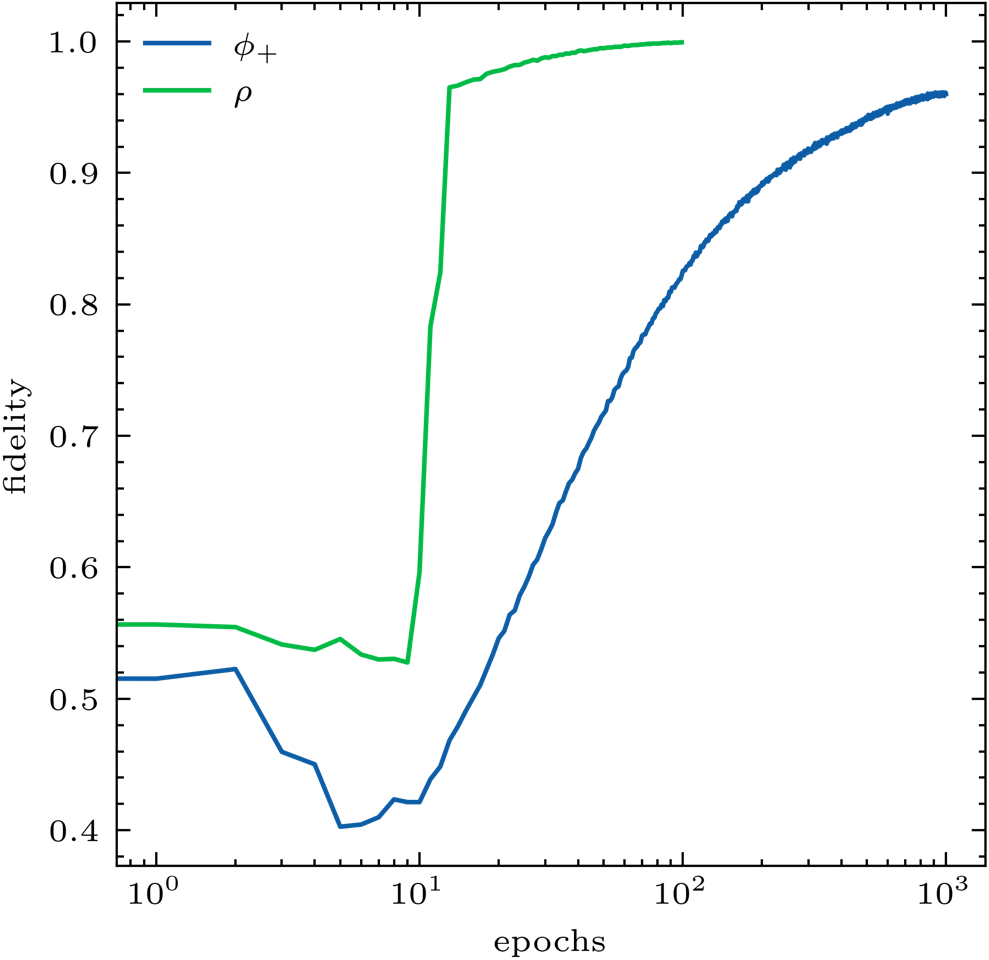}
\caption{Learning density matrices by a stochastic state of neurons. We plot the
fidelity $F(\rho,\sigma)$ by comparing the density matrix $\rho$ extracted from
the expectation values and correlations of two-level neurons to a given density
matrix $\sigma$ that is to be learned. The fidelity monitors the progress after
a given number of training epochs. We compare two density matrices: a maximally
entangled one ($\psi_+$) and a random one ($\rho$). The figure is taken from
ref.\,\cite{PW}.}
\label{fig:dm_learning}
\end{figure}

Fig.\,\ref{fig:dm_learning} demonstrates the learning of two particular density
matrices\,\cite{PW}. The precision of the agreement of the matrix obtained as a
result of a given number of learning steps (epochs) with the wanted density
matrix is measured by the ``fidelity", a concept generally used to measure
precision in quantum computations. The fidelity compares two density matrices
$\rho$ and $\sigma$. It is defined by
\begin{equation}
F(\rho,\sigma) = \left( \tr \left\{ \sqrt{ \sqrt{\rho}\sigma \sqrt{\rho} }
\right\} \right)^2.
\label{eq:FFF1}
\end{equation}
We demonstrate the learning of the probability distribution which realizes the
density matrix of the pure maximally entangled state with wave function
\begin{equation}
\psi_+ = \frac{1}{\sqrt{2}} ( \ket{\uparrow\uparrow} +
\ket{\downarrow\downarrow} ),
\label{eq:FFF2}
\end{equation}
as well as the one for a randomly generated density matrix $\rho$.
Both can be learned after a sufficient number of epochs, where we observe that
the learning of a highly entangled state stakes somewhat longer.

\begin{figure}[h]
\includegraphics[width=\linewidth]{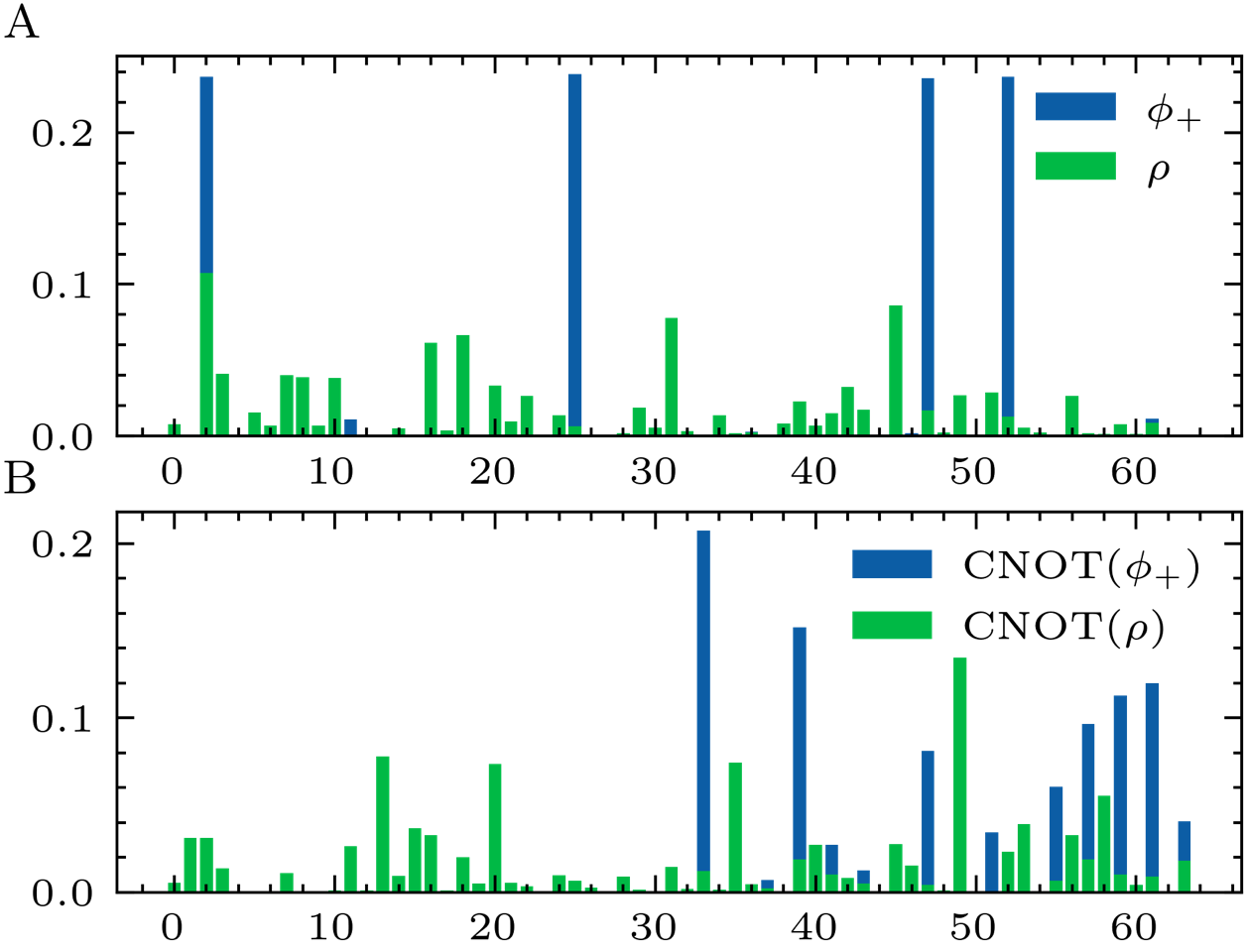}
\caption{Classical probabilities for quantum states. In part $A$ we plot the
probability distributions which are mapped to the density matrix of the
maximally entangled pure state
$\psi_+=\frac{1}{\sqrt{2}}(|00\rangle+|11\rangle)$ (blue) and for a randomly
chosen density matrix (green). In part $B$ we display the probability
distributions corresponding to the density matrices which are obtained from the
ones of part $A$ by applying the CNOT-gate. The labels $0\dots63$ refer to the
configurations of six classical Ising spins $s_k^{(i)}$, $i=1,2$, $k=1\dots3$.
They can be though as binary numbers constructed from the bits associated to
Ising spins. For example the label three corresponds to the spin configuration
$(-1,-1,-1,-1,1,1)$. This figure demonstrates that entangled quantum states can
be realized by classical probability distributions, and quantum gates by changes
of these probability distributions. The figure is taken from ref.~\cite{PW}.}
\label{fig:sixIsing}
\end{figure}

Once the stochastic dynamics realizing a given density matrix has been learned,
one can measure all correlations of the six Ising spins and construct the
probability distribution. In Fig.~\ref{fig:sixIsing} we display the
probabilities for the $64$ configurations of the six Ising spins that result
from this learning.
We do this for both the state $\psi_+$ and the randomly chosen density matrix.
We also perform the learning of the density matrices which obtain by applying
the CNOT-gate to $\psi_+$ and the randomly chosen $\rho$. The result is also
shown in Fig.~\ref{fig:sixIsing}. This demonstrates that the system can learn
the density matrices that obtain by applying basis quantum gates to a given
density matrix.

\begin{figure}[h]
\includegraphics[width=\linewidth]{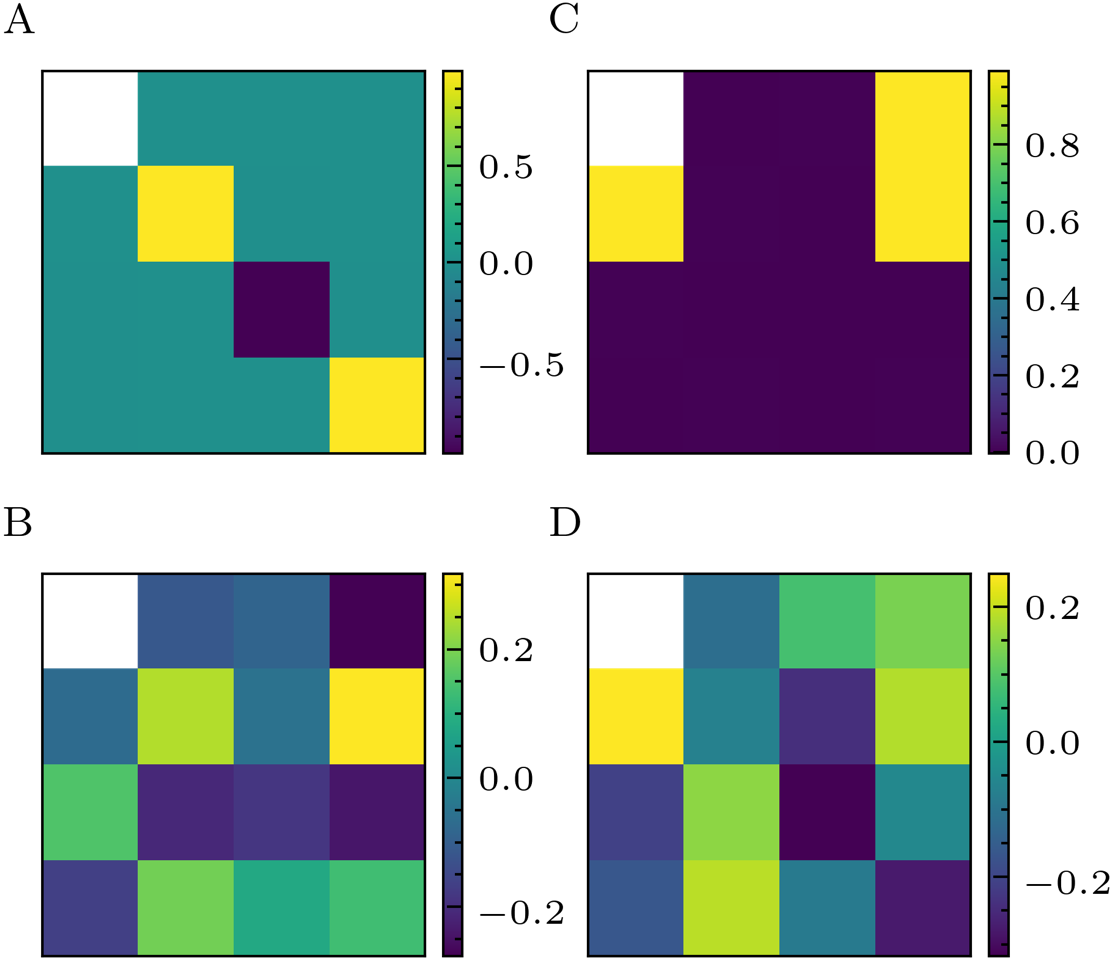}
\caption{Expectation values and correlations of classical Ising spins which are
used for the quantum density matrix. The first row indicates the three
expectation values $\langle s_k^{(1)}\rangle$, while the first column indicates
$\langle s_k^{(2)}\rangle$. The remaining $3\times3$ matrices display the correlations
$\langle s_k^{(1)}s_k^{(2)}\rangle$. We display the expectation values and
correlations learned by the neuromorphic stochastic system for the pure
entangled quantum state in part $A$, and for a randomly chosen density matrix in
part $B$. The parts $C$ and $D$ display the results for density matrices that
obtain from $A$ and $B$ by applying the CNOT gate. Similar results obtain by
analytic computation from the probability distributions in
Fig.~\ref{fig:sixIsing}. The figure is taken from ref.~\cite{PW}.}
\label{fig:probabilities}
\end{figure}

Fig.~\ref{fig:probabilities} displays the expectation values and correlations
used for the quantum density matrix.
They can directly be extracted from measurement. On the other hand, we can
equivalently compute them from the classical probability distributions shown in
Fig.~\ref{fig:sixIsing}. This can be used as a check that the probability
distributions found are indeed correct. We emphasize that the probability
distributions for the $64$ configurations realizing a given quantum density
matrix are not unique. Only $15$ linear combinations of the $64$ probabilities 
enter the density matrix. Therefore a different training typically finds a
different probability distribution for the configurations realizing a given
quantum density matrix.

If the system would have to learn how to perform the basic quantum gates for
arbitrary probability distributions realizing a given quantum density matrix, it
has to learn how to update the parameters of the stochastic dynamics for
arbitrary probability distributions realizing a given density matrix. Such a
procedure would learn a large amount of redundant information. For learning a
given quantum gate it is sufficient that a map between stochastic parameter sets
for two consecutive time periods is learned such that the $15$ expectation
values and correlations are changed according to the unitary transformation.
This task is similar to the one discussed in the preceding section. The other
$63-15=48$ correlations for the six Ising spins simply play no role. Together
with the neurons not used for the quantum density matrix they constitute the
environment for the quantum subsystem.

Already the implementation of the quantumness gate can be used to answer a
computational question which is not easily accessible otherwise: Is the minimal
correlation map a complete bit-quantum map or not? The answer is positive for
two qubits and negative for three or more qubits. Completeness of the
correlation map means that arbitrary quantum density matrices can be realized
for suitable probability distributions of the classical bits. For this purpose
one investigates a very large number of randomly chosen density matrices. For
two qubits the learning of the associated stochastic dynamics and corresponding
probability distribution has always been successful. Without being a formal
proof this constitutes a rather strong argument for completeness. For three
qubits we have found no successful learning for a certain class of density
matrices. This is again not a formal proof of incompleteness of the minimal
correlation map. In principle, it could be a shortcoming of the learning
algorithm. Having the indication of incompleteness it was possible, however, to
find analytically obstructions for the minimal correlation map to work for
GHZ-states~\cite{PW}.

\paragraph*{Outlook on correlated computing}

We have presented several examples for correlated computing. They are
necessarily based on some type of probabilistic computing since only non-trivial
probability distributions can realize the necessary correlations between
two-level observables. Quantum computation by real atoms or photons implement
these correlations by the very nature of quantum systems. For example, the
correlations between different directions of the quantum spin, which may be
regarded as independent two-level observables, follow automatically from the
uncertainty relations or similar properties of the quantum formalism. In our
example the two-level observables are represented by some type of classical
bits, as neurons in two different possible states. Then the quantum correlations
have to be prepared by the initialization through a quantumness gate. It seems
very unlikely that the high precision of quantum correlations for sufficiently
isolated real quantum particles can be reached by the classical spins. Also the
extension to many entangled qubits is, in principle, straightforward for real
quantum particles, while it is far from obvious for classical bits. This points
to a clear superiority of real quantum computers as compared to realizations by
classical bits.

What is then the point of our very simple correlated computers realized by
classical bits? It is the demonstration that the performance of tasks by
correlated computing does not need the extreme isolation of microscopic qubits.
The Ising spins can be macroscopic observables, and the issue of decoherence
does not seem to play an important role. This implies that correlated computing
may be used in practice by nature for performing certain tasks. The evolution
may have taught animals to employ correlations between the states of neurons in
order to recognize patterns or to store memory. In the same way deep artificial
neural networks may learn to employ correlations between different building
blocks. Our simple examples demonstrate that the presence of probabilities needs
not to reduce the computational power. In contrast, probabilities open the door
to the use of correlations. Much more information can be stored in correlations
between configurations of Ising spins than in the sharp independent
configurations. And, most important, for a correlated system the manipulation of
one observable can simultaneously affect many other observables as well.

\section{Conditional probabilities and measurements}
\label{sec:conditional_probabilities_4_7}

Probabilistic realism~\cite{CWPW} is based on the concept of an overall
probability distribution, describing the whole Universe from the infinite past
to the infinite future. In practice, one is often interested, however, in
subsystems that are local in time and space. A typical physicists question asks:
If I have prepared certain initial experimental conditions, what will be the
outcome? This type of questions concerns conditional probabilities.

Conditional probabilities are the key concept for understanding the outcomes of
sequences of measurements. One needs the conditional probabilities $(w_a^A)_b^B$
to find for an observable $A(t_2)$ the value $a$ under the condition that
another observable $B(t_1)$ has been found previously to have the value $b$.
They determine the correlations found in sequences of measurements -- the
measurement correlations. Conditional probabilities and measurement correlations
are not unique -- they depend on the details how measurements are performed.
Conditional probabilities cannot be computed from the probabilistic information
for a subsystem. They involve the specification of additional input on the
circumstances how a sequence of two measurements for the subsystem is performed.
We define criteria for ideal measurements for subsystems. For ideal measurements
in typical subsystems the measurement correlations do not correspond to the
classical correlations in the overall probabilistic system. This is particularly
apparent for the continuum limit of time-local subsystems. The classical
correlation functions depend on precise details of the environment and
measurement apparatus, as well as on details of averaging procedures. They do
not correspond to ideal measurements. In contrast, there exist other robust
measurement correlations obeying the criteria for ideal measurements. They are
based on operator products in our formalism for time-local subsystems. 

We discuss the connection between ideal measurements and conditional
probabilities in detail since many misconceptions for measurements in quantum
mechanics arise from an insufficient consideration of conditional probabilities.
In particular, the reduction of the wave function is nothing else than an
appropriate formalism for the description of conditional probabilities for
decoherent ideal measurements. There is no need to relate the reduction of the
wave function to a non-unitary physical process or to a ``many world
interpretation'' of quantum mechanics. We will find that for most cases the
classical correlation function is not an acceptable measurement correlation for
ideal measurements in subsystems. Confounding measurement correlations and
classical correlations is at the root of some other ``paradoxes of quantum
mechanics'' and ``no-go theorems'' that we will discuss in
sect.\,\ref{sec:the_paradoxes_of_quantum_mechanics}.

\subsection{Conditional probabilities}
\label{sec:conditional_probabilities}

Conditional probabilities concern sequences of events or observations. These are
typically time sequences, but not necessarily so. Consider
two Ising spins $A$ and $B$. One wants to make statements about the probability
for the event $A=1$, given that the event
$B=1$ has happened. For a corresponding sequence of two measurements the
question asks for the conditional probability 
$(w_+^A)_+^B$ to find for $A$ the value $+1$ if $B$ is measured to be $B=1$.
Similarly, the conditional probability to find
$A = +1$ given that $B = -1$ is denoted by $(w_+^A)_-^B$ and so on. For an Ising
spin $A$ either $A=1$ or $A=-1$ has to happen 
independently of the outcome of the measurement of $B$, such that the
conditional probabilities obey the rule
\begin{equation}
\label{M1}
(w_+^A)_+^B + (w_-^A)_+^B = 1\,, \quad (w_+^A)_-^B + (w_-^A)_-^B = 1\,.
\end{equation}
The generalization of conditional probabilities to observables with more than
two possible measurement values is straightforward.

Most statements in physics concern conditional probabilities, rather than the
probabilities for events $A$ or $B$. Imagine that a person
holds a pen between two fingers one meter above the floor and opens the hand. A
physicist would predict that after some time the probability for the pen to be
on the floor in a radius of $20\,$m around this person ($A=1$) is close to one
if the hand is open
($B=1$), while the probability that the pen is not on the floor at this location
($A=-1$) is almost zero. This is a statement about
conditional probabilities, $(w_+^A)_+^B \approx 1$, $(w_-^A)_+^B \approx 0$. In
contrast, from the point of view of the overall 
probability distribution for the whole Universe the probability for a pen to be
on the floor at the given time and place 
$w_+^A = w(A=1)$ is almost zero. Given initial conditions at the time of the
emission CMB fluctuations the pen on the floor requires
1) that a galaxy has formed in the vicinity of the position $x$ on the floor, 2)
that a star with a planet is there, 3) that a
civilization with pens has developed and so on. If $A=1$ is the event that there
is a pen in some interval of time and space 
around $t$ and $x$, the probability $w(A=1)$ is extremely close to zero, in
contrast to the conditional probability $(w_+^A)_+^B$ which
is very close to one. It is rather obvious from this simple example that the
interest lies in the conditional probabilities, not in 
the probabilities themselves. For most practical purposes one uses conditional
probabilities without naming them in this way. The
condition that certain initial conditions have been prepared is not mentioned
explicitly. 

It is important to distinguish between the conditional probabilities
$(w_+^A)_+^B$ and the probabilities $w_{++}^{(AB)}$ to find the 
events $A=1$, $B=1$ in the overall probabilistic system. The probability
$w_{++}^{(AB)}$ to find a sequence $B=1$, $A=1$ can be 
expressed as the product of the probability for the event $B=1$ and the
conditional probability to find $A=1$ for $B=1$ given 
\begin{equation}
\label{M2}
w_{++}^{(AB)} = (w_+^A)_+^B w(B=1) = (w_+^A)_+^B w_+^B\,.
\end{equation}
While both $w(B=1) = w_+^B$ and $w_{++}^{(AB)}$ may be very small, the
conditional probability $(w_+^A)_+^B$ can be large. 
This is precisely what happens in our example with the pen. From the point of
view of the whole Universe the probability for a hand
with a pen opening at $t_1$ and $x_1$, e.g. $w(B=1)$, is tiny. Also the
probability $w_{++}^{(AB)}$ for the two events, a hand with a
pen opening at ($t_1,x_1$), ($B=1$), and a pen at ($t_2, x_2$), ($A=1$), is
extremely small. Nevertheless, the conditional probability,
given formally by
\begin{equation}
\label{M3}
 (w_+^A)_+^B = \frac{w_{++}^{(AB)}}{w_+^B}\,,
\end{equation}
is close to one. We note that conditional probabilities can be determined by
eq.~\eqref{M3} for arbitrarily small non-zero $w_+^B$.
Eq.~\eqref{M3} is not meaningful, however, if the probability for the event
$B=1$ is precisely zero, e.g. $w_+^B = w_{++}^{(AB)} = w_{-+}^{(AB)} = 0$.

In the following we will often forget about the overall probabilistic system and
rather concentrate on subsystems as quantum systems prepared by some
experimental conditions. The importance of conditional probabilities for
sequences of measurements does not change. Only the probabilistic information of
the subsystem replaces the overall probability distribution. We will denote in
this case by $w_+^A$ the probability to find the value $+1$ for the observable
$A$, as computed from the probabilistic information of the subsystem. The
meaning and notation of conditional probabilities remains the same as before.

Let us consider subsystems with incomplete statistics, where $A$ and $B$ are
system observables but the classical correlation function 
$\braket{AB}_\text{cl}$ is not available. The probabilistic information of the
subsystem permits the computation of $w_\pm^A$ and
$w_\pm^B$, but provides no direct prescription how to compute $w_{++}^{(AB)}$
etc. At this stage neither $w_{++}^{(AB)}$ nor the 
conditional probability $(w_+^A)_+^B$ is fixed. For subsystems with incomplete
statistics one needs some type of independent information
that defines the conditional probabilities. In other words, the conditional
probabilities are not simply properties of the 
probabilistic information of the subsystem. They need additional input how a
sequence of measurements of $A$ and $B$ is done. This will
lead us to the concept of ideal measurements.

\subsection{Sequence of measurements}
\label{sec:sequence_of_measurements}

A conceptual understanding of measurements is a rather complex issue. From the
point of view of the overall probabilistic description
of the world, measurements and the humans or apparatus performing them are part
of the world and included in the overall probability
distribution. We do not aim here for a systematic discussion of the measurement
process. We rather highlight a few aspects~\cite{CWB} that are
crucial for the conceptual understanding of a ``classical'' probabilistic
description. 

In particular, we emphasize that the correlations found in sequences of
measurements are not unique. They depend on the precise way how measurements are
performed. Correspondingly, there are many different product structures of
observables that correspond to sequences of measurements. In general, they do
not correspond to the ``classical product'' of observables, which is often not
available for a subsystem. The products of observables relevant for measurements
are often non-commutative. The order in a sequence of measurement matters.

\paragraph*{Different types of measurements}

Every student in physics learns that the outcome of a measurement or a sequence
of measurements depends on how the measurement is done. 
There are good measurements that provide valuable information about a system,
and other measurements that depend on rather random 
circumstances of the environment of a system. In the case of two Ising spins or
yes/no decisions $A$ and $B$, every sequence of first 
measuring $B$ and subsequently $A$ will give one of the four possible results
($++$), ($+-$), ($-+$), ($--$). Imagine a physics class
where each student should perform the sequence of measurements of $A$ and $B$
with his own constructed apparatus. An experienced 
researcher may be able to estimate the outcomes of the different measurement
devices. She may concentrate on the measurements where the
first measurement has found $B=1$. Knowing the physics law behind the
experiment, she predicts that an ideally constructed apparatus
will find $A=1$. For this ideal apparatus the conditional probability is
$(w_+^A)_+^B = 1$. For some other apparatus she may 
judge that the outcome will be random $(w_+^A)_+^B = (w_-^A)_+^B = 1/2$. And
still for others there will be conditional probabilities
in between. 

The lesson from this simple example is that conditional probabilities for a
sequence of measurements depend on how the measurement is
done. The conditional probabilities do actually not depend on the judgement of
the experienced researcher. In view of a probabilistic
description of the whole process they are properties of the measurement
apparatus employed. Only for an ideal measurement apparatus 
one has a conditional probability, $(w_+^A)_+^B = 1$, independently if this has
been recognized by the experienced researcher or not.

\paragraph*{Measurement correlation}

Depending on the precise way how a measurement is done one will find different
``measurement correlations". For the two
observables $A$ and $B$ the outcomes depend on the conditional probabilities.
Those depend, in turn, on the way how the measurements
are done. A basic rule for measurements associates the probabilities $w^{(AB)}$
for the outcome of a sequence to the conditional
probabilities and the probabilities to find a given value for the first
measurement, 
\begin{align}
\label{M4}
w_{++}^{(AB)} &= (w_+^A)_+^B w_+^B\,, \quad w_{+-}^{(AB)} = (w_+^A)_-^B w_-^B\,,
\nonumber \\
w_{-+}^{(AB)} &= (w_-^A)_+^B w_+^B\,, \quad w_{--}^{(AB)} = (w_-^A)_-^B w_-^B\,.
\end{align}
This can be used in order to define the measurement correlation
\begin{equation}
\label{M5}
\braket{AB}_m = w_{++}^{(AB)} + w_{--}^{(AB)} - w_{+-}^{(AB)} -w_{-+}^{(AB)}\,.
\end{equation}
The measurement correlation is not a universal quantity. It depends on how the
measurement is done, as expressed by the conditional probabilities. 
In general, the measurement correlation is not the classical correlation
function or any other universally defined correlation function.
It always involves the particular realisation of a measurement by a given
apparatus or observation.

By using the same symbol $w_{++}^{(AB)}$ for the probability of a sequence of
events in some time- and space-local subsystem as the one used for the overall
probability distribution of the Universe in eq.\,\eqref{M2} we follow a commonly
used procedure. We treat the subsystem as if it would be the whole Universe and
take it for this particular purpose as a replacement of the overall
probabilistic system. From the point of view of the whole Universe
$w_{++}^{(AB)}$ is related to some sort of conditional probability, namely under
the condition that a suitable subsystem is realized. This condition is here
tacitly assumed, and the context makes the meaning of $w_{++}^{(AB)}$ clear.

Still, the time- and space-local subsystem associated to the new overall system
can be much larger than the subsystem actually employed for the description of
the sequence of measurements. It may contain an environment which may influence
the outcome of the sequence of measurements. Typically, the precise details of
the measurement apparatus are part of this environment. The subsystem for $A$
and $B$ does not include the probabilistic information related to these details.
Subsystems occur here on different levels. For example, the quantum subsystem
for the observables $A$ and $B$ is a subsystem of the experiment-subsystem which
includes the measurement apparatus. This in turn is a subsystem of the overall
probabilistic system. We also deal with classical statistical subsystems of the
overall probabilistic system, which have in turn quantum systems as subsystems.

\paragraph*{Different products of observables for different\\measurements}

Within the subsystem with observables $A$ and $B$ one can formally define a
product of the two observables $(A \circ B)_m$ such that its expectation value
is the measurement correlation,
\begin{equation}
\label{M6}
\braket{(A \circ B)_m} = \braket{AB}_m\,.
\end{equation}
Indeed, for any given apparatus the sequence of measurements of $A$ and $B$ is a
new combined observable with possible measurement values $\pm 1$. 
Different types of apparatus correspond to different products $(A \circ B)_m$.
We conclude that for subsystems the product of two observables is not 
unique. There exist many different definitions of observable products $C =
(A\circ B)_m$, since there are many different ways to
perform measurements. The classical observable product in the overall
probabilistic system, $C_\tau = A_\tau B_\tau$, is only
one out of many possibilities. We will see that for many subsystems, in
particular for subsystems with incomplete statistics, it plays
no role.

The product $A\circ B$ is, in general, not commutative. The order in the
sequence of two measurements can matter. It makes a difference
if $A$ or $B$ are measured first. Thus the measurement correlation can depend on
the order of the two factors
\begin{equation}
\label{M6A}
\braket{AB}_m \neq \braket{BA}_m\,,
\end{equation}
in distinction to the classical correlation. This can be seen by the different
expressions in terms of the conditional probabilities
\begin{equation}
\label{M6B}
\braket{BA}_m = w_{++}^{(BA)} + w_{--}^{(BA)} - w_{+-}^{(BA)} -w_{-+}^{(BA)}\,,
\end{equation}
where
\begin{align}
\label{M6C}
w_{++}^{(BA)} &= (w_+^B)_+^A w_+^A\,, \quad w_{+-}^{(BA)} = (w_+^B)_-^A w_-^A\,,
\nonumber \\
w_{-+}^{(BA)} &= (w_-^B)_+^A w_+^A\,, \quad w_{--}^{(BA)} = (w_-^B)_-^A w_-^A\,.
\end{align}
There is no a priori direct relation between $w^{(AB)}$ in eq.~\eqref{M4} and
$w^{(BA)}$ in eq.~\eqref{M6C}. One has to 
find this relation for each concrete sequence of two measurements.

The expectation value of the observable that is measured first does not depend
on the conditional probabilities for the sequence of
measurements. Measuring first $B(t_1)$ one has
\begin{equation}
\label{M6D}
\braket{B(t_1)} = \braket{B} = w_+^B - w_-^B\,,
\end{equation}
where the probabilities $w_\pm^B$ to find $B=\pm 1$ are part of the
probabilistic information of the subsystem. For the expectation
value of the second observable $A(t_2)$ the way how the measurement is performed
matters, however. Indeed, $\braket{A(t_2)}$ involves
the conditional probabilities
\begin{multline}
\label{M6E}
\braket{A(t_2)}_B = \braket{A}_B = w_{++}^{(AB)} + w_{--}^{(AB)} - w_{+-}^{(AB)}
-w_{-+}^{(AB)} \\
= \left[ (w_+^A)_+^B - (w_-^A)_+^B \right] w_+^B + \left[ (w_+^A)_-^B -
(w_-^A)_-^B \right] w_-^B \,.
\end{multline}

Performing first a measurement of $B(t_1)$ can influence the expectation value
for $A(t_2)$. The expectation value $\braket{A(t_2)}_B$
can differ from the expectation value obtained without the measurement of $B$,
i.e.
\begin{equation}
\label{M6F}
\braket{A(t_2)} = w_+^A - w_-^A\,.
\end{equation}
The expectation value \eqref{M6F} describes a measurement in the subsystem which
evolves without any disturbance. In contrast, 
$\braket{A}_B$ in eq.~\eqref{M6E} takes into account that the subsystem may be
influenced by the measurement of $B(t_1)$. The
measurement brings a subsystem into contact with its environment. A closed
subsystem follows its evolution law, as formulated in terms
of the probabilistic information for the subsystem, only for the time between
measurements. The interaction with the environment due
to the measurement of $B$ at $t_1$ can influence the state of the subsystem at
$t_1$, which serves as initial condition for the 
evolution at $t > t_1$. It is this influence that is responsible for a possible
difference between $\braket{A(t_2)}_B$ and $\braket{A(t_2)}$.

\paragraph*{Conditional probabilities from measurement\\correlations}

The relation between the conditional probabilities and the measurement
correlation can be inverted. Whenever $\braket{AB}_m$,
$\braket{A}_B$ and $\braket{B}$ are known, one can reconstruct the conditional
probabilities if they are defined. With
\begin{equation}
\label{M7}
w_\pm^B = \frac{1}{2} \left( 1 \pm \braket{B} \right)
\end{equation}
and
\begin{align}
\label{M8}
w_{++}^{(AB)} = \frac{1}{4} (1+\braket{A}_B+\braket{B}+ \braket{AB}_m)\,,
\nonumber \\
w_{+-}^{(AB)} = \frac{1}{4} (1+\braket{A}_B-\braket{B}- \braket{AB}_m)\,,
\nonumber \\
w_{-+}^{(AB)} = \frac{1}{4} (1-\braket{A}_B+\braket{B}- \braket{AB}_m)\,,
\nonumber \\
w_{--}^{(AB)} = \frac{1}{4} (1-\braket{A}_B-\braket{B}+ \braket{AB}_m)\,,
\nonumber \\
\end{align}
the conditional probabilities obtain by inverting the relations \eqref{M4}. For
the system of two Ising spins we
observe a one to one correspondence between the measurement correlation and
expectation values $\braket{B}$, $\braket{A}_B$ on one side
and the conditional probabilities on the other side.


\subsection{Ideal measurements for subsystems}
\label{sec:ideal_measurements_for_subsystems}

Not all measurements are equivalent -- some are better than others. Physicists
have developed the concept of ``ideal measurements"
in order to find out the properties of subsystems. An ideal measurement
apparatus is one that is best suited to measure the
properties of the subsystem rather than its environment. The concept of an ideal
measurement may single out a particular set of 
conditional probabilities or a particular measurement correlation among the many
possibilities. In turn, it may single out a specific
ideal observable product $A \circ B$ among the many possible choices of products
$(A \circ B)_m$. (If we discuss ideal measurements we often will omit the
subscript $m$ for measurement.)

Ideal measurements should be as insensitive as possible to the state of the
environment of a subsystem, and we develop criteria for this property. An
important finding is that the measurement correlations for ideal measurements
are not given by the classical correlation function. We distinguish between
coherent and decoherent ideal measurements. For the particular case of quantum
subsystems we discuss in detail the different outcomes for correlation functions
for these two types of ideal measurements. For coherent ideal measurements the
measurement correlation is the quantum correlation as defined by the product of
Heisenberg operators. For decoherent ideal measurements the state of the quantum
subsystem is influenced by the interaction with the environment during the
measurement process. This is related to the ``reduction of the wave function".

\paragraph*{Criteria for ideal measurements}

Ideal measurements for subsystems should obey five criteria:
\begin{enumerate}[wide=0pt,listparindent=1.25em]
\item\label{item:ci} \textit{Measurement of subsystems properties}

The measurement should measure properties of the subsystem, not of its
environment. The outcome of a given ideal measurement should only depend on the
probabilistic information of the subsystem. This means that the conditional
probabilities and the measurement correlation should be computable from the
variables characterizing the subsystem.
 
\item\label{item:cii} \textit{Independence of environment}

In other words, the outcome of an ideal measurement for a subsystem should not
depend on the state of its environment. This is a type of ``common sense
criterion'' that is used in practice. The influence of the state of the
environment is considered as ``noise'' which has to be minimized for an ideal
measurement. The criterion \ref{item:cii} does not state that the environment plays no
role for the measurement. Only the outcome of the measurement should not depend
on the particular state of the environment.

Typically, the measurement apparatus is part of the environment of a subsystem.
The measurement process necessarily involves an interaction between the
subsystem and the measurement apparatus, and therefore an interaction between
the subsystem and its environment. Nevertheless, ideal measurements should not
introduce additional probabilistic information from the environment into the
subsystem, or at least should restrict such additional information to a minimum.
Despite the interaction with the environment during the measurement process and
a possible change of state of the subsystem induced by this interaction, the
outcome of the sequences of ideal measurements should not depend on the state of
the measurement apparatus. A possible change of state of the subsystem during
the measurement process should be computable from the probabilistic information
of the subsystem.

\item\label{item:ciii} \textit{Non-intrusiveness}

The outcome of a sequence of ideal measurements should be computable with the
time evolution of the subsystem between two measurements. We distinguish
coherent and decoherent ideal measurements. For coherent ideal measurements the
first measurement of $B(t_1)$ should not alter the subsystem. The probabilistic
information of the subsystem after the measurement is the same as before the
measurement. This implies, in particular, that $\braket{A(t_2)}$ is the same
with or without the measurement of $B(t_1)$. If the probabilistic information in
a subsystem is sufficient to compute for two probabilistic observables $A(t_2)$
and $B(t_1)$ the expectation value $\braket{A(t_2)B(t_1)}$, $t_2 > t_1$, this
correlation should coincide with the measurement correlation for coherent ideal
measurements. For probabilistic observables $\braket{A(t_2)B(t_1)}$ often
differs from the classical correlation function.

 For ``decoherent ideal measurements'' the measurement of $B(t_1)$ may change
the state of the subsystem. This change should be computable from the
probabilistic information of the subsystem. The non-intrusiveness of the
measurement procedure is now limited to the time inbetween measurements.

\item\label{item:civ} \textit{Repetition of identical measurements}

If the second observable $A(t_2)$ is identical to $B(t_1)$ and $t_2 \to t_1$,
one measures twice the observable $B(t_1)$. If the first measurement finds the
value $b_m$, the second measurement should find this value again. In the
continuum limit for time this property should hold if $| t_2 - t_1 |$ is much
smaller than the characteristic time for the evolution of the subsystem.

\item\label{item:cv} \textit{Compatibility with equivalence classes}

Probabilistic observables for a subsystem correspond to equivalence classes of
observables of the overall probabilistic system. Many different observables of
the overall system are mapped to the same probabilistic observable of the
subsystem. They form an equivalence class of observables. From the point of view
of the subsystem the differences between overall observables belonging to the
same equivalence class should be regarded as properties of the environment.
Ideal measurements in a subsystem should be compatible with the notion of the
equivalence class. The outcome should only depend on the equivalence class, not
on the specific member. If two observables $A$ and $A'$ belong to the same
equivalence class they may still be different observables in the overall system.
This difference concerns properties of the environment of the subsystem. Ideal
measurements in a subsystem should not be sensitive to this difference. The
outcome should be the same for all members of a given equivalence class.

\end{enumerate}

The five criteria are not independent. They reflect different facets of the
basic requirement
that any ideal measurement in a subsystem should be as independent as possible
from the
state of the environment.

\paragraph*{Time-local subsystem}

For an understanding of time-sequences of measurements we employ the time-local
subsystem as a crucial concept. It is characterized by the probabilistic
information at a given time $t$. Time-local subsystems can be defined for every
overall probabilistic system for all events in time and space. They are
discussed in detail in ref.~\cite{CWPW}. We summarize here the results relevant
for measurement sequences.

By integrating in the overall probability distribution over the probabilities
for all configurations in the past, $t'<t$, and focusing on a given
configuration $\tau$ at $t$, one obtains the classical wave function $\tilde
q(t)$ with components $\tilde q_\tau(t)$. The wave function $\tilde q(t+\eps)$
at the next time step is determined by a linear evolution law involving the step
evolution operator $\Shat(t)$,
\begin{equation}
\label{eq:LT1}
\tilde q_\tau(t+\eps) = \Shat_{\tau\rho}(t)\tilde q_\rho(t)\,.
\end{equation}
Similarly, integrating out the future, $t'>t$, yields the conjugate wave
function $\bar q(t)$. We concentrate here on orthogonal step evolution operators
$\Shat(t)$. In this case one can identify $\bar q(t)=\tilde q(t)=q(t)$, once one
factors out an overall transition amplitude~\cite{CWPW}. The time-local
classical probabilities $p_\tau(t)$ are then given by the square of the
components of the classical wave function
\begin{equation}
\label{eq:LT2}
p_\tau(t)=q_\tau^2(t)\,.
\end{equation}
For probabilistic automata we recover the setting of
sect.~\ref{sec:classical_wave_function_and_step_evolution_operator}. The
classical wave function is a probability amplitude. In contrast to the general
case for local probabilities it obeys a linear evolution law~\eqref{eq:LT1}. As
compared to the probabilities the classical wave function contains some
redundancy, namely the signs of $q_\tau(t)$. The corresponding local
$\mathbb{Z}_2$ gauge symmetry can be fixed by choosing a sign convention for
$q_\tau(t)$. Due to the linear evolution law and the possibility of basis
changes familiar from quantum mechanics the classical wave function is more
suitable for the understanding of the time evolution than the classical
time-local probability distribution.

The classical density matrix $\rho'(t)$ is a bilinear of the wave function
\begin{equation}
\label{eq:LT3}
\rho'_{\tau\rho}(t) = q_\tau(t)q_\rho(t)\,.
\end{equation}
It contains more time-local probabilistic information than the probabilities
which are given by its diagonal elements, $p_\tau(t)=\rho_{\tau\tau}(t)$. It
obeys again a linear evolution law, $\Shat^{-1}(t) = \Shat^T(t)$,
\begin{equation}
\label{eq:LT4}
\rho'(t+\eps) = \Shat(t)\rho'(t)\Shat^{-1}(t)\,.
\end{equation}
This can be extended to an arbitrary time $t_1>t$
\begin{equation}
\label{eq:LT5}
\rho'(t_1) = U(t_1,t)\rho'(t)U^{-1}(t_1,t)\,,
\end{equation}
with
\begin{equation}
\label{eq:LT6}
U(t_1,t) = \Shat(t_1-\eps)\Shat(t_1-2\eps)\dots\Shat(t+\eps)\Shat(t)\,.
\end{equation}

For a time-local classical observable $A(t)$ the possible measurement values
$A_\tau(t)$ depend only on the configuration $\tau$ at $t$. Its expectation
value obeys
\begin{equation}
\label{eq:LT7}
\langle A(t)\rangle = \sum_\tau p_\tau(t)A_\tau(t) =
\tr\big\{\rho'(t)\hat A(t)\big\}\,,
\end{equation}
with diagonal operator $\hat A_{\tau\rho}(t) = A_\tau(t)\delta_{\tau\rho}$. This
generalizes eq.~\eqref{QM6}. The probabilistic information of the density matrix
at $t$ is sufficient for a computation of the expectation value of a time-local
classical observable at $t_1$ different from $t$,
\begin{align}
\label{eq:LT8}
\langle A(t_1)\rangle =& \tr\big\{\hat A(t_1)\rho'(t_1)\big\}\nonumber\\
=& \tr\big\{\hat A(t_1) U(t_1,t)\rho'(t)U^{-1}(t_1,t)\big\}\nonumber\\
=& \tr\big\{\hat A_H(t_1,t)\rho'(t)\big\}\,,
\end{align}
where we introduce the Heisenberg operator
\begin{equation}
\label{eq:LT9}
\hat A_H(t_1,t) = U(t,t_1)\hat A(t_1)U(t_1,t)\,.
\end{equation}
With
\begin{equation}
\label{eq:LT10}
U(t,t_1) = U^{-1}(t_1,t)\,,\quad U(t,t) = 1\,,
\end{equation}
the relation~\eqref{eq:LT8} is valid for arbitrary $t_1$.

Let us consider next two time-local classical observables $A(t_2)$ and $B(t_1)$.
Their classical correlation function $\langle A(t_2)B(t_1)\rangle\subt{cl}$ can
be computed from the overall probability distribution. Equivalently it can be
expressed in terms of the time ordered product of Heisenberg operators
\begin{equation}
\label{eq:LT11}
\langle A(t_2)B(t_1)\rangle\subt{cl} = \tr\big\{\rho'(t)T\{\hat A_H(t_2,t)\hat
B_H(t_1,t)\}\big\}\,,
\end{equation}
with
\begin{equation}
\label{eq:LT12}
T\{\hat A_H(t_2,t)\hat B_H(t_1,t)\} = \begin{cases} \hat A_H(t_2,t)\hat
B(t_1,t)\ \text{for}\ t_2>t_1\\ \hat B_H(t_1,t)\hat A_H(t_2,t)\ \text{for}\
t_2<t_1\,.\end{cases}
\end{equation}
The time ordering is important since in general the operators $\hat A_H(t_2,t)$
and $\hat B(t_1,t)$ do not commute. The time ordered product is commutative, as
appropriate for the classical correlation function. We will see that it is
precisely this time ordering that makes the classical correlation inappropriate
for a measurement correlation in the continuum limit.

\paragraph*{Coherent ideal measurements}

The measurement correlation depends on the type of local observables. For the
sake of simplicity we focus on two-level observables $A, B$. The simplest case
are Ising spins at neighboring sites, as $A=s(t_2), B=s(t_1), t_2 > t_1$.
In this case the expectation value $\braket{AB}$ can be computed from the
probabilistic information of the subsystem. For coherent ideal measurements one
has according to the criterion~\ref{item:ciii}
\begin{equation}
 \label{M9}
 \braket{A B}_m = \braket{s(t_2)s(t_1)} =
 \tr \left\{ \rho'(t) \hat{A}_H(t_2,t) \hat{B}_H(t_1,t) \right\}\,,
\end{equation}
with $\hat{A}_H(t_2,t)$ and $ \hat{B}_H(t_1,t)$ the Heisenberg operators
associated
to $A = s(t_2)$ and $B = s(t_1)$. The density matrix $\rho'$ and
Heisenberg operators for observables are defined for the time-local ``classical"
probabilistic subsystem. Since $\rho'$ is a symmetric matrix only the symmetric
part of the operator product $\hat A_H(t_2)\hat B_H(t_1)$ contributes to
$\langle AB\rangle_m$. We may therefore use equivalently an expression in terms
of the anticommutator $\{\hat A_H, \hat B_H\}$,
\begin{equation}
\label{7.3.1A}
\langle AB\rangle_m = \frac12\tr\Big(\rho'(t)\{\hat A_H(t_2,t),\hat
B_H(t_1,t)\}\Big)\,,
\end{equation}
where we employ that $\hat A_H$ and $\hat B_H$ are symmetric matrices for
orthogonal $\Shat$ and $U$.

The expectation value
\begin{equation}
 \label{M9A}
 \braket{A(t_2)} = \tr \left\{ \rho'(t) \hat{A}_H(t_2,t) \right\}\,,
\end{equation}
is the same if $B$ is measured or not. The conditional probabilities can be
inferred from this measurement correlation and the expectation values
$\braket{A(t_2)}$ and $\braket{B(t_1)}$ according to eq.~\eqref{M8}, \eqref{M4}.
For this particular case the measurement correlation \eqref{M9} coincides with
the classical correlation in the overall probabilistic system~\cite{CWPW}.

The prescription for ideal coherent measurements by Heisenberg
operators~\eqref{M9} can be extended to arbitrary observables $A(t_2)$ and
$B(t_1)$. This formulation in terms of the Heisenberg operators $\hat{A}_H,
\hat{B}_H$ is compatible with the notion of equivalence classes. Two members of
a given equivalence class are mapped to the same Heisenberg operator. By the
criterion~\ref{item:civ} the same measurement correlation should hold for all local
probabilistic observables $A'$ and $ B'$ that are represented by the operators
$\hat{A}_H$ and $\hat{B}_H$, respectively. As we have discussed in
ref.~\cite{CWPW}, the classical correlations in the overall probabilistic
system often differ for different representatives in the same equivalence class.
Thus the relation \eqref{M9} for the measurement correlation implies that, in
general, the measurement correlation differs from the classical correlation. The
robust object that respects the equivalence class is an observable product $A
\circ B$ based on the operator product $\hat{A}_H\hat{B}_H$. It is typically
non-commutative.

\paragraph*{Continuum limit in time}

These more formal considerations become particularly relevant for the continuum
limit in time. We will argue that in this limit the classical correlation
function based on the time-ordered operator product is no longer compatible with
the criteria for ideal measurements. The reason is a clash between time ordering
and time averaging. In contrast, the operator product without time ordering
remains compatible with an ideal measurement.

In the continuum limit the relevant observables typically involve an averaging
over infinitesimal time steps. In ref.~\cite{CWPW} we have introduced a
time-averaged spin observable
\begin{equation}
\label{eq:CT1}
\sigma(\bar{t}) = \sum_{t'}a^{(\sigma)}(\bar t + t')s(\bar t + t')\,,
\end{equation}
with averaging function $a^{(\sigma)}$ taking a suitable shape such that only
some time-region around $\bar t$ contributes effectively to the averaged spin.
Since $\sigma(\bar t)$ is linear in the spins $s(\bar t + t’)$ we can introduce
the associated Heisenberg operator
\begin{equation}
\label{eq:CT2}
\hat \sigma_H(\bar t,t) = \sum_{t'}a^{(\sigma)}(\bar t + t') U(t,\bar t +
t')\hat s(\bar t + t')U(\bar t + t',t)
\end{equation}
and compute the expectation value as
\begin{equation}
\label{eq:CT3}
\langle\sigma(\bar t)\rangle = \tr\big\{\rho'(t)\hat\sigma_H(\bar t,t)\big\}\,.
\end{equation}
There is a large family of different averaging functions $a^{(\sigma)}$ which
lead to the same Heisenberg operator $\hat\sigma_H(\bar t,t)$. If this
difference cannot be resolved in the continuum limit for time the corresponding
average observables $\sigma(\bar t)$ for this family should be considered as
equivalent.

One can define a possible correlation function for $t_2 > t_1$ as
\begin{equation}
 \label{M11}
 \braket{\sigma(t_2) \sigma(t_1)} =
 \tr \left\{ \rho'(t) \hat{\sigma}_H(t_2,t) \hat{\sigma}_H(t_1,t) \right\}\,.
\end{equation}
This is a good candidate for the measurement correlation for coherent ideal
measurements in the subsystem. It respects the structure of equivalence classes,
is compatible with the time evolution of the subsystem and only uses the
probabilistic information in the subsystem as encoded in the classical density
matrix $\rho'(t)$.

The correlation \eqref{M11} is, in general, no longer a classical correlation
function. The classical correlation function for $\sigma(t_2) \sigma(t_1)$
involves in eq.~\eqref{M11} the time ordered operator $\TO \left\{
\hat{\sigma}_H(t_2,t) \hat{\sigma}_H(t_1,t) \right\}$ instead of the product
$\hat{\sigma}_H(t_2,t) \hat{\sigma}_H(t_1,t)$. The two expressions only coincide
if $t_2-t_1$ is sufficiently large as compared to the interval used for the
averaging $\Delta t$, and the Heisenberg operators $\hat{s}_H(t_2+t',t)$ and
$\hat{s}_H(t_1+t',t)$ in the definition of $\hat{\sigma}_H(t_2,t)$ and
$\hat{\sigma}_H(t_1,t)$ have no overlapping time region. Whenever $t_2-t_1$
becomes of the order $\Delta t$ or smaller the time ordered product becomes very
complicated. It involves microscopic details not available in the continuum
limit. It does not respect the notion of equivalence classes since different
average procedures that lead to the same operators $\hat{\sigma}_H(t_2,t)$ or
$\hat{\sigma}_H(t_1,t)$ do not yield the same time ordered products.
Furthermore, no simple time evolution law exists for the time ordered product.

We conclude that the classical correlation function is not suitable for the
measurement correlation for ideal measurements in the classical time-local
subsystem. We generalize these findings by postulating that for two local
observables $A(t_2), B(t_1), t_2 > t_1$ the measurement correlation for coherent
ideal measurements is given by
\begin{equation}
 \label{M12}
 \braket{A(t_2)B(t_1)}_m =
 \tr \left\{ \rho'(t) \hat{A}_H(t_2,t) \hat{B}_H(t_1,t) \right\}\,.
\end{equation}
Here $\hat{A}_H(t_2,t)$ and $\hat{B}_H(t_1,t)$ are the Heisenberg operators
associated to $A(t_2)$ and $B(t_1)$. This measurement correlation obeys all
criteria for ideal measurements in a subsystem. It equals suitable classical
correlation functions in certain limiting cases, but is a much more robust
object. If $A(t_2)$ and $B(t_1)$ are two-level observables, the measurement
correlation fixes the conditional probabilities for ideal measurements. For
more general cases also higher order measurement correlations will be needed for
the determination of the conditional probabilities.

The measurement correlation \eqref{M12} in terms of the operator product does
not employ particular properties of quantum subsystems. It holds for all
local-time subsystems, both for quantum systems and more general probabilistic
subsystems. The central motivation arises from the continuum limit for
the time-local subsystem. Still, the measurement correlation \eqref{M12} remains
a postulate for coherent ideal measurements. This is necessarily so and there is
no direct way to derive conditional probabilities from the probability
distribution of the subsystem or overall system. One has to \textit{define} what
is an ideal measurement -- this is done in the form of a postulate for
measurement correlations. There may be other possible definitions for ideal
measurements in time-local subsystems. What should be clear at this stage is
that the classical correlation function is not a viable candidate.

It is not always guaranteed that a measurement process exists which leaves the
probabilistic information of the time-local subsystem the same before and after
the measurement of $B(t_1)$. If not, we will have to deal with decoherent ideal
measurements. We will discuss below such decoherent ideal measurements for
quantum subsystems, which are particular local-time subsystems. At the end it
remains an experimental question if a measurement apparatus can be constructed
whose results come close to coherent ideal measurements.

\paragraph*{Quantum subsystems and quantum correlation}

Quantum subsystems are time-local subsystems with incomplete statistics. We may
therefore try to take over the measurement correlation \eqref{M12} for coherent ideal measurements in
time-local subsystems and employ the quantum correlation $\langle
AB\rangle_q$,
\begin{equation}
\label{M13} 
\braket{A(t_2)B(t_1)}_q = \tr \left\{ \rho(t) \hat{A}_H(t_2,t) \hat{B}_H(t_1,t)
\right\}\,.
\end{equation}
Here $\hat{A}_H(t_2,t)$ and $\hat{B}_H(t_1,t)$ are the Heisenberg operators in
the quantum subsystem associated to the observables $A(t_2)$ and $B(t_1)$.
In the real formulation $U(t_2,t_1)$ is an orthogonal matrix and
the density matrix $\rho'$ is symmetric. In the presence of a complex structure
the density matrix $\rho$ becomes a Hermitian complex matrix, and the evolution
operators $U(t_2,t_1)$ are unitary matrices. As a result the quantum correlation
has an imaginary part if the operators $\hat A_H(t_2)$ and $\hat B_H(t_1)$ do
not commute. For a real measurement correlation for coherent ideal measurements
we propose to use the anticommutator
\begin{equation}
\label{7.3.6A}
\langle A(t_2)B(t_1)\rangle_m = \frac12\tr\Big(\rho(t)\{\hat A_H(t_2),\hat
B_H(t_1)\}\Big)\,,
\end{equation}
similar to eq.~\eqref{7.3.1A} for the real formulation. This correlation is
compatible with the notion of equivalence classes of observables and therefore
robust. It obeys all criteria for ideal measurements. Translated to the real
formulation of quantum mechanics it is equivalent to eq.~\eqref{M9}.
Combining the measurement correlation~\eqref{7.3.6A} with the expectation value
as
\begin{equation}
 \label{M13A}
 \braket{A(t_2)} = \tr \left\{ \hat{A}_H(t_2,t) \rho(t) \right\}\,,
\end{equation}
one can extract the conditional probabilities for coherent ideal measurements of
two-level observables from eq.~\eqref{M8}. The conditional probabilities for
coherent ideal measurements can also be obtained directly from a suitable
projection of $\rho$~\cite{CWB}.

\paragraph*{Decoherent ideal measurements}

Not all ideal measurements in quantum subsystems are coherent ideal
measurements. Often a measurement apparatus cannot preserve the coherence of the
quantum information. For this case we define the notion of decoherent ideal
measurements. Bell-type experiments typically assume coherent ideal
measurements, while sequences of Stern-Gerlach type experiments employ
decoherent ideal measurements.

We discuss the concept of decoherent ideal measurements for one qubit quantum
mechanics with $A(t_2)$ and $B(t_1)$ having possible measurement values $\pm 1$.
For a decoherent ideal measurement the measurement of $B(t_1)$ can change the
state of the subsystem. Let us work in a basis where $\hat{B}_H(t_1,t_1)$ is
diagonal, $\hat{B}_H(t_1,t_1) = \tau_3$. The complex density matrix $\rho(t_1)$
takes the general form
\begin{equation}
\label{M18}
\rho(t_1) = \begin{pmatrix}
w_+^B & c \\
c^* & w_-^B
\end{pmatrix},
\end{equation}
with $w_\pm^B$ the probabilities to find $B=\pm1$ and 
\begin{align}
\label{M19}
\braket{B(t_1)} &= \tr \left\{ \rho(t_1) \hat{B}_H(t_1,t_1) \right\} = \tr
\left\{ \rho(t_1) \tau_3 \right\} \nonumber \\
&= w_+^B - w_-^B\,.
\end{align}
For a pure state one has $|c|^2 = w_+^B w_-^B$, while an incoherent mixed state
is characterized by $c=0$.

A decoherent measurement can change $\rho(t_1)$ to $\rho'(t_1)$,
\begin{equation}
\label{M20}
\rho'(t_1) = \begin{pmatrix}
w_+^B & c' \\ c'^* & w_-^B
\end{pmatrix}.
\end{equation}
The diagonal elements of $\rho'(t_1)$ and $\rho(t_1)$ have to be the same in
order to guarantee the criterion (iv) for ideal measurements. For a repetition
of the same measurement the conditional probabilities have to obey
\begin{equation}
\label{M21}
(w_+^B)_+^B = (w_-^B)_-^B = 1\,, \quad (w_+^B)_-^B = (w_-^B)_+^B = 0\,.
\end{equation}
This means that for the second measurement of $B$ one has
\begin{equation}
\label{M22}
{w'}_+^B = (w_+^B)_+^B w_+^B + (w_+^B)_-^B w_-^B = w_+^B\,,
\end{equation}
and similarly for ${w'}_-^B$. The probabilities to find $B= \pm 1$ should not
change by the first measurement of $B$. 
In contrast, the off-diagonal elements $c'$ in $\rho'(t_1)$ are not constrained
by this requirement. They play no role for 
$\braket{B(t_1)}$ or $w_\pm^B$.

For decoherent ideal measurements we assume that the coherent information is
lost by the measurement, as we will discuss in
sect.~\ref{sec:decoherence_and_syncoherence} in more detail. After the
measurement of $B(t_1)$ the state of the quantum subsystem is
described by the incoherent ``reduced density matrix"
\begin{equation}
\label{M23}
\rho_r(t_1) = \begin{pmatrix}
w_+^B & 0 \\ 0 & w_-^B
\end{pmatrix}.
\end{equation}
With $\rho_+(t_1)$ and $\rho_-(t_1)$ pure state density matrices for the
eigenstates with $B(t_1) = \pm 1$,
\begin{equation}
\label{M24}
\hat{B}_H(t_1,t_1) \rho_\pm(t_1) = \pm \rho_\pm(t_1)\,,
\end{equation}
the reduced density matrix can be written as a linear combination of $\rho_\pm$,
\begin{equation}
\label{M25}
\rho_r(t_1) = w_+^B \rho_+(t_1) + w_-^B \rho_-(t_1)\,.
\end{equation}
The relations \eqref{M24}, \eqref{M25} are independent of the basis chosen for
the quantum subsystem. The subsequent evolution 
of $\rho_r(t)$, $t > t_1$ is given by the unitary evolution of the quantum
system
\begin{equation}
\label{M26}
\rho_r(t) = U(t,t_1) \rho_r(t_1) U^\dagger(t,t_1)\,.
\end{equation}
Criterion~\ref{item:ciii} for ideal measurements will be obeyed if we define
conditional probabilities in terms of $\rho_r(t)$. 

For decoherent ideal measurements we postulate the conditional probabilities
\begin{equation}
\label{M27}
(w_+^A)_\pm^B = \tr \left\{ \frac{1}{2} \left(1+\hat{A}_H(t_2,t_1) \right)
\rho_\pm(t_1) \right\}\,,
\end{equation}
and
\begin{equation}
\label{M27A}
(w_-^A)_\pm^B = \tr \left\{ \frac{1}{2} \left(1-\hat{A}_H(t_2,t_1) \right)
\rho_\pm(t_1) \right\}\,.
\end{equation}
This implies for the expectation value of $A(t_2)$ in the presence of a first
measurement of $B(t_1)$ the relation
\begin{align}
\label{M28}
\braket{A(t_2)}_B &= (w_+^A)_+^B w_+^B + (w_+^A)_-^B w_-^B \nonumber \\
& \quad - (w_-^A)_+^B w_+^B - (w_-^A)_-^B w_-^B \nonumber \\
&= w_+^B \tr \left\{ \hat{A}_H(t_2,t_1) \rho_+(t_1) \right\} \nonumber \\
& \quad + w_-^B \tr \left\{ \hat{A}_H(t_2,t_1) \rho_-(t_1) \right\} \nonumber \\
&= \tr \left\{ \hat{A}_H(t_2,t_1) \rho_r(t_1) \right\}\ .
\end{align}
In other words, the decoherent ideal measurement assumes that coherence is lost by
the interaction with the apparatus at the first measurement. After the
measurement the reduced density matrix $\rho_r$ evolves according to the
von-Neumann equation without further disturbance by the environment. At $t_2$
one evaluates $\braket{A(t_2)}$ using $\rho_r(t_2)$,
\begin{equation}
\label{7.37A}
\braket{A(t_2)}_B=\tr\big\{A(t_2)\rho_r(t_2)\}\ ,
\end{equation}
which coincides with the Heisenberg picture~\eqref{M28}.

For the measurement correlation one finds
\begin{align}
\label{M29}
\braket{A(t_2)B(t_1)}_m =&\, (w_+^A)_+^B w_+^B + (w_-^A)_-^B w_-^B \nonumber \\
&- (w_-^A)_+^B w_+^B - (w_+^A)_-^B w_-^B \nonumber \\
=&\, w_+^B \tr \left\{ \hat{A}_H(t_2,t_1) \rho_+(t_1) \right\} \nonumber \\
&- w_-^B \tr \left\{\hat{A}_H(t_2,t_1) \rho_-(t_1) \right\} \nonumber \\
=&\, w_+^B \tr \left\{ \hat{A}_H(t_2,t_1) \hat{B}_H(t_1,t_1) \rho_+(t_1) \right\}
\nonumber \\
&+ w_-^B \tr \left\{ \hat{A}_H(t_2,t_1) \hat{B}_H(t_1,t_1) \rho_-(t_1)
\right\} \nonumber \\
=&\, \tr \left\{ \hat{A}_H(t_2,t_1) \hat{B}_H(t_1,t_1) \rho_r(t_1) \right\}\,.
\end{align}
In comparison with the expressions \eqref{M13}, \eqref{M13A} for coherent ideal
measurements the decoherent ideal measurements replace
$\rho(t)$ by $\rho_r(t)$, and $\braket{A}$ by $\braket{A}_B$.

The reduced density matrix $\rho_r(t_1)$ can be computed from $\rho(t_1)$ by an
appropriate projection
\begin{equation}
\label{M30}
\rho_r = P_+ \rho P_+ + P_- \rho P_-\,,
\end{equation}
where 
\begin{equation}
P_+ = \begin{pmatrix}
1 & 0 \\ 0 & 0 
\end{pmatrix}, \quad 
P_- = \begin{pmatrix}
0 &0 \\ 0 & 1
\end{pmatrix}.
\end{equation}
Thus $\rho_r$ can be computed from the probabilistic information of the
subsystem which is contained in $\rho(t_1)$. This extends
to the expectation value
\begin{equation}
\label{M32}
\braket{B(t_1)} = \tr \left\{ \hat{B}_H(t_1,t_1) \rho_r(t_1) \right\}\,,
\end{equation}
as well as $\braket{A(t_2)}_B$ in eq.~\eqref{M28} and the measurement
correlation \eqref{M29}. In turn, the conditional probabilities \eqref{M27} are
computable from the information in the subsystem and the
criteria~\ref{item:ci},~\ref{item:cii} for ideal measurements are obeyed. We
observe that for decoherent ideal measurements the measurement affects the
subsystem. This happens, however, in a universal way which does not depend on
the particular state of the environment. One easily verifies that also the
criteria~\ref{item:ciii}--\ref{item:cv} for ideal measurement are obeyed.
Decoherent ideal measurements are a reasonable definition for ideal measurements
for cases where decoherence of quantum subsystems plays an important role.

\paragraph*{Coherent and decoherent ideal measurements}

In contrast to $\braket{B(t_1)}$, which does not depend on the particular type
of measurement, the expectation value $\braket{A(t_2)}$ for coherent ideal
measurements differs from $\braket{A(t_2)}_B$ for decoherent ideal measurements.
This is easily seen in a basis of eigenstates of $\hat{B}_H(t_1,t_1)$ where
$\rho(t_1)$ and $\rho_r(t_1)$ are given by eqs.~\eqref{M18}, \eqref{M23}. One
finds
\begin{equation}
\label{M33}
\braket{A} - \braket{A}_B = \tr \left\{ \hat{A}_H(t_2,t_1) (\rho(t_1) -
\rho_r(t_1)) \right\}\,,
\end{equation}
where $\rho(t_1) - \rho_r(t_1)$ involves the off diagonal elements of
$\rho(t_1)$
\begin{equation}
\label{M34}
\rho(t_1) - \rho_r(t_1) = \begin{pmatrix}
0 & c \\ c^* & 0
\end{pmatrix}.
\end{equation}
The expression \eqref{M33} differs from zero for many cases where
$\hat{A}_H(t_2,t_1)$ has off-diagonal elements, which occur for
\begin{equation}
\label{M35}
\left[ \hat{A}_H(t_2,t_1), \hat{B}_H(t_1,t_1) \right] \neq 0\,.
\end{equation}
This is the general case. By the same argument the measurement correlations can
differ for coherent and decoherent ideal measurements.

\paragraph*{Sequence of three measurements}

The difference between coherent and decoherent ideal measurements can be seen
easily for a sequence of measurements of three spin observables. We consider a
one qubit quantum system with an evolution operator \begin{equation} \label{M36}
U(t_2,t_1) = \exp \left\{ i \omega \tau_3 (t_2 - t_1) \right\}\,, \end{equation}
and measurements of the spins $S_z(0)$, $S_x(\pi/\omega)$ and $S_z(2\pi,
\omega)$. In a basis where $\hat{S}_{z,H}(0,0) = \tau_3$ one has
$\hat{S}_{x,H}(\pi/\omega,0) = \tau_1$ and $\hat{S}_{z,H}(\pi/2\omega,0) =
\tau_3$. We consider a pure initial state with $\rho(0) = \rho_+(0)$. The first
measurement of $S_z(0)$ only confirms that at $t=0$ the system is in an
eigenstate of $S_z$. The probability to find $S_z(0) = 1$ equals one. 

Consider first coherent ideal measurements. In this case the expectation value
of $S_z(2\pi/\omega)$ equals one and one is certain to find for the third
measurement the value $S_z(2\pi/\omega) = 1$. For $S_x(\pi/\omega)$ the
expectation value vanishes,
\begin{equation}
\label{M37}
\braket{S_z(0)} = 1\,, \quad \braket{S_x(\pi/\omega)} = 0\,, \quad
\braket{S_z(2\pi/\omega)} = 1\,.
\end{equation}
We denote by $w_{+++}$ the probabilities to find for the sequence of
measurements the values $(+1,+1,+1)$, and similar for the other combinations.
For the coherent ideal measurements one has
\begin{equation}
w_{+++} = w_{+-+} = \frac{1}{2}\,,
\end{equation}
while all other combinations with either $S_z(0) = -1$ or $S_z(2\pi/\omega) =
-1$ vanish. The different correlations are easily obtained from these
probabilities.

The outcome differs for a sequence of decoherent ideal measurements. The first
measurement of $S_z(0)$ does not change the state of the quantum system. The
second measurement of $S_x(\pi/\omega)$ yields with equal probability
$S_x(\pi/\omega)=1$ or $S_x(\pi/\omega)=-1$. After this measurement the quantum
state is characterized by a reduced density matrix
\begin{equation}
\label{M39}
\rho_r\left( \frac{\pi}{\omega} \right) = \frac{1}{2} \begin{pmatrix}
1 & 0 \\ 0 & 1
\end{pmatrix},
\end{equation}
for which the two eigenstates of $S_x(\pi/\omega)$ have equal probability $1/2$.
This state does not change by the evolution from $t=\pi/\omega$ to
$t=2\pi/\omega$. The expectation value of $S_z(2\pi/\omega)$ in this state is
therefore zero,
\begin{equation}
\label{M40}
\braket{S_z(0)} = 1\,, \quad \braket{S_x(\pi/\omega)} = 0\,, \quad
\braket{S_z(2\pi/\omega)} = 0\,.
\end{equation}
The third expectation differs from eq.~\eqref{M37} for coherent ideal
measurements. The non-zero probabilities for sequences of different results are
now given by
\begin{equation}
\label{M41}
w_{+++} = w_{+-+} = w_{-++} = w_{--+} = \frac{1}{4}\,.
\end{equation}

One may realize the sequence of measurements by a series of Stern-Gerlach
apparatus for which beams are split, going upwards for $S_z = 1$ and downwards
for $S_z = -1$, and left for $S_x = 1$ and right for $S_x = -1$. The apparatus
are positioned in all the possible beam directions, and at distances such that
the time sequence of measurements described above is realized. Coherent ideal
measurements would predict a final outcome of two beams, both going upwards, one
left and one right. Decoherent ideal measurements predict four beams, two up and
two down, and in each pair one left and one right. Experiments will typically
find the latter situation with four beams. We will discuss in
sect.~\ref{sec:decoherence_and_syncoherence} why decoherent ideal measurements
are appropriate for this setting. 

With a sufficient effort an experimenter may also be able to perform a sequence
of measurements that come close to coherent ideal measurements. This supposes
that she can limit the loss of quantum correlations by decoherence. This
demonstrates that the issue which type of ideal measurement is realized is not
given a priori. The conditional probabilities for subsystems always require
additional information how measurements are performed. They are not properties
of the subsystem alone, even though for ideal measurements the outcome can be
predicted only based on the probabilistic information of the subsystem. 

\subsection{Reduction of the wave function}
\label{sec:reduction_of_the_wave_function}

The ``reduction of the wave function" is often considered as one of the
mysteries of quantum mechanics. At some given time $t_1$ the
quantum system is characterized by a density matrix $\rho(t_1)$. Consider a
first measurement of the observable $B(t_1)$. The outcome
of the measurement is one of the eigenvalues $b_m$ of the operator
$\hat{B}_H(t_1,t_1)$. The ``reduction of the wave function" states
that after this measurement the quantum system is in a new state, namely a pure
state with wave function $\psi_m(t_1)$, which is an
eigenstate of the operator $\hat{B}_H(t_1,t_1)$ corresponding to the measured
eigenvalue $b_m$. Subsequently, the system will continue
its unitary quantum evolution, now with initial value $\psi_m(t_1)$. At some
later time $t_2$ one can measure another observable 
$A(t_2)$. The expectation value is then given by
\begin{equation}
\label{RW1}
\braket{A(t_2)}_m = \braket{\psi_m(t_2) \hat{A}(t_2,t_2) \psi_m(t_2)}\,,
\end{equation}
where $\psi_m(t_2)$ obtains from $\psi_m(t_1)$ by a unitary evolution,
\begin{equation}
\label{RW2}
\psi_m(t_2) = U(t_2,t_1) \psi_m(t_1)\,.
\end{equation}

This simple prescription for the computation of $\braket{A(t_2)}$ seems to lead
to a conceptual problem. The jump from the 
density matrix $\rho(t_1)$ to the new pure state density matrix $\rho_m(t_1)$,
$\rho_{m,\alpha \beta}(t_1) = \psi_{m,\alpha}(t_1) \psi^*_{m,\beta}(t_1),$ is
not unitary if $\rho(t_1)$ is not a pure state
density matrix. Even if $\rho(t_1)$ is a pure state density matrix,
$\rho_{\alpha \beta}(t_1) = \psi_\alpha(t_1) \psi^*_\beta(t_1)$,
the jump from the associated wave function $\psi(t_1)$ to $\psi_m(t_1)$ is
discontinuous. Such a jump cannot be accounted for by 
the continuous unitary evolution of the quantum subsystem. This has led to many
proposals for modifications of quantum mechanics in order
to account for such discontinuous jumps.

We will show that the reduction of the wave function is simply a convenient
mathematical identity, or ``technical trick", for the 
computation of conditional probabilities for decoherent ideal measurements. As
such it does not need to correspond to a continuous
unitary evolution of the quantum subsystem. Measurements involve the interaction
of the subsystem with the measurement apparatus.

\paragraph*{Reduction of wave function for one qubit quantum subsystem}

Let us demonstrate our statement first for a one-qubit quantum system. We
consider two-level observables $A(t_2)$ and $B(t_1)$, with
possible measurement values $\pm 1$ and associated Heisenberg operators
$\hat{A}_H(t_2,t)$ and $\hat{B}_H(t_1,t)$. The reduction 
of the wave function defines conditional probabilities by the rule
\begin{align}
\begin{split}
(w_\pm^A)_+^B = \frac{1}{2} \left( 1 \pm \braket{A}_{B=1} \right)\,, \\
(w_\pm^A)_-^B = \frac{1}{2} \left( 1 \pm \braket{A}_{B=-1} \right)\,,
\end{split}
\label{RW3}
\end{align}
such that 
\begin{align}
\begin{split}
\braket{A}_{B=1} = (w_+^A)_+^B - (w_-^A)_+^B\,, \\
\braket{A}_{B=-1} = (w_+^A)_-^B - (w_-^A)_-^B\,.
\end{split}
\label{RW3b}
\end{align}

The expression 
\begin{equation}
\label{RW4}
\braket{A}_{B=\pm 1} = \braket{\psi_\pm(t_2) | \hat{A}_H(t_2,t_2) |
\psi_\pm(t_2)}
\end{equation}
corresponds to the rule \eqref{RW1} according to the reduction of the wave
function. It is the expectation value of $A(t_2)$
evaluated in the pure quantum state
\begin{equation}
\label{RW5}
\psi_\pm(t_2) = U(t_2,t_1) \psi_\pm(t_1)\,,
\end{equation}
with $\psi_\pm(t_1)$ corresponding to the reduced wave functions obeying
\begin{equation}
\label{RW6}
\hat{B}_H(t_1,t_1) \psi_\pm(t_1) = \pm \psi_\pm(t_1)\,.
\end{equation}
The evolution operator $U(t_2,t_1)$ describes the evolution of the quantum
subsystem from $t_1$ to $t_2$, without any
disturbance. We may call $\braket{A}_{B=1}$ the ``conditional expectation
value", i.e. the expectation value of 
$A(t_2)$ under the condition that $B(t_1)=1$ is found previously, and similarly
for $\braket{A}_{B=-1}$.
We will show that the expression \eqref{RW4} coincides with the expression
\eqref{RW3b} in terms of conditional probabilities.

For a proof of this statement we define at $t_1$ the pure state density matrices
$\rho_\pm(t_1)$ in terms of the reduced wave function
\begin{equation}
\label{RW7}
\rho_\pm(t_1)_{\alpha \beta} = \psi_{\pm, \alpha}(t_1)
\psi^*_{\pm,\beta}(t_1)\,,
\end{equation}
such that
\begin{equation}
\begin{split}
\braket{A}_{B=\pm 1} &= \tr \left\{ \hat{A}_H(t_2,t_1) \rho_\pm(t_1) \right\} \\
&= \tr \left\{ \hat{A}_H(t_2,t_2) \rho_\pm(t_2) \right\}\,.
\end{split}
\label{RW8}
\end{equation}
Here we employ the standard unitary evolution law for density matrices
\begin{equation}
\label{RW9}
\rho_\pm(t_2) = U(t_2,t_1) \rho_\pm(t_1) U^\dagger(t_2,t_1)\,,
\end{equation}
in order to establish the equivalence of eqs.\,\eqref{RW4} and \eqref{RW8}.
Insertion of eq.~\eqref{RW8} into eq.~\eqref{RW3} establishes
that the conditional probabilities computed from the reduction of the wave
function equal the conditional probabilities \eqref{M27}
for decoherent ideal measurements. 

The expectation value for $B(t_1)$,
\begin{equation}
\braket{B(t_1)} = \tr \{ \rho(t_1) \hat{B}_H(t_1,t_1) \},
\label{eq:RW10}
\end{equation}
and the associated probabilities to find $B(t_1)=1$ or $B(t_1)=-1$,
\begin{equation}
w_\pm^B = \frac{1}{2} (1+ \braket{B(t_1)} ),
\label{eq:RW11}
\end{equation}
do not depend on the reduction of the wave function and the way how ideal
measurements are defined. These quantities are independent of a possible later
measurement of $A(t_2)$ and involve only the probabilistic information in the
local time subsystem at $t_1$. From the conditional probabilities and the
probabilities $w_\pm^B$ we can compute the probabilities $w^{(AB)}$ according to
eq.\,\eqref{M4}, and infer the expectation values $\braket{A(t_2)}$ and
$\braket{AB}_m$ from eq.\,\eqref{M5}, \eqref{M6E}. All these quantities are the
same if determined from the conditional probabilities for decoherent ideal
measurements or from the reduction of the wave function. In particular, one has
for the measurement correlation $\braket{AB}_m$ and the expectation values
$\braket{A}$ and $\braket{B}$ for a sequence of decoherent ideal measurements
the simple identities
\begin{align}
\begin{split}
1 \pm \braket{A} + \braket{B} \pm \braket{AB}_m &= (1+\braket{B})(1\pm
\braket{A}_{B=1}),\\
1 \pm \braket{A} - \braket{B} \mp \braket{AB}_m &= (1-\braket{B})(1\pm
\braket{A}_{B=-1}).
\end{split}
\label{eq:RW12}
\end{align}
The computation of the conditional expectation values $\braket{A}_{B=\pm 1}$
according to the rule \eqref{RW4} for the reduction of the wave function is
indeed a convenient tool for the computation of the values on the r.\,h.\,s.\ of
eq.\,\eqref{eq:RW12}.

There is, however, no input from the reduction of the wave function beyond the
rules for conditional probabilities for decoherent ideal measurements. There is
no need to employ the reduction of the wave function. Everything can be computed
from the conditional probabilities \eqref{RW3}. In particular, no specification
of a physical process that achieves the reduction of the wave function is
needed. It is sufficient that the measurement apparatus performs a decoherent
ideal measurement, independently of all details how this is done. We emphasize
that the reduction of the wave function accounts specifically for decoherent
ideal measurements. It is not valid for other types of measurements as, for
example, the coherent ideal measurements. It is not a general property of the
evolution of quantum systems but rather describes a particular type of ideal
measurements in a subsystem.\

\paragraph*{Reduction of wave function for two and more qubits}

For a spin measurement in a one qubit system the reduction of the wave function
is unique. There is a unique eigenfunction to any given eigenvalue of the spin
operator. This does not hold for systems of two or more qubits. The spectrum of
eigenvalues of a spin operator is now degenerate. The space of eigenfunctions is
therefore multi-dimensional. There is no unique eigenfunction, such that
additional information is needed in order to specify to which eigenfunction the
wave function should be reduced after the measurement. This is in line with our
general argument that conditional probabilities for a sequence of measurements
need additional information on how an experiment is performed.

Consider a system of two qubits and spin observables $S_k^{(1)}$ and $S_k^{(2)}$
for the Cartesian spin directions of the two spins. The spin observable
$S_z^{(1)}$ has the possible measurement values $\pm 1$. The corresponding
operator $\hat{S}_z^{(1)}$ is a $4\times 4$ matrix with two eigenvalues $+1$ and
two eigenvalues $-1$. If $S_z^{(1)}=1$ is measured, the state with respect to
the second spin is not specified. One could have a pure state, say an eigenstate
to one of the spin operators $\hat{S}_l^{(2)}$. One could also take a linear
superposition of such states, or even a mixed state with density matrix obeying
\begin{equation}
\hat{S}_z^{(1)} \rho = \rho \hat{S}_z^{(1)} = \rho.
\label{eq:RW13}
\end{equation}
The outcome depends on what happens to the second spin during the measurement of
$S_z^{(1)}$. The apparatus could simultaneously measure $S_l^{(2)}$ in some
direction given by $l$. With a measurement of a complete set of commuting
operators the eigenfunction for a given outcome of the measurement would be
unique and the reduction of the wave function unambiguous. The measurement could
also not affect the second spin at all. Then one may suppose that the
measurement of $S_z^{(1)}$ keeps as much previous information on the second spin
as possible. For systems with many quantum spins a simultaneous measurement of
all spins is typically not realistic. A unique reduction of the wave function is
then not given.

One could formulate decoherent ideal measurements for situations with more than
one quantum spin or, more generally, for incomplete quantum measurements where
the measurement does not determine a maximal set of commuting operators. This is
a more basic conceptual framework from which effective rules similar to the
reduction of the wave function can be derived. A possible rule for decoherent
ideal measurements is the generalization of eq.\,\eqref{M30}, where the
projectors $P_\pm$ are replaced by projectors on the possible measurement values
of the observable that is actually measured. For the example of a two-qubit
system in a basis where $\hat{S}_z^{(1)} = \diag(1,1,-1,-1)$ one has $P_+ =
\diag(1,1,0,0)$, $P_- = \diag(0,0,1,1)$. The matrix
\begin{equation}
\tilde{\rho}_+ = P_+ \rho P_+,\quad 
\hat{S}_z^{(1)}\tilde{\rho}_+ = \tilde{\rho}_+ \hat{S}_z^{(1)} = \tilde{\rho}_+,
\label{eq:RW14}
\end{equation}
can be renormalized by defining
\begin{equation}
\rho_+ = \frac{\tilde{\rho}_+}{\tr \{\tilde{\rho}_+\}},\quad \tr \rho_+ = 1.
\label{eq:RW15}
\end{equation}
This generalizes the pure state density matrix $\rho_+$ for the single qubit
system. A projection on $\rho_+$ after the measurement of $S_z^{(1)}$ with
result $S_z^{(1)}=1$ replaces the reduction of the wave function. It keeps a
maximum amount of information on the second spin since it is insensitive to the
properties of $\rho$ with respect to the second spin. The generalization of the
rule for decoherent ideal measurements of a single observable with possible
measurement values $\pm 1$, for which $+1$ is found after the first measurement,
would be a ``reduction of the density matrix''. After the measurement, the new
state of the system is given by $\rho_+$. 

In general, $\rho_+$ will not be a pure state density matrix, however. Let us
write a general $4\times 4$ density matrix in terms of $2\times 2$ matrices
$\hat{\rho}_+$, $\hat{\rho}_-$, $c$ as
\begin{equation}
\rho = \begin{pmatrix}
\hat{\rho}_+ & c \\
c^\dagger & \hat{\rho}_-
\end{pmatrix},\quad
\hat{\rho}_\pm^\dagger = \hat{\rho}_\pm.
\label{eq:RW16}
\end{equation}
The density matrix $\rho_+$ reads
\begin{equation}
\rho_+ = \frac{1}{\tr \{ \hat{\rho}_+ \} }
\begin{pmatrix}
\hat{\rho}_+ & 0 \\
0 & 0
\end{pmatrix}.
\label{eq:RW17}
\end{equation}
This is a pure state density matrix only if one of the eigenvalues of
$\hat{\rho}_+$ vanishes, which is not the general case. 

The conditional probabilities for a sequence of two measurements are again
defined by eq.\,\eqref{M27}, which does not assume that $\rho_\pm$ are pure
state density matrices. With the reduced density matrix $\rho_\mathrm{r}$
defined by eq.\,\eqref{M30}, one has again
\begin{align}
\begin{split}
\braket{A(t_2)}_{B} &= \tr \{ \hat{A}_H(t_2,t_1) \rho_\mathrm{r}(t_1) \}, \\
\braket{A(t_2) B(t_1)}_m &= \tr \{ \hat{A}_H(t_2,t_1) \hat{B}_H(t_1,t_1)
\rho_\mathrm{r}(t_1) \}.
\end{split}
\label{eq:RW18}
\end{align}
This setting is easily generalized to simultaneous measurements of two two-level
observables. According to the possible outcomes $(++)$, $(+-)$, $(-+)$, $(- -)$
one defines projectors $P_{++}$ etc., and
\begin{equation}
\rho_\mathrm{r} = P_{++} \rho P_{++} + P_{+-}\rho P_{+-} + P_{-+} \rho P_{-+} +
P_{- -} \rho P_{- -}.
\label{eq:RW19}
\end{equation}
For a simultaneous measurement of a maximal set of commuting operators the
different pieces in the sum \eqref{eq:RW19} are pure state density matrices up
to normalization.

\subsection{Decoherence and syncoherence}
\label{sec:decoherence_and_syncoherence}

Decoherence and syncoherence are possible properties of the time evolution of
subsystems. Decoherence\,\cite{ZEH,JZ,ZUR,JZKG} describes how a pure state can
become a mixed state, and syncoherence\,\cite{CWQM} accounts for a mixed state
evolving to a pure state. For the full local-time subsystem a pure state remains
a pure state during the evolution. This is a direct consequence of the evolution
laws for the classical density matrix $\rho'$. With 
\begin{equation}\label{D51}
\rho'(t+\epsilon)=\hat{S}(t)\rho'(t) \hat{S}^{-1}(t)
\end{equation}
the eigenvalues of $\rho'(t+\epsilon)$ are the same as for $\rho'(t)$. 
The same properties hold for closed quantum systems. In a complex formulation
they correspond to the replacement $\rho' \rightarrow \rho$, $\hat{S}
\rightarrow U$. Full local-time subsystems or closed quantum subsystems do not
admit decoherence and syncoherence.

\paragraph*{Decoherence for a two-qubit quantum system}

The situation changes if we consider the evolution of subsystems. We may
describe the main issues within a two-qubit quantum system in a complex
formulation. The subsystem is given by the first qubit, and the environment,
which is very simple in this case, consists of the second qubit and its possible
correlation with the first qubit. For a pure quantum state we denote the four
complex components of the wave function for the two-qubit system by
$\psi_{\alpha\gamma} (t)$, with $(\alpha,\gamma)$ a double index where
$\alpha=1,2$ refers to the first qubit and $\gamma=1,2$ to the second qubit.
Correspondingly, a density matrix is described by a Hermitian positive matrix
$\rho_{\alpha\gamma,\beta\delta}(t)$, with a pure state density matrix given by
$\rho_{\alpha\gamma,\beta\delta}(t)=\psi_{\alpha\gamma}(t)\psi^*_{\beta\delta}(t)$.
The density matrix $\bar\rho$ of the subsystem for the first qubit obtains by
taking a trace over the degrees of freedom of the environment
\begin{equation}\label{D52}
\bar\rho_{\alpha\beta} (t)=\rho_{\alpha\gamma,\beta\delta}(t)
\delta^{\gamma\delta}.
\end{equation}

A pure quantum state of the full system can be a mixed state of the subsystem.
This may be demonstrated by comparing two different pure states. The first state
is given by
\begin{equation}\label{D53}
\psi^{(1)}=
\begin{pmatrix}1\\0\end{pmatrix}\otimes\begin{pmatrix}1\\0\end{pmatrix}, \,
\rho^{(1)}=\begin{pmatrix}1&0&0&0\\0&0&0&0\\0&0&0&0\\0&0&0&0\end{pmatrix},
\end{equation} 
or
\begin{equation}\label{D54}
\psi_{11}=1, \, \psi_{12}=\psi_{21}=\psi_{22}=0.
\end{equation}
The second state is an entangled state
\begin{align}\label{D55}
\psi^{(2)}= \frac{1}{\sqrt{2}}\left[ \begin{pmatrix}1\\0\end{pmatrix}
\otimes\begin{pmatrix}1\\0\end{pmatrix}\,-\,\begin{pmatrix}0\\1\end{pmatrix}\otimes\begin{pmatrix}0\\1\end{pmatrix}\right]\
, 
\nonumber\\
\psi^{(2)}=\frac{1}{\sqrt{2}} \begin{pmatrix} 1\\0\\0\\ -1\end{pmatrix} , \,
\rho^{(2)}=
\frac{1}{2}\begin{pmatrix}1&0&0&-1\\0&0&0&0\\0&0&0&0\\-1&0&0&1\end{pmatrix},
\end{align}
or
\begin{equation}\label{D55A}
\psi_{11}=\frac{1}{\sqrt{2}} \; , \; \psi_{12}=\psi_{21}=0 \; , \;
\psi_{22}=-\frac{1}{\sqrt{2}}. 
\end{equation}
The density matrix for the subsystem is a pure state density matrix for
$\rho^{(1)}$, and a mixed state density matrix for $\rho^{(2)}$,
\begin{equation}\label{D56}
\bar{\rho}^{(1)}=\begin{pmatrix}
1&0\\0&0
\end{pmatrix}, \, \bar{\rho}^{(2)}=\frac{1}{2}\begin{pmatrix}
1&0\\0&1
\end{pmatrix}.
\end{equation}

We next want to describe a unitary time evolution which turns a pure state of
the subsystem into a mixed state. For this purpose we consider a unitary
evolution of the full quantum system, 
\begin{equation}\label{D57} 
U(t)=\exp\{{i}{\omega}{t}{T}\},\quad 
T=\frac{1}{\sqrt{2}}
\begin{pmatrix}1&0&0&-1\\0&1&0&0\\0&0&1&0\\ -1&0&0&-1\end{pmatrix}.
\end{equation} With
\begin{equation}\label{D58} 
T^\dagger = T, \quad T^2=1,
\end{equation} 
we can write 
\begin{equation}\label{D59} 
U(t)=\cos{({\omega}{t})}+i\sin{({\omega}{t})}T.
\end{equation} 
In particular, for $t=\pi/(2\omega)$ one has
\begin{equation}\label{D510} 
U(\frac{\pi}{2\omega})=iT.
\end{equation} 

Let us start at $t=0$ with the pure state $\psi^{(1)}$,
\begin{equation}\label{D511}
\psi(0)=\psi^{(1)}, \quad \rho(0)=\rho^{(1)}.
\end{equation} With
\begin{equation}\label{D512}
T \psi^{(1)}=\psi^{(2)},
\end{equation} one has
\begin{equation}\label{D513}
\psi(\frac{\pi}{2\omega})=i\psi^{(2)}, \quad
\rho(\frac{\pi}{2\omega})=\rho^{(2)}.
\end{equation}
Correspondingly, the density matrix for the subsystem evolves from the pure
state density matrix $\bar{\rho}^{(1)}$ to the mixed state density matrix
$\bar{\rho}^{(2)}$, 
\begin{equation}\label{D514}
\bar{\rho}(0)=\bar{\rho}^{(1)}, \quad
\bar{\rho}({\frac{\pi}{\omega}})=\bar{\rho}^{(2)}.
\end{equation}
This is a simple example of decoherence. Syncoherence, the change from a mixed
state to a pure state, is encountered for
\begin{equation}\label{D515}
\rho(0)=\rho^{(2)}, \quad \rho({\frac{\pi}{{2}{\omega}}})=\rho^{(1)}.
\end{equation}

\paragraph*{Decoherent evolution equation}

From the unitary evolution equation for the two-qubit system
\begin{equation}\label{D516}
\partial_{t}\rho=-i[{H},{\rho}], \quad H=-{\omega}{T},
\end{equation} with $T$ given by eq.\,\eqref{D57}, and the definition
\eqref{D52} of the one-qubit subsystem, one can infer the evolution equation for
the density matrix of the subsystem, 
\begin{equation}\label{D517}
\partial_{t}\bar{\rho}=-i[{\bar{H}},{\bar{\rho}}]+ \bar{F},
\end{equation} 
with
\begin{equation}\label{D518}
\bar{H}=-\frac{\omega}{{2}{\sqrt{2}}} \tau_{3} .
\end{equation}
The term $\bar{F}$ involves the properties of the environment
\begin{equation}\label{D519}\bar{F}={\begin{pmatrix}
	A&B\\B^{*}& - A
	\end{pmatrix}}, 
\end{equation}
where
\begin{align}
\begin{split}
A&= -\sqrt{2}\omega \,{\mathrm{Im}(\rho_{1122})}, \\
B&={\frac{i\omega}{\sqrt{2}}}(\rho_{1211}+\rho_{1222}-\rho_{1121}-\rho_{2221}).
\end{split}\label{D520}
\end{align} 
The evolution equation \eqref{D517} is the general evolution equation for
subsystems obtained by taking a subtrace,
with
\begin{equation}\label{D520A}
\bar{H}^\dagger = \bar{H},\quad \bar{F}^\dagger = \bar{F},\quad \tr \bar{F}=0.
\end{equation}
The particular form \eqref{D518} \eqref{D519} \eqref{D520} is valid for the
particular unitary evolution with $U$ given by eq.\eqref{D57}. 

For $ \bar{F}\neq 0$ the evolution of the subsystem is no longer closed. It
cannot be computed from the probabilistic information of the subsystem alone,
but also involves properties of the environment. It is the interaction with the
environment that is responsible for decoherence or syncoherence in the
subsystem. This can be be seen by the evolution of the purity $P$, defined by
\begin{equation}\label{D521}
P = \rho_k \rho_k ,\quad \bar{\rho}_{\alpha\beta} =
\frac{1}{2}(1+{\rho_k}(\tau_k)_{\alpha\beta}).
\end{equation}
A pure quantum state of the subsystem has $P=1$. In terms of the matrix elements
$\bar{\rho}_{{\alpha}{\beta}}$ one has
\begin{align}
\begin{split}
\rho_{1} &= 2\Re(\bar{\rho}_{12}), \quad \rho_{2}=-2\Im(\bar{\rho}_{12}), \\
\rho_{3} &= \bar{\rho}_{11}-\bar{\rho}_{22}
\end{split}\label{D522}
\end{align}
or
\begin{equation}\label{D523}
P = 4 |\bar{\rho}_{12}|^{2}+(\bar{\rho}_{11}-\bar{\rho}_{22})^{2}.
\end{equation}
The purity is conserved by a unitary evolution and therefore for $\bar{F}=0$. A
change of the purity is directly reflecting the coupling to the environment
\begin{equation}\label{D524}
\partial_{t}P=4(\rho_{1}Re(B)-\rho_{2}Im(B)+\rho_{3}A) .
\end{equation} 

Due to the coupling to the environment the purity of the subsystem can decrease,
accounting for decoherence, or increase, corresponding to syncoherence.We
observe that there are particular states of the subsystem and environment for
which the purity remains constant despite the coupling to the environment. For
example, for $\rho_{3}=0$ and a coupling to the environment with $B=0$, the
purity is conserved even for $A\neq0$. This is compatible with a non-trivial
unitary evolution of the subsystem which may correspond to a rotation in the
$(\rho_{1},\rho_{2})$-plane, with $\rho_{3}=0$.

\paragraph*{Decoherence for macroscopic environment} 

Our two-qubit system is a rather extreme case for a subsystem coupled to its
environment. Typically, the environment may involve many more degrees of
freedom, as for the coupling of the quantum subsystem to a macroscopic
measurement apparatus. For the simple two-qubit system the overall unitary
evolution is periodic with period $2\pi/\omega$ - or $\pi/\omega$ if we consider
the density matrix. Phases of decoherence and syncoherence follow each other.
This is a simple example of ``recurrence''. One may separate the characteristic
time scales for the unitary evolution of the subsystem and for decoherence or
syncoherence by adding to $\bar{H}$ in eq.\,\eqref{D517} a term with a period
much shorter than $\pi/\omega$. This is easily done on the level of the
two-qubit system by adding to $H$ in eq.\,\eqref{D516} a piece acting only on
the first qubit. With eigenvalues $\bar{E}$ of $\bar{H}$ we may consider the
limit of a small ratio $\omega/\bar{E}$. On the time scale of the unitary
evolution given by $1/\bar{E}$ the decoherence or syncoherence is very slow. The
subsystem almost performs a unitary evolution, with only minor corrections due
to the decoherence. Nevertheless, after a ``recurrence time'' $\pi/(2\omega)$
decoherence stops and changes to syncoherence.

Recurrence occurs because the matrix $\bar{F}$ in eq.\,\eqref{D517}
``remembers'' the unitary evolution of the overall system. For a macroscopic
environment this memory is effectively lost. For an increasing number of degrees
of freedom in the environment the recurrence time becomes rapidly very long,
much longer than the typical time scale of decoherence or syncoherence. In
practice, the recurrence time can be taken to infinity. The subsystem may then
undergo decoherence until minimal purity $P=0$ is reached, or until it reaches
some of the states for which $\partial_t P = 0$ at nonzero $P$. If there is no
subsequent syncoherence, the state with constant purity is typically reached
asymptotically for $t\rightarrow\infty$. After fast initial decoherence the
phenomenon of decoherence can effectively stop. This is analogous to
thermalization. The same can hold in the opposite direction for syncoherence.
We note that for an environment with many degrees of freedom the time reflection
symmetry can be effectively lost for the evolution of the subsystem.

\paragraph*{Decoherent ideal measurements}

A measurement couples a quantum subsystem to the measurement apparatus, which is
typically a macroscopic system with many degrees of freedom. We may consider a
one-qubit quantum subsystem and measure the spin observable in the 3-direction
$S_3$. The measurement apparatus is assumed to have two pointer positions
$B=\pm1$. For an ideal measurement one will find $B=1$ whenever $S_{3}=1$, and
$B=-1$ whenever $S_{3}=-1$. An example is the ``Schrödinger cat'' system, where
a decaying nucleus triggers the emission of poison which kills the cat. The
decaying nucleus corresponds to $S_{3}=1$, and the non-decaying nucleus to
$S_{3}=-1$. For $B=1$ the cat is dead, for $B=-1$ it is alive. 

Let us consider some subsystem which contains the probabilistic observables
$S_3$ and $B$. We may call it the ``pointer-probe subsystem". For definiteness
we consider a two-qubit quantum system, for which $S_{3}^{(1)}=S_{3}$
corresponds to the yes/no decision if the nucleus has decayed or not, and
$S_{3}^{(2)}=B$ indicates if the cat is dead or alive. The two-qubit quantum
subsystem contains further observables as $S_{1}^{(1)}$ or $S_{1}^{(2)}$ that
will play no particular role here. The reason why we have chosen a quantum
subsystem is a demonstration that the decoherent ideal measurement can be fully
described within quantum mechanics. More general probabilistic systems could be
used as well. The density matrix $\bar{\rho}$ for the pointer-probe subsystem is
a Hermitian positive $4\times 4$ matrix, obeying the evolution law \eqref{D517}.
It is not a closed subsystem, since the two ``pointer states" $B=\pm 1$ are
connected to many other states of the measurement apparatus which act as an
environment for the subsystem. The Hermitian traceless $4\times 4$ matrix
$\bar{F}$ in eq. \eqref{D517} accounts for the coupling to this environment and
does not vanish. The evolution of the pointer-probe systems is not unitary and
can admit decoherence or syncoherence.

An ideal measurement correlates the values of $S_{3}^{(1)}$ and $S_{3}^{(2)}$,
\begin{equation}\label{D525}
\braket{S_3^{(1)} \, S_3^{(2)}}=1.
\end{equation}
This correlation should be achieved during the measurement. Once achieved, it
should not change anymore during the measurement process. If we employ the
direct product basis \eqref{E11} for the two-qubit system
\begin{equation}\label{D526}
\bar{\rho}=\dfrac{1}{4}(\rho_{\mu\nu} \, L_{\mu\nu}),\quad 
\bar{F}=f_{\mu\nu} \,L_{\mu\nu} \,,
\end{equation}
the correlation \eqref{D525} is realised for 
\begin{equation}\label{D527}
\rho_{33}=1.
\end{equation}
Any ideal measurement has to establish the condition \eqref{D527} in early
stages of the measurement when the pointer adapts its value to the value of the
measured observable. After this initial stage the correlation \eqref{D525} has
to remain stable. In the ending stage of the measurement $\rho_{33}$ has to be
conserved, and the evolution has to obey \begin{equation}\label{D528}
[ \bar{H}, \, \tau_{3}\otimes\tau_{3})]=0,\quad f_{33}=0.
\end{equation}

A second requirement for an ideal measurement is that the expectation value
$\braket{S_{3}^{(1)}}$ is not changed during the measurement. The relative
probabilities for the nucleus having decayed or not should not be affected by
the measurement. In our notation this requires that $\rho_{30}$ is invariant,
and the time evolution should obey during the whole measurement 
\begin{equation}\label{D529}
[ \bar{H}, \, \tau_{3}\otimes1)]=0,\quad f_{30}=0.
\end{equation}
One concludes that during the ending stage of any ideal measurement both
$\rho_{33}$ and $\rho_{30}$ should not depend on time. 

We have not made any assumption on the time evolution of $\rho_{03}(t)$. In a
basis of eigenstates to $S_{3}^{(1)}$ and $S_{3}^{(2)}$ with double indices
refering to the two qubits, the diagonal elements of the density matrix during
the ending stage of the measurement are given by \begin{align}\label{D530}
\bar{\rho}_{11,11}=\frac{1}{2}+\frac{1}{4}\bigl(\rho_{30}+\rho_{03}(t)\bigr),
\nonumber\\
\bar{\rho}_{12,12}=-\bar{\rho}_{21,21}=\frac{1}{4}\bigl(\rho_{30}-\rho_{03}(t)\bigr),
\nonumber\\
\bar{\rho}_{22,22}=\frac{1}{2}-\frac{1}{4}\bigl(\rho_{30}+\rho_{03}(t)\bigr).
\end{align}
(There should be no confusion between the elements
$\bar{\rho}_{\alpha\gamma\,,\beta\delta}$ of the density matrix and the
coefficients $\rho_{\mu\nu}$ of the expansion \eqref{D526}.)

Consider now the one-qubit subsystem whose properties are measured. We denote
its density matrix by $\bar{\rho}_{\alpha\beta}^{(1)}$. According to
eq.\,\eqref{D52} its diagonal elements are given by
\begin{align}\label{D531}
\bar{\rho}_{11}^{(1)}=\bar{\rho}_{1111}+\bar{\rho}_{1212}=\frac{1}{2}(1+\rho_{30})
\, , \nonumber\\
\bar{\rho}_{22}^{(1)}=\bar{\rho}_{2121}+\bar{\rho}_{2222}=\frac{1}{2}(1-\rho_{30})
\, ,
\end{align}
independently of $\rho_{03}(t)$. This reflects that the expectation value
$\braket{S_{3}^{(1)}}$ is not affected by the measurement. We are interested in
$\bar{\rho}^{(1)}(t_{f})$ at the time $t_{f}$ at the end of the measurement.

The difference between coherent and decoherent ideal measurements concerns the
off-diagonal elements $\bar{\rho}_{12}^{(1)}$ and
$\bar{\rho}_{21}^{(1)}=(\bar{\rho}_{12}^{(1)})^{*}$ at $t_{f}$. A decoherent
ideal measurement assumes that the only probabilistic information in the
one-qubit subsystem at the end of the measurement is given by $<S_{3}^{(1)}>$.
This amounts to vanishing off-diagonal elements $\bar{\rho}_{12}(t_{f})=0$. For
a decoherent ideal measurement the one-qubit subsystem at the end of the
measurement is a mixed state whenever $\left|\rho_{30}\right|\neq1$,
\begin{equation}\label{D532}
\bar{\rho}^{(1)}=\dfrac{1}{2}\begin{pmatrix}
{1+\rho_{30}}&0\\0&{1-\rho_{30}}
\end{pmatrix}.
\end{equation}
This corresponds to the ``reduction of the wave function" discussed previously.
In contrast, for a coherent ideal measurement the off-diagonal elements of
$\bar{\rho}_{1}(t_{f})$ at the end of the measurement are the same as the ones
before the measurement, $\bar{\rho}_{1}(t_{f})=\bar{\rho}_{1}(t_{in})$. 

A rough picture of the evolution corresponding to ideal measurements can be
depicted as follows. Before the measurement the total system of the measured
subsystem and the measurement apparatus is a direct product system, for which
the subsystem and the apparatus follow their separate evolution. During the
measurement between $t_{in}$ and $t_{f}$ the interactions between the measured
system and the apparatus play a role. This is the range for which
eq.\eqref{D517} describes the evolution of the ``pointer-probe subsystem". After
the measurement the probe and the apparatus are separated and follow again a
separate evolution. The measured one-qubit subsystem follows its own unitary
evolution, starting from $\bar{\rho}_{2}(t_{f})$ at the end of the measurement.
A second measurement in a sequence of two measurements can be performed
afterwards at some time $t_{2}>t_{f}$.

For an understanding why decoherent ideal measurements are realistic for many
macroscopic measurements we need to investigate the off-diagonal elements of
$\bar{\rho}^{(1)}$,
\begin{equation}\label{D533}
\bar{\rho}_{12}^{(1)}=\bigl(\bar{\rho}_{21}^{(1)}\bigr)^{*}=\bar{\rho}_{1121}+\bar{\rho}_{1222}.
\end{equation}
The part of $\bar{\rho}_{\alpha\gamma,\beta\delta}$ contributing to the
off-diagonal part of $\bar{\rho}_{nd}^{(1)}$,
\begin{equation}\label{D534}
\bar{\rho}_{nd}^{(1)}=\begin{pmatrix}
0&g\\g^*&0
\end{pmatrix} , \quad g=\bar{\rho}_{12}^{(1)},
\end{equation}
is given by
\begin{equation}\label{D535}
\bar{\rho}_{nd}=\dfrac{1}{2}\bar{\rho}_{nd}^{(1)}\otimes1=\dfrac{1}{4}\lbrace\rho_{10}(\tau_{1}\otimes1)+\rho_{20}(\tau_{2}\otimes1)\rbrace,
\end{equation}
and involves among the $\rho_{\mu\nu}$ the coefficients $\rho_{10}$ and
$\rho_{20}$. Only off-diagonal elements of the two-qubit density matrix
$\bar{\rho}$ contribute to $\bar{\rho}_{12}^{(1)}$.

As every density matrix, the two-qubit density matrix can be interpreted as a
linear combination of pure state density matrices $\bar{\rho}^{(i)}$
\begin{equation}\label{D536}
\bar{\rho}=\sum\limits_{i} w_{i}\bar{\rho}^{(i)},
\end{equation}
with $w_{i}$ the probabilities to ``realise'' $\bar{\rho}^{(i)}$, i.e. 
$\sum_{i}w_{i}=1, w_{i}\geq0$. Assume now that the probabilities vanish for all
pure states that do not either have $B=1$ or $B=-1$. In other words, only
eigenstates of $B$ contribute in the sum \eqref{D536}. This is the statement
that no superposition states of dead and living cats can be realized. This
assumption restricts the possible form of
$\rho_{\tau\rho}^{(i)}=\psi_{\tau}^{(i)}\psi_{\rho}^{(i)}$, with $\psi^{(i)}$
taking the possible forms
\begin{equation}\label{G537}
\psi_{+}^{(i)}=\begin{pmatrix} a\\0\\c\\0 \end{pmatrix}, \,
\psi_{-}^{(i)}=\begin{pmatrix} 0\\b\\0\\d \end{pmatrix},
\end{equation}
For an ideal measurement in the ending stage of the measurement the probability
for eigenstates with opposite values of $S_{3}^{(1)}$ and $S_{3}^{(2)}$
has to vanish by virtue of eq. \eqref{D525}. This implies that only pure states
with $c=0$, $b=0$ can contribute the sum \eqref{D536}. As a consequence, the
pointer-probe density matrix in $\bar{\rho}(t_{f})$ is diagonal. This translates
to diagonal $\bar{\rho}^{(1)}(t_{f})$.
The selection of decoherent ideal measurements therefore follows from the
vanishing probability of superposition states with $B=1$ and $B=-1$.

The absence of superposition states for different positions of the pointer (dead
and living cat) is a property of the apparatus that does not depend on the
presence of the probe to be measured.
The interaction between the probe to be measured and the apparatus is not
relevant for this issue. The formal reason for the absence of the superposition
of the different pointer states resides in the fact that the pointer subsystem
-- the one-qubit subsystem corresponding to the observables $S_{k}^{(2)}$ -- is
itself a subsystem of the macroscopic apparatus. The term $\bar{F}$ in eq.
\eqref{D517} can produce the decoherence of any superposition state. Even if one
would start with a superposition state of the pointer subsystem it will end in a
mixed state after some characteristic time $\tau_{dc}$.

As we have seen above, decoherence in the pointer subsystem is perfectly
compatible with a unitary evolution of a quantum system for the whole apparatus.
The ``rest of the apparatus'' is the environment for the pointer subsystem. The
decoherence time $\tau_{dc}$ is typically a property of the apparatus. There is
no need to put the apparatus in a further environment and to invoke, for
example, its interaction with the cosmic microwave radiation or similar effects.
Using a cat as a measurement apparatus, $\tau_{dc}$ is typically some
``biological time". Dying is a complex issue and not instantaneous. The final
stage for $t\gg \tau_{dc}$ is either dead or alive, however. A rather long
biological $\tau_{dc}$ does not mean that other superposition states do not
decohere much faster. The decoherence time is not universal - it depends on the
particular selection of a pointer subsystem used for the measurement. 


Whenever the typical time interval for the measurement $\Delta{t}=t_{f}-t_{in}$
is much longer than the decoherence time, ideal measurements are decoherent
ideal measurements. A coherent ideal measurement could be realised in the
opposite limit $\Delta{t}\ll\tau_{dc}$. It needs a pointer subsystem with a
sufficiently long decoherence time.

\paragraph*{Syncoherence} 

Syncoherence\,\cite{CWQM} in subsystems is a frequent phenomenon in Nature. We
typically find isolated atoms in a unique pure quantum state, namely the ground
state. This would not happen without syncoherence. If the time evolution of
subsystems would be either unitary or decoherent, quantities as the purity could
not increase. Once smaller than one at $t_{1}$, the purity would have to be
smaller than one for all $t_{2}>t_{1}$. There is no need, however, for the
purity to be monotonically decreasing or constant. 
The general evolution equation for subsystems \eqref{D517} is perfectly
compatible with increasing purity or syncoherence.


As an example, consider a single atom emitted from a hot region where it has
been in thermal equilibrium. At the time $t_{in}$ when it leaves the hot region
its state is characterized by a thermal density matrix, with energy levels
occupied according to Boltzmann factors. This is a mixed state. Away from the
hot region the atom subsystem follows a new evolution law for which the thermal
environment does no longer play a role. The time evolution of the atom subsystem
is not closed, however. 

The atom still interacts with its environment, e.g. with the photon states of
the vacuum. In a quantum field theory the atom can emit photons, until it
reaches its ground state. The corresponding evolution is characterized by
syncoherence. The term $\bar{F}$ in eq. \eqref{D517} leads to increasing purity
of the atom subsystem. Starting from a mixed state at $t_{in}$, the atom
subsystem reaches a pure state for sufficiently large $t-t_{in}$ for many
situations.

\section{The ``paradoxes" of quantum\\mechanics}
\label{sec:the_paradoxes_of_quantum_mechanics}

The literature is full of statements that quantum mechanics cannot be described
by classical probabilistic systems, that quantum mechanics has to be incomplete,
or that quantum mechanics is not compatible with a single world. These arguments
are based on no-go theorems or ``paradoxes" for quantum mechanics. We have
described quantum mechanics as particular local-time subsystems of an overall
probabilistic description of one single world. Our description is based only on
the fundamental laws for ``classical" probabilities. Our explicit constructions
are counter examples for no-go theorems forbidding the embedding of quantum
mechanics in classical statistics. These no-go theorems cannot be complete.
Still we should explain why there is no conflict with no-go theorems, and how
the paradoxes can be understood. As usual, the no-go theorems are not wrong.
Only the assumptions, often implicit, for the applicability of the no-go
theorems do not hold for quantum subsystems. Most of the time the apparent
conflicts and paradoxes arise from a too narrow view on subsystems of
probabilistic systems. Key properties such as incompleteness, the equivalence
classes of probabilistic observables or the correct choice of the measurement
correlation are often not taken into account. The structure of possible
subsystems is much richer than for simple direct product subsystems.
Correlations of the subsystem with its environment play an important role.

We have argued that for arbitrary quantum systems there is no obstruction to
embed them as appropriate subsystems in a probabilistic overall description of
the world. We should therefore find out at what point the assumptions of
specific no-go theorems fail to be realized. We will discuss Bell's inequalities
in sect.\,\ref{sec:classical_correlation_functions_and_bells_inequalities} and
the Kochen-Specker theorem\,\cite{KOSP,MER,PER,STRA} in
sect.\,\ref{sec:kochen-specker_theorem}. In
sect.\,\ref{sec:einstein-podolski-rosen_paradox} we turn to the
Einstein-Podolski-Rosen (EPR)-paradox. We have already discussed the reduction
of the wave function in sect.\,\ref{sec:reduction_of_the_wave_function}.

\subsection{Classical correlation functions and Bell's
inequalities}\label{sec:classical_correlation_functions_and_bells_inequalities}

Bell's inequalities are powerful constraints that classical correlation
functions have to obey. Measured correlation functions in quantum systems are
found to violate these constraints. Statements that this implies the
impossibility to embed quantum mechanics into a classical statistical system
make one important implicit assumption, namely that the measured correlations
are described by classical correlation functions. As we have seen in
sect.~\ref{sec:conditional_probabilities_4_7} this assumption typically does not
hold for measurement correlations in quantum subsystems. The classical
correlation functions cannot describe the correlation functions for ideal
measurements in many circumstances. The reasons are the incomplete statistics of
the quantum subsystem and the incompatibility of the classical correlations with
the structure of equivalence classes for observables.

In short, classical correlations obey Bell's inequalities but are not
appropriate for a description of the outcome of measurements. There exist other
correlation functions describing ideal measurements in subsystems. These are
typically the quantum correlations based on operator products. These
``measurement correlations'' can violate Bell's inequalities.

In the first part of this work~\cite{CWPW} we have encountered observables for
which the classical correlation functions simply do not exist. One example is
the momentum observable for a simple probabilistic automaton describing free
massless fermions in two dimensions. It does not take a definite value for a
given configuration of the overall probabilistic system. It rather measures
properties of the time-local probabilistic information. It is a ``statistical
observable", with a status similar to temperature in classical equilibrium
systems. Nevertheless, it is a conserved quantity which is crucial for the
dynamics of particles. Since there are no simultaneous values for momentum and
occupation numbers for the overall configurations, a classical correlation for
such pairs of observables does not exist. The energy and momentum observables
for the quantum subsystem for a particle in a harmonic potential discussed in
sect.~\ref{subsec:quantum_particle_in_harmonic_potential} are of a similar
nature.

Another example are the time-derivative observables. The classical correlation
function for the time-derivative observables has been found to be incompatible
with the continuum limit. This means that whenever a continuum limit is possible
a classical correlation between position and the time-derivative of the position
cannot be defined. Constraints on classical correlation functions do not apply
for such cases.

Bell's inequalities apply, however, also for simple spin systems. This is where
important experiments have been done. We have to discuss why the classical
correlation functions are inappropriate for measurements in such systems.

\paragraph*{Bell type inequalities}

Bell-type inequalities\,\cite{BELL2,BELL,CHSH,CLSH,CLHO} are constraints on
systems of classical correlation functions. By a classical correlation function
for a pair of two observables $A$ and $B$ we understand for this discussion any
correlation function that can be written in the form
\begin{equation}
\braket{AB}_\mathrm{cl} = \sum_{i,j} w_{ij}^{(AB)} A_i B_j,
\label{eq:pq1}
\end{equation}%
where $A_i$ and $B_j$ are the possible measurement values of the observables $A$
and $B$, and $w_{ij}^{(AB)}$ are the \textit{simultaneous probabilities} to find
$A_i$ for $A$ and $B_j$ for $B$. They have to obey
\begin{equation}
w_{ij}^{(AB)} \geq 0,\quad \sum_{i,j} w_{ij}^{(AB)} =1.
\label{eq:pq2}
\end{equation}%

A system of classical correlations for three observables $A$, $B$, $C$ consists
of the classical correlation functions $\braket{AB}_\mathrm{cl}$,
$\braket{AC}_\mathrm{cl}$, $\braket{BC}_\mathrm{cl}$, obeying
eqs.\,(\ref{eq:pq1}), (\ref{eq:pq2}). For a system of classical correlations for
three observables we further require that the simultaneous probabilities to find
$A_i$ for $A$, $B_j$ for $B$ and $C_k$ for $C$ are defined
\begin{equation}
w_{ijk}^{(ABC)} \geq 0,\quad \sum_{i,j,k} w_{ijk}^{(ABC)} = 1.
\label{eq:pq3}
\end{equation}%
The simultaneous probabilities for pairs (\ref{eq:pq1}), (\ref{eq:pq2}) follow
by partial summation, e.\,g.
\begin{equation}
w_{ij}^{(AB)} = \sum_k w_{ijk}^{(ABC)}.
\label{eq:pq4}
\end{equation}%

If these simultaneous probabilities are available we can define new observables
by linear combinations, as $B+C$ with possible measurement values given by the
sums of $B_j$ and $C_{k}$
\begin{equation}
D= B+C,\quad D_l = D_{(jk)} = B_j + C_k.
\label{eq:pq5}
\end{equation}%
Classical correlations involving $D$ obey
\begin{equation}
\braket{AD}_{\mathrm{cl}} = \sum_{i,l} A_i\, D_l\,w_{il}^{(AD)},
\label{eq:pq6}
\end{equation}%
where $l=(jk)$,
\begin{equation}
w_{il}^{(AD)} = w_{i(jk)}^{(AD)} = w_{ijk}^{(ABC)}.
\label{eq:pq7}
\end{equation}%
In case of a degenerate spectrum, where a given $D_l$ can be reached by more
than one combination $B_j+C_k$, the probability $w_{il}^{(AD)}$ obtains by
summing $w_{ijk}^{(ABC)}$ over all pairs $(jk)$ that correspond to a given $l$.
This generalizes to systems of classical correlations for more than three
observables.

Bell type inequalities concern systems of classical correlation functions for
three or more observables. A crucial assumption for these inequalities is the
existence of the simultaneous probabilities $w_{ijk}^{(ABC)}$. For a subsystem
with complete statistics the probabilities $w_{ijk}^{(ABC)}$ are available,
while this is typically not the case for subsystems characterized by incomplete
statistics. A central assumption for these inequalities (that is often not
stated) is that all relevant measurement correlations are classical correlations
that obey eqs. (\ref{eq:pq1}) and (\ref{eq:pq2}). Furthermore, it is assumed
that the system is characterized by complete statistics for which the
simultaneous probabilities $w_{ijk}^{(ABC)}$ are defined. With this assumption
Bell type inequalities follow as constraints on combinations of classical
correlations belonging to a system of classical correlations for three or more
observables.

We have already discussed in sect.\,\ref{sec:bells_inequalities} the
CHSH-inequalities\,\cite{CHSH,CLSH,CLHO}. They concern combinations of
correlation functions for a system of classical correlations for four
observables $A$, $A'$, $B$, $B'$. As a special case they include Bell's original
inequality if two out of the four observables are identified. The
CHSH-inequalities apply if the simultaneous probabilities $w_{ijkl}^{(AA'BB')}$
are defined and used for the definition of the correlation functions. For
comparison with observation one further assumes that the measurement
correlations coincide with the classical correlations of this system.

For complete statistics the assumption for the CHSH- or Bell-inequalities are
obeyed. In turn, if correlations are found to violate the CHSH-inequalities,
complete statistics are not possible. The issue concerns the existence of
simultaneous probabilities as $w_{ijk}^{(ABC)}$ which could be used for the
prediction of outcomes of ideal measurements. As we have discussed, they are
often not available for measurements in subsystems. In this case the
CHSH-inequalities dot not need to hold. It may happen that a system of
classical correlations for three or more observables exists, but cannot be used
for ideal measurements in subsystems. In this case the CHSH-inequalities dot not
apply to the measurement correlation found in this type of measurements. If the
CHSH-inequalities are violated by a measurement of correlations, either the
corresponding subsystem has incomplete statistics, or classical correlations
cannot be used.

The observation that classical correlation functions may not be available or not
be appropriate for a description of the outcome of measurements in subsystems
does not constitute a problem. As we have seen in
sect.\,\ref{sec:conditional_probabilities_4_7}, other correlation functions
based on conditional probabilities are available and well adapted for ideal
measurements in subsystems. These measurement correlations do not have to obey
Bell's inequalities. In the following we will work out in more detail why
classical correlation functions are not appropriate.

\paragraph*{Classical correlations of overall probabilistic\\systems}

For observables that take fixed values for the configurations of the overall
probabilistic system the classical correlation function (\ref{eq:pq1}) always
exists. The probabilities $w_{ij}^{(AB)}$ obtain by summing the probabilities of
all states for which $A$ takes the value $A_i$ and $B$ takes the value $B_j$.
They obey the relations (\ref{eq:pq2}). This extends to systems of classical
correlation functions. The simultaneous probabilities $w_{ijk}^{(ABC)}$ are all
available as sums of the probabilities for appropriate states. The overall
probabilistic system has complete statistics for all observables that take fixed
values for each given overall configuration.

Often these classical correlations are, however, not the correlations appearing
in ideal measurements for subsystems. For subsystems characterized by incomplete
statistics not all simultaneous probabilities are accessible by the
probabilistic information of the subsystem. Typically, classical correlations
depend on properties of the environment of the subsystem. They take different
values for two observables that belong to the same equivalence class of
probabilistic observables for the subsystem, but differ in ``environment
properties". This excludes a use of classical correlation functions for ideal
measurements in a subsystem, since the latter should not measure properties of
the environment. We have discussed in
sect.\,\ref{sec:conditional_probabilities_4_7} the measurement correlations that
reflect ideal measurements in a subsystem. They do not need to obey the
CHSH-inequalities.

We have also encountered probabilistic observables in subsystems that do not
take fixed values for the configurations of the overall probabilistic system. An
example is the momentum observable. For such observables the classical
correlation functions are not defined at all.

\paragraph*{Coherent ideal measurements}

For local-time subsystems we have advocated that ideal measurements should use a
measurement correlation based on the product of associated local operators. This
holds, in different ways, for decoherent and coherent ideal measurements. For
experiments testing the CHSH-inequalities one typically measures two parts of a
subsystem, with observables $A$, $A'$ for the first part and $B$, $B'$ for the
second part. Since the two sets of operators for these observables commute with
each other,
\begin{equation}
[\hat{A},\hat{B}] = [\hat{A},\hat{B}'] = [\hat{A}',\hat{B}] =
[\hat{A}',\hat{B}'] = 0,
\label{eq:pq8}
\end{equation}
the precise time sequence of the measurements does not matter for correlations
of the type $\braket{AB}$, $\braket{AB'}$. One typically tries to measure both
observables simultaneously in order to exclude signals sent from one part of
the subsystem to the other. With eq.\,(\ref{eq:pq8}) we can extend our
discussion of sequences of ideal measurements to this case. The appropriate
setting are coherent ideal measurements since the measurement of $B$ has no
influence on the simultaneous measurement of $A$ and $B$.

The measured correlations have been found to violate Bell's inequalities. This
possibly may be anticipated because the measurement correlations are not the
classical correlations of the overall probabilistic system, and the local-time
subsystem is characterized by incomplete statistics. It is instructive to
understand at which point the logic leading to CHSH-inequalities does not apply.

\paragraph*{Simultaneous probabilities}

For the measurement correlation \eqref{M9} of coherent ideal measurements the
simultaneous probabilities for the pairs $w_{ij}^{(AB)}$, $w_{ij}^{(AB')}$,
$w_{ij}^{(A'B)}$ and $w_{ij}^{(A'B')}$ can still be computed. This follows from
the definition of conditional probabilities and the relations \eqref{M6C},
\eqref{M8}, which imply the relations (\ref{eq:pq2}). The assumptions
(\ref{eq:pq1}), (\ref{eq:pq2}) for the derivation of the CHSH-inequalities are
therefore obeyed. This holds independently of the property if the measurement
correlations can be associated with classical correlations of the overall system
or not. For the correlation map in sect.\,\ref{sec:correlation_map} some of the
measurement correlations can be associated to classical correlations, while this
is not the case for the average spin map \eqref{E31}. For both bit-quantum maps
the assumptions (\ref{eq:pq1}), (\ref{eq:pq2}) hold, since they are only based
on the relations for conditional probabilities.

The point where a proof of the CHSH-inequalities fails for general two-level
observables represented by operators $\hat{A}$, $\hat{A}'$, $\hat{B}$,
$\hat{B}'$ with eigenvalues $\pm 1$ is the absence of simultaneous probabilities
$w_{ijk}^{(ABB')}$ etc. Typically, $\hat{B}$ and $\hat{B}'$ do not commute for
the interesting cases, and similarly for $\hat{A}$ and $\hat{A}'$. The violation
of the CHSH-inequalities for measurement correlations concerns the case where
not all the four observables are Cartesian spins.

A crucial point in the simple proof of the CHSH-inequality in
sect.\,\ref{sec:bells_inequalities} is the relation
\begin{equation}
\braket{AB} + \braket{AB'} + \braket{A'B} - \braket{A'B'} = \braket{AD_+} +
\braket{A'D_-},
\label{eq:pq9}
\end{equation}%
where
\begin{equation}
D_+ = B + B',\quad D_- = B - B'
\label{eq:pq10}\end{equation}%
are observables with possible measurement values $\pm 2,\,0$. If $B$ and $B'$
are represented by non-commuting operators, $[\hat{B},\hat{B}']\neq0$, the
simultaneous probabilities $w_{ijk}^{(ABB')}$ are not available for the quantum
subsystem. As a consequence, simultaneous probabilities as $w_{il}^{(AD_\pm)}$
are not available either, and a proof of the CHSH-inequality is no longer
possible.

To be more concrete we take $B = S_1^{(2)}$ and $B' = S_3^{(2)}$. The
corresponding local operators are
\begin{equation}
\hat{B} = (1 \otimes \tau_1),\quad \hat{B}' = (1 \otimes \tau_3).
\label{eq:pq11}
\end{equation}%
For the observable $D_+$ there is no associated local-observable operator,
however. The measurement correlation $\braket{AD_+}_m$ is not defined. We can,
of course, define the sums and products of operators, as
\begin{equation}
\hat{D}_+ = \hat{B} + \hat{B}' = (1 \otimes (\tau_1 + \tau_3)).
\label{eq:pq12}
\end{equation}%
This operator has eigenvalues $\pm \sqrt{2}$. It is not the local-observable
operator associated to the observable $D_+$, which has possible measurement
values $\pm 2,\,0$. It is at this point where the proof of the CHSH-inequalities
fails for the measurement correlation based on operator products.

\paragraph*{CHSH-inequalities for special cases of\\measurement correlations}

For general spin observables $A$, $A'$, $B$, $B'$ the CHSH-inequalities dot not
have to hold for measurement correlations. There are special cases, however, for
which these inequalities can be proven, nevertheless. This holds whenever a
given system of measurement correlations can be expressed as an equivalent
system of classical correlations. An example are the Cartesian spin observables
in the two-qubit quantum system. The existence of the correlation map tells us
that the measurement correlations for the Cartesian spin observables can be
associated to a system of classical correlations computed from a local
probability distribution. If the correlation map is complete there exists a
probability distribution for every arbitrary quantum state or every density
matrix. For arbitrary quantum states a classical probability distribution can
therefore represent the measurement correlations for Cartesian spins by a system
of classical correlations. As a consequence, the measurement correlations for
Cartesian spins have to obey the CHSH-inequalities. This is indeed the case. The
violations of the CHSH-inequalities only occur for angles between spins
different from $\pi /2$. The proof of the CHSH-inequalities for Cartesian spins
is independent of the fact if the correlation map is used or not for the
definition of the quantum subsystem. The existence of a complete map is
sufficient. There is also no need that the local probability distribution
defining the system of classical correlation functions is unique. Typically,
this in not the case.

This argument can be inverted. For any system of measurement correlations that
violates the CHSH-inequalities there cannot be a classical probability
distribution such that all measurement correlations of this system can be
associated to classical correlations.

\subsection{Kochen-Specker theorem}
\label{sec:kochen-specker_theorem}

The Kochen-Specker no-go theorem\,\cite{KOSP,MER,PER,STRA} concerns the possible
associations between quantum operators and classical observables. It makes the
(generally implicit) assumption that one can associate to a quantum operator a
unique ``classical observable" whose expectation value can be computed from a
probability distribution according to the standard rule of classical statistics.
With this assumption of uniqueness it establishes contradictions.

For local-time subsystems, including quantum subsystems, we have shown that one
can associate to each system observable an operator, such that its expectation
value, as defined in the overall probabilistic ensemble, can equivalently be
computed by the quantum rule using the associated operator. The map from system
observables to operators associates to each system observable a unique operator.
The inverse is not given. There are equivalence classes of system observables
for which all members are mapped to the same operator. Such equivalence classes
have more than a single member. There is therefore no inverse map from quantum
operators to classical observables. The central assumption of uniqueness for the
Kochen-Specker theorem is not obeyed for quantum subsystems.

We briefly describe the Kochen-Specker theorem and show how the non-uniqueness
of the classical observables which are mapped to a given quantum operator avoids
the applicability of the no-go theorem.

\paragraph*{Commuting operators and observables}

Let us consider two observables $A$, $B$ that are represented by two different
commuting quantum operators $\hat{A}$, $\hat{B}$. Two such observables may be
called ``comeasurable''. For comeasurable observables it is possible to
represent the classical product observable $AB$ by the operator product
$\hat{A}\hat{B}$. The simultaneous probabilities $w_{ij}^{(AB)}$ to find for $A$
the value $A_i$, and for $B$ the value $B_j$, can be part of the probabilistic
information of the quantum subsystem. We will consider pairs of comeasurable
observables for which the classical observable product $AB$ is mapped to the
operator product $\hat{A}\hat{B}$.

This does not mean that the associative classical product of observables is
isomorphic to the associative operator product. As a simple example we consider
observables and operators in a two-qubit system. We associate
\begin{align}
\begin{split}
A&\rightarrow \hat{A}=(\tau_1\otimes 1),\quad 
B\rightarrow \hat{B}=(\tau_1\otimes \tau_3),\\
C&\rightarrow \hat{C}=(\tau_3\otimes \tau_1),
\end{split}
\label{eq:pq13}
\end{align}%
where
\begin{equation}
\hat{A}\hat{B}=\hat{B}\hat{A}=(1\otimes\tau_3),\quad 
\hat{B}\hat{C}=\hat{C}\hat{B}=(\tau_2\otimes\tau_2).
\label{eq:pq14}
\end{equation}%
While $\left[\hat{A},\hat{B}\right]=0$, $\left[\hat{B},\hat{C}\right]=0$, the
operators $\hat{A}$ and $\hat{C}$ do not commute,
\begin{equation}
\hat{A}\hat{C}=-\hat{C}\hat{A}=-i(\tau_2\otimes\tau_1).
\label{eq:pq15}
\end{equation}%
In contrast, the classical observable product is always commutative.

Let us now assume that the inverse map would exist for all pairs of commuting
operators
\begin{equation}
\hat{A}\rightarrow A,\quad \hat{B}\rightarrow B,\quad \hat{A}\hat{B}\rightarrow
AB.
\label{eq:pq16}
\end{equation}%
We define the operator $\hat{D} = \hat{A}\hat{B}$, and assume a further operator
$\hat{E}$ that commutes with $\hat{D}$. With
\begin{equation}
\hat{F}=\hat{D}\hat{E}\rightarrow F=DE,
\label{eq:pq17}
\end{equation}%
this implies
\begin{equation}
\hat{A}\hat{B}\hat{E} = \hat{D}\hat{E}=\hat{F}\rightarrow F=ABE.
\label{eq:pq18}
\end{equation}%
We can in this way construct chains of operators that are mapped to multiple
classical products of observables. This construction contradicts the
non-commuting structure of operator products, as we will show next.

\paragraph*{Complete comeasurable bit chains}

Consider a number of Ising spins or bits that are represented by a set of
commuting operators. They form a ``comeasurable bit chain." For a given number
$Q$ of qubits there are maximally $2^Q-1$ mutually commuting two-level
operators. A set of Ising spins that is mapped to a maximal set of commuting
operators is called a ``complete comeasurable bit chain."

As an example we take a three-qubit quantum system. Complete comeasurable bit
chains consist each of seven different Ising spins. These seven Ising spins
contain ``composite Ising spins" as products of Ising spins. Let us consider
four different complete comeasurable bit chains that we specify by the commuting
sets of operators used:\newline\newline
F-chain:
\begin{align}
\begin{split}
&\hat{F}_1 = (\tau_3 \otimes 1 \otimes 1),\;
\hat{F}_2 = (1 \otimes \tau_1 \otimes 1),\;
\hat{F}_3 = (1 \otimes 1 \otimes \tau_1),\\
&\hat{F}_{12} = (\tau_3 \otimes \tau_1 \otimes 1),\;
\hat{F}_{13} = (\tau_3 \otimes 1 \otimes \tau_1),\\
&\hat{F}_{23} = (1 \otimes \tau_1 \otimes \tau_1),\;
\hat{F}_{123} = (\tau_3 \otimes \tau_1 \otimes \tau_1),
\end{split}
\label{eq:pq19}
\end{align}%
G-chain:
\begin{align}
\begin{split}
&\hat{G}_1 = (\tau_1 \otimes 1 \otimes 1),\;
\hat{G}_2 = (1 \otimes \tau_3 \otimes 1),\;
\hat{G}_3 = (1 \otimes 1 \otimes \tau_1),\\
&\hat{G}_{12} = (\tau_1 \otimes \tau_3 \otimes 1),\;
\hat{G}_{13} = (\tau_1 \otimes 1 \otimes \tau_1),\\
&\hat{G}_{23} = (1 \otimes \tau_3 \otimes \tau_1),\;
\hat{G}_{123} = (\tau_1 \otimes \tau_3 \otimes \tau_1),
\end{split}
\label{eq:pq20}
\end{align}%
H-chain:
\begin{align}
\begin{split}
&\hat{H}_1 = (\tau_1 \otimes 1 \otimes 1),\;
\hat{H}_2 = (1 \otimes \tau_1 \otimes 1),\;
\hat{H}_3 = (1 \otimes 1 \otimes \tau_3),\\
&\hat{H}_{12} = (\tau_1 \otimes \tau_1 \otimes 1),\;
\hat{H}_{13} = (\tau_1 \otimes 1 \otimes \tau_3),\\
&\hat{H}_{23} = (1 \otimes \tau_1 \otimes \tau_3),\;
\hat{H}_{123} = (\tau_1 \otimes \tau_1 \otimes \tau_3),
\end{split}
\label{eq:pq21}
\end{align}%
Q-chain:
\begin{align}
\begin{split}
&\hat{Q}_1 = \hat{F}_{123} = (\tau_3 \otimes \tau_1 \otimes \tau_1),\;
\hat{Q}_2 = \hat{G}_{123} = (\tau_1 \otimes \tau_3 \otimes \tau_1),\\
&\hat{Q}_3 = \hat{H}_{123} = (\tau_1 \otimes \tau_1 \otimes \tau_3),\;
\hat{Q}_{12} = (\tau_2 \otimes \tau_2 \otimes 1),\\
&\hat{Q}_{13} = (\tau_2 \otimes 1 \otimes \tau_2),\;
\hat{Q}_{23} = (1 \otimes \tau_2 \otimes \tau_2),\\
&\hat{Q}_{123} = -(\tau_3 \otimes \tau_3 \otimes \tau_3),
\end{split}
\label{eq:pq22}
\end{align}%
If we can associate to each operator a unique Ising spin, e.\,g.
\begin{equation}
\hat{F}_{12} = \hat{F}_1\hat{F}_2 \rightarrow F_{12} = F_1 F_2,
\label{eq:pq23}
\end{equation}%
one finds
\begin{equation}
\hat{Q}_{123} = \hat{F}_{123}\hat{G}_{123}\hat{H}_{123} = F_1 F_2 F_3 G_1 G_2
G_3 H_1 H_2 H_3.
\label{eq:pq24}
\end{equation}%
With
\begin{equation}
F_2 = H_2,\quad F_3 = G_3,\quad G_1 = H_1,
\label{eq:pq25}
\end{equation}%
one has for Ising spins
\begin{equation}
F_2 H_2 = 1,\quad F_3 G_3 =1,\quad G_1 H_1 =1,
\label{eq:pq26}
\end{equation}%
and therefore the map
\begin{equation}
\hat{Q}_{123} \rightarrow F_1 G_2 H_3.
\label{eq:pq27}
\end{equation}%

On the other hand we may construct one more complete comeasurable bit
chain:\newline\newline
C-chain:
\begin{align}
\begin{split}
&\hat{C}_1 = \hat{F}_{1} = (\tau_3 \otimes 1 \otimes 1),\;
\hat{C}_2 = \hat{G}_{2} = (1 \otimes \tau_3 \otimes 1),\\
&\hat{C}_3 = \hat{H}_{3} = (1 \otimes 1 \otimes \tau_3),\;
\hat{C}_{12} = (\tau_3 \otimes \tau_3 \otimes 1),\\
&\hat{C}_{13} = (\tau_3 \otimes 1 \otimes \tau_3),\;
\hat{C}_{23} = (1 \otimes \tau_3 \otimes \tau_3),\\
&\hat{C}_{123} = (\tau_3 \otimes \tau_3 \otimes \tau_3).
\end{split}
\label{eq:pq28}
\end{align}%
With
\begin{equation}
\hat{Q}_{123} = -\hat{C}_{123}
\label{eq:pq29}
\end{equation}%
a unique map from operators to observables implies
\begin{equation}
\hat{Q}_{123}\rightarrow -C_1 C_2 C_3.
\label{eq:pq30}
\end{equation}%
On the other hand one has
\begin{equation}
F_1 = C_1,\quad G_2 =C_2,\quad H_3 =C_3,
\label{eq:pq31}
\end{equation}%
such that eq.\,(\ref{eq:pq27}) reads
\begin{equation}
\hat{Q}_{123} \rightarrow C_1 C_2 C_3.
\label{eq:pq32}
\end{equation}%
The signs in eqs.\,(\ref{eq:pq30}) and (\ref{eq:pq32}) contradict each other.
One concludes that no map from quantum operators to observables is possible.
This particular, rather simple version of the Kochen-Specker theorem follows the
elegant derivation by N.\ Straumann \cite{STRA}.

The Kochen-Specker no-go theorem has often been misinterpreted by stating that
it is not possible to associate quantum operators and classical observables. The
correct interpretation tells us that one can map classical observables to
quantum operators, but that this map is not invertible. Different classical
observables in the same equivalence class are mapped to the same quantum
operator. The Kochen-Specker theorem is not applicable and no contradiction for
the embedding of quantum mechanics in classical mechanics arises.

\paragraph*{Minimal correlation map for three qubits}

The minimal correlation map for three qubits maps 9 classical Ising spins
$s_k^{(i)}$, $k=1..3$, $i=1..3$, plus 27 products for two different Ising spins
$s_k^{(i)} s_l^{(j)}$, $i \neq j$, and 27 products of three different Ising
spins $s_k^{(1)} s_l^{(2)} s_m^{(3)}$, to the corresponding quantum spin
operators $\hat{S}_k^{(i)}$ and products thereof. The expectation values of
these 63 classical spin observables can be equivalently computed as classical
expectation values and correlations or by the quantum rule with the associated
operators, using the density matrix
\begin{equation}
\rho = \frac{1}{8}(\braket{s_{\mu\nu\rho}} \tau_\mu \otimes \tau_\nu \otimes
\tau_\rho),
\label{eq:pq33}
\end{equation}%
with
\begin{align}
\begin{split}
&s_{000}=1,\; s_{k00}=s_k^{(1)},\; s_{0k0}=s_k^{(2)},\; s_{00k}=s_k^{(3)},\\
&s_{kl0}= s_k^{(1)}s_l^{(2)},\; s_{k0l}= s_k^{(1)}s_l^{(3)},\; s_{0kl}=
s_k^{(2)}s_l^{(3)},\\ 
&s_{klm}= s_k^{(1)}s_l^{(2)}s_m^{(3)}.
\end{split}
\label{eq:pq34}
\end{align}%

The two level operators
\begin{equation}
\hat{S}_{\mu\nu\rho} = \tau_\mu \otimes \tau_\nu \otimes \tau_\rho
\label{eq:pq35}
\end{equation}%
are of the type of the operators associated to the complete comeasurable bit
chains in eqs.\,(\ref{eq:pq19})-(\ref{eq:pq22}). Thus the minimal correlation
map maps the classical spin observables to quantum operators. This includes
products of spins with different ``flavor" $i$. The correlation map does not
involve classical correlation functions with four or more factors, or
correlations of spins $s_k^{(i)}$ with different $k$ but the same $i$. These
quantities are not accessible from the probabilistic information of the quantum
subsystem. The map from the observables to operators is not invertible. For
example, the product $\hat{F}_{123}\hat{G}_{123} = (\tau_2 \otimes \tau_2
\otimes 1)$ is not uniquely associated to the product
$s_3^{(1)}s_1^{(2)}s_1^{(1)}s_3^{(2)}$ which would follow from identities of the
type (\ref{eq:pq18}) for an invertible map. The classical observable
$s_2^{(1)}s_2^{(2)}$ is mapped to the operator $\hat{F}_{123}\hat{G}_{123}$, but
many other observables, for example observables at different times, are
typically mapped to this observable as well. The Kochen-Specker theorem is not
relevant for the correlation map. In particular, it does not impose restrictions
for the completeness of this bit-quantum map.

\subsection{Einstein-Podolski-Rosen paradox}
\label{sec:einstein-podolski-rosen_paradox}

While Bell's inequalities and the Kochen-Specker theorem have often been invoked
for an argument that there cannot be a ``classical probabilistic system"
underlying quantum mechanics, the Einstein-Podolski-Rosen (EPR) argument
\cite{EPR} tries to show that some extension of quantum mechanics is
conceptually necessary. It argues in favor of some type of ``hidden variables"
that contain information beyond quantum mechanics. Our embedding of quantum
systems as subsystems of the overall probabilistic system provides for such
hidden variables. In our view, the additional probabilistic information in the
overall system is, however, not necessary to understand the dynamics of closed
quantum subsystems and ideal measurements which are compatible with these
subsystems.

The overall probabilistic system provides for a satisfactory conceptual
framework for understanding the origin of the rules of quantum mechanics. Once
one accepts that subsystems are characterized by probabilistic observables and
incomplete statistics, and admits the concept of ideal measurements, the quantum
subsystems are self-contained logical systems without inherent contradictions.

\paragraph*{EPR-type experiments}

A typical EPR-type experiment considers the decay of a spinless particle into
two fermions with spin. Spin conservation requires that the spins of the two
decay products have to be opposite. (We neglect here spin-nonconservation by a
coupling to angular momentum or magnetic fields. We also omit position or
momentum degrees of freedom.) After the decay, the two particles are treated as
two qubits in a spin singlet state, with quantum wave function
\begin{equation}
\psi = \frac{1}{\sqrt{2}} \left( \braket{\uparrow\downarrow} -
\braket{\downarrow\uparrow}\right).
\label{eq:pq36}
\end{equation}%
This is the maximally entangled state~\eqref{E1} discussed in
sect.\,\ref{sec:entanglement_in_classical_and_quantum_statistics}. The spins in
all directions are maximally anticorrelated. For example, the correlation
function for Cartesian spin directions obey eq.~\eqref{E2}, while the expectation
values $\braket{s_k^{(i)}}$ vanish.

After the decay the two fermions may fly to regions that are no longer causally
connected. No event happening in the region of the first particle at time $t_1$
can send signals to the region of the second particle which could influence the
behavior of the second particle in a finite time interval $\Delta t$ around
$t_1$. Assume that two observers situated in these causally disconnected regions
both measure the spin $S_3$ of the fermions for a series of decays. If they
later come together and compare their results they will find out that each
observer sees in average as many events with $S_3$ up or down, corresponding to
the vanishing expectation values $\braket{S_3^{(i)}}=0$. Whenever for a given
decay $S_3^{(1)}=1$ is measured by the first observer, the second observer finds
precisely $S_3^{(2)}=-1$, as predicted by the maximal anticorrelation
$\braket{S_3^{(1)}S_3^{(2)}}=-1$, or more basically, by the conservation of
total spin.

\paragraph*{Reality of correlations}

The EPR-argument in favor of ``incompleteness of quantum mechanics" or the
equivalent necessity of additional information (hidden variables) for a complete
description of physics goes in several steps.
\begin{enumerate*}[label={(\arabic*)}]
\item\label{item:s1} Assume that $S_3^{(1)} = 1$ is measured at $t_1$. After the measurement it
is certain that $S_3^{(1)}$ has the value one. 
\item\label{item:s2} Whenever some event is
certain a piece of physical reality is associated to it. (This concept of
reality concerns the notion of ``restricted reality" discussed in the
introduction of ref.~\cite{CWPW}.) 
\item\label{item:s3} For $t>t_1$ the value $S_3^{(1)}$ is real. 
\item\label{item:s4} It is also
certain that a measurement of $S_3^{(2)}$ at $t_2>t_1$, $(t_2-t_1)<\Delta t$,
will find $S_3^{(2)}=-1$. 
\item\label{item:s5} For $t>t_2$ the value $S_3^{(2)}=-1$ is real.

\item\label{item:s6} Since no signal has affected the region of the second particle, the spin of
the second particle cannot have changed in the interval $t_1 - \Delta t/2 < t_1
< t_1 + \Delta t/2$. 
\item\label{item:s7} Therefore $S_3^{(2)}$ has with certainty the value $-1$
already for some time $t<t_1$. 
\item\label{item:s8} The value $S_3^{(2)} = -1$ is real for
$t<t_1$. 
\item\label{item:s9} This information is not given by quantum mechanics since without
the measurement of $S_3^{(1)}$ at $t_1$ the probability to find $S_3^{(2)} =-1$
is only one half. Quantum mechanics is therefore incomplete.
\end{enumerate*}

The shortcoming of this argument is the assignment of reality to the individual
spins. What is certain in this setting, and therefore real, is the maximal
anticorrelation between the spins of the two fermions, not the individual spins.
A possible description of the world predicts for this situation a probability
one half for $S_3^{(1)}=1$, $S_3^{(2)}=-1$, and one half for $S_3^{(1)}=-1$,
$S_3^{(2)}=1$. What is certain, and directly expected by spin conservation with
$S_3^{(1)}+S_3^{(2)}=0$, is the opposite value of the two spins. There is no
reason why certainty or reality should only be attributed to individual spins.
Correlations can be real in the restricted sense for situations where individual
spin values are not real.

Of course, if one believes in a deterministic world, the event $S_3^{(1)}=1$,
$S_3^{(2)}=-1$ may be associated with fixed values of these spins before the
measurement. From a deterministic point of view any probabilistic setting for
subsystems, and in particular quantum mechanics, is incomplete in the sense that
knowledge of the deterministic full system would contain additional information
beyond the subsystem. There is, however, no necessity for such a deterministic
description. The probabilistic description is fully self-consistent.

\paragraph*{Indivisibility of correlated systems}

It is often felt as counter-intuitive that a measurement in one system can
provide information about the state of another system that is not in causal
contact with the first system. The mistake in this intuition is the
consideration of the two fermions after the decay as separate systems. They are,
however, only parts of a common system. In the presence of correlations between
two parts of a system these parts of the system cannot be treated as separate
systems, even if no signals can be exchanged between the parts after some time.
The correlation does not disappear because of the separation in space. The
system has always to be considered as a whole. Any measurement, even if done
only on one of the spins, provides information on the whole system of the spins
for both fermions.

The simple intuition that for total vanishing spin a measurement of one of the
spins provides automatically information about the other spin having the
opposite value is correct and does not lead to any contradiction. It is only
based on the sum of both spins being zero. Only measurements that change the
spin $S_3^{(1)}$ could destroy the anticorrelation between the two spins by
introducing spin nonconservation into the system. Such measurements are not
ideal measurements of $S_3^{(1)}$. Any ideal measurement has to respect spin
conservation and therefore preserves the correlation
$\braket{S_3^{(1)}S_3^{(2)}}=-1$. It is actually causality that implies that
only an ideal measurement of $S_3^{(1)}$ which does not change its value does
also not change the anticorrelation relation $S_3^{(1)}+S_3^{(2)}=0$, simply
because it cannot influence $S_3^{(2)}$.

In summary, the discussion of the EPR-paradox confirms an old wisdom: The whole
is more than the sum of its parts.

\section{Embedding quantum mechanics in classical statistics}
\label{sec:Embedding_quantum_mechanics_in_classical_statistics}

Numerous statements have asserted that an embedding of quantum mechanics in
classical statistics is not possible. The present work demonstrates by explicit
examples that this claim is not justified. We have presented a complete
classical statistical description of two simple quantum systems:
\begin{itemize}
\item[--] a single qubit with an arbitrary time-dependent Hamiltonian,
\item[--] a quantum particle in a harmonic potential.
\end{itemize}
For both systems we have found a family of classical overall probability
distributions and specified how they are mapped to the quantum subsystem. We
further have described how a rather simple ``classical" neuromorphic computer
can learn to perform unitary transformations for the entangled states of two
qubits. We have indicated an evolution law for classical time-local probability
distributions in phase space which accounts for a quantum particle in an
arbitrary potential as a subsystem. We have argued that all probabilistic
automata with deterministic updating are actually discrete quantum systems.

It is important to stress that all these examples realize both sides of quantum
mechanics: the dynamics in terms of a unitary evolution of a continuous wave
function and the realization of observables which may have a discrete or
continuous spectrum of possible measurement values. For example, every spin
direction for single qubit quantum mechanics corresponds to a classical
two-level observable based on a yes/no decision. Observables are mapped to
operators whose spectrum of eigenvalues coincides with the possible measurement
values. Classical observables have a well defined value for every configuration
of the overall statistical ensemble. These values coincide with the possible
measurement values of the quantum system. In addition, we have encountered
statistical observables which describe properties of the probabilistic
information without taking fixed values for the configurations of the overall
statistical ensemble.

Our final goal is the construction of an overall probability distribution for
events at all times and locations which can describe the dynamics of a realistic
model for fundamental particles and their interactions. This overall probability
distribution is the equivalent to the functional integral for a quantum field
theory, with the important specification that the (euclidean) action $S$ is
real, such that $Z^{-1}\exp(-S)$ describes a distribution of real positive
probabilities.

A unitary evolution of the time-local probabilistic information from one
time-layer to the next does not need a complex functional integral. It is possible
for a real action as well. A simple way -- not necessarily the only way -- to
realize this unitary evolution are probabilistic automata. The deterministic
updating from one time-layer to the next guarantees the unitary evolution for
which no information is lost. Probabilistic initial conditions induce the
probabilistic aspects crucial for quantum systems. The time-local probabilistic
information is encoded in a continuous real wave function. The squares of the
components of this wave function are the time-local probabilities. The evolution
law for the wave function is linear, such that the superposition principle
holds. In consequence, all probabilistic automata are quantum systems in a
discrete and real formulation. For a suitable complex structure the real wave
function is mapped to a complex wave function. The standard continuous
description of the quantum evolution in terms of a Schrödinger or von-Neumann
equation follows if a continuum limit exists.

Probabilistic cellular automata with cells associated to positions in space
implement the locality and causality structures of quantum field theories.
Suitable probabilistic cellular automata for Ising spins describe fermionic
quantum field theories in an occupation number basis~\cite{CWFCS, CWQFFT, CWCA},
with occupation numbers and Ising spins in direct correspondence. A large
variety of discrete quantum field theories for fermions have been constructed in
this way~\cite{CWQFCB, CWFQPCA, CWFPCA, CWSG, CWQPP}. They include models with
local gauge symmetry or local Lorentz symmetry. A key next issue will be the
establishment of a continuum limit for these models~\cite{AKCW}.

The present part of this work does not focus on the construction of quantum
field theories. It rather supplements this approach by a complementary
``bottom-up" approach based on simple constructions of quantum systems. The light
that these examples shed on basic questions of an embedding of quantum mechanics
in classical statistics may be summarized by simple answers to a few quantum
questions.

\subsection{Short answers to quantum questions}
\label{subsec:Short_answers_to_quantum_questions}

We conclude this part of our work by a list of short answers to questions that
are typically asked for the understanding of quantum mechanics. While these
questions are certainly not exhaustive, our answers should summarize in a
concise form important lessons from an embedding of quantum mechanics in a
classical probabilistic description of our world.

\begin{enumerate}
\item\textit{Can one understand particle-wave duality?} The positive answer is
rooted in the probabilistic description of the world. On the one side
observables have often discrete possible measurement values, as yes-no answers
to the question if at a given time a particle is located in a certain space
interval or not. On the other side the dynamics describes the evolution of
continuous probabilistic information. This wave aspect is particularly apparent
if one uses classical wave functions in order to encode the time-local
probabilistic information. The classical wave functions are probability
amplitudes which obey a linear evolution law. This realizes the superposition
principle characteristic for waves.

\item\textit{Why is the time-local probabilistic information described by wave
functions?} The time-local subsystem ``integrates out" the past and the future of
the overall probability distribution. Each one of the two integrations leaves a
wave function, which obeys a linear evolution law. Time-local probabilities, or
more generally the classical density matrix, are bilinear in the wave functions
since both the past and the future are integrated out.

\item\textit{Does quantum mechanics require a complex functional integral?} In
Feynman's approach to quantum mechanics the functional integral is complex, and
quantum field theories are typically based on complex functional integrals with
a Minkowski signature for the metric. The steps from the functional integral to
wave functions can be done as well for real functional integrals which describe
an overall probability distribution for events at all times. The evolution of
the wave function can nevertheless be unitary. A real ``classical" or
``microscopic" action defining the functional integral can lead under certain
conditions to a unitary time evolution of the wave function. A simple example
are probabilistic automata.

\item\textit{Where do the phases of the wave function come from?} Classical wave
functions which are computed from the overall probability distribution or
functional integral with real action are real functions. Nevertheless, a
suitable complex structure is often compatible with the evolution. This maps the
real wave function to a complex wave function. In this complex formulation for
quantum subsystems the phases play the usual role. We have given several
explicit examples.

\item\textit{Where do non-commuting operators come from?} Contrary to widespread
prejudice non-commuting operators play a role in classical statistics. A prime
example is the transfer matrix. It commutes with operators for observables only
for conserved quantities. The basic structure of operators describing
observables arises from the projection of the overall probability distribution
or functional integral to the time-local subsystem. The issue if operators
commute or not in a given subsystem depends on the completeness of the
probabilistic information in the subsystem. Operators for observables commute if
the simultaneous probabilities for their measurement values are available in the
probabilistic information of the subsystem. This cannot be realized for
statistical observables. Also for classical observables a given subsystem can be
characterized by incomplete statistics, with a representation of these
observables by non-commuting operators.

\item\textit{What is the origin of Planck's constant $\hbar$?} Actually, $\hbar$
is only a conversion factor for units. It does not appear if the units of
momentum are inverse length and the units of energy are inverse time. The role
of $\hbar$ in quantum mechanics reflects the non-commutativity of operators. It
appears in the commutator relations if other units for momentum or energy are
used. We choose units for which $\hbar$ is set to one.

\item\textit{How can one explain the quantum rule for expectation values of
observables?} For classical observables these quantum rules follow directly from
the basic classical law for expectation values in terms of the classical
probabilities of the overall probability distribution. They are a result of the
projection to the time-local subsystem and involve only the standard laws for
probabilities. The generalization to expectation values of statistical
observables arises from the observation that for subsystems the statistical
observables are typically represented by operators very similar to the classical
observables. The possible measurement values of statistical observables
correspond to the eigenvalues of the operators for suitable eigenstates. These
eigenstates correspond to particular forms of the time-local probabilistic
information. Simple examples are observables for momentum or energy.

\item\textit{What is the origin of the quantum mechanical uncertainty?} This
uncertainty is related to the incomplete statistics of subsystems. Observables
are then represented by non-commuting operators. Uncertainty relations follow
from the non-vanishing commutator. The incomplete statistics can be of a basic
nature as for statistical observables. It can also be the result of the
projection to a subsystem. Some subsystems are possible only if the classical
time-local probability distribution obeys certain constraints. These constraints
-- the quantum constraints for our examples -- can enforce directly the
uncertainty relations.

\item\textit{How can the unitary time evolution of quantum mechanics be realized
in classical probabilistic systems?} For many classical systems the evolution is
not unitary, but rather describes the approach to some equilibrium state. The
rate of this approach is given by some correlation time or length. There exist,
however, classical statistical systems for which the step evolution operator is
orthogonal, or its projection to a subsystem is orthogonal. This realizes a
unitary evolution for which no information is lost as time proceeds. A simple
example are probabilistic automata. Here the deterministic updating guarantees
the conservation of the time-local probabilistic information.

\item\textit{Why is the world described by quantum mechanics?} An overall
probability distribution which approaches an equilibrium state as time
progresses cannot describe the complexity of our world. Any realistic
description of the universe has to be based on an overall probability
distribution for which the step evolution operator is orthogonal, at least once
projected on a suitable subsystem. This subsystem may have an environment which
equilibrates. The universe is then described by the subsystem. With an
orthogonal step evolution operator the subsystem is a quantum system in a real
or complex formulation.

\item\textit{Does on expect deviations from quantum mechanics?} Quantum
mechanics is exact for subsystems with a unitary evolution. Possible deviations
from quantum mechanics can only arise from the interaction of the quantum
subsystem with its environment. One the fundamental level the step evolution
operator can be made block-diagonal in a part with orthogonal evolution for the
subsystem and a part for the environment. The absolute size of the eigenvalues
of the step evolution operator for the environment is smaller than one. As time
progresses, this erases the probabilistic information in the environment. Due to
the very long history of the universe in terms of fundamental time units --
Planck time or smaller -- the probabilistic information in the environment is
completely lost at present time. With an equilibrated environment only the
subsystem remains. On the fundamental level the quantum laws are exact without
any deviations. The situation can be different for particular quantum
subsystems of the time-local subsystem for the whole world. They may not be
completely decoupled from their environment. For example, this may happen if
quantum constraints are not obeyed exactly. The effective deviations from
quantum mechanics for such subsystems can be understood within the exact quantum
laws of the overall exact quantum mechanics for the whole world~\cite{KOS, LIN,
ZOL}.

\item\textit{Does entanglement distinguish quantum mechanics from classical
statistics?} No. Entanglement is a statement about correlations. Correlation
functions are defined for all probabilistic systems. Correlations between parts
imply that the system has to be regarded as a whole and cannot be separated into
independent parts.

\item\textit{Why are observables probabilistic?} For a given state in quantum
mechanics only probabilities are available for finding one of the possible
measurement values for an observable. This feature typically also occurs for
subsystems in a classical statistical setting. Even observables which have
definite values for the configurations of the overall probabilistic system are
often mapped to probabilistic observables for the subsystem. Constraints on the
overall probability distribution which can realize a given subsystem may not
allow simultaneously sharp values for all observables of the subsystem.

\item\textit{Can one use the classical correlation function for a sequence of
measurements?} In general, the correlations for sequences of measurements are
described by measurement correlations based on conditional probabilities. For
particular cases, that may be called ideal classical measurements, the
measurement correlation coincides with the classical correlation function. In
case of incomplete statistics the classical correlation function is not
available for arbitrary pairs of observables. This is the case for statistical
observables for which classical correlation functions are not defined. It can
also happen for subsystems if the probabilistic information available for the
subsystem does not contain simultaneous probabilities for the possible
measurement values of a pair of observables. Even if available, the classical
correlation function is often not appropriate for measurements because of a
lack of robustness.

\item\textit{Why do Bell's inequalities not prevent an embedding of quantum
mechanics into classical statistics?} Bell's inequalities concern classical
correlation functions. The measurement correlation in quantum systems has been
found experimentally to violate Bell's inequalities for certain pairs of
observables. The conclusion is that classical correlations cannot be used for
the measurement correlations of such pairs of observables. Since for classical
statistical subsystems the measurement correlation for ideal measurements often
differs from the classical correlation function, the non-applicability of the
classical correlation function is no argument why quantum mechanics cannot be
embedded in classical statistics. Our examples of explicit constructions of
quantum subsystems from an overall classical statistical probability
distribution prove that such an embedding is possible.

\item\textit{What is the reduction of the wave function?} The reduction of the
wave function is a convenient mathematical procedure to describe the conditional
probabilities for a sequence of incoherent ideal measurements. It does not apply
to sequences of arbitrary measurements. It is also not well defined for quantum
systems of many qubits. No violation of quantum laws or the unitarity of the
evolution are necessary for the realization of the conditional probabilities of
incoherent ideal measurements.

\item\textit{Which classical probability distribution can describe a quantum
particle in an arbitrary potential?} One can formulate a time-evolution law for
the time local probabilities which describe the quantum particle as an
appropriate subsystem. An overall probability distribution which realizes this
evolution law is not known at present, with the exception of harmonic
potentials. Given the fact that particles are excitations of a complex vacuum of
a quantum field theory, a final answer to this question may have to follow the
route of a quantum field theory and its vacuum first.
\end{enumerate}

When one comes to more concrete questions as the last one, one realizes that the
way is still long for a quantitative understanding of interesting quantum
systems from an overall classical probability distribution. The examples of this
part of our work mainly help to clarify conceptual issues since all steps can be
followed explicitly in a simple way. A promising route to understand real
physical quantum systems from an overall ``classical" probability distribution
of the world may proceed by probabilistic cellular automata. They guarantee a
unitary time evolution and implement the locality and causality of quantum field
theories. Suitable probabilistic cellular automata for Ising spins are
equivalent to discretized quantum field theories for fermions. A key task is the
establishment of a continuum limit and an understanding of the vacuum of these
quantum field theories. First advances are described in the first part of this
work~\cite{CWPW}. A good part of these tasks still needs to be done.


	


\vspace{2.0cm}\noindent

\nocite{*}

\vspace{2.0cm}\noindent

\bibliography{probabilistic_world}

\end{document}